\documentclass[%
 reprint,
superscriptaddress,
 amsmath,amssymb,
 aps,
]{revtex4-2}

\usepackage[utf8]{inputenc}
\usepackage{graphicx}
\usepackage{color}
\usepackage{physics}
\usepackage{braket}
\usepackage{comment}
\usepackage{mathrsfs}
\usepackage{enumerate}

\usepackage[version=4]{mhchem}
\usepackage[breaklinks=true]{hyperref}
\usepackage{lipsum}
\usepackage{algorithmic}
\usepackage{algorithm}

\newcommand{\mr}[1]{\mathrm{#1}}

\newcommand{\mcl}[1]{\mathcal{#1}}
\newcommand{\bbC}{\mathbb{C}}
\newcommand{\bbR}{\mathbb{R}}
\newcommand{\bbZ}{\mathbb{Z}}
\newcommand{\bbN}{\mathbb{N}}

\newcommand{\ad}{\mathrm{ad}}

\newcommand{\poly}[1]{\mathrm{poly} \left( #1 \right)}
\newcommand{\Otilde}[1]{\tilde{\mathcal{O}} \left( #1 \right)}
\date{\today}
\usepackage{amsthm}
\theoremstyle{definition}
\newtheorem{theorem}{Theorem}[]
\newtheorem{definition}[theorem]{Definition}

\newtheorem{lemma}[theorem]{Lemma}
\newtheorem{corollary}[theorem]{Corollary}

\begin{document}
\title{Explicit error bounds with commutator scaling \\ for time-dependent product and multi-product formulas}

\author{Kaoru Mizuta}
\email{mizuta@qi.t.u-tokyo.ac.jp}
\affiliation{Department of Applied Physics, Graduate School of Engineering, The University of Tokyo, Hongo 7-3-1, Bunkyo, Tokyo 113-8656, Japan}
\affiliation{Photon Science Center, Graduate School of Engineering, The University of Tokyo, Hongo 7-3-1, Bunkyo, Tokyo 113-8656, Japan}
\affiliation{RIKEN Center for Quantum Computing (RQC), Hirosawa 2-1, Wako, Saitama 351-0198, Japan}

\author{Tatsuhiko N. Ikeda}
\affiliation{Faculty of Social Informatics, ZEN University, Zushi, Kanagawa, 249-0007, Japan}
\affiliation{RIKEN Center for Quantum Computing (RQC), Hirosawa 2-1, Wako, Saitama 351-0198, Japan}
\affiliation{Department of Applied Physics, Hokkaido University, Sapporo, Hokkaido, 060-8628, Japan}

\author{Keisuke Fujii}
\affiliation{Graduate School of Engineering Science, Osaka University,
1-3 Machikaneyama, Toyonaka, Osaka 560-8531, Japan.}
\affiliation{Center for Quantum Information and Quantum Biology, Osaka University, 560-0043, Japan.}
\affiliation{RIKEN Center for Quantum Computing (RQC), Hirosawa 2-1, Wako, Saitama 351-0198, Japan}
\affiliation{Fujitsu Quantum Computing Joint Research Division at QIQB,
Osaka University, 1-2 Machikaneyama, Toyonaka 560-0043, Japan}

\begin{abstract}
Product formula (PF), which approximates the time evolution under a many-body Hamiltonian by the product of local time-evolution operators, is one of the central approaches for simulating quantum dynamics by quantum computers.
It has been of great interest whether the errors of PFs can be expressed by commutators among local terms (called \textit{commutator scaling}), since it brings the substantial suppression of the computational cost in the system size.
Although recent studies have revealed the error bounds with commutator scaling for generic time-independent systems, those for time-dependent Hamiltonians remain to be a difficult problem except for low-order PFs.
In this paper, we derive an error bound of generic PFs for smooth time-dependent Hamiltonians, which is expressed explicitly by nested commutators and their time derivatives.
It indicates the substantial error suppression in the system size for generic local Hamiltonians, thereby giving a much better estimate of gate counts. 
Our derivation employs Floquet theory originally developed for time-periodic systems, with which we analyze the time-dependent PF errors by mapping them to those for time-independent PFs defined on an infinite-dimensional space.
This approach is also applicable to a family of time-dependent PFs, such as the one for non-unitary dynamics.
In particular, we clarify the error bound of a time-dependent multi-product formula, with which we can achieve exponentially better scaling with respect to the target accuracy than the PF.
Our results will shed light on applications of near-term and future quantum computers, ranging from quantum simulation of nonequilibrium materials to faster algorithms exploiting time-dependent Hamiltonians such as adiabatic state preparation.

\end{abstract}

\maketitle

\section{Introduction}\label{Sec:Introduction}

Simulating time evolution of quantum many-body systems, called Hamiltonian simulation, is one of the most promising tasks of quantum computers with the essential quantum speedup.
Owing to its broad application to physics and chemistry, various quantum algorithms such as Trotterization \cite{Lloyd1996-ko} and quantum singular value transformation (QSVT) \cite{Low2017-qsp,Low2019-qubitization,Gilyen2019-qsvt} have been vigorously developed.
Trotterization has been one of the central approaches for a long time, which constructs an approximate time-evolution operator under a Hamiltonian $H=\sum_\gamma H_\gamma$ by the product of those under simple Hamiltonians $\{ H_\gamma \}$ based on the product formulas (PFs)\cite{Lloyd1996-ko}.
It guarantees accuracy and also provides a simple construction of quantum circuits by local quantum gates without using any ancilla qubits, making it being a promising approach both for near-term and long-term quantum computers.

The error of the PFs from the exact time-evolution operator is the most significant issue that determines the computational cost of Hamiltonian simulation.
For time-independent Hamiltonians, there have been much development in understanding the error bounds of the PFs for generic orders.
Firstly, the $p$-th order PF has been proven to have an error bound $\order{(\sum_\gamma \norm{H_\gamma} t)^{p+1}}$, which is called the $1$-norm scaling \cite{Berry2007-ie-L1norm,Papageorgiou2012-bs-L1norm,Hadfield2018-mu-L1norm}.
However, there exists a better bound having so-called the commutator scaling, which states that the coefficient of the $p$-th order PF error $\order{t^{p+1}}$ is proportional to the summation of $(p+1)$-fold commutators among $\{ H_\gamma \}$.
The commutator scaling substantially suppresses the error and thereby the gate counts in terms of the system size for generic local Hamiltonians compared to the $1$-norm scaling.
It has been well known for the low-order PFs from the context of Lie algebra \cite{Suzuki1985-io-Lie}, and a series of studies have revealed it also for higher-orders \cite{Somma2015-fs-commutator,childs-prl2019-pf}.
Importantly, Ref.~\cite{childs2021-trotter} has derived the error bound of generic time-independent PFs explicitly expressed by the commutators among the terms $\{H_\gamma\}$, which is applicable to various local Hamiltonians having finite- or long-ranged interactions.
Such commutator scaling contributes to better gate count estimates for simulating a variety of spin or electronic systems \cite{Wecker-PRA2015-t-dep-pf,Babbush-2015-models,Reiher2017-un-model,Childs2018-eu-model}.

On the other hand, it is contrasting for time-dependent Hamiltonians.
While the commutator-scaling error has been understood for low order time-dependent PF errors as well \cite{Huyghebaert1990-zb-tpf,Osborne-prl2010-lb,Poulin-prl2011_time_dep,Wecker-PRA2015-t-dep-pf,An2021-tpf,Ikeda_2023_time-dep}, the error bounds for generic time-dependent PFs remain to be a difficult problem.
Ref.~\cite{Wiebe2010-mu} shows an error bound for smooth time-dependent Hamiltonians including higher orders, which corresponds to the error with the $1$-norm scaling.
Ref. \cite{childs-prl2019-pf} shows a bound involving commutators in a similar manner to time-independent cases, but the commutators include a fraction of the time-dependent PFs, which are difficult to compute except for finite-ranged Hamiltonians.

In this paper, we derive the explicit error bound of generic time-dependent PFs.
We consider generic smooth time-dependent Hamiltonians in the form of $H(t)=\sum_\gamma H_\gamma (t)$.
We show that the error bound for the $p$-th order, which amounts to $\order{t^{p+1}}$, have a coefficient expressed by commutators among the set of terms $\{ H_\gamma (t)\}$ and their derivatives in time $\{ H_\gamma^{(q)}(t) \}$ for $q=1,\cdots,p$.
This scaling hosts the substantial suppression with respect to the system size for local Hamiltonians with finite-, short-, and long-ranged interactions like time-independent cases.
In contrast to previous studies, our error bound simultaneously achieves the commutator scaling for generic orders and its applicability to various time-dependent local Hamiltonians. 
We also estimate the gate count for Hamiltonian simulations based on the time-dependent PFs.
They are substantially improved in the system size for generic local finite-, short-, and long-ranged Hamiltonians, which becomes almost as small as time-independent cases. 

Our strategy makes use of Floquet theory, originally developed for time-periodic Hamiltonians, as a mathematical tool to analyze general smooth time-dependent Hamiltonians.
By mapping a generic time-dependent Hamiltonian to a time-independent one acting on an infinite-dimensional extended space \cite{Oka2019review}, we can derive the error bound in three steps.
First, based on the Floquet mapping, we prove the existence of a time-independent PF on an infinite-dimensional space, which can reproduce a generic time-dependent PF.
Next, using a technique developed for time-independent cases \cite{childs2021-trotter}, we obtain the error representation for the time-independent PF on the infinite-dimensional space.
Finally, we reduce this error to the one for the time-dependent PF in the original space.
By carefully showing the convergence of the error by the inverse Floquet mapping, we arrive at the error bound explicitly represented by nested commutators and their time derivatives.

Our formalism also applies to various families of time-dependent PFs, such as the time-dependent multi-product formula (MPF) and the time-dependent PF for non-unitary dynamics.
In particular, MPF is a promising approach that can exponentially improve the gate counts with respect to  the allowable error $\varepsilon$ \cite{Chin2010-mpf,low2019-mpf}.
While the commutator scaling of MPF error has recently been proven for time-independent Hamiltonians \cite{zhuk2024-mpf,aftab2024-mpf,mizuta2025-mpf}, that for time-dependent ones has been an open problem \cite{watkins-2024-clock}.
Our approach reveals the commutator-scaling-type error of the time-dependent MPFs, resulting in their efficiency with respect to the system size and the accuracy.

Our error bound will directly contribute to offering better estimations of costs for various tasks of quantum computation, such as simulation of nonequilibrium materials, adiabatic state preparation \cite{Wu-PRL2002-adiabatic,Aspuru-Guzik2005-adiabatic,Albash-RMP2018-adiabatic}, and even the exploration of time-independent systems using the interaction picture \cite{Low2018-dyson,bosse-2024_int_pic,Sharma-2024-int_pic}.
We also expect that our derivation based on Floquet theory enables us to explore various families of time-dependent PFs.
Our results will shed light on the application of near-term and future fault-tolerant quantum computers toward simulation of quantum physics and chemistry.

The rest of this paper is organized as follows.
In Section \ref{Sec:Preliminary}, we briefly review time-independent PFs and their commutator-scaling errors.
We also provide the construction of time-dependent PFs.
In Section \ref{Sec:Summary}, we summarize our main results on the error bound of the time-dependent PFs.
Its derivation ranges from Section \ref{Sec:relation} to Section \ref{Sec:Error}.
In Section \ref{Sec:Examples}, we derive the gate count estimates for generic local Hamiltonians, in which we confirm the efficiency brought by the commutator scaling for time-dependent cases.
Section \ref{Sec:MPF} extends our result to time-independent MPFs.
We conclude our paper in Section \ref{Sec:Discussion}.
We provide Appendices that include known results for completeness and detailed proofs of our results.

\section{Preliminary}\label{Sec:Preliminary}

In this section, we introduce some notations used in this paper (Section~\ref{Subsec:Notation}), and briefly review the errors of the time-independent and time-dependent PFs (Sections~\ref{Subsec:time_indep}) and \ref{Subsec:time_dep}.
\subsection{Notation}\label{Subsec:Notation}

\begin{itemize}
    \item Norm: $\norm{A}$ for an operator $A$ denotes its operator norm.
    $\norm{\vec{c}}_1$ for a vector $\vec{c}=(c_1,c_2,\cdots)$ represents its $1$-norm $\sum_j |c_j|$.
    
    \item Product of operators: For a series of operators $U_1,U_2,\cdots, U_K$, we define their product with arrows by
    \begin{equation}
    \qquad \prod_{k=1}^K U_k = U_K \cdots U_1, \quad \prod_{k=K}^1 U_k = U_1 \cdots U_K.
    \end{equation}

    \item Tuple of indices: An index $v\gamma$ denotes a pair of indices $(v,\gamma)$.
    When $v$ and $\gamma$ are respectively chosen from $\{1,2,\cdots,V\}$ and $\{1,2,\cdots,\Gamma\}$, we align the indices in the order, $11,12,\cdots,1\Gamma,21,22,\cdots,V\Gamma$.
    The notation $v\gamma \prec v'\gamma'$ means that $v\gamma$ precedes $v'\gamma'$ in this order, and the symbols $\preceq, \succ, \succeq$ are defined similarly.
    The index $v(\gamma+1)$ in the case of $\gamma=\Gamma+1$ means $(v+1)1$.
    The index $v(\gamma-1)$ for $\gamma=1$ means $(v-1)\Gamma$ as well.
    
    \item Commutators: For operators $A$ and $B$, we use $[A,B]$ or $\mr{ad}_A B$ to represent their commutator $AB-BA$.
    We also note that $e^{\mr{ad}_A} B=\sum_{n=0}^\infty\frac{1}{n!} (\mr{ad}_A)^n B= e^A B e^{-A}$ holds if $(\mr{ad}_A)^n B$ is bounded for each $n \in \bbN$.
    \item Derivative: For a time-dependent operator $A(t)$, its derivative $A^{(p)}(t)$ means $\dv[p]{t} A(t)$.
    \item Time evolution: For a time-dependent Hamiltonian $H(t)$, its time-evolution operator from $t_0$ to $t+t_0$ is
    \begin{equation}
    \quad U(t+t_0,t_0) = \mcl{T} \exp \left( -i \int_{t_0}^{t+t_0} \dd t^\prime H(t^\prime)\right).
    \end{equation}
    The symbol $\mcl{T}$ denotes a time-ordered product.

    \item
    Landau symbol: The symbols $\order{\cdot}$, $o(\cdot)$, and $\Omega (\cdot)$ denote the Landau symbols defined by
    \begin{eqnarray}
        \order{f(x)} &=& \left\{ g(x) \, |  \, \lim_{x \to \text{$0$ or $\infty$}} \left| \frac{g(x)}{f(x)}\right| < \infty \right\}, \\
        o(f(x)) &=&  \left\{ g(x) \, |  \, \lim_{x \to \text{$0$ or $\infty$}} \left| \frac{g(x)}{f(x)}\right| =0 \right\}, \\
        \Omega (f(x)) &=& \left\{ g(x) \, |  \, \lim_{x \to \text{$0$ or $\infty$}} \left| \frac{f(x)}{g(x)}\right| < \infty \right\}.
    \end{eqnarray}
    The symbols $\Otilde{\cdot}$, $\tilde{o}(\cdot)$, and $\tilde{\Omega}(\cdot)$ allow poly-logarithmic corrections in the variables.
    
\end{itemize}
\subsection{Time-independent product formula}\label{Subsec:time_indep}

We briefly review the PF for time-independent systems and its commutator scaling.
Consider a time-independent Hamiltonian $H$ given by
\begin{equation}
    H = \sum_{\gamma = 1}^\Gamma H_\gamma.
\end{equation}
The partition $\gamma$ is determined so that the time-evolution operators $\{ e^{-i H_\gamma \tau} \}$ are easily implemented by local quantum gates for any $\tau$.
The aim of the PF is to approximate $e^{-iHt}$ under sufficiently small time $t$ by the product of $\{ e^{-i H_\gamma \tau} \}$.
To be precise, the $p$th-order PF is defined by
\begin{equation}\label{Eq_Pre:PF_t_indep}
    T(t) = \prod_{v=1}^{V_p} \prod_{\gamma=1}^\Gamma e^{-i H_{\pi_v(\gamma)} \alpha_{v\gamma} t},
\end{equation}
satisfying the order condition $T(t)=e^{-iHt}+\order{t^{p+1}}$ in $t \to 0$.
The symbol $\pi_v$ denotes the permutation of the indices $\{1,2,\cdots,\Gamma\}$ depending on the stage $v \in \{1,2,\cdots, V_p \}$.
The number of stages $V_p$ is an $\order{1}$ constant under $p \in \order{1}$, while it is usually large exponentially in $p$.

The most famous example is so-called Lie-Suzuki-Trotter formula \cite{Suzuki1991-za}.
The first- and second-order formulas are given by
\begin{equation}
    T_1(t) = \prod_{\gamma=1}^\Gamma e^{-iH_\gamma t}, \quad T_2(t) = T_1(-t/2)^\dagger T_1(t/2).
\end{equation}
Higher-order formulas $T_{2q}(t)$ for $q \in \bbN$ are obtained by the recursive relation,
\begin{equation}
    T_{2q}(t) = T_{2q-2} (a_q t)^2 T_{2q-2}(b_q t)^\dagger T_{2q-2} (a_q t)^2.
\end{equation}
The positive factors $a_q$ and $b_q$ are respectively defined by $a_q = (4-4^{1/(2q-1)})^{-1}$ and $b_q = 4a_q-1$. 
We have $V_p=1$ (for $p=1$) and $V_p=2 \cdot 5^{q-1}$ (for $p=2q$).
The permutation $\pi_v$ is designated by
\begin{equation}\label{Eq_Pre:permutation}
    \pi_v(\gamma) = \begin{cases}
        \gamma & (\text{$v$: odd}), \\
        \Gamma+1-\gamma & (\text{$v$: even}).
    \end{cases}
\end{equation}
Throughout this work, $T(t)$ without a subscript represents an arbitrary time-independent PF.
On the other hand, $T_p(t)$ having a subscript $p$ denotes the $p$th-order Lie-Suzuki-Trotter formula.

The error of the PF, $\norm{e^{-iHt}-T(t)}$, is of great interest since it determines the gate counts for Hamiltonian simulation.
The prototypical form of the PF error is so-called the $1$-norm scaling \cite{Berry2007-ie-L1norm,Papageorgiou2012-bs-L1norm,Hadfield2018-mu-L1norm}, expressed by
\begin{equation}\label{Eq_Pre:1_norm_scaling}
    \norm{e^{-iHt}-T(t)} \in \order{\left( \sum_\gamma \norm{H_\gamma} t\right)^{p+1}}.
\end{equation}
However, this formula overestimates the error since it becomes nonzero even when $\{ H_\gamma \}$ mutually commutes.
There exists a better bound resolving such a problem, given by
\begin{equation}\label{Eq_Pre:com_scaling}
    \norm{e^{-iHt}-T(t)} \in \order{\alpha_{\mr{com},p} t^{p+1}}.
\end{equation}
It is called commutator scaling since the factor $\alpha_{\mr{com},p}$ is given by
\begin{equation}\label{Eq_Pre:commutator}
    \alpha_{\mr{com},p} = \sum_{\gamma_0,\cdots,\gamma_p=1}^\Gamma \norm{[H_{\gamma_p},[H_{\gamma_{p-1}},\cdots,[H_{\gamma_1},H_{\gamma_0}]]]}.
\end{equation}
While this better bound has been well known for low order cases over the past decades \cite{Suzuki1985-io-Lie}, the one for generic time-independent formulas has been proven recently \cite{Somma2015-fs-commutator,childs-prl2019-pf,childs2021-trotter}.

The commutator scaling contributes to reducing the computational cost.
When considering large evolution time $t$, the query complexity $r$ in the PF is determined so that the desirable error $\varepsilon \in (0,1)$ can be achieved by
\begin{equation}
    \norm{e^{-iHt}-\{T(t/r)\}^r} \leq \varepsilon.
\end{equation}
The number $r$ is called the Trotter number.
The commutator scaling error by Eq. (\ref{Eq_Pre:com_scaling}) brings substantially better dependency in the system size than the one based on the $1$-norm scaling Eq.~(\ref{Eq_Pre:1_norm_scaling}) and those of post-Trotter algorithms like LCU and QSVT for various local Hamiltonians \cite{Wecker-PRA2015-t-dep-pf,Babbush-2015-models,Reiher2017-un-model,Childs2018-eu-model}.
The commutator scaling is essential for PFs, but it has not been fully understood in time-dependent cases.

\subsection{Time-dependent product formula}\label{Subsec:time_dep}

In this section, we introduce time-dependent PFs.
Let us consider a time-dependent Hamiltonian $H(t)$, decomposed by
\begin{equation}\label{Eq_Pre:H_t_decomp}
    H(t) = \sum_{\gamma = 1}^\Gamma H_\gamma(t),
\end{equation}
and construct the approximation of the time-evolution operator $U(t+t_0,t_0)$.
We set $t_0=0$ without loss of generality since we can reproduce the results for $t_0 \neq 0$ by shifting the time by $t_0$.
In contrast to time-independent systems, there are various candidates for time-dependent PFs, originating from the existence of the time-ordered product.
Throughout this paper, we focus on the following two formulations.

\textbf{Standard PFs.---}
The standard PF is composed of time-evolution operators under instantaneous Hamiltonians as
\begin{equation}\label{Eq_Pre:Standard_PF}
    \bar{S}(t,0) = \prod_{v=1}^{V_p} \prod_{\gamma=1}^\Gamma e^{-iH_{\pi_v(\gamma)}(\beta_{v\gamma} t) \alpha_{v\gamma} t},
\end{equation}
where the coefficients $\alpha_{v\gamma},\beta_{v\gamma} \in \bbR$ are determined by the order condition $\bar{S}(t,0)=U(t,0)+\order{t^{p+1}}$ \cite{Suzuki1991-za}.
Each time evolution under the instantaneous Hamiltonian $H_{\pi_v(\gamma)}(\beta_{v\gamma} t)$ is implemented by local gates as well as time-independent cases.
As the simplest construction, the Lie-Suzuki-Trotter formulas are given by
\begin{equation}\label{Eq_Pre:LST_Std_1_2}
    \bar{S}_1(t,0) = \prod_{\gamma=1}^\Gamma e^{-iH_\gamma(0)t}, \quad \bar{S}_2(t,0) = \bar{S}_1(t/2,t)^\dagger \bar{S}_1(t/2,0).
\end{equation}
The higher-order formulas are obtained by the recursive relation,
\begin{eqnarray}\label{Eq_Pre:LST_recursive_t_dep}
    \bar{S}_{2q}(t,0) &=& \bar{S}_{2q-2}(t,t-a_qt) \bar{S}_{2q-2}(t-a_qt,t-2a_qt) \nonumber \\
    && \quad \times \bar{S}_{2q-2}(2a_qt,t-2a_qt)^\dagger \nonumber \\
    && \qquad \times \bar{S}_{2q-2}(2a_qt,a_qt)\bar{S}_{2q-2}(a_qt,0),
\end{eqnarray}
with $a_q=[4-4^{1/(2q-1)}]^{-1}$.
The error bound of the standard PFs has been vigorously investigated in the past decade, while the one for low-orders are well-established \cite{Suzuki1991-za,Suzuki1993-general}.
For generic orders $p \in \bbN$, it is shown to have the error bound,
\begin{eqnarray}
    && \norm{U(t,0)-\bar{S}(t,0)} \in \order{\left\{\max_{\tau \in [0,t]} \left[ \alpha_{\text{1-norm},p}(\tau)\right] t\right\}^{p+1}}, \nonumber \\
    && \label{Eq_Pre:1_norm_error_t_dep}\\
    && \alpha_{\text{1-norm},p}(\tau)= \sum_\gamma \max_{q=0,\cdots,p} \left(\norm{\dv[q]{\tau} H_\gamma(\tau)}^{\frac{1}{q+1}}\right), \label{Eq_Pre:1_norm_factor_t_dep}
\end{eqnarray}
which is a counterpart of the $1$-norm scaling by Eq. (\ref{Eq_Pre:1_norm_scaling}) \cite{Wiebe2010-mu}.
Another study \cite{childs-prl2019-pf} has revealed the error bound expressed by nested commutators.
While it can provide size-efficient cost for generic local Hamiltonians with finite-range interactions, it includes matrix exponentials which are difficult to be evaluated for Hamiltonians with long-range interactions.

\textbf{Generalized PFs.---}
The other time-dependent PF is called the generalized PF, which is organized by the time-evolution operators,
\begin{equation}
    U_\gamma(t+t_0,t_0) = \mcl{T} \exp \left( -i \int_{t_0}^{t+t_0} \dd t^\prime H_\gamma(t^\prime)\right).
\end{equation}
The $p$th-order generalized PF $S(t,0)$ is defined by
\begin{equation}\label{Eq_Pre:Generalized_PF}
    S(t,0) = \prod_{v=1}^{V_p} \prod_{\gamma=1}^\Gamma U_{\pi_v(\gamma)}(\beta_{v\gamma} t + \alpha_{v\gamma} t, \beta_{v\gamma} t)
\end{equation}
satisfying the order condition, $S(t,0)=U(t,0)+\order{t^{p+1}}$.
The factors $\{\alpha_{v\gamma},\beta_{v\gamma}\}$ are not necessarily the same as those for the standard PF in Eq. (\ref{Eq_Pre:Standard_PF}).
The time-ordered integral in each time evolution $U_\gamma(t,0)$ can be efficiently computed by classical computers when the partition $\gamma$ is properly chosen.
Thus, the generalized PF $S(t,0)$ can be implemented by local gates as simply as the standard PF $\bar{S}(t,0)$.
The Lie-Suzuki-Trotter formulas in this case are given by
\begin{equation}
    S_1(t,0) = \prod_{\gamma=1}^\Gamma U_\gamma(t,0), \quad S_2(t,0) = S_1(t/2,t)^\dagger S_1(t/2,0),
\end{equation}
for the orders $p=1,2$, and the higher-orders are obtained recursively in the same way as Eq. (\ref{Eq_Pre:LST_recursive_t_dep}).
The error bound of the generalized PFs is investigated only for low-orders.
For instance, the first-order Lie-Suzuki-Trotter formula $S_1(t,0)$ has the commutator-scaling error in the form of
\begin{equation}\label{Eq_Pre:LST_error_t_dep_1}
    \norm{U(t,0)-S_1(t,0)} \leq \int_0^t \dd t_2 \int_0^{t_2} \dd t_1 \norm{[H_1(t_1),H_2(t_2)]},
\end{equation}
for a Hamiltonian $H(t)=H_1(t)+H_2(t)$.

Throughout this paper, we make several assumptions on the coefficients $\{\alpha_{v\gamma},\beta_{v\gamma}\}$, which do not essentially lose the generality.
First, we assume that the relations
\begin{equation}\label{Eq_Pre:coef_1}
    |\alpha_{v\gamma}| \leq 1, \quad \sum_{v=1}^{V_p} \sum_{\gamma': \pi_v(\gamma')=\gamma} \alpha_{v\gamma'} = 1
\end{equation}
are satisfied for every $\gamma=1,2,\cdots,\Gamma$.
The former one comes from that each evolution time $|\alpha_{v\gamma}|t$ is smaller than $t$.
The latter one means that the total evolution time under $H_\gamma(\cdot)$ is $t$ for every $\gamma$.
In addition, we assume 
\begin{equation}\label{Eq_Pre:coef_3}
    \begin{cases}
        0 \leq \beta_{v\gamma} \leq 1 & \text{for $\bar{S}(t,0)$} \\
        0 \leq \beta_{v\gamma}, \beta_{v\gamma}+\alpha_{v\gamma} \leq 1 & \text{for $S(t,0)$}
    \end{cases},
\end{equation}
which means a natural expectation that the time evolution $U(t,0)$ can be reproduced by the Hamiltonian $H_\gamma(\tau)$ during $\tau \in [0,t]$.
Finally, for making the calculation simple, we assume
\begin{equation}
    \beta_{11} = 0
\end{equation}
throughout this paper, which means that the approximation begins with the origin of the time.
Although $\beta_{11}$ can be nonzero in general, the results for $\beta_{11} \neq 0$ are essentially the same as those for $\beta_{11} = 0$ (See Appendix \ref{SubsecA:Error_beta_nonzero}).

As a special case, we also define the PF with the uniform coefficients as follows.
\begin{definition}\label{Def_Pre:uniform}
\textbf{(PF with uniform coefficients)}

If a standard or generalized PF has the coefficients $\{ \beta_{v\gamma} \}$ independent of $\gamma$, i.e.,
\begin{equation}
    \beta_{v\gamma} = \beta_v \in \bbR, \quad \text{for $v=1,\cdots,V_p$},
\end{equation}
it has uniform coefficients.
\end{definition}
Since each $\gamma$ equally contributes the Hamiltonian, many of the existing PFs such as the Lie-Suzuki-Trotter formula and the Forest-Ruth-Suzuki formula \cite{Forest1990-bg,Suzuki1990-fractal} satisfy the uniformity. 
Like time-independent cases, $\bar{S}(t,0)$ and $S(t,0)$ denote arbitrary standard and generalized PFs satisfying the order condition.
On the other hand, $\bar{S}_p(t,0)$ and $S_p(t,0)$ mean those with the uniform coefficients, while we often focus on the Lie-Suzuki-Trotter formulas as their particular examples without loss of generality.

It is an important but unresolved problem whether they have an error bound expressed by nested commutators, which is applicable to generic orders $q$ and generic local Hamiltonians.
It plays a central role for determining whether PFs can simulate time-dependent systems with the cost favorable in the system size like the time-independent cases \cite{childs-prl2019-pf}.

\subsection{Floquet theory}\label{Subsec:Floquet_theory}

We briefly review Floquet theory and pick up some relation which are frequently used in the proof.
Floquet theory is developed for analyzing time-periodic Hamiltonians in the form of $H(t)= \sum_m H_m e^{-im\omega t}$.
Each operator $H_m$ means the Fourier coefficient and is assumed to be $\norm{H_m} \in o(|m|^{-1})$ for the absolute and uniform convergence of the Fourier series.
The constant $\omega \in \bbR$ denotes the frequency.

In Floquet theory, we consider an infinite-dimensional space (Floquet-Hilbert space), defined by
\begin{equation}
    \mcl{H}_\mr{FH} = l^2 (\bbZ) \otimes \mcl{H},
\end{equation}
where $l^2 (\bbZ)$ denotes a square-summable space over $\bbZ$.
A Floquet Hamiltonian $H^F$ is an operator acting on $\mcl{H}_\mr{FH}$, which is defined by
\begin{equation}\label{Eq_Pre:H_F}
    H^F = \sum_{l \in \bbZ}\left( \sum_{m} \ket{l+m}\bra{l} \otimes H_m -l\omega \ket{l}\bra{l} \otimes I \right). 
\end{equation}
The ancilla system $\ket{l}$ contains the Fourier index $l \in \bbZ$.
It is often separated into two terms as $H^F = H^\mr{Add} - H^\mr{LP}$ with
\begin{equation}\label{Eq_Pre:H_add}
    H^\mr{Add} = \sum_m \mr{Add}_m\otimes H_m, \quad \mr{Add}_m = \sum_{l\in \bbZ} \ket{l+m}\bra{l},
\end{equation}
and the linear-potential term,
\begin{equation}\label{Eq_Pre:H_LP}
    H^\mr{LP} = \sum_{l \in \bbZ} l\omega \ket{l}\bra{l} \otimes I.
\end{equation}
Floquet theory dictates that the time evolution under the time-periodic Hamiltonian $H(t)$ can be expressed by the one under the time-independent operator $H^F$ as
\begin{equation}\label{Eq_Pre:Time_Evol_Floquet}
    U(t+t_0,t_0) = \sum_{l \in \bbZ} e^{-il\omega(t+t_0)} \braket{l|e^{-iH^Ft}|0},
\end{equation}
which is absolutely and uniformly convergent under $\norm{H_m} \in o(|m|^{-1})$ \cite{Levante1995-tx,Mizuta_Quantum_2023}.
Instead of discarding the difficulty by time dependence, $H^F$ is an unbounded operator on the infinite-dimensional system.

We introduce the translation symmetry in Floquet theory, which will be frequently used in the proof.
We can immediately check the invariance under the shift of the Fourier index,
\begin{eqnarray}
    && \braket{l+l'|H^\mr{Add}|l'} = \braket{l|H^\mr{Add}|0}, \label{Eq_Pre:tr_sym_H_Add}\\
    && \braket{l+l'|H^\mr{LP}|l'} = \braket{l|H^\mr{LP}|0} + l'\omega, \label{Eq_Pre:tr_sym_H_LP}
\end{eqnarray}
for any $l,l' \in \bbZ$ from their definitions, Eqs. (\ref{Eq_Pre:H_add}) and (\ref{Eq_Pre:H_LP}).
Using these relations, we also have the symmetry,
\begin{equation}\label{Eq_Pre:tr_sym_H_F}
    \braket{l+l'|e^{-iH^F t}|l'} = e^{il'\omega t} \braket{l|e^{-iH^F t}|0},
\end{equation}
for any $l,l' \in \bbZ$ and any time $t \in \bbR$.

\section{Summary of results}\label{Sec:Summary}

\subsection{Setup}\label{Subsec:Setup}

We first specify the setup in this section.
The central assumption is that the time-dependent Hamiltonian $H(t)=\sum_\gamma H_\gamma(t)$, defined on a finite-dimensional Hilbert space $\mcl{H}$, smoothly varies in time.
To be concrete, when we are interested in a $p$-th order PF during the time interval $[0,t]$, we assume that each $H_\gamma(\tau)$ is of class $C^{p+2}$ in $\tau \in [0,t]$.
Therefore, our results are applicable to broad types of time-dependence as well as some previous studies \cite{Wiebe2010-mu,childs-prl2019-pf}.

When each Hamiltonian $H_\gamma (t)$ is smooth, we can embed it into a time-periodic Hamiltonian.
Namely, there are series of time-periodic Hamiltonians $\{H^\mr{ex}_\gamma (t)\}$ satisfying the following conditions (See Appendix \ref{A_Sec:Smooth}).
\begin{itemize}
    \item Periodicity: There exists a $\gamma$-independent period $T$, larger than the time of interest $t$, such that $H_\gamma^\mr{ex}(\tau) = H_\gamma^\mr{ex}(\tau+T)$ for any $\tau \in \bbR$.
    Each $H_\gamma^\mr{ex}$ is expressed by a Fourier series,
    \begin{equation}
        H_\gamma^\mr{ex}(\tau) = \sum_{m \in \bbZ} H_{\gamma m}^\mr{ex} e^{-im\omega \tau}, \quad \omega \equiv \frac{2\pi}T,
    \end{equation}
    where the Fourier coefficient $H_{\gamma m}^\mr{ex}$ satisfies
    \begin{equation}\label{Eq_Sum:ex_decay}
        \norm{H_{\gamma m}^\mr{ex}} \in \order{|m|^{-p-2}}.
    \end{equation}

    \item Coincidence: Each $H_\gamma^\mr{ex}(\tau)$ is equivalent to $H_\gamma(\tau)$ in the interval $[0,t]$ by
    \begin{equation}
        H_\gamma^{\mr{ex} (q)}(\tau) = H_\gamma^{(q)}(\tau), \quad \text{for $\tau \in [0,t]$},
    \end{equation}
    for each $q=0,1,\cdots,p+2$.
\end{itemize}
The smoothness of $H(t)$ appears as the decay of $H_{\gamma m}^\mr{ex}$ by Eq.~(\ref{Eq_Sum:ex_decay}), which is required for the absolute and uniform convergence of the Fourier series of $H_\gamma^{\mr{ex}(q)}(\tau)$ for every $q=0,1,\cdots,p$.
This embedding enables us to employ the theory for analysing time-periodic Hamiltonians, called Floquet theory~\cite{Oka2019review}, in order to obtain the error bound of the time-dependent PFs as we will discuss below.

While we will use the time-periodic Hamiltonian $H^\mr{ex}(\tau) = \sum_\gamma H_\gamma^\mr{ex}(\tau)$ during the derivation, the resulting error bound includes the information only in the interval $[0,t]$, which is equivalent to that for the original Hamiltonian $H(t)$.
Therefore, without loss of generality, we omit the superscript ``ex" and deal with a generic smooth Hamiltonian $H(t)$ as if each $H_\gamma(t)$ is a time-periodic Hamiltonian satisfying
\begin{equation}\label{Eq_Sum:Fourier_decay}
    H_\gamma(t) = \sum_{m \in \bbZ} H_{\gamma m} e^{-im\omega t}, \quad \norm{H_{\gamma m}} \in \order{|m|^{-p-2}}
\end{equation}
with some period $T = 2\pi/\omega$.
We note the Fourier series for its time derivatives,
\begin{equation}
    H_\gamma^{(q)}(t) = \sum_{m \in \bbZ}  (-im\omega)^q H_{\gamma m} e^{-im\omega t},
\end{equation}
is absolutely and uniformly convergent for any finite $t$ up to $q=p$.
We use the notations for Floquet theory in Section \ref{Subsec:Notation}.

\begin{table*}
    \centering
    
    \begin{tabular}{|c|c|c|c|}
        \begin{tabular}{c}
            Algorithm \\
            (Error type)
        \end{tabular} &
             \begin{tabular}{c} Finite- / Short-ranged \\ Hamiltonians \end{tabular} &
             Long-ranged Hamiltonians & Ancilla qubits \\ \hline \hline
       \begin{tabular}{c} Standard PF, uniform \\
      ($1$-norm, Ref.~\cite{Wiebe2010-mu}) \end{tabular} & \(\displaystyle \order{N^2t\left[ \frac{(g+f)^pgt}{\varepsilon} \right]^{\frac1p}} \) & \(\displaystyle \order{N^{k+1}t\left[ \frac{(g+f)^pgt}{\varepsilon} \right]^{\frac1p}} \) & $0$ 
       \\ \hline
       \begin{tabular}{c} Standard PF \\
      (Commutator, Ref.~\cite{childs-prl2019-pf}) \end{tabular} & \(\displaystyle \order{Nt\left[ \frac{(g+f)^pgNt}{\varepsilon} \right]^{\frac1p}} \) & --- & $0$
       \\ \hline
       \begin{tabular}{c} Standard PF \\
       \& Generalized PF \\
       (Commutator, this work) \end{tabular} & 
        \(\displaystyle \order{Nt\left[ \frac{(g+f)^pgNt}{\varepsilon} \right]^{\frac1p}} \)
         & \(\displaystyle \order{N^kt\left[ \frac{(g+\Gamma f)^pgNt}{\varepsilon} \right]^{\frac1p}} \) &$0$ 
       \\ \hline
       \begin{tabular}{c} Standard PF, uniform \\
       (Commutator, this work) \end{tabular} & 
        \(\displaystyle \order{Nt\left[ \frac{(g+f)^pgNt}{\varepsilon} \right]^{\frac1p}}\)
         & \(\displaystyle \order{N^kt\left[ \frac{(g+f)^pgNt}{\varepsilon} \right]^{\frac1p}}\) &$0$ 
       \\ \hline
       \begin{tabular}{c} Standard MPF, uniform \\ ($1$-norm, Ref. \cite{watkins-2024-clock}) \end{tabular} &  \( \displaystyle \Otilde{N^2 t\left[ (g+f)^p g\right]^{\frac1{p+1}}} \) &  \( \displaystyle \Otilde{N^{k+1}t\left[(g+f)^pg\right]^{\frac1{p+1}}} \) & \( \displaystyle \order{\log \log  \frac{N(g+f)t}{\varepsilon}} \) \\ \hline
      
       \begin{tabular}{c} Standard MPF \\
       \& Generalized MPF \\
       (Commutator, this work) \end{tabular} & \( \displaystyle \Otilde{Nt \left\{ \left[ (g+f)^p gN \right]^{\frac1{p+1}} + f \right\}} \) &  \( \displaystyle \Otilde{N^k t \left\{ \left[ (g+\Gamma f)^p gN\right]^{\frac1{p+1}} +\Gamma f \right\} } \) & \( \displaystyle \order{\log \log \frac{N(g+\Gamma f)t}{\varepsilon} }\) \\ \hline
       \begin{tabular}{c} Standard MPF, uniform \\
       (Commutator, this work) \end{tabular} &  \( \displaystyle \Otilde{Nt\left\{ \left[ (g+f)^p gN \right]^{\frac1{p+1}} +f \right\}} \) &  \( \displaystyle \Otilde{N^k t\left\{ \left[ (g+f)^p gN\right]^{\frac1{p+1}} +f \right\}} \) & \( \displaystyle \order{\log \log \frac{N(g+f)t}{\varepsilon} }\) \\ \hline
      \begin{tabular}{c} LCU \cite{Low2018-dyson,Kieferova2019-dyson}  \end{tabular} & \(\displaystyle \order{N^2 gt \frac{\log (Ngt/\varepsilon)}{\log \log (Ngt/\varepsilon)}} \) & \(\displaystyle \order{ N^{k+1} gt \frac{\log (Ngt/\varepsilon)}{\log \log (Ngt/\varepsilon)}} \) & $\order{\log (Ngt/\varepsilon)}$ \\ \hline
       QSVT                           \cite{Mizuta_Quantum_2023,Mizuta_2023_multi} & \(\displaystyle \order{ N [Ngt+ft \log (ft/\varepsilon)] } \)  & \(\displaystyle \order{ N^k [Ngt+ft \log (ft/\varepsilon)] }\) & $\order{\log (Ngt/\varepsilon)}$ \\ \hline
      HHKL \cite{Haah2021-time-dep,Tran-PRX2019-hhkl} & $\Otilde{Ngt}$ & \begin{tabular}{c} \(\displaystyle \Otilde{Ngt \left( Ngt/\varepsilon \right)^{\frac{2d}{\nu-d}}} \) \\  (Only for $\nu>2d$) \end{tabular}& \( \displaystyle \order{\log \log (Ngt/\varepsilon)) }\) \\ \hline
    \end{tabular}
    
    \caption{Number of $\order{1}$-local quantum gates for Hamiltonian simulation with the system size $N$, the evolution time $t$, and the error tolerance $\varepsilon$. The extensiveness $g$ defined by Eq. (\ref{Eq_Cost:extensiveness}) satisfies $g \in \order{1}$ for finite-range interactions or Eq. (\ref{Eq_Cost:g_scaling}) for long-range interactions.
    The quantity $f$ represent the local scale of time-dependency by Eq. (\ref{Eq_Cost:time_dependency}), which can be $\order{1}$ for a broad class of local Hamiltonians in practice. The number of partitions in PFs, $\Gamma$, depends on the range of interactions as discussed in Section \ref{Subsec:gate_counts}. The notation $\Otilde{\cdot}$ contains polylogarithmic corrections with respect to $N$, $t$, or $1/\varepsilon$. See Definition~\ref{Def_Pre:uniform} for the definition of uniform PFs.}
    \label{Table:comparison_algorithms}
\end{table*}

\subsection{Explicit error bound and improved cost}\label{Subsec:commutator_scaling}

We summarize the central results of this paper, i.e., the error bound for time-dependent PFs explicitly organized by commutators among $\{H_\gamma(t)\}$ and their time derivatives, and the improved cost reflecting the locality based on it.

We begin with the derivation of the time-dependent PF error.
The central strategy is to employ Floquet theory developed for time-periodic Hamiltonians.
The proof is divided into three steps.
First, we find a mapping of the time-dependent PF error to a time-independent one.
Floquet theory states that the time evolution $U(t,0)$ can be expressed simply by the one under the time-independent Hamiltonian $H^F$ [See Eq. (\ref{Eq_Pre:Time_Evol_Floquet})].
We find that a similar relation holds for the time-dependent PF errors in general.
For instance, we prove that the error of the standard PF $\bar{S}(t,0)$ is expressed by
\begin{equation}\label{Eq_Sum:Error_Std}
    U(t,0)-\bar{S}(t,0) = \sum_{l \in \bbZ} e^{-il\omega t} \braket{l|[e^{-iH^Ft}-\bar{T}^F(t)]|0},
\end{equation}
where $\bar{T}^F(t)$ is a time-independent PF for $e^{-iH^Ft}$ defined on the infinite-dimensional space $\mcl{H}_\mr{FH}$.
The same goes also for the generalized PF $S(t,0)$.
As the second step, we employ the theory of the time-independent PF errors \cite{childs2021-trotter} for the mapped error like $e^{-iH^Ft}-\bar{T}^F(t)$. 
We note that this does not immediately imply the commutator scaling due to the summation over $l \in \bbZ$ in Eq.~(\ref{Eq_Sum:Error_Std}) and the unboundedness of $H^F$.
In the last step, we reduce the time-independent PF error on the infinite-dimensional space to the time-dependent PF error by carefully taking this summation.
We obtain the following error bounds as a first main result.

\begin{theorem}\label{Thm:main_theorem_informal}
\textbf{(Commutator scaling, informal)}

The $p$th-order standard and generalized PFs $\bar{S}(t,0)$ and $S(t,0)$ have the following error bounds, 
\begin{eqnarray}
    \norm{U(t,0)-\bar{S}(t,0)} &\in&  \order{\max_{\tau \in [0,t]} \bar{\alpha}_{\mr{com},p}(\tau) \cdot t^{p+1}}, \\
    \norm{U(t,0)-S(t,0)} &\in& \order{\max_{\tau \in [0,t]} \alpha_{\mr{com},p}(\tau)  \cdot t^{p+1}},
\end{eqnarray}
where the factors $\bar{\alpha}_{\mr{com},p}(\tau)$ and $\alpha_{\mr{com},p} (\tau)$ are defined by
\begin{widetext}
\begin{eqnarray}
    \bar{\alpha}_{\mr{com},p} (\tau) &=& \sum_{\gamma_0=1}^\Gamma \sum_{\gamma_1,\cdots,\gamma_p=1}^{\Gamma+1}   \norm{\left[\prod_{q=1}^p \bar{\mcl{D}}_{\gamma_q}(\tau) \right] H_{\gamma_0}(\tau)}, \quad \bar{\mcl{D}}_\gamma(\tau) = \begin{cases}
        \mr{ad}_{H_\gamma(\tau)} & (\gamma=1,\cdots,\Gamma) \\
        \Gamma \dv{\tau} & (\gamma=\Gamma+1)
    \end{cases}, \label{Eq_Sum:alpha_com_Std}\\
    \alpha_{\mr{com},p} (\tau) &=& \sum_{\gamma_0=1}^\Gamma \sum_{\gamma_1,\cdots,\gamma_p=1}^{\Gamma+1}   \norm{\left[\prod_{q=1}^p \mcl{D}_{\gamma_q}(\tau) \right] H_{\gamma_0}(\tau)}, \quad \mcl{D}_\gamma(\tau) = \begin{cases}
        \mr{ad}_{H_\gamma(\tau)} + i \dv{\tau} & (\gamma=1,\cdots,\Gamma) \\
        \Gamma \dv{\tau} & (\gamma=\Gamma+1)
    \end{cases}. \label{Eq_Sum:alpha_com_Gen}
\end{eqnarray}
\end{widetext}
\end{theorem}

These error bounds are composed of nested commutators among $\{ H_\gamma (t)\}$ and their time derivatives, while the latter one is inherent in time-dependent cases.
The advantage of our result is that they have many vanishing terms in the nested commutators when the Hamiltonian has the locality of interactions.
Indeed, we show that our error bounds achieve much better scaling in the system size $N$ for generic time-dependent local Hamiltonians than the previous one with the $1$-norm scaling \cite{Wiebe2010-mu}.

A better error bound of a quantum algorithm implies better complexity of running it with a target accuracy.
We prove the improved cost of the time-dependent PFs based on the above error bounds as the second part of our main results, which is summarized in Table \ref{Table:comparison_algorithms}.
Our result achieves better scaling in the system size $N$ than the known results by the $1$-norm scaling error \cite{Wiebe2010-mu}.
We also note that it can be applied to generic local time-dependent Hamiltonians with finite-, short-, and long-ranged interactions in contrast to Ref. \cite{childs-prl2019-pf}, which shows the commutator-scaling error valid for finite-range interactions.
Furthermore, they can reproduce the best known results for the time-independent PFs obtained by the commutator-scaling errors \cite{childs2021-trotter}.

Finally, we also emphasize that our approach based on Floquet theory applies to a family of various time-dependent PFs.
The one for non-hermitian time-dependent Hamiltonians is the simplest example (See Appendix \ref{A_Sec:non_unitary}).
In particular, the time-dependent multi-product formula (MPF), which uses a linear combination of time-dependent PFs, is a promising extension, with which we can exponentially improve the scaling in the desirable accuracy \cite{low2019-mpf,aftab2024-mpf,mizuta2025-mpf}.
As one of the main results, we derive the error bounds of the time-dependent MPFs which can reflect the locality of the Hamiltonian via nested commutators, and we prove the improved cost based on them as shown in Table \ref{Table:comparison_algorithms}.
The time-dependent MPFs achieve the good $N$-scaling as small as the PFs and also achieve the good scaling in $t$ and $\varepsilon$ like the post-Trotter algorithms such as LCU and QSVT.

\section{Relationship between time-independent and -dependent Product formulas
}\label{Sec:relation}

In this section, we map the time-dependent PFs into the time-independent one in the infinite-dimensional Floquet-Hilbert space.

\subsection{PF in Floquet-Hilbert space}\label{Subsec:formula_FH}

We begin by defining the time-independent PF in the Floquet-Hilbert space which can reproduce the time-dependent PF.
As introduced in Section \ref{Sec:Summary}, the time evolution under $H(t)$ can be organized by the Floquet Hamiltonian $H^F$ [See Eq.~(\ref{Eq_Pre:Time_Evol_Floquet})].
We note that, for smoothly varying Hamiltonians $H(t)$ discussed in Sec.~\ref{Subsec:Setup}, the infinite series for $U(t,t_0)$ in Eq.~(\ref{Eq_Pre:Time_Evol_Floquet}) is uniformly and absolutely convergent \cite{Mizuta_Quantum_2023}.

We similarly construct a Floquet Hamiltonian $H_\gamma^F$ for each term $H_\gamma (t)$ by
\begin{equation}\label{Eq_Rel:H_Add_gamma}
    H_\gamma^F =  H^\mr{Add}_\gamma - H^\mr{LP}, \quad H^\mr{Add}_\gamma = \sum_m \mr{Add}_m \otimes H_{\gamma m},
\end{equation}
and then, the time evolution under $H_\gamma (t)$ is expressed by
\begin{equation}\label{Eq_Rel:Time_Evol_Floquet_gamma}
    U_\gamma(t+t_0,t_0) = \sum_{l \in \bbZ} e^{-il\omega (t+t_0)} \braket{l|e^{-iH_\gamma^Ft}|0}.
\end{equation}
We define two kinds of time-independent PFs for $H^F$, which correspond to the standard and generalized PFs, respectively.

\begin{definition}\label{Def_Rel:PF_Floquet_Hamiltonian}
\textbf{(PFs for Floquet Hamiltonian)}

We define time-independent PFs $\bar{T}^F(t)$ and $T^F(t)$, acting on the Floquet-Hilbert space $\mcl{H}_\mr{FH}$, respectively by
\begin{eqnarray}
    \bar{T}^F(t) &=& \prod_{v=1}^{V_p}  \prod_{\gamma=1}^\Gamma \left( e^{iH^\mr{LP} (\beta_{v(\gamma+1)}-\beta_{v\gamma})t} e^{-iH_{\pi_v(\gamma)}^\mr{Add} \alpha_{v\gamma} t}\right), \label{Eq_Rel:T_F_Std}\\
    T^F(t) &=&  \prod_{v=1}^{V_p}  \prod_{\gamma=1}^\Gamma \left( e^{iH^\mr{LP} (\beta_{v(\gamma+1)}-\beta_{v\gamma}-\alpha_{v\gamma})t} e^{-iH_{\pi_v(\gamma)}^F \alpha_{v\gamma}t} \right), \nonumber \\
    && \label{Eq_Rel:T_F_Gen}
\end{eqnarray}
where we set $\beta_{v(\Gamma+1)}=\beta_{(v+1)1}$ and $\beta_{V_p(\Gamma+1)}=1$.
\end{definition}

We show that the time-independent PFs $\bar{T}^F(t)$ and $T^F(t)$ are related to the time-dependent PFs as follows.

\begin{theorem}\label{Thm_Rel:Relation}
\textbf{(Bridging PFs for original and Floquet Hamiltonians)}

The standard and generalized PFs $\bar{S}(t,t_0)$ and $S(t,t_0)$ by Eqs. (\ref{Eq_Pre:Standard_PF}) and (\ref{Eq_Pre:Generalized_PF}) are related to the time-independent PFs, Eqs. (\ref{Eq_Rel:T_F_Std}) and (\ref{Eq_Rel:T_F_Gen}), by
\begin{eqnarray}
    \bar{S}(t,0) = \sum_{l \in \bbZ} e^{-il\omega t} \braket{l|\bar{T}^F(t)|0}, \label{Eq_Rel:Rel_PF_Std} \\
    S(t,0) = \sum_{l \in \bbZ} e^{-il\omega t} \braket{l|T^F(t)|0}, \label{Eq_Rel:Rel_PF_Gen}
\end{eqnarray}
when the coefficients $\{\alpha_{v\gamma}, \beta_{v\gamma} \}$ satisfy Eqs. (\ref{Eq_Pre:coef_1})-(\ref{Eq_Pre:coef_3}).
\end{theorem}

\textbf{Proof.---}
We focus on the standard PF.
We begin with showing that the time-evolution operator $e^{-iH_\gamma(\beta t) \alpha t}$ for real $\alpha,\beta$ is expressed by 
\begin{equation}\label{Eq_Rel:TE_Floquet_inst}
    e^{-i H_\gamma (\beta t) \alpha t} = \sum_{l \in \bbZ} e^{-il\omega\beta t}\braket{l|e^{-i H_\gamma^\mr{Add} \alpha t}|0}.
\end{equation}
We note that the right-hand side is absolutely and uniformly convergent, which immediately follows from $\| \braket{l|e^{-i H_\gamma^\mr{Add} \alpha t}|0}\| \in o(|l|^{-1})$ by Lemma~\ref{LemA_smooth:evolution_decay}, which is shown in Appendix \ref{A_Sec:Smooth}.
Regarding $\alpha t$ and $\beta t$ as independent variables and fixing $\beta t$, we obtain
\begin{eqnarray}
    && \dv{(\alpha t)} \sum_{l \in \bbZ} e^{-il\omega\beta t}\braket{l|e^{-i H_\gamma^\mr{Add} \alpha t}|0} \nonumber \\
    && \quad = -i \sum_{l \in \bbZ} e^{-il\omega\beta t}\braket{l|H_\gamma^\mr{Add} e^{-i H_\gamma^\mr{Add} \alpha t}|0} \nonumber \\
    && \quad = -i \sum_{l,m \in \bbZ} H_{\gamma m} e^{-i\{(l-m)+m\}\omega \beta t} \braket{l-m|e^{-iH_\gamma^\mr{Add}\alpha t}|0} \nonumber \\
    && \quad = -i H_\gamma (\beta t) \sum_{l \in \bbZ} e^{-il\omega\beta t}\braket{l|e^{-i H_\gamma^\mr{Add} \alpha t}|0}, \label{Eq_Re:Differential_Eq_inst}
\end{eqnarray}
in which the exchange of the summation and the differentiation is allowed by the uniform convergence.
The second equality comes from Eq. (\ref{Eq_Rel:H_Add_gamma}).
In the last equality, we use the Fourier series Eq. (\ref{Eq_Sum:Fourier_decay}) and its absolute convergence, which leads to $\sum_{l,m \in \bbZ}=\sum_{m \in \bbZ} \sum_{l-m \in \bbZ}$.
We arrive at Eq. (\ref{Eq_Rel:TE_Floquet_inst}) by solving the differential equation Eq. (\ref{Eq_Re:Differential_Eq_inst}).
Using the translation symmetry of $H_\gamma^\mr{Add}$ by Eq. (\ref{Eq_Pre:tr_sym_H_Add}), we obtain
\begin{widetext}
\begin{eqnarray}
    \bar{S}(t,0) &=& \prod_{v=1}^{V_p} \prod_{\gamma=1}^\Gamma \left( \sum_{l_{v\gamma} \in \bbZ} e^{-il_{v\gamma} \omega \beta_{v\gamma} t}\Braket{\sum_{v'\gamma' \preceq v\gamma} l_{v'\gamma'}|e^{-iH_{\pi_v(\gamma)}^\mr{Add}\alpha_{v\gamma}t}| \sum_{v'\gamma' \prec v\gamma} l_{v'\gamma'}} \right) \nonumber \\
    &=& \prod_{v=1}^{V_p} \prod_{\gamma=1}^\Gamma \left( \sum_{l_{v\gamma} \in \bbZ} \Braket{\sum_{v'\gamma' \preceq v\gamma} l_{v'\gamma'}| e^{-iH^\mr{LP} \beta_{v\gamma} t} e^{-iH_{\pi_v(\gamma)}^\mr{Add}\alpha_{v\gamma}t} e^{iH^\mr{LP}\beta_{v\gamma} t}| \sum_{v'\gamma' \prec v\gamma} l_{v'\gamma'}} \right) \nonumber \\
    &=& \sum_{l \in \bbZ} \Braket{l|e^{-iH^\mr{LP}\beta_{V_p(\Gamma+1)}t}\prod_{v=1}^{V_p} \prod_{\gamma=1}^\Gamma \left(  e^{iH^\mr{LP} (\beta_{v(\gamma+1)}-\beta_{v\gamma})t} e^{-iH_{\pi_v(\gamma)}^\mr{Add} \alpha_{v\gamma} t} \right) e^{iH^\mr{LP}\beta_{11}t}|0},
\end{eqnarray}
\end{widetext}
where we used the completeness $\sum_{l \in \bbZ} \ket{l}\bra{l} = I$ in the last equality.
We note that this calculation is validated by the absolute convergence of the infinite series, Eq. (\ref{Eq_Rel:TE_Floquet_inst}), which allows us to change the way of taking the summation over $\{ l_{v'} \in \bbZ \}$.
This completes the proof of Eq. (\ref{Eq_Rel:Rel_PF_Std}) by the definition of $\bar{T}^F(t)$, Eq. (\ref{Eq_Rel:T_F_Std}).

The proof of Eq. (\ref{Eq_Rel:Rel_PF_Gen}) is parallel to the above calculation.
We substitute the expression of $U_\gamma(t,0)$ by Eq. (\ref{Eq_Rel:Time_Evol_Floquet_gamma}), and then we use the translation symmetry of $e^{-iH^Ft}$ by Eq. (\ref{Eq_Pre:tr_sym_H_F}).
We arrive at Eq. (\ref{Eq_Rel:Rel_PF_Gen}) as the expression of the generalized PF $S(t,0)$. $\quad \square$

Theorem \ref{Thm_Rel:Relation} shows a simple relation between time-independent and -dependent PFs.
Instead of discarding the time dependence, the PFs acquire infinite dimensionality as well as the exact time evolution, Eq.~(\ref{Eq_Pre:Time_Evol_Floquet}).
Consequently, the errors of the time-dependent PFs can be attributed to those of the time-independent PFs $e^{-iH^Ft}-\bar{T}^F(t)$ or $e^{-iH^Ft}-T^F(t)$ like Eq. (\ref{Eq_Sum:Error_Std}).
This is the origin of the commutator scaling of the time-dependent PFs.
However, we emphasize that the presence of the commutator scaling in the time-dependent PF errors is nontrivial at the present stage due to the infinite-dimensionality.
The Hamiltonians $H_\gamma^F$ and $H^\mr{LP}$ are both unbounded, and the summation over $l \in \bbZ$ does not necessarily provides a meaningful bound.
The explicit computation of Eq.~(\ref{Eq_Sum:Error_Std}) in Section \ref{Sec:Error} is essential for concluding the commutator scaling.

We remark several points before finishing this section.
First, in order to satisfy the order condition $\bar{S}(t,0)=U(t,0)+\order{t^{p+1}}$ or $S(t,0)=U(t,0)+\order{t^{p+1}}$, we will demand that $\bar{T}^F(t)$ or $T^F(t)$ become $p$th-order time-independent PFs for $e^{-iH^Ft}$.
The difference between the standard and generalized PFs lies in the way of the decomposition of the Floquet Hamiltonian,
\begin{eqnarray}
    H^F &=& \sum_{\gamma=1}^\Gamma H_\gamma^\mr{Add} - H^\mr{LP} \label{Eq_Rel:H_F_decomp_Std}\\
    &=& \sum_{\gamma=1}^{\Gamma} H_\gamma^F + (\Gamma-1)H^\mr{LP}. \label{Eq_Rel:H_F_decomp_Gen}
\end{eqnarray}
The decomposition by the first line corresponds to the standard PF, which employs the time-evolution operators under $H_\gamma^\mr{Add}$ and $H^\mr{LP}$ in $\bar{T}^F(t)$.
Similarly, the one by the second line corresponds to the generalized PF.
These decompositions will be used for determining the proper coefficients $\{\alpha_{v\gamma}, \beta_{v\gamma} \}$.

We also note that the assumption $\beta_{11}=0$ affects the construction of the operators $\bar{T}^F(t)$ and $T^F(t)$ via this order condition.
Equations (\ref{Eq_Rel:T_F_Std}) and (\ref{Eq_Rel:T_F_Gen}) cannot provide a $p$th-order time-independent PF for $e^{-iH^Ft}$ under $\beta_{11}=0$.
When we have $\beta_{11} \neq 0$, the operator $\bar{T}^F(t)$ is replaced by
\begin{equation}
    \prod_{v=1}^{V_p}  \prod_{\gamma=1}^\Gamma \left( e^{iH^\mr{LP} (\beta_{v(\gamma+1)}-\beta_{v\gamma})t} e^{-iH_{\pi_v(\gamma)}^\mr{Add} \alpha_{v\gamma} t}\right) e^{iH^\mr{LP} \beta_{11}t}
\end{equation}
and so does the operator $T^F(t)$ similarly.
The calculation in the following sections is a little bit modified, while the scalings of the error bounds and the computational cost do not essentially alter.
See Appendix \ref{SubsecA:Error_beta_nonzero} for the detailed discussion.

Finally, we provide the translation symmetry of the time-independent PFs $\bar{T}^F(t)$ and $T^F(t)$ like Eq.~(\ref{Eq_Pre:tr_sym_H_F}), which will be necessary for the calculation in the next section. 

\begin{lemma}\label{Lem_Rel:transl_sym_PF}
\textbf{(Translation symmetry of PFs)}

The time-independent PFs $\bar{T}^F(t)$ and $T^F(t)$ by Definition \ref{Def_Rel:PF_Floquet_Hamiltonian} satisfy the translation symmetry,
\begin{eqnarray}
    \braket{l|\bar{T}^F(t)|l^\prime} &=& e^{il^{\prime \prime}\omega t} \braket{l-l^{\prime \prime}|\bar{T}^F(t)|l^\prime-l^{\prime \prime}}, \\
    \braket{l|T^F(t)|l^\prime} &=& e^{il^{\prime \prime}\omega t} \braket{l-l^{\prime \prime}|T^F(t)|l^\prime-l^{\prime \prime}},
\end{eqnarray}
for every $l,l^\prime, l^{\prime \prime} \in \bbZ$.
\end{lemma}

\textbf{Proof.---}
Suppose that operators $A^F$ and $B^F$, acting on the Floquet-Hilbert space $\mcl{H}_\mr{FH}$, satisfy the symmetry,
\begin{eqnarray}
    \braket{l|A^F|l^\prime} &=& e^{il^{\prime \prime}a} \braket{l-l^{\prime \prime}|A^F|l^\prime-l^{\prime \prime}}, \\
    \braket{l|B^F|l^\prime} &=& e^{il^{\prime \prime}b} \braket{l-l^{\prime \prime}|B^F|l^\prime-l^{\prime \prime}},
\end{eqnarray}
for every $l,l^\prime, l^{\prime \prime} \in \bbZ$ with certain values $a,b \in \bbC$.
Then, we have
\begin{eqnarray}
    \braket{l|A^FB^F|l'} &=& \sum_{l''' \in \bbZ} \braket{l|A^F|l'''}\braket{l'''|B^F|l'} \nonumber \\
    &=& \sum_{l''' \in \bbZ} e^{il''(a+b)}\braket{l-l''|A^F|l'''-l''} \nonumber \\
    && \qquad \qquad \qquad \times \braket{l'''-l''|B^F|l'-l''} \nonumber \\
    &=& e^{il''(a+b)} \braket{l-l''|A^F B^F|l'-l''}.
\end{eqnarray}
We repeat this relation based on the symmetry of $H^\mr{Add}_\gamma$ and $H^\mr{LP}$ by Eqs. (\ref{Eq_Pre:tr_sym_H_Add}) and (\ref{Eq_Pre:tr_sym_H_LP}) for the standard PF $\bar{T}^F(t)$, which results in 
\begin{eqnarray}
    \braket{l|\bar{T}^F(t)|l'} &=& e^{i l''\omega \sum_{v=1}^{V_p} \sum_{\gamma=1}^\Gamma (\beta_{v(\gamma+1)}-\beta_{v\gamma}) t} \nonumber \\
    && \qquad \qquad \qquad  \times \braket{l-l''|\bar{T}^F(t)|l'-l''} \nonumber \\
    &=& e^{il''\omega t} \braket{l-l''|\bar{T}^F(t)|l'-l''}.
\end{eqnarray}
We used $\beta_{11}=0$ and $\beta_{V_p(\Gamma+1)}=1$ in the last equality.
The symmetry for the generalized PF $T^F(t)$ is shown by using that for $e^{-iH^Ft}$ by Eq. (\ref{Eq_Pre:tr_sym_H_F}) in a similar manner. $\quad \square$

While the PFs $\bar{T}^F(t)$ and $T^F(t)$ are constructed as approximate formulas of $e^{-iH^Ft}$, Lemma \ref{Lem_Rel:transl_sym_PF} implies that the translation symmetry of $e^{-iH^Ft}$ is exactly inherited.
This relation will be used to obtain the error representation of the time-dependent PF errors.

\subsection{Example: Lie-Suzuki-Trotter formula}\label{Subsec:Lie-Suzuki-Trotter}

In this section, we consider the Lie-Suzuki-Trotter formulas as a concrete example, and we show that those for time-dependent systems are completely expressed by those for time-independent cases via Theorem \ref{Thm_Rel:Relation}.
We begin with the first-order formulas $\bar{S}_1(t,0)$ and $S(t,0)$, in which we have $V_1 = 1$, $\alpha_{v\gamma}= 1$, and $\beta_{v\gamma}=0$.
The corresponding time-independent PFs based on Definition \ref{Def_Rel:PF_Floquet_Hamiltonian} are given by
\begin{eqnarray}
    \bar{T}^F(t) &=& e^{iH^\mr{LP}t} \prod_{\gamma=1}^\Gamma e^{-iH_\gamma^\mr{Add}t}, \label{Eq_Rel:LST_Std_1}\\ 
    T^F(t) &=& e^{-iH_\Gamma^Ft} \prod_{\gamma=1}^\Gamma \left( e^{-iH^\mr{LP}t} e^{-iH_\gamma^Ft} \right). \label{Eq_Rel:LST_Gen_1}
\end{eqnarray}
The former one is just the first-order Lie-Suzuki-Trotter formula with the decomposition, $H^F=H_1^\mr{Add}+\cdots+H_\Gamma^\mr{Add}-H^\mr{LP}$. 
The latter one comes from the decomposition, $H^F = H_1^F + H^\mr{LP} + H_2^F + \cdots + H^\mr{LP} + H_\Gamma^F$.
These relations hold for generic orders $p$, which is described by the following theorem.

\begin{theorem}\label{Thm_Rel:Coincidence_LST}
\textbf{}

We define the time-independent PF $\bar{T}_p^F(t)$ by Eq. (\ref{Eq_Rel:LST_Std_1}) for $p=1$, $\bar{T}_2^F(t)=\bar{T}_1^F(-t/2)^\dagger \bar{T}_1^F(t/2)$ for $p=2$, and
\begin{equation}\label{Eq_Rel:Recursive_T_F_LST}
    T_{2q}^F(t) = [T_{2q-2}^F(a_qt) ]^2 T_{2q-2}^F (b_qt)^\dagger [T_{2q-2}^F(a_qt) ]^2
\end{equation}
for $q \in \bbN + 1$.
The $p$th-order Lie-Suzuki-Trotter formula $\bar{S}_p(t,0)$ is expressed by
\begin{equation}\label{Eq_Rel:Rel_LST_Std}
    \bar{S}_p(t,0) = \sum_{l \in \bbZ} e^{-il\omega t} \braket{l|\bar{T}_p^F(t)|0}.
\end{equation}
We also have
\begin{equation}
    S_p(t,0) = \sum_{l \in \bbZ} e^{-il\omega t} \braket{l|T_p^F(t)|0},
\end{equation}
where $T_p^F(t)$ is constructed recursively beginning with Eq. (\ref{Eq_Rel:LST_Gen_1}) in a similar manner.
\end{theorem}

\textbf{Proof.---}
We focus on the standard PF with the order $p \geq 2$.
Based on the definition of $\bar{S}_2(t,0)$ by Eqs. (\ref{Eq_Pre:Standard_PF}) and (\ref{Eq_Pre:LST_Std_1_2}), we have $V_2=2$, $\alpha_{1\gamma}=\alpha_{2\gamma}=1/2$, $\beta_{1\gamma}=0$, and $\beta_{2\gamma}=1$.
The permutation $\pi_v$ is given by Eq. (\ref{Eq_Pre:permutation}).
The time-independent PF defined by Eq. (\ref{Eq_Rel:T_F_Std}) is equal to
\begin{eqnarray}
    \bar{T}^F(t) &=& \left( \prod_{\gamma=\Gamma}^1 e^{-iH_\gamma^\mr{Add}t/2} \right) e^{iH^\mr{LP}t} \left( \prod_{\gamma=1}^\Gamma e^{-iH_\gamma^\mr{Add}t/2}\right) \nonumber \\
    &=& \bar{T}_1^F(-t/2)^\dagger \bar{T}_1^F(t/2).
\end{eqnarray}
Thus, Theorem \ref{Thm_Rel:Relation} implies the satisfaction of Eq. (\ref{Eq_Rel:Rel_LST_Std}) for $p=2$.
Next, we assume Eq. (\ref{Eq_Rel:Rel_LST_Std}) for $p=2q-2$.
The recursive relation Eq. (\ref{Eq_Rel:Recursive_T_F_LST}) implies
\begin{widetext}
\begin{eqnarray}
    && [\text{r.h.s. of Eq. (\ref{Eq_Rel:Rel_LST_Std})}] \nonumber \\
    && \quad = \sum_{l,l_1,l_2,l_3,l_4 \in \bbZ} e^{-il\omega t} \braket{l|\bar{T}_{2q-2}^F(a_qt)|l_1} \braket{l_1|\bar{T}_{2q-2}^F(a_qt)|l_2} \braket{l_2|\bar{T}_{2q-2}^F (b_qt)^\dagger|l_3} \braket{l_3|\bar{T}_{2q-2}^F(a_qt)|l_4}\braket{l_4|\bar{T}_{2q-2}^F(a_qt)|l_0} \nonumber \\
    && \quad = \sum_{l,l_1,l_2,l_3,l_4 \in \bbZ} e^{-i(l-l_1)\omega t} \braket{l-l_1|\bar{T}_{2q-2}^F(a_qt)|0} e^{-i(l_1-l_2)\omega (1-a_q)t} \braket{l_1-l_2|\bar{T}_{2q-2}^F(a_qt)|0} \nonumber \\
    && \qquad \quad \times \left( e^{-i(l_3-l_2)\omega (1-2a_q)t}\braket{l_3-l_2|\bar{T}_{2q-2}^F (b_qt)|0}\right)^\dagger  e^{-i(l_3-l_4)\omega (2a_q) t} \braket{l_3-l_4|\bar{T}_{2q-2}^F(a_qt)|0} e^{-il_4\omega a_q t} \braket{l_4|\bar{T}_{2q-2}^F(a_qt)|0} \nonumber \\
    && \quad = \bar{S}_{2q-2}(t,t-a_qt) \bar{S}_{2q-2}(t-a_qt,t-2a_qt) \bar{S}_{2q-2}(2a_qt,t-2a_qt)^\dagger \bar{S}_{2q-2}(2a_qt,a_qt) \bar{S}_{2q-2}(a_qt,0), 
\end{eqnarray}
\end{widetext}
where we used the completeness in the first equality, and translation symmetry by Lemma \ref{Lem_Rel:transl_sym_PF} in the second equality.
The definition of $\bar{S}_{2q}(t,0)$ by Eq. (\ref{Eq_Pre:LST_recursive_t_dep}) completes the proof of Eq. (\ref{Eq_Rel:Rel_LST_Std}) by induction.
The same discussion goes also for the generalized PF. $\quad \square$

To summarize, Theorem \ref{Thm_Rel:Coincidence_LST} implies that the Lie-Suzuki-Trotter formulas for time-dependent Hamiltonian are completely translated into those for time-independent Hamiltonians.
The appearance of the different types of the PFs, i,e., the standard PF $\bar{S}(t,0)$ and the generalized PF $S(t,0)$, can be attributed to the ways of the decomposition of $H^F$.


\section{Error bound of time-dependent product formulas}\label{Sec:Error}

We derive the error bounds for the time-dependent PFs, which shows the commutator scaling.
As shown in the previous section, the time-dependent PF error can be translated into the time-independent one for the Floquet Hamiltonian.
We first clarify the error representation by applying the technique for time-independent systems \cite{childs2021-trotter} in Section \ref{Subsec:Error_repr}.
We have to be aware that the time-independent counterpart is defined on the infinite-dimensional Floquet-Hilbert space, and that this does not immediately provide a meaningful error bound.
In Section \ref{Subsec:explicit_error_bound}, we explicitly compute the summation over $l \in \bbZ$ and show a convergent error bound, which results in the meaningful commutator scaling of the time-dependent PF error.
Section~\ref{Subsec:form_commutators} is devoted to computing commutators appearing in the errors in Section \ref{Subsec:explicit_error_bound}, which gives the origin of the errors expressed by the nested commutators and their time derivatives.
Throughout this section, we concentrate on deriving the error bound of the standard PF $\bar{S}(t,0)$.
While we briefly discuss the results for the generalized PF $S(t,0)$, we provide their detailed derivation in Appendix \ref{A_Sec:Generalized_PF}.

\subsection{Error representation}\label{Subsec:Error_repr}

Section \ref{Sec:relation} proves that the error of the time-dependent PF can be translated into the one of the time-independent counterpart by Eq.~(\ref{Eq_Sum:Error_Std}) defined on $\mcl{H}_\mr{FH}$.
Next, we apply the theory of PF errors for time-independent Hamiltonians \cite{childs2021-trotter} to this.
Focusing on the standard PF $\bar{S}(t,0)$, we assume that its counterpart $\bar{T}^F(t)$ by Eq. (\ref{Eq_Rel:T_F_Std}) formally gives a $p$th-order time-independent PF for $e^{-iH^Ft}$.
We associate $H_\gamma^\mr{Add}$ and $-H^\mr{LP}$ respectively with $H_\gamma$ and $H_0$ of a Hamiltonian $H = \sum_{\gamma=0}^\Gamma H_\gamma$, and then it is equivalent to consider the time-independent PF,
\begin{equation}\label{Eq_Err:PF_equiv}
    \bar{T}(t) = \prod_{v=1}^{V_p}  \prod_{\gamma=1}^\Gamma \left( e^{-iH_0 (\beta_{v(\gamma+1)}-\beta_{v\gamma})t} e^{-iH_{\pi_v(\gamma)} \alpha_{v\gamma} t}\right),
\end{equation}
satisfying the order condition $\bar{T}(t)=e^{-iHt}+\order{t^{p+1}}$.
The error $e^{-iH^Ft}-\bar{T}^F(t)$ can be represented completely in the same way as $e^{-iH^Ft}-\bar{T}(t)$.

We introduce some useful notations before give the error representation based on the above strategy.
The time-independent PF $\bar{T}^F(t)$ by Eq.~(\ref{Eq_Rel:T_F_Std}) can be rewritten as
\begin{equation}\label{Eq_Err:T_F_rewrite}
    \bar{T}^F(t) \equiv \prod_{v=1}^{V_p} \prod_{\tilde{\gamma}=1}^{2\Gamma} e^{-i \tilde{H}_{v \tilde{\gamma}} \tilde{\alpha}_{v \tilde{\gamma}} t},
\end{equation}
where $\tilde{H}_{v \tilde{\gamma}}$ and $\tilde{\alpha}_{v \tilde{\gamma}}$ are respectively given by
\begin{equation}\label{Eq_Err:H_tilde_def}
    (\tilde{H}_{v\tilde{\gamma}}, \tilde{\alpha}_{v \tilde{\gamma}}) = \begin{cases}
        (H_{\pi_v(\gamma)}^\mr{Add},\alpha_{v \gamma}) & (\tilde{\gamma}=2\gamma-1) \\
        (-H^\mr{LP},\beta_{v(\gamma+1)}-\beta_{v\gamma}) & (\tilde{\gamma}=2\gamma)
    \end{cases}.
\end{equation}
The coefficient $\tilde{\alpha}_{v \tilde{\gamma}}$ has an absolute value smaller than $1$ owing to Eqs. (\ref{Eq_Pre:coef_1})-(\ref{Eq_Pre:coef_3}).
We also define the partial product of $\bar{T}^F(t)$ by
\begin{equation}\label{Eq_Err:Truncated_T_F}
    \bar{T}^F_{\prec v \tilde{\gamma}} (t) \equiv \prod_{v'\tilde{\gamma}'=11}^{v(\tilde{\gamma}-1)} e^{-i \tilde{H}_{v' \tilde{\gamma}'} \tilde{\alpha}_{v'\tilde{\gamma}'} t}.
\end{equation}
With these notations, we obtain the following error representation.

\begin{lemma}
\textbf{(Error representation)}

Suppose that the coefficients $\{\alpha_{v\gamma},\beta_{v\gamma}\}$ are chosen so that the time-independent PF $\bar{T}(t)$ by Eq. (\ref{Eq_Rel:T_F_Std}) can satisfy the order condition $\bar{T}(t)=e^{-iHt}+\order{t^{p+1}}$ for a Hamiltonian $H=\sum_{\gamma=0}^\Gamma H_\gamma$.
The difference $e^{-iH^Ft}-\bar{T}^F(t)$ is given by
\begin{equation}\label{Eq_Err:Error_repr_PF_Floquet}
    e^{-iH^Ft}-\bar{T}^F(t) = -i \int_0^t \dd \tau e^{-iH^F(t-\tau)} \bar{T}^F(\tau) \bar{\Delta}^F(\tau),
\end{equation}
where the operator $\bar{\Delta}^F(\tau)$ is defined by
\begin{widetext}
\begin{eqnarray}
    \bar{\Delta}^F(\tau) &=&  \sum_{v=1}^{V_p} \sum_{\tilde{\gamma}=1}^{2\Gamma}
    \sum_{\substack{q_{v\tilde{\gamma}}, \cdots,q_{V_p 2\Gamma} \geq 0: \\ q_{v\tilde{\gamma}} + \cdots + q_{V_p 2\Gamma} = p,
    \\ q_{v\tilde{\gamma}} \neq 0}} \int_0^\tau \dd \tau_1 \frac{i^p (\tau-\tau_1)^{q_{v\tilde{\gamma}}-1} \tau^{p-q_{v\tilde{\gamma}}}}{(q_{v\tilde{\gamma}}-1)!q_{v(\tilde{\gamma}+1)}! \cdots q_{V_p 2\Gamma} !} C_1(H^F) \nonumber \\
    && \qquad  - \sum_{v'=1}^{V_p} \sum_{\tilde{\gamma}'=1}^{2\Gamma} \sum_{v\tilde{\gamma} \preceq v'\tilde{\gamma}'} \tilde{\alpha}_{v'\tilde{\gamma}'}
    \sum_{\substack{q_{v\tilde{\gamma}}, \cdots,q_{v'\tilde{\gamma}'} \geq 0: \\ q_{v\tilde{\gamma}} + \cdots + q_{v'\tilde{\gamma}'} = p,
    \\ q_{v\tilde{\gamma}} \neq 0}} \int_0^\tau \dd \tau_1 \frac{i^p (\tau-\tau_1)^{q_{v\tilde{\gamma}}-1} \tau^{p-q_{v\tilde{\gamma}}}}{(q_{v\tilde{\gamma}}-1)!q_{v(\tilde{\gamma}+1)}! \cdots q_{v'\tilde{\gamma}'} !} C_2(\tilde{H}_{v'\tilde{\gamma}'}), \label{Eq_Err:Delta_F_expansion}
\end{eqnarray}
\begin{eqnarray}
    C_1 (H^F) &=& \bar{T}^F_{\prec v \tilde{\gamma}}(\tau)^\dagger e^{i \tilde{H}_{v\tilde{\gamma}} \tilde{\alpha}_{v\tilde{\gamma}} \tau_1} \left[ \prod_{v''\tilde{\gamma}''=V_p 2\Gamma}^{v\tilde{\gamma}} \left( \tilde{\alpha}_{v''\tilde{\gamma}''} \ad_{\tilde{H}_{v''\tilde{\gamma}''}}\right)^{q_{v''\tilde{\gamma}''}} H^F \right] e^{-i \tilde{H}_{v\tilde{\gamma}} \tilde{\alpha}_{v\tilde{\gamma}} \tau_1} \bar{T}^F_{\prec v \tilde{\gamma}}(\tau), \label{Eq_Err:C_1}\\
    C_2 (\tilde{H}_{v'\tilde{\gamma}'}) &=& \bar{T}^F_{\prec v \tilde{\gamma}}(\tau)^\dagger e^{i \tilde{H}_{v\tilde{\gamma}} \tilde{\alpha}_{v\tilde{\gamma}} \tau_1} \left[ \prod_{v''\tilde{\gamma}''=v'\tilde{\gamma}'}^{v\tilde{\gamma}} \left( \tilde{\alpha}_{v''\tilde{\gamma}''} \ad_{\tilde{H}_{v''\tilde{\gamma}''}}\right)^{q_{v''\tilde{\gamma}''}} \tilde{H}_{v' \tilde{\gamma}'} \right] e^{-i \tilde{H}_{v\tilde{\gamma}} \tilde{\alpha}_{v\tilde{\gamma}} \tau_1} \bar{T}^F_{\prec v \tilde{\gamma}}(\tau). \label{Eq_Err:C_2}
\end{eqnarray}
\end{widetext}
\end{lemma}

\textbf{Proof.---}
Since the representation Eq. (\ref{Eq_Err:T_F_rewrite}) is essentially the same as Eq. (\ref{Eq_Pre:PF_t_indep}), we can directly apply the approach for time-independent cases (See Theorem 5 in Ref. \cite{childs2021-trotter}).
We begin by noticing that the difference is represented by
\begin{equation}
    e^{-iH^Ft}-\bar{T}^F(t) = -e^{-iH^Ft}\int_0^t \dd \tau \dv{\tau} e^{iH^F\tau} \bar{T}^F(\tau).
\end{equation}
This implies that we have Eq.~\eqref{Eq_Err:Error_repr_PF_Floquet} with
\begin{eqnarray}
    \bar{\Delta}^F(\tau) &=& -i \bar{T}^F(\tau)^\dagger e^{-iH^F\tau} \dv{\tau} e^{iH^F\tau} \bar{T}^F(\tau) \nonumber \\
    &=& \bar{T}^F(\tau)^\dagger H^F \bar{T}^F(\tau) \nonumber \\
    && \quad - \sum_{v'=1}^{V_p}\sum_{\tilde{\gamma}'=1}^{2\Gamma} \tilde{\alpha}_{v'\tilde{\gamma}'} \bar{T}_{v' \tilde{\gamma}'}^F(\tau)^\dagger \tilde{H}_{v'\tilde{\gamma}'} \bar{T}_{v'\tilde{\gamma}'}^F(\tau). \nonumber \\
    && \label{Eq_Err:Delta_F_def}
\end{eqnarray}

In the above equation, both of the two terms have the form,
\begin{equation}
    \left( \prod_{k=1}^K e^{\tau \ad_{A_k}} \right) B = e^{A_K \tau}  \cdots e^{A_1 \tau} B e^{-A_1 \tau} \cdots e^{-A_K \tau},
\end{equation}
with some operators $\{A_k\}$ and $B$.
We repeatedly use the Taylor's theorem,
\begin{eqnarray}
    e^{\tau \ad_{A_k}} B &=& \sum_{q=0}^{p-1} \frac{(\tau \ad_{A_k})^q}{q!} B \nonumber \\
    && \qquad + \int_0^\tau \dd \tau_1 e^{\tau_1 \ad_{A_k}}\frac{(\tau_1 \ad_{A_k})^p}{p!} B, \label{Eq_Err:Taylor_theorem}
\end{eqnarray}
for $k=1,\cdots,K$ so that the Taylor's reminder can be $\order{\tau^p}$.
This results in the following formula, which is used for time-independent PFs \cite{childs2021-trotter}, 
\begin{widetext}
\begin{equation}\label{Eq_Err:adjoint_product_exp}
    \left( \prod_{k=1}^K e^{\tau \ad_{A_k}} \right) B = \sum_{q=0}^{p-1} \mcl{A}_q \tau^q + \sum_{k=1}^K \sum_{\substack{q_1+\cdots+q_k=p \\ q_{k} \neq 0}} \int_0^\tau \dd \tau_1 \frac{(\tau-\tau_1)^{q_k-1} \tau^{p-q_k}}{(q_k-1)!q_{k-1}!\cdots q_1!} \left( \prod_{k^\prime =k+1}^K e^{\tau \ad_{H_{k^\prime}} }\right) e^{\tau_1 \ad_{H_k}} \left( \prod_{k^\prime=1}^k (\ad_{H_{k^\prime}})^{q_{k^\prime}} B \right).
\end{equation}
\end{widetext}

When we apply this formula to the first and the second terms in Eq. (\ref{Eq_Err:Delta_F_def}), the first term of Eq. (\ref{Eq_Err:adjoint_product_exp}) vanishes due to the order condition $\tilde{T}(t)=e^{-iH^Ft}+\order{t^{p+1}}$ just like the time-independent PF.
The remaining second term of Eq. (\ref{Eq_Err:adjoint_product_exp}) give Eq. (\ref{Eq_Err:Delta_F_expansion}) as the form of $\bar{\Delta}^F(\tau)$, which completes the proof. $\quad \square$

In the case of the time-independent PFs, the counterparts of Eqs.~(\ref{Eq_Err:C_1}) and (\ref{Eq_Err:C_2}) directly give the commutator scaling of the errors~\cite{childs2021-trotter}.
We have a certain operator $\Delta(\tau)$ similar to Eq.~(\ref{Eq_Err:Delta_F_expansion}), such that
\begin{equation}\label{Eq_Err:t_indep_error_Delta}
    e^{-iHt}-T(t) = -i \int_0^t \dd \tau e^{-iH(t-\tau)} T(\tau) \Delta(\tau).
\end{equation}
This immediately implies the error bound,
\begin{equation}\label{Eq_Err:t_indep_error_bound_Delta}
    \norm{e^{-iHt}-T(t)} \leq \int_0^t \dd \tau \norm{\Delta(\tau)},
\end{equation}
and the operator norm $\norm{\Delta(\tau)}$ characterizes the error bound.
The representation of $\Delta(\tau)$ similar to Eq.~(\ref{Eq_Err:Delta_F_expansion}) indicates the scaling $\Delta(\tau) \in \order{\tau^p}$ and the commutator scaling by the counterpart of Eqs.~(\ref{Eq_Err:C_1}) and (\ref{Eq_Err:C_2}).
As a result, the time-independent PF error of $\order{\alpha_{\mr{com},p} t^{p+1}}$ is obtained.

In time-dependent cases, Eqs.~(\ref{Eq_Err:C_1}) and (\ref{Eq_Err:C_2}) are the origins of the commutator scaling for the time-dependent PF errors, although evaluating its error bound is not so simple.
The difficulty mainly comes from the facts that the error $e^{-iH^Ft}-\bar{T}^F(t)$ is defined on the infinite-dimensional space. 
We have to consider the infinite summation over $l \in \bbZ$ to reproduce the error in the original space as follows.

\begin{theorem}\label{Thm_Err:t_dep_error_Delta}
\textbf{}

The additive error of the time-dependent PF error is expressed by
\begin{eqnarray}
    && U(t,0) - \bar{S}(t,0) = \nonumber \\
    && \qquad -i \int_0^t \dd \tau U(t,\tau) \bar{S}(\tau,0) \left( \sum_{l \in \bbZ} \braket{l|\bar{\Delta}^F(\tau)|0}\right), \nonumber \\
    && \label{Eq_Err:t_dep_error_Delta}
\end{eqnarray}
where the operator $\bar{\Delta}^F(\tau)$ is given by Eqs. (\ref{Eq_Err:Delta_F_expansion})-(\ref{Eq_Err:C_2}).
\end{theorem}

\textbf{Proof.---} 
Using the error representation by Eq.~(\ref{Eq_Err:Error_repr_PF_Floquet}), the error of interest is computed as follows;
\begin{eqnarray}
    && U(t,0) - \bar{S}(t,0) \nonumber \\
    && \quad = -i \sum_{l \in \bbZ} e^{-il\omega t} \int_0^t \dd \tau \braket{l|e^{-iH^F(t-\tau)}\bar{T}^F(\tau)\bar{\Delta}^F(\tau)|0} \nonumber \\
    && \quad = -i \sum_{l,l_1,l_2 \in \bbZ} \int_0^t \dd \tau e^{-il\omega t} \braket{l|e^{-iH^F(t-\tau)}|l_1} \nonumber \\
    && \qquad \qquad  \qquad \qquad  \qquad \, \times \braket{l_1|\bar{T}^F(\tau)|l_2} \braket{l_2|\bar{\Delta}^F(\tau)|0} \nonumber \\
    && \quad  = -i \sum_{l,l_1,l_2 \in \bbZ} \int_0^t \dd \tau e^{-i(l-l_1)\omega t} \braket{l-l_1|e^{-iH^F(t-\tau)}|0} \nonumber \\
    && \qquad \qquad \times e^{-i(l_1-l_2)\omega \tau} \braket{l_1-l_2|\bar{T}^F(\tau)|0} \braket{l_2|\bar{\Delta}^F(\tau)|0}, \nonumber \\
    && \label{Eq_Err:t_dep_error_Delta_1}
\end{eqnarray}
where we use the translation symmetry [Eq.~(\ref{Eq_Pre:tr_sym_H_F})] and Lemma \ref{Lem_Rel:transl_sym_PF}] in the third equality.
The integrand of the last line forms a Cauchy product of the infinite series for $U(t,\tau)$ [Eq.~(\ref{Eq_Pre:Time_Evol_Floquet})], $\bar{S}(\tau,0)$ [Eq.~(\ref{Eq_Rel:Rel_PF_Std})], and $\sum_{l \in \bbZ} \braket{l|\bar{\Delta}^F(\tau)|0}$.
While the former two series are absolutely convergent, the last one $\sum_{l \in \bbZ} \braket{l|\bar{\Delta}^F(\tau)|0}$ is convergent as we will compute in Sections \ref{Subsec:form_commutators}-\ref{Subsec:explicit_error_bound}.
Then, we can use Mertens' theorem~\cite{Knopp1956-vi}, which implies the convergence of Eq.~(\ref{Eq_Err:t_dep_error_Delta_1}) to the right-hand side of Eq.~(\ref{Eq_Err:t_dep_error_Delta}). $\quad \square$

As a consequence, the operator norm of the PF error is bounded by
\begin{equation}\label{Eq_Err:t_dep_error_norm_Delta}
    \norm{U(t,0)-\bar{S}(t,0)} \leq \int_0^t \dd \tau \norm{\sum_{l \in \bbZ} \braket{l|\bar{\Delta}^F(\tau)|0}}
\end{equation}
for time-dependent cases instead of Eq.~(\ref{Eq_Err:t_indep_error_bound_Delta}), and this difference brings the difficulty of dealing with time dependence.
While each matrix element of $\Delta^F(\tau)$ is expected to be $\order{\tau^p}$ and to have a coefficient with the commutator scaling, it is nontrivial whether these properties are maintained after infinite summation over $l \in \bbZ$.
We also note that the time-dependent PF error is an infinite series composed of the nested commutators among $\{ H_\gamma^\mr{Add} \}_\gamma$ and $H^\mr{LP}$ via Eqs. (\ref{Eq_Err:C_1}) and (\ref{Eq_Err:C_2}).
It should be examined whether it provides a meaningful bounded value since the latter one is unbounded, and it is also of importance how it is related to nested commutators among $\{ H_\gamma (t) \}_\gamma$.
To solve these problems, we have to compute the explicit form of $\sum_{l \in \bbZ} \braket{l|\bar{\Delta}^F(\tau)|0}$.


\subsection{Origin of commutator scaling}\label{Subsec:form_commutators}

The next step for the explicit error bound is to compute the operator $\sum_{l \in \bbZ} \braket{l|\bar{\Delta}^F(\tau)|0}$ in detail.
In this section, we calculate nested commutators appearing in $\bar{\Delta}^F(\tau)$, which will be proven to be the origins of nested commutators among $\{ H_\gamma (t)\}_\gamma$ and their time derivatives in the error of time-dependent PFs.

Let us consider the following term appearing in Eqs. (\ref{Eq_Err:C_1}) and (\ref{Eq_Err:C_2}),
\begin{equation}\label{Eq_Err:Multi_com}
    \prod_{v''\tilde{\gamma}''=v'\tilde{\gamma}'}^{v\tilde{\gamma}} \left( \tilde{\alpha}_{v''\tilde{\gamma}''} \ad_{\tilde{H}_{v''\tilde{\gamma}''}}\right)^{q_{v''\tilde{\gamma}''}} \tilde{H},
\end{equation}
for $v\tilde{\gamma} \preceq v'\tilde{\gamma}'$, where $\tilde{H}$ is chosen from either $\{H_\gamma^\mr{Add} \}_\gamma$ or $-H^\mr{LP}$. 
Since each $\tilde{H}_{v''\tilde{\gamma}''}$ is chosen from the same set by Eq. (\ref{Eq_Err:H_tilde_def}), it is the nested commutator among $\{ H_\gamma^\mr{Add} \}_\gamma$ or $-H^\mr{LP}$.
We show that it can be related to nested commutators among $\{ H_\gamma (t)\}_\gamma$ and their time derivatives as follows.

\begin{lemma}\label{Lem_Err:Form_commutators}
\textbf{}

We define an operation $\bar{D}_{v\tilde{\gamma}}(t)$ by
\begin{equation}\label{Eq_Err:Dbar_v_gamma_def}
    \bar{D}_{v\tilde{\gamma}}(t) = \begin{cases}
        \ad_{H_{\pi_v(\gamma)}(t)} & (\tilde{\gamma}=2\gamma-1) \\
        i \dv{t} & (\tilde{\gamma} = 2\gamma)
    \end{cases},
\end{equation}
for $\gamma=1,\cdots,\Gamma$.
We also use $(f(t))_m$ to represent the Fourier coefficient of a time-dependent function $f(t)$, given by
\begin{equation}\label{Eq_Err:Fourier_coef}
    (f(t))_m = \frac1T \int_0^t \dd t^\prime f(t^\prime) e^{im\omega t^\prime}.
\end{equation}
The commutators with $\tilde{H}_{v\tilde{\gamma}}$ are given by
\begin{eqnarray}
    && \ad_{\tilde{H}_{v\tilde{\gamma}}} H_{\gamma'}^\mr{Add}  = \sum_{m \in \bbZ} \mr{Add}_m \otimes \left( \bar{D}_{v\tilde{\gamma}}(t) H_{\gamma'}(t)\right)_m, \label{Eq_Err:Com_gamma_Add} \\
    && \ad_{\tilde{H}_{v\tilde{\gamma}}} H^\mr{LP} \nonumber \\
    && \quad =
    \begin{cases}
        \displaystyle \sum_{m \in \bbZ} \mr{Add}_m \otimes \left( i \dv{t} H_{\pi_{v}(\gamma)}(t) \right)_m & (\tilde{\gamma}=2\gamma-1)\\
        \displaystyle 0 & (\text{otherwise})
    \end{cases}, \nonumber \\
    && \label{Eq_Err:Com_gamma_LP}
\end{eqnarray}
for $\gamma=1,\cdots,\Gamma$.
\end{lemma}

\textbf{Proof.---}
We first consider Eq. (\ref{Eq_Err:Com_gamma_Add}).
For an odd integer $\tilde{\gamma} = 2\gamma-1$, we have
\begin{eqnarray}
    \ad_{\tilde{H}_{v\tilde{\gamma}}} H_{\gamma'}^\mr{Add} &=& [H_{\pi_{v}(\gamma)}^\mr{Add},H_{\gamma'}^\mr{Add}] \nonumber \\
    &=& \sum_{m,m' \in \bbZ} \mr{Add}_{m+m'} \otimes [H_{\pi_{v}(\gamma)m},H_{\gamma' m'}] \nonumber \\
    &=& \sum_{m} \mr{Add}_m \otimes \sum_{m'} [H_{\pi_{v}(\gamma)(m-m')},H_{\gamma'm'}]. \nonumber \\
    &=& \sum_{m} \mr{Add}_m \otimes \left(\ad_{H_{\pi_{v}(\gamma)}(t)}H_{\gamma'}(t)\right)_m. 
\end{eqnarray}
The last equality is based on the fact that the Fourier coefficient of the commutator $[H_{\gamma}(t),H_{\gamma'}(t)]$ is expressed by
\begin{eqnarray}
    ([H_{\gamma}(t),H_{\gamma'}(t)])_m &=& \frac1T \int_0^T [H_{\gamma}(t),H_{\gamma'}(t)] e^{im\omega t} \dd t \nonumber \\
    &=& \sum_{m'} [H_{\gamma (m-m')}, H_{\gamma'm'}].
    \label{Eq_Err:Fourier_Com}
\end{eqnarray}
For an even integer $\tilde{\gamma}=2\gamma$, we have
\begin{eqnarray}
    \ad_{\tilde{H}_{v\tilde{\gamma}}} H_{\gamma'}^\mr{Add} &=&
    [-H^\mr{LP}, H_{\gamma'}^\mr{Add}] \nonumber \\
    &=& - \sum_m \sum_{l \in \bbZ} l \omega [\ket{l}\bra{l},\mr{Add}_m] \otimes H_{\gamma' m} \nonumber \nonumber \\
    &=& \sum_m \mr{Add}_m \otimes (m \omega H_{\gamma' m}) \nonumber \\
    &=& \sum_m \mr{Add}_m \otimes \left( i \dv{t} H_{\gamma'} (t) \right)_m. \label{Eq_Err:Com_LP_Add}
\end{eqnarray}
This completes the proof of Eq. (\ref{Eq_Err:Com_gamma_Add}).
It is easy to confirm Eq. (\ref{Eq_Err:Com_gamma_LP}).
We have $\ad_{\tilde{H}_{v\tilde{\gamma}}} H^\mr{LP} = - \ad_{H^\mr{LP}} H_{\pi_{v}(\gamma)}^\mr{Add}$ for odd $\tilde{\gamma}=2\gamma-1$, which is similar to Eq. (\ref{Eq_Err:Com_LP_Add}).
It is trivially $0$ for even $\tilde{\gamma}$ due to $\tilde{H}_{v\tilde{\gamma}}=-H^\mr{LP}$. $\quad \square$

This lemma indicates that the commutator with $H^\mr{Add}_\gamma$ in the Floquet-Hilbert space is replaced by the one with $H_\gamma(t)$, defined on the original space.
The commutator with $H^\mr{LP}$ is converted to the time derivative.
We next consider the expression of Eq. (\ref{Eq_Err:Multi_com}).
Lemma \ref{Lem_Err:Form_commutators} states that the application of $\ad_{\tilde{H}_{v\tilde{\gamma}}}$ to $H_\gamma^\mr{Add}=\sum_m \mr{Add}_m \otimes (H_\gamma(t))_m$ merely transforms its Fourier coefficient $(H_\gamma(t))_m$.
Thus, we can use this lemma again when further applying $\ad_{\tilde{H}_{v\tilde{\gamma}}}$.
Equation (\ref{Eq_Err:Delta_F_expansion}) for $\tilde{H}=H_\gamma^\mr{Add}$ is expressed by
\begin{eqnarray}
    && \prod_{v''\tilde{\gamma}''=v'\tilde{\gamma}'}^{v\tilde{\gamma}} \left( \tilde{\alpha}_{v''\tilde{\gamma}''} \ad_{\tilde{H}_{v''\tilde{\gamma}''}}\right)^{q_{v''\tilde{\gamma}''}} H_\gamma^\mr{Add} \nonumber \\
    && \qquad \qquad \qquad \qquad  = \sum_{m \in \bbZ} \mr{Add}_m \otimes \left( C_{v\tilde{\gamma},v'\tilde{\gamma}'}^{\{q\},\gamma}(t) \right)_m, \label{Eq_Err:Multi_Com_repr}\\
    && C_{v\tilde{\gamma},v'\tilde{\gamma}'}^{\{q\},\gamma}(t) = \prod_{v''\tilde{\gamma}''=v'\tilde{\gamma}'}^{v\tilde{\gamma}}\left( \tilde{\alpha}_{v''\tilde{\gamma}''} \bar{D}_{v''\tilde{\gamma}''}(t)\right)^{q_{v''\tilde{\gamma}''}} H_\gamma (t). \nonumber \\
    && \label{Eq_Err:Multi_Com_repr_time}
\end{eqnarray}
In the case of $\tilde{H}=-H^\mr{LP}$, let us define the maximal nontrivial index by
\begin{equation}\label{Eq_Err:max_index}
    (v_\mr{M},\tilde{\gamma}_\mr{M}) = \max \left[ (v'',\tilde{\gamma}'') \text{ s.t. } q_{v''\tilde{\gamma}''} \neq 0 \right].
\end{equation}
Using Eq. (\ref{Eq_Err:Com_gamma_LP}), we arrive at
\begin{eqnarray}
    && \prod_{v''\tilde{\gamma}''=v'\tilde{\gamma}'}^{v\tilde{\gamma}} \left( \tilde{\alpha}_{v''\tilde{\gamma}''} \ad_{\tilde{H}_{v''\tilde{\gamma}''}}\right)^{q_{v''\tilde{\gamma}''}} (-H^\mr{LP}) \nonumber \\
   && \qquad \qquad \qquad = \sum_{m \in \bbZ} \mr{Add}_m \otimes \left( C_{v\tilde{\gamma},v'\tilde{\gamma}'}^{\{q\},\mr{LP}}(t)\right)_m,
\end{eqnarray}
where the operator $C_{v\tilde{\gamma},v'\tilde{\gamma}'}^{\{q\},\mr{LP}}(t)$ is defined by
\begin{eqnarray}
    && C_{v\tilde{\gamma},v'\tilde{\gamma}'}^{\{q\},\mr{LP}}(t) \nonumber \\
    && \quad = \prod_{v''\tilde{\gamma}''=v_\mr{M}(\tilde{\gamma}_\mr{M}+1)}^{v\tilde{\gamma}}\left( \tilde{\alpha}_{v''\tilde{\gamma}''} \bar{D}_{v''\tilde{\gamma}''}(t)\right)^{q_{v''\tilde{\gamma}''}} \nonumber \\
    && \qquad \times \left( \tilde{\alpha}_{v_\mr{M}\tilde{\gamma}_\mr{M}} \bar{D}_{v_\mr{M}\tilde{\gamma}_\mr{M}}(t)\right)^{q_{v_\mr{M}\tilde{\gamma}_\mr{M}}-1}\left(-i \dv{t}H_{\gamma_\mr{M}}(t)\right) \nonumber \\
    &&
\end{eqnarray}
for an odd integer $\tilde{\gamma}_\mr{M}=2\gamma_\mr{M}-1$ and $C_{v\tilde{\gamma},v'\tilde{\gamma}'}^{\{q\},\mr{LP}}(t)=0$ for an even integer $\tilde{\gamma}_\mr{M}=2\gamma_\mr{M}$.

We next show that the operators $C_{v\tilde{\gamma},v'\tilde{\gamma}'}^{\{q\},\gamma}(t)$ and $C_{v\tilde{\gamma},v'\tilde{\gamma}'}^{\{q\},\mr{LP}}(t)$ directly appear the error bounds of the time-dependent PFs.
Namely, they are the origins of the commutator scaling involving the nested commutators $\{H_\gamma(t)\}$ and their time derivative.
First, we prove the following lemma for calculation.

\begin{lemma}\label{Lem_Err:truncated}
\textbf{}

Let $\bar{S}_{\prec v\gamma}(t,0)$ be the partial standard PF defined by
\begin{equation}\label{Eq_Err:truncated_Standard_PF}
    \bar{S}_{\prec v\gamma}(t,0) = \prod_{v'\gamma'=11}^{v(\gamma-1)} e^{-i H_{\pi_{v'}(\gamma')}(\beta_{v'\gamma'}t) \alpha_{v'\gamma'}t}.
\end{equation}
When the time $\tau_1$ is in $[0,\tau]$, there exists certain time $\tau_{v\tilde{\gamma}}(\tau,\tau_1) \in [0,\tau]$ such that 
\begin{eqnarray}
    && \sum_{l \in \bbZ} e^{-il \omega \tau_{v\tilde{\gamma}}(\tau,\tau_1)} \braket{l|e^{-i \tilde{H}_{v\tilde{\gamma}} \tilde{\alpha}_{v\tilde{\gamma}} \tau_1} \bar{T}^F_{\prec v \tilde{\gamma}}(\tau)|0} \nonumber \\
    && \quad = \begin{cases}
        e^{-iH_{\pi_v(\gamma)}(\beta_{v\gamma}\tau) \alpha_{v\gamma} \tau_1} \bar{S}_{\prec v\tilde{\gamma}}(\tau,0) & (\tilde{\gamma}=2\gamma-1) \\
        \bar{S}_{\prec v \tilde{\gamma}} (\tau,0) & (\tilde{\gamma}=2\gamma)
    \end{cases}
    \nonumber \\
    && \label{Eq_Err:truncatd_PF_relation}
\end{eqnarray}
is satisfied.
In addition, it has the translation symmetry, 
\begin{eqnarray}
    &&\braket{l-l^{\prime\prime}|e^{-i \tilde{H}_{v\tilde{\gamma}} \tilde{\alpha}_{v\tilde{\gamma}} \tau_1} \bar{T}^F_{\prec v \tilde{\gamma}}(\tau)|l^\prime-l^{\prime \prime}} \nonumber \\
    && \qquad = e^{il^{\prime \prime} \omega \tau_{v\tilde{\gamma}}(\tau,\tau_1)} \braket{l|e^{-i \tilde{H}_{v\tilde{\gamma}} \tilde{\alpha}_{v\tilde{\gamma}} \tau_1} \bar{T}^F_{\prec v \tilde{\gamma}}(\tau)|l^\prime},
    \label{Eq_Err:truncated_PF_symmetry}
\end{eqnarray}
satisfied for every $l,l',l'' \in \bbZ$.
\end{lemma}

\textbf{Proof.---}
We consider an odd integer $\tilde{\gamma}=2\gamma-1$.
In a similar manner to the proof of Theorem \ref{Thm_Rel:Relation}, we obtain the relation,
\begin{equation}
    \bar{S}_{\prec v\gamma}(\tau,0) = \sum_{l \in \bbZ} e^{-il\omega \beta_{v\gamma}\tau}\braket{l|\bar{T}^F_{\prec v\tilde{\gamma}}(\tau)|0}.
\end{equation}
We set the time $\tau_{v\tilde{\gamma}}(\tau,\tau_1)=\beta_{v\gamma}\tau \in [0,\tau]$.
Using the translation symmetry by Eq. (\ref{Eq_Pre:tr_sym_H_Add}) and the relation Eq. (\ref{Eq_Rel:TE_Floquet_inst}) results in
\begin{eqnarray}
    && \sum_{l \in \bbZ} e^{-il \omega \tau_{v\tilde{\gamma}}(\tau,\tau_1)} \braket{l|e^{-i \tilde{H}_{v\tilde{\gamma}} \tilde{\alpha}_{v\tilde{\gamma}} \tau_1} \bar{T}^F_{\prec v \tilde{\gamma}}(\tau)|0} \nonumber \\
    && \quad = \sum_{l,l_1 \in \bbZ} e^{-il\omega \beta_{v\gamma} \tau} \braket{l-l_1|e^{-iH_{\pi_v(\gamma)}^\mr{Add} \alpha_{v\gamma}\tau_1}|0}\braket{l_1|\bar{T}^F_{\prec v\tilde{\gamma}}(\tau)|0} \nonumber \\
    && \quad = e^{-iH_{\pi_v(\gamma)}(\beta_{v\gamma}\tau) \alpha_{v\gamma} \tau_1} \bar{S}_{\prec v\tilde{\gamma}}(\tau,0).
\end{eqnarray}
We next consider the case where $\tilde{\gamma}$ is an even integer by $\tilde{\gamma}=2\gamma$.
We set time $\tau_{v\tilde{\gamma}}(\tau,\tau_1) = \beta_{v\gamma}\tau+(\beta_{v(\gamma+1)}-\beta_{v\gamma})\tau_1$, which lies between $\beta_{v\gamma}\tau$ and $\beta_{v(\gamma+1)}\tau$, and hence it is included in $[0,\tau]$.
The same calculation leads to the satisfaction of Eq. (\ref{Eq_Err:truncatd_PF_relation}).
The symmetry by Eq. (\ref{Eq_Err:truncated_PF_symmetry}) immediately follows from Lemma \ref{Lem_Rel:transl_sym_PF}. $\quad \square$

As discussed in Eq. (\ref{Eq_Err:t_dep_error_norm_Delta}), the error bound is determined by $\sum_{l \in \bbZ} \braket{l|\bar{\Delta}^F(\tau)|0}$.
Using the explicit formula of  $\bar{\Delta}^F(\tau)$ by Eq. (\ref{Eq_Err:Delta_F_expansion}), it is further represented by
\begin{eqnarray}
    && \sum_{l \in \bbZ} \braket{l|C_1(H^F)|0} \nonumber \\
    && \quad = \sum_{\gamma'=1}^\Gamma \sum_{l \in \bbZ} \braket{l|C_1(H_{\gamma'}^\mr{Add})|0} - \sum_{l \in \bbZ} \braket{l|C_1(H^\mr{LP})|0}, \nonumber \\
    && \label{Eq_Err:C_1_decomp}
\end{eqnarray}
and also by $\sum_{l\in \bbZ} \braket{l|C_2(\tilde{H}_{v'\tilde{\gamma}'})|0}$ similarly.
The first term can be calculated as follows.
For an odd integer $\tilde{\gamma}=2\gamma-1$, we have
\begin{widetext}
\begin{eqnarray}
    \sum_{l \in \bbZ} \braket{l|C_1(H_{\gamma'}^\mr{Add})|0} &=&  \sum_{l,l_1 \in \bbZ} \bra{l} \bar{T}^F_{\prec v \tilde{\gamma}}(\tau)^\dagger e^{i \tilde{H}_{v\tilde{\gamma}} \tilde{\alpha}_{v\tilde{\gamma}} \tau_1} \left[\sum_m \ket{l_1+m}\bra{l_1} \otimes \left( C_{v\tilde{\gamma},V_p2\Gamma}^{\{q\},\gamma'}(t)\right)_m \right] e^{-i \tilde{H}_{v\tilde{\gamma}} \tilde{\alpha}_{v\tilde{\gamma}} \tau_1} \bar{T}^F_{\prec v \tilde{\gamma}}(\tau) \ket{0} \nonumber \\
    &=& \sum_{l,l_1,m} \bra{l-m}\bar{T}^F_{\prec v \tilde{\gamma}}(\tau)^\dagger e^{i \tilde{H}_{v\tilde{\gamma}} \tilde{\alpha}_{v\tilde{\gamma}} \tau_1} \ket{l_1} \left( C_{v\tilde{\gamma},V_p 2\Gamma}^{\{q\},\gamma'}(t)\right)_m e^{-im\omega \tau_{v\tilde{\gamma}} (\tau,\tau_1)} \bra{l_1}  e^{-i \tilde{H}_{v\tilde{\gamma}} \tilde{\alpha}_{v\tilde{\gamma}} \tau_1} \bar{T}^F_{\prec v \tilde{\gamma}}(\tau) \ket{0}\nonumber \\
    &=& \sum_{l,l_1 \in \bbZ} e^{-il \omega \tau_{v\tilde{\gamma}} (\tau,\tau_1)} \bra{0} \bar{T}^F_{\prec v \tilde{\gamma}}(\tau)^\dagger e^{i \tilde{H}_{v\tilde{\gamma}} \tilde{\alpha}_{v\tilde{\gamma}} \tau_1}  \ket{l_1-l} C_{v\tilde{\gamma},V_p 2\Gamma}^{\{q\},\gamma'}(\tau_{v\tilde{\gamma}} (\tau,\tau_1)) \bra{l_1}  e^{-i \tilde{H}_{v\tilde{\gamma}} \tilde{\alpha}_{v\tilde{\gamma}} \tau_1} \bar{T}^F_{\prec v \tilde{\gamma}}(\tau) \ket{0} \nonumber \\
    &=& \bar{S}_{\prec v\tilde{\gamma}}(\tau,0)^\dagger e^{iH_{\pi_v(\gamma)}(\beta_{v\gamma}\tau) \alpha_{v\gamma} \tau_1}  C_{v\tilde{\gamma},V_p 2\Gamma}^{\{q\},\gamma'}(\tau_{v\tilde{\gamma}} (\tau,\tau_1)) e^{-iH_{\pi_v(\gamma)}(\beta_{v\gamma}\tau) \alpha_{v\gamma} \tau_1}  \bar{S}_{\prec v\tilde{\gamma}}(\tau,0). \label{Eq_Err:Proof_Thm_1v}
\end{eqnarray} 
\end{widetext}
The first equality comes from Eq. (\ref{Eq_Err:Multi_Com_repr}).
We use Lemma \ref{Lem_Err:truncated} for the third and last equalities.
As a result, the norm of Eq. (\ref{Eq_Err:Proof_Thm_1v}) is bounded by
\begin{equation}
    \norm{\sum_{l \in \bbZ} \braket{l|C_1(H_{\gamma'}^\mr{Add})|0}} \leq \norm{C_{v\tilde{\gamma},V_p 2\Gamma}^{\{q\},\gamma'}(\tau_{v\tilde{\gamma}} (\tau,\tau_1))},
\end{equation}
which is valid also for even $\tilde{\gamma}$ due to Lemma \ref{Lem_Err:truncated}.
We also obtain
\begin{equation}\label{Eq_Err:C_1_LP_bound}
    \norm{\sum_{l \in \bbZ} \braket{l|C_1(H^\mr{LP})|0}} \leq \norm{C_{v\tilde{\gamma},V_p 2\Gamma}^{\{q\},\mr{LP}}(\tau_{v\tilde{\gamma}} (\tau,\tau_1))},
\end{equation}
and the similar relations hold for the norm of $\sum_{l\in \bbZ} \braket{l|C_2(\tilde{H}_{v'\tilde{\gamma}'})|0}$ as well.

To summarize, all the terms included in the norm of $\sum_{l \in \bbZ}\braket{l|\bar{\Delta}^F(\tau)|0}$, which determines the error bound, can be expressed by those of $C_{v\tilde{\gamma},v'\tilde{\gamma}'}^{\{q\},\gamma'}(\tau_{v\tilde{\gamma}} (\tau,\tau_1))$ and $C_{v\tilde{\gamma},v'\tilde{\gamma}'}^{\{q\},\mr{LP}}(\tau_{v\tilde{\gamma}} (\tau,\tau_1))$.
This is why the nested commutators among $H_\gamma(t)$ and their time derivative appear in the error bound of the time-dependent PFs.
We remark that the time $\tau_{v\tilde{\gamma}} (\tau,\tau_1)$ is always included in $[0,t]$.
We extrapolate the Hamiltonian $H(t)$ for $\bbR \backslash [0,t]$ as discussed in Section \ref{Subsec:Setup}, and its Floquet Hamiltonian $H^F$ contains the information for all the time including the extrapolated domain $\bbR \backslash [0,t]$, which is irrelevant for the time evolution $U(t,0)$.
The fact $\tau_{v\tilde{\gamma}} (\tau,\tau_1) \in [0,t]$ indicates that the nested commutators and their time derivatives only during $[0,t]$ determine the PF errors independent of the extrapolation.

\subsection{Explicit error bound}\label{Subsec:explicit_error_bound}

Now, we are ready to derive the concrete error bound for generic time-dependent PFs by summarizing the results so far.
The error bound of the standard PF $\bar{S}(t,0)$ is given by the following theorem.

\begin{theorem}\label{Thm_Err:Main_theorem}
\textbf{(Error of the standard PF)}

Suppose that the coefficients $\{\alpha_{v\gamma},\beta_{v\gamma}\}$ are chosen so that the time-independent PF $\bar{T}(t)$ by Eq. (\ref{Eq_Err:PF_equiv}) can satisfy the order condition $\bar{T}(t)=e^{-iHt}+\order{t^{p+1}}$.
The standard PF error has an upper bound by
\begin{equation}\label{Eq_Err:PF_bound_theorem}
    \norm{U(t,0)-\bar{S}(t,0)} \leq 4 (V_p t)^{p+1} \max_{\tau \in [0,t]} \bar{\alpha}_{\mr{com},p}(\tau), 
\end{equation}
where the quantity $\bar{\alpha}_{\mr{com},p}(\tau)$ is defined by Eq. (\ref{Eq_Sum:alpha_com_Std}).
\end{theorem}

\textbf{Proof.---}
Combining the discussion in Eqs. (\ref{Eq_Err:C_1_decomp})-(\ref{Eq_Err:C_1_LP_bound}) with the expression of $\bar{\Delta}^F(\tau)$ by Eq. (\ref{Eq_Err:Delta_F_expansion}), we have
\begin{widetext}
\begin{eqnarray}
\norm{\sum_{l \in \bbZ} \braket{l|\bar{\Delta}^F(\tau)|0}} &\leq&  \tau^{p-1} \int_0^\tau \dd \tau_1 \sum_{v=1}^{V_p} \sum_{\tilde{\gamma}=1}^{2\Gamma}
    \sum_{\substack{q_{v\tilde{\gamma}}, \cdots,q_{V_p 2\Gamma} \geq 0: \\ q_{v\tilde{\gamma}} + \cdots + q_{V_p 2\Gamma} = p,
    \\ q_{v\tilde{\gamma}} \neq 0}} \left( \sum_{\gamma'=1}^\Gamma \norm{C_{v\tilde{\gamma},V_p 2\Gamma}^{\{q\},\gamma'}(\tau_{v\tilde{\gamma}} (\tau,\tau_1))} + \norm{C_{v\tilde{\gamma},V_p 2\Gamma}^{\{q\},\mr{LP}}(\tau_{v\tilde{\gamma}} (\tau,\tau_1))}\right) \nonumber \\
&& \qquad + \tau^{p-1} \int_0^\tau \dd \tau_1 \sum_{v'=1}^{V_p} \sum_{\tilde{\gamma}':\text{odd}} \sum_{v\tilde{\gamma} \preceq v'\tilde{\gamma}'} 
    \sum_{\substack{q_{v\tilde{\gamma}}, \cdots,q_{v'\tilde{\gamma}'} \geq 0: \\ q_{v\tilde{\gamma}} + \cdots + q_{v'\tilde{\gamma}'} = p,
    \\ q_{v\tilde{\gamma}} \neq 0}}   \norm{C_{v\tilde{\gamma},v'\tilde{\gamma}'}^{\{q\},\frac{\tilde{\gamma}'+1}2}(\tau_{v\tilde{\gamma}} (\tau,\tau_1))} \nonumber \\
&& \qquad \qquad + \tau^{p-1}\int_0^\tau \dd \tau_1 \sum_{v'=1}^{V_p} \sum_{\tilde{\gamma}':\text{even}} \sum_{v\tilde{\gamma} \preceq v'\tilde{\gamma}'} 
    \sum_{\substack{q_{v\tilde{\gamma}}, \cdots,q_{v'\tilde{\gamma}'} \geq 0: \\ q_{v\tilde{\gamma}} + \cdots + q_{v'\tilde{\gamma}'} = p,
    \\ q_{v\tilde{\gamma}} \neq 0}}  \norm{C_{v\tilde{\gamma},v'\tilde{\gamma}'}^{\{q\},\mr{LP}}(\tau_{v\tilde{\gamma}} (\tau,\tau_1))}. \label{Eq_Err:Delta_F_bound_start}
\end{eqnarray}
\end{widetext}
Owing to $\tau_{v\tilde{\gamma}}(\tau,\tau_1) \in [0,\tau]$, the integral in the above equation can be replaced by
\begin{equation}
    \tau^{p-1} \int_0^\tau \dd \tau_1 f(\tau_{v\tilde{\gamma}}(\tau,\tau_1)) \leq \tau^p \max_{\tau_1 \in [0,\tau]} f(\tau_1),
\end{equation}
where $f(\tau_1)$ denotes the summations appearing in the integrand.

We first focus on the summation of $C_{v\tilde{\gamma},V_p2\Gamma}^{\{q\},\gamma'}(\tau_1)$, which appears in the first term of Eq. (\ref{Eq_Err:Delta_F_bound_start}).
Since the indices $q_{v\tilde{\gamma}},\cdots,q_{V_p 2\Gamma}$ satisfy the constraint $q_{v\tilde{\gamma}}+\cdots+q_{V_p 2\Gamma}=p$, each of $C_{v\tilde{\gamma},V_p2\Gamma}^{\{q\},\gamma'}(\tau_1)$ by Eq. (\ref{Eq_Err:Multi_Com_repr_time}) has the form
\begin{equation}
    \left[ \prod_{q=1}^p \tilde{\alpha}_{v_q \tilde{\gamma}_q} \bar{D}_{v_q \tilde{\gamma}_q}(\tau_1)  \right] H_{\gamma'} (\tau_1)
\end{equation}
with some indices $\{v_q\tilde{\gamma}_q\}_{q=1}^p$.
Since all the terms appearing in the summation over the indices $q_{v\tilde{\gamma}}, \cdots, q_{V_p 2\Gamma}$ are covered by the one over all the possible indices $\{v_q\tilde{\gamma}_q\}$, we obtain
\begin{eqnarray}
    && \sum_{v=1}^{V_p} \sum_{\tilde{\gamma}=1}^{2\Gamma}
    \sum_{\substack{q_{v\tilde{\gamma}}, \cdots,q_{V_p 2\Gamma} \geq 0: \\ q_{v\tilde{\gamma}} + \cdots + q_{V_p 2\Gamma} = p,
    \\ q_{v\tilde{\gamma}} \neq 0}} \norm{C_{v\tilde{\gamma},V_p2\Gamma}^{\{q\},\gamma'}(\tau_1)} \nonumber \\
    && \quad = \sum_{\substack{q_{11}, \cdots,q_{V_p 2\Gamma} \geq 0: \\ q_{11} + \cdots + q_{V_p 2\Gamma} = p}} \norm{C_{11,V_p2\Gamma}^{\{q\},\gamma'}(\tau_1)}  \nonumber \\
    && \quad \leq \sum_{v_1,\cdots,v_p=1}^{V_p}\sum_{\tilde{\gamma}_1,\cdots,\tilde{\gamma}_p=1}^{2\Gamma} \norm{\left[\prod_{q=1}^p  \bar{D}_{v_q\tilde{\gamma}_q}(\tau_1) \right] H_{\gamma'}(\tau_1)}, \nonumber \\
    &&
\end{eqnarray}
where we use $|\tilde{\alpha}_{v\tilde{\gamma}}| \leq 1$ as discussed in Eq. (\ref{Eq_Err:H_tilde_def}).
Since the $\Gamma$ copies of the derivative $\dv{\tau_1}$ by even $\tilde{\gamma}$ can be replaced by $\bar{\mcl{D}}_{\Gamma+1}(\tau_1)=\Gamma \dv{\tau_1}$, we obtain
\begin{eqnarray}
    && \sum_{v=1}^{V_p} \sum_{\tilde{\gamma}=1}^{2\Gamma}
    \sum_{\substack{q_{v\tilde{\gamma}}, \cdots,q_{V_p 2\Gamma} \geq 0: \\ q_{v\tilde{\gamma}} + \cdots + q_{V_p 2\Gamma} = p,
    \\ q_{v\tilde{\gamma}} \neq 0}} \sum_{\gamma'=1}^\Gamma \norm{C_{v\tilde{\gamma},V_p2\Gamma}^{\{q\},\gamma'}(\tau_1)} \nonumber \\
    && \quad \leq (V_p)^p \bar{\alpha}_{\mr{com},p} (\tau_1),
\end{eqnarray}
according to the definition, Eq. (\ref{Eq_Sum:alpha_com_Std}).
The same goes also for the summation of $C_{v\tilde{\gamma},v'\tilde{\gamma}'}^{\{q\},\frac{\tilde{\gamma}'+1}2}(\tau_1)$, which results in
\begin{eqnarray}
    && \sum_{v'=1}^{V_p} \sum_{\tilde{\gamma}':\text{odd}}\sum_{\substack{v\tilde{\gamma} \\ \, \preceq v'\tilde{\gamma}'}} 
    \sum_{\substack{q_{v\tilde{\gamma}}, \cdots,q_{v'\tilde{\gamma}'} \geq 0: \\ q_{v\tilde{\gamma}} + \cdots + q_{v'\tilde{\gamma}'} = p,
    \\ q_{v\tilde{\gamma}} \neq 0}} \norm{C_{v\tilde{\gamma},v'\tilde{\gamma}'}^{\{q\},\frac{\tilde{\gamma}'+1}2}(\tau_1)} \nonumber \\
    && \quad \leq (V_p)^{p+1} \bar{\alpha}_{\mr{com},p} (\tau_1).
\end{eqnarray}
The additional factor $V_p$ comes from the summation over $v'$ from $1$ to $V_p$.

The quantity $C_{v\tilde{\gamma},V_p 2\Gamma}^{\{q\},\mr{LP}}(\tau_1)$ in Eq. (\ref{Eq_Err:Delta_F_bound_start}) is evaluated in a similar manner, whose summation over the indices $q_{v\tilde{\gamma}}, \cdots, q_{V_p 2\Gamma}$ can be replaced by the one over $\{v_q\tilde{\gamma}_q\}_{q=2}^{p}$ as well.
While the last index is fixed to $\gamma_\mr{M}$ defined by Eq. (\ref{Eq_Err:max_index}), we can delete it by using a trivial relation,
\begin{equation}
    \norm{(\cdots) \dv{t}H_{\gamma_\mr{M}}(\tau_1)} \leq \sum_{\gamma_0=1}^\Gamma \norm{(\cdots) \dv{t}H_{\gamma_0}(\tau_1)}.
\end{equation}
As a result, the summation of $C_{v\tilde{\gamma},V_p 2\Gamma}^{\{q\},\mr{LP}}(\tau_1)$ appearing in the first term of Eq. (\ref{Eq_Err:Delta_F_bound_start}) is bounded by
\begin{eqnarray}
    && \sum_{v=1}^{V_p} \sum_{\tilde{\gamma}=1}^{2\Gamma}
    \sum_{\substack{q_{v\tilde{\gamma}}, \cdots,q_{V_p 2\Gamma} \geq 0: \\ q_{v\tilde{\gamma}} + \cdots + q_{V_p 2\Gamma} = p,
    \\ q_{v\tilde{\gamma}} \neq 0}} \norm{C_{v\tilde{\gamma},V_p 2\Gamma}^{\{q\},\mr{LP}}(\tau_1)} \nonumber \\
    && \quad \leq \sum_{v_2\tilde{\gamma}_2,\cdots,v_p\tilde{\gamma}_p=11}^{V_p2\Gamma}\sum_{\gamma_0=1}^\Gamma \norm{\left[\prod_{q=2}^p  \bar{D}_{v_q\tilde{\gamma}_q}(\tau_1) \right] \dv{t} H_{\gamma_0}(\tau_1)}, \nonumber \\
    && \quad \leq \frac{(V_p)^{p-1}}{\Gamma} \bar{\alpha}_{\mr{com},p}(\tau_1). \label{Eq_Err:sum_LP_a}
\end{eqnarray}
In the second inequality, we use $\dv{\tau} = \Gamma^{-1} \bar{\mcl{D}}_{\Gamma+1}(\tau)$ and the fact that it is covered by the summation of $\bar{\mcl{D}}_{\gamma_1}(\tau)$ over $\gamma_1$ from $1$ to $\Gamma+1$.
Similarly, the summation of $C_{v\tilde{\gamma},v'\tilde{\gamma}'}^{\{q\},\mr{LP}}(\tau_1)$ is bounded by
\begin{eqnarray}
    && \sum_{v'=1}^{V_p} \sum_{\tilde{\gamma}':\text{even}}\sum_{v\tilde{\gamma} = 11}^{v'\tilde{\gamma}'}
    \sum_{\substack{q_{v\tilde{\gamma}}, \cdots,q_{v'2\gamma'} \geq 0: \\ q_{v\tilde{\gamma}} + \cdots + q_{v'2\gamma'} = p,
    \\ q_{v\tilde{\gamma}} \neq 0}} \norm{C_{v\tilde{\gamma},v'\tilde{\gamma}'}^{\{q\},\mr{LP}}(\tau_1)} \nonumber \\
    && \quad \leq (V_p)^p \bar{\alpha}_{\mr{com},p}(\tau_1), \label{Eq_Err:sum_LP_b}
\end{eqnarray}
where the additional factor $V_p\Gamma$ compared to Eq. (\ref{Eq_Err:sum_LP_a}) comes from the summation over $v'$ and even $\tilde{\gamma}'$.

Finally, summarizing the relations, Eqs. (\ref{Eq_Err:Delta_F_bound_start})-(\ref{Eq_Err:sum_LP_b}), we obtain
\begin{equation}
    \norm{\sum_{l \in \bbZ} \braket{l|\bar{\Delta}^F(\tau)|0}} \leq 4 (V_p \tau)^{p+1} \max_{\tau_1 \in [0,\tau]} \bar{\alpha}_{\mr{com},p}(\tau_1).
\end{equation}
Equation (\ref{Eq_Err:t_dep_error_norm_Delta}) immediately leads to Eq. (\ref{Eq_Err:PF_bound_theorem}), which completes the proof.
$\quad \square$

Theorem \ref{Thm_Err:Main_theorem} is the central result of this paper, which gives the time-dependent PF error explicitly composed of the commutators among $\{ H_\gamma (t) \}$ and their time-derivatives.
Similarly, the error bound of the generalized PF $S(t,0)$ is given by the following theorem, which completes Theorem \ref{Thm:main_theorem_informal} in Section \ref{Sec:Summary} with Theorem \ref{Thm_Err:Main_theorem}.

\begin{theorem}\label{Thm_Err:main_thm_Gen}
\textbf{(Error of the generalized PF)}

Suppose that the coefficients $\{\alpha_{v\gamma}\}$ and $\{ \beta_{v\gamma} \}$ are chosen so that the time-independent PF,
\begin{equation}\label{Eq_Err:T_F_ctrpart_Gen}
    T(t) = \prod_{v=1}^{V_p}  \prod_{\gamma=1}^\Gamma \left( e^{iH_0 (\beta_{v(\gamma+1)}-\beta_{v\gamma}-\alpha_{v\gamma})t} e^{-iH_{\pi_v(\gamma)} \alpha_{v\gamma}t} \right)
\end{equation}
satisfies the order condition $T(t)=e^{-iHt}+\order{t^{p+1}}$ for $H=\sum_{\gamma=1}^\Gamma H_\gamma + (\Gamma-1)H_0$.
The generalized PF error has an upper bound by
\begin{equation}
    \norm{U(t,0)-S(t,0)} \leq 5 (V_pt)^{p+1} \max_{\tau \in [0,t]} \alpha_{\mr{com},p}(\tau),
\end{equation}
where the quantity $\alpha_{\mr{com},p}(\tau)$ is defined by Eq. (\ref{Eq_Sum:alpha_com_Gen}).
\end{theorem}

We provide the detailed proof in Appendix \ref{A_Sec:Generalized_PF}.
The difference from the standard PF is the usage of $T^F(t)$ by Eq. (\ref{Eq_Rel:T_F_Gen}) rather than $\bar{T}^F(t)$.
In terms of the order condition, we demand that $T^F(t)$ becomes a $p$th-order PF of $e^{-iH^Ft}$ with the Floquet Hamiltonian $H^F = \sum_{\gamma=1}^\Gamma H_\gamma^F + (\Gamma-1)H^\mr{LP}$ as shown in Eq. (\ref{Eq_Rel:H_F_decomp_Gen}).
Associating $H_0$ and $H_\gamma$ respectively with $H^\mr{LP}$ and $H_\gamma^F$, we arrive at the order condition of Eq. (\ref{Eq_Err:T_F_ctrpart_Gen}) for $H=\sum_{\gamma=1}^\Gamma H_\gamma + (\Gamma-1)H_0$.
In terms of the error bound, we observe commutators with $H_\gamma^F=H_\gamma^\mr{Add}-H^\mr{LP}$ instead of those with $H_\gamma^\mr{Add}$.
Remembering that $\ad_{H_\gamma^\mr{Add}}$ and $\ad_{H^\mr{LP}}$ respectively correspond to $\ad_{H_\gamma(t)}$ and $-i\dv{t}$ in the original space, this is why the operator $\bar{\mcl{D}}_{\gamma}(\tau)$ is replaced by $\mcl{D}_\gamma(t)=\ad_{H_\gamma(t)}+i\dv{t}$ for each $\gamma=1,\cdots,\Gamma$.

We also consider the standard PF $\bar{S}_p(t,0)$ with the uniform coefficients as Definition \ref{Def_Pre:uniform}.
We can obtain a better error bound in that case as a corollary of Theorem \ref{Thm_Err:Main_theorem}.
Since we have $\beta_{v\gamma}=\beta_v$, the coefficient $\tilde{\alpha}_{v\tilde{\gamma}}$ in Eq. (\ref{Eq_Err:H_tilde_def}) for even $\tilde{\gamma}$ is equal to $0$ except for $\tilde{\gamma}=2\Gamma$.
Namely, the $\Gamma$ copies of the derivative $\dv{\tau}$ included in the summation over even integers $\tilde{\gamma}$ is replaced by a single copy.
This results in the following error bound in the uniform case.

\begin{corollary}\label{Cor_Err:error_uniform}
\textbf{(Error of the uniform standard PF)}

Let $\bar{S}_p(t,0)$ be the $p$th-order standard PF having the uniform coefficients as Definition \ref{Def_Pre:uniform}.
Its error is bounded by
\begin{equation}
    \norm{U(t,0)-\bar{S}_p(t,0)} \leq 4 (V_pt)^{p+1} \max_{\tau \in [0,t]} \bar{\alpha}_{\mr{com},p}^\mr{uni}(\tau).
\end{equation}
The quantity $\bar{\alpha}_{\mr{com},p}^\mr{uni}(\tau)$ is defined by
\begin{eqnarray}
    && \bar{\alpha}_{\mr{com},p}^\mr{uni}(\tau) \nonumber \\
    && \quad = \sum_{v_1,\cdots,v_p=1}^{V_p} \sum_{\gamma_1,\cdots,\gamma_p=1}^{\Gamma+1} \sum_{\gamma_0=1}^\Gamma \norm{\prod_{q=1}^p \mcl{\bar{D}}_{\gamma_q}^\mr{uni}(\tau) H_{\gamma_0}(\tau)}, \nonumber \\
    && \label{Eq_Err:alpha_com_uni}
\end{eqnarray}
where the operator $\mcl{\bar{D}}_{\gamma}^\mr{uni}(\tau)$ is given by $\mcl{\bar{D}}_{\gamma}^\mr{uni}(\tau) = \ad_{H_\gamma(\tau)}$ for $\gamma=1,\cdots,\Gamma$ and $\mcl{\bar{D}}_{\Gamma+1}^\mr{uni}(\tau)=\dv{\tau}$.
\end{corollary}

In contrast to Theorems \ref{Thm_Err:Main_theorem} and \ref{Thm_Err:main_thm_Gen}, the strength of the above error bound is directly independent of the number of terms $\Gamma$.
As discussed in Section \ref{Subsec:time_dep}, it can be applied to the common examples of the standard PFs such as the Lie-Suzuki-Trotter formula and the Forest-Ruth-Suzuki formula.
The independence from $\Gamma$ is important when considering long-range interactions, which will be discussed in Section \ref{Subsec:gate_counts}.

Finally, we emphasize that better error bounds for time-independent PFs lead to better error bounds for time-dependent PFs.
We employ the time-independent PF error by Ref. \cite{childs2021-trotter} so that the error bounds for generic time-dependent PFs can be obtained, which are expected to be loose.
On the other hand, we can exploit any type of time-independent PF errors for evaluating $e^{-iH^Ft}-\bar{T}^F(t)$ or $e^{-iH^Ft}-T^F(t)$ in Section \ref{Subsec:Error_repr}. 
If we know better time-independent PFs, we can improve the error bounds for the time-dependent PFs.
For instance, since a smaller coefficient of the error bounds is derived for the time-independent Lie-Suzuki-Trotter formulas $T_p(t)$ up to $p=4$ \cite{childs2021-trotter}, the same goes also for the time-dependent cases.

\subsection{Comparison with previous error bounds}

There have been various attempts to obtain error bounds of time-dependent PFs so far, and we compare our results with them.
For low-order PFs, the explicit formulas for the errors have been known up to the second order \cite{Huyghebaert1990-zb-tpf,An2021-tpf}, as far as we know.
For instance, the first order Lie-Suzuki-Trotter formula has an error bounded by Eq.~(\ref{Eq_Pre:LST_error_t_dep_1}).
Our result does not strictly reproduce it while they are qualitatively the same.
As we can see in Eq.~(\ref{Eq_Err:PF_bound_theorem}), the former contains commutators of Hamiltonians at equal time, and the latter has those at different time.
Their error scaling in the system size coincides with each other owing to their commutator scaling in many of local Hamiltonians (see Section~\ref{Sec:Examples}), although this difference brings the difference of their prefactors.

On the analysis of higher-order PF errors, Refs.~\cite{Mechthild-SIAM2008_ho,Wiebe2010-mu,childs-prl2019-pf} deal with the standard PFs for generic smooth time-dependent Hamiltonians.
Reference~\cite{Wiebe2010-mu} proves an error bound called the $1$-norm scaling by Eqs. (\ref{Eq_Pre:1_norm_error_t_dep}) and (\ref{Eq_Pre:1_norm_factor_t_dep}), which does not explicitly contain commutators.
Its system-size scaling for local Hamiltonians is worse than ours, which will be shown in Section \ref{Subsec:bound_commutator}.
Reference~\cite{childs-prl2019-pf} shows an error bound whose scaling in the system size can decrease for local Hamiltonians with short-ranged Hamiltonians.
However, the explicit formula for the error is expressed by commutators involving many matrix exponentials $\exp (-i \alpha_{v\gamma} t\, \mr{ad}_{H_{\pi_v(\gamma)} (\beta_{v\gamma} t)})$ in the middle.
Such a form makes it difficult to evaluate the system-size scaling when the interactions are long-ranged.
By contrast, our error bound is directly expressed by nested commutators among  $\{ H_\gamma(t) \}$ and their derivatives.
It enables us to evaluate the system-size dependence of PF errors and to obtain its reduction for generic local Hamiltonians, as discussed in Section \ref{Sec:Examples}.

In terms of the way of derivation, we remark the approach based on time-shift operators~\cite{Suzuki1993-general,Hatano2005-uk}.
The time-shift operator is generated by the time derivative, which corresponds to $H^\mr{LP}$ in the Floquet formalism, and it enables us to formally map the time-dependent PFs and MPFs to those for time-independent cases.
Approximating these relations with introducing an auxiliary system that labels the discretized time~\cite{watkins-2024-clock}, as we do for the Fourier index $l \in \bbZ$, we can reproduce the time-dependent PF or MPF errors by those in time-independent cases under the continuous time limit.
However, it is rather complicated to examine the convergence in that case compared to the Floquet formalism, in which the number of the discrete frequencies labeled by $\{ \ket{l} \}_{l \in \bbZ}$ goes to infinity instead.
Indeed, any error bound explicitly organized by commutators has not been shown by this approach for either the time-dependent PFs or MPFs.

\section{Computational cost with improved $N$-scaling}\label{Sec:Examples}

Commutator scaling plays a central role in reducing the computational cost with respect to the size $N$ for local Hamiltonians.
In this section, we evaluate the cost for simulating time-dependent local Hamiltonians with the standard and generalized PFs.
While their errors by Theorem  \ref{Thm_Err:Main_theorem}, Theorem \ref{Thm_Err:main_thm_Gen}, and Corollary \ref{Cor_Err:error_uniform} involve the time derivatives, we will show that time-dependent systems are simulated as efficiently as time-independent ones.

\subsection{Local time-dependent Hamiltonians}

First of all, we define the locality and extensiveness for time-dependent Hamiltonians.
We say that $H(t)$ is $k$-local if it is expressed by
\begin{equation}
    H (t) = \sum_{X \subset \Lambda: |X| \leq k} h_X(t),
\end{equation}
where each $h_X(t)$ denotes at most $k$-body interactions acting on a domain $X \subset \Lambda$.
We consider local Hamiltonians having the locality $k \in \order{1}$.
When constructing the time-dependent PFs, we split $H(t)$ into $\Gamma$ terms by Eq. (\ref{Eq_Pre:H_t_decomp}).
Each $H_\gamma(t)$ can be also expanded by
\begin{equation}
    H_\gamma(t) = \sum_{X \subset \Lambda: |X| \leq k} h_X^\gamma(t),
\end{equation}
where the set of $\{h_X^\gamma(t)\}$ satisfies $[h_X^\gamma(t),h_{X'}^\gamma(t)]=0$ and reproduces $h_X(t)$ by $h_X(t)=\sum_\gamma h_X^\gamma(t)$.
The extensiveness $g$ is defined by a number such that
\begin{equation}\label{Eq_Cost:extensiveness}
    \max_{i \in \Lambda} \sup_{\tau \in [0,t]}\sum_{X \subset \Lambda; X \ni i} \sum_{\gamma=1}^\Gamma \norm{h_X^\gamma(\tau)} \leq g
\end{equation}
holds.
It means the maximal energy scale per site \cite{Kuwahara2016-yn}, and depends on the range of interactions of the Hamiltonian.
We consider finite-range interactions (those working within finite distance), short-range interactions (those exponentially decaying in distance), and long-range interactions (those polynomially decaying in distance $r$ as $r^{-\nu}$).
We can set the extensiveness $g$ whose scaling is given by
\begin{equation}\label{Eq_Cost:g_scaling}
    g \in \begin{cases}
        \order{1} & (\text{finite- or short-range}) \\
        \order{1} & (\text{long-range, $\nu > d$}) \\
        \order{\log N} & (\text{long-range, $\nu =d$}) \\
        \order{N^{1-\frac{\nu}d}} & (\text{long-range, $\nu<d$})
    \end{cases},
\end{equation}
where $d$ denotes the spatial dimension of the lattice $\Lambda$.

As a characteristic quantity for time-dependent Hamiltonians, we additionally define the following value measuring the degree of time-dependency, $f$, by a number satisfying
\begin{equation}\label{Eq_Cost:time_dependency}
    \max_{i \in \Lambda} \sup_{\tau \in [0,t]} \sum_{X \subset \Lambda; X \ni i} \sum_{\gamma=1}^\Gamma \norm{\dv[q]{\tau} h_X^\gamma(\tau)} \leq f^qg 
\end{equation}
for every $q=1,2,\cdots,p$.
Each time derivative $H^{(q)}(t)$ has the extensiveness smaller than $f^q g$ by definition.
The $1$-norm factor $\alpha_{\text{1-norm},p}(\tau)$ by Eq. (\ref{Eq_Pre:1_norm_factor_t_dep}) is bounded by
\begin{equation}
    \alpha_{\text{1-norm},p}(\tau) \leq N g \max \left[ 1, (f/g)^{\frac{p}{p+1}} \right].
\end{equation}
For standard local Hamiltonians in condensed matter physics and quantum chemistry, where each local term $h_X^\gamma (t)$ is driven locally, the strength of time-dependency $f$ is usually size-independent.
For example, let us consider the case where each term varies in time as $h_X^\gamma(t)=h_X^\gamma \sin (\omega_X^\gamma t)$, which describes typical quantum systems subject to light with frequency $\omega_X^\gamma$.
The strength $f \in \order{\max_{X,\gamma} (\omega_X^\gamma)}$ is usually a size-independent constant as long as we do not consider highly-oscillatory dynamics such that the frequency of the light $\omega_X^\gamma$ is macroscopically large in the system size $N$.
We assume $f \in \order{N^0}$ below considering standard local Hamiltonians in physics and chemistry.

\subsection{Bounds on time-dependent nested commutators}\label{Subsec:bound_commutator}

We next evaluate the upper bounds on $\bar{\alpha}_{\mr{com},p}(\tau)$ and $\alpha_{\mr{com},p}(\tau)$ by Eqs. (\ref{Eq_Sum:alpha_com_Std}) and (\ref{Eq_Sum:alpha_com_Gen}).
They are composed of nested commutators and their time derivatives, characterizing the error of the time-dependent PFs.
In order to conclude the substantial improvement by the commutator scaling, they should have scaling favorable with respect to the size $N$.
Based on the locality and the extensiveness, we obtain the following bounds.

\begin{theorem}\label{Thm_Cost:bound_commutators}
\textbf{}

Local and extensive Hamiltonians $H(t)$ satisfy the inequalities,
\begin{eqnarray}
    \max_{\tau \in [0,t]} \bar{\alpha}_{\mr{com},p}(\tau) &\leq& p! (2kg+\Gamma f)^p Ng, \label{Eq_Cost:alpha_Std_bound} \\
    \max_{\tau \in [0,t]} \alpha_{\mr{com},p}(\tau) &\leq& p! (2kg+2\Gamma f)^p Ng, \label{Eq_Cost:alpha_Gen_bound} \\
    \max_{\tau \in [0,t]} \bar{\alpha}_{\mr{com},p}^\mr{uni}(\tau) &\leq& p! (2kg+f)^p Ng, \label{Eq_Cost:alpha_uni_bound}
\end{eqnarray}
where the commutator factors $\bar{\alpha}_{\mr{com},p}(\tau)$,  $\alpha_{\mr{com},p}(\tau)$, and $\bar{\alpha}_{\mr{com},p}^\mr{uni}(\tau)$ are defined respectively by Eqs. (\ref{Eq_Sum:alpha_com_Std}), (\ref{Eq_Sum:alpha_com_Gen}), and (\ref{Eq_Err:alpha_com_uni}).
\end{theorem}

\textbf{Proof.---} We focus on the quantity $\bar{\alpha}_{\mr{com},p}(\tau)$.
We first evaluate the upper bound on
\begin{equation}\label{Eq_Cost:p_com_r_deriv}
    \Gamma^{p-r} \sum_{\gamma_0,\cdots,\gamma_r=1}^\Gamma \norm{\left[ \prod_{r'=1}^r \ad_{H_{\gamma_{r'}}^{(q_{r'})}(\tau)}\right] H_{\gamma_0}^{(q_0)}(\tau)}
\end{equation}
for $r,q_0,\cdots,q_r$ such that $r+q_0+\cdots+q_r=p$.
The integer $p-r$ mean the number of times that the derivative $\bar{D}_{\Gamma+1}(\tau)=\Gamma \dv{\tau}$ is chosen in $\bar{\alpha}_{\mr{com},p}(\tau)$.
We decompose each $H^{(q_{r'})}_{\gamma_{r'}}$ into local terms by
\begin{equation}
    H^{(q_{r'})}_{\gamma_{r'}}(\tau) = \sum_{X_{r'} \subset \Lambda: |X_{r'}| \leq k} h_{X_{r'}}^{\gamma_{r'}(q_{r'})}(\tau).
\end{equation}
Let us define the union $Y_{r'}=X_1 \cup \cdots \cup X_{r'-1}$ for $r'\geq 2$.
The nested commutator among the local terms $\{ h_{X_{r'}}^{\gamma_{r'}(q_{r'})}\}$ vanishes if there exists an index $r'$ such that $X_{r'} \cap Y_{r'} = \phi$.
As a result, with repeating the relation $\norm{[A,B]} \leq 2 \norm{A} \norm{B}$, we obtain
\begin{eqnarray}
    && [\text{Eq. (\ref{Eq_Cost:p_com_r_deriv})}] \nonumber \\
    && \quad \leq 2^r \Gamma^{p-r} \sum_{\gamma_0,\cdots,\gamma_r} \sum_{\substack{X_0,\cdots,X_r: \\ X_{r'} \cap Y_{r'}\neq\phi}} \prod_{r'=0}^r \norm{h_{X_{r'}}^{\gamma_{r'} (q_{r'})}(\tau)} \nonumber \\
    && \quad \leq 2^r \Gamma^{p-r} \sum_{\gamma_0} \sum_{X_0} \norm{h_{X_0}^{\gamma_0 (q_0)}(\tau)} \prod_{r'=1}^r \left( r'k f^{q_{r'}} g\right) \nonumber \\
    && \quad \leq r! (2kg)^r (\Gamma f)^{p-r} Ng.
\end{eqnarray}
The second inequality comes from the fact that each $h_{X_{r'}}^{\gamma_{r'} (q_{r'})}(\tau)$ has the locality $k$ and the extensiveness $f^{q_{r'}} g$ by Eq. (\ref{Eq_Cost:time_dependency}), which leads to
\begin{eqnarray}
    \sum_{\substack{X_{r'}: |X_{r'}| \leq k, \\ X_{r'} \cap Y_{r'} \neq\phi}} \norm{h_{X_{r'}}^{\gamma_{r'} (q_{r'})}(\tau)} &\leq& \sum_{i \in Y_{r'}} \sum_{X_{r'}: X_{r'} \ni i} \norm{h_{X_{r'}}^{\gamma_{r'} (q_{r'})}(\tau)} \nonumber \\
    &\leq& r'kf^{q_{r'}}g.
\end{eqnarray}
We use the relation $|Y_{r'}|=|X_0\cup\cdots\cup X_{r'-1}| \leq r' k$ in the second inequality.

The quantity $\bar{\alpha}_{\mr{com},p}(\tau)$ can be decomposed into terms having the form of Eq. (\ref{Eq_Cost:p_com_r_deriv}).
Let $Q$ denote a possible subset of $(1,\cdots,p)$ such that $\gamma_{q'} = \Gamma + 1$ for $q' \in Q$.
This leads to the expression of $\bar{\alpha}_{\mr{com},p}(\tau)$ by
\begin{eqnarray}
    && \bar{\alpha}_{\mr{com},p} (\tau) \nonumber \\
    && \quad =  \sum_{r=0}^p \sum_{\substack{Q \subset (1,\cdots,p):\\ |Q|=p-r}} \underbrace{\sum_{\gamma_0,\cdots,\gamma_q=1}^\Gamma}_{\text{$q' \notin Q$}} \norm{ \left[ \prod_{q=1}^p \underbrace{\bar{\mcl{D}}_{\gamma_{q'}}(\tau)}_{\text{$\Gamma \dv{\tau}$ if $q' \in Q$}}\right] H_{\gamma_0}(\tau)}. \nonumber \\
    &&
\end{eqnarray}
Considering the Leibnitz rule, the above equation contains at most $(r+1)^{p-r}$ copies of Eq. (\ref{Eq_Cost:p_com_r_deriv}) for each $r$ and $Q$ such that $|Q|=p-r$.
As a result, we obtain
\begin{eqnarray}
    \bar{\alpha}_{\mr{com},p} (\tau) &\leq& \sum_{r=0}^p \sum_{\substack{Q \subset (1,\cdots,p):\\ |Q|=p-r}} (r+1)^{p-r} r! (2kg)^r (\Gamma f)^{p-r} Ng \nonumber \\
    &\leq& p! Ng \sum_{r=0}^p \frac{p!}{(p-r)!r!} (2kg)^r (\Gamma f)^{p-r} \nonumber \\
    &=& p! (2kg+\Gamma f)^p Ng.
\end{eqnarray}
This completes the proof of Eq. (\ref{Eq_Cost:alpha_Std_bound}).

The inequalities for $\alpha_{\mr{com},p}(\tau)$ and $\bar{\alpha}_{\mr{com},p}^\mr{uni}(\tau)$ are obtained in a similar manner.
Using the triangle inequality, the quantity $\alpha_{\mr{com},p}(\tau)$ defined by Eq. (\ref{Eq_Sum:alpha_com_Gen}) is further bounded by
\begin{eqnarray}
    && \alpha_{\mr{com},p}(\tau) \nonumber \\
    && \quad \leq \sum_{v_1,\cdots,v_p=1}^{V_p} \sum_{\gamma_1,\cdots,\gamma_p=1}^{\Gamma+1} \sum_{\gamma_0=1}^\Gamma \norm{\prod_{q=1}^p \mcl{D}_{\gamma_q}'(\tau) H_{\gamma_0}(\tau)}, \nonumber \\
    &&
\end{eqnarray}
with $\mcl{D}_\gamma'(\tau)=\ad_{H_\gamma(\tau)}$ for $\gamma=1,\cdots,\Gamma$ and $\mcl{D}_{\Gamma+1}'(\tau)=2\Gamma \dv{\tau}$.
Thus, Eq. (\ref{Eq_Cost:alpha_Gen_bound}) is obtained by replacing $\Gamma f$ in Eq. (\ref{Eq_Cost:alpha_Std_bound}) by $2\Gamma f$.
Similarly, replacing $\Gamma f$ by $f$ results in Eq. (\ref{Eq_Cost:alpha_uni_bound}). $\quad \square$

These bounds reproduce the bound on the nested commutators for time-independent cases, Eq. (\ref{Eq_Pre:commutator}) \cite{childs2021-trotter},
\begin{equation}
    \alpha_{\mr{com},p} \leq p! (2kg)^p Ng,
\end{equation}
at $f=0$.
Errors of the time-dependent PFs for small time $t$ are bounded scale as 
\begin{eqnarray}
    \norm{U(t,0)-\bar{S}(t,0)} &\in& \order{(gt+\Gamma ft)^p Ngt}, \label{Eq_Cost:error_scaling_com_Std}\\
    \norm{U(t,0)-S(t,0)} &\in& \order{(gt+\Gamma ft)^p Ngt}, \label{Eq_Cost:error_scaling_com_Gen}
\end{eqnarray}
for orders $p \in \order{1}$.
When the standard PF is uniform, the error scaling becomes
\begin{equation}\label{Eq_Cost:error_scaling_com_uni}
    \norm{U(t,0)-\bar{S}_p(t,0)} \in \order{(gt+ft)^p Ngt}.
\end{equation}
They are substantially improved from the previous result hosting the $1$-norm scaling \cite{Wiebe2010-mu},
\begin{eqnarray}
    \norm{U(t,0)-\bar{S}_p(t,0)} &\in& \order{\left( \max_{\tau \in [0,t]} \alpha_{\text{1-norm},p} (\tau) t\right)^{p+1}} \nonumber \\
    &\subset& \order{(Ngt+Nft)^p Ngt}, \label{Eq_Cost:error_scaling_1_norm}
\end{eqnarray}
with respect to the size $N$ [See Eq. (\ref{Eq_Pre:1_norm_error_t_dep})].

\subsection{Gate counts for time-dependent PFs}\label{Subsec:gate_counts}

We are ready for evaluating the computational cost.
For simulating sufficiently large evolution time $t$, we split it into $r$ parts and approximate the time evolution $U(t,0)$ by the product of time-dependent PFs for small time $t/r$.
Namely, the number $r$, often called the Trotter number, is the query complexity in PFs.
The Trotter number $r$ is determined so that the desirable error $\varepsilon$ can be achieved.
For instance, the requirement when using the standard PF is
\begin{eqnarray}
     \norm{ U(t,0)-\prod_{l=1}^r \bar{S}(lt/r,(l-1)t/r)} \leq \varepsilon.
\end{eqnarray}
Since the left-hand side is bounded by
\begin{eqnarray}
    && \norm{ U(t,0)-\prod_{l=1}^r \bar{S}(lt/r,(l-1)t/r)} \nonumber \\
    && \quad \leq \sum_{l=1}^r \norm{U(lt/r,(l-1)t/r)-\bar{S}_p(lt/r,(l-1)t/r)} \nonumber \\
    && \quad \leq 4\left(\frac{V_p t}r\right)^{p+1}\sum_{l=1}^r \max_{\tau \in [(l-1)t/r,lt/r]} \bar{\alpha}_{\mr{com},p}(\tau)\nonumber \\
    && \quad \leq \frac{4(2V_p kgt + V_p \Gamma ft)^p V_p Ngt}{r^p}
\end{eqnarray}
by Theorem \ref{Thm_Err:Main_theorem} and Theorem \ref{Thm_Cost:bound_commutators}, it is sufficient to choose $r$ by
\begin{equation}
    r = \left\lceil V_p (2kg+\Gamma f)t \left( \frac{4V_p Ngt}{\varepsilon} \right)^{\frac1p}\right\rceil.
\end{equation}
Similar analysis applies to the generalized PF and the standard PF in the uniform case.
The scaling of the Trotter number $r$ is summarized as follows,
\begin{equation}
    r \in \begin{cases}
        \displaystyle \Theta \left( (g+\Gamma f)t \left( \frac{Ngt}{\varepsilon} \right)^{\frac1p} \right) & \text{for $\bar{S}(t,0)$, $S(t,0)$} \\
        \displaystyle \Theta \left( (g+f)t \left( \frac{Ngt}{\varepsilon} \right)^{\frac1p} \right) & \text{for $\bar{S}_p(t,0)$}
    \end{cases}.
\end{equation}

We estimate the gate counts for Hamiltonian simulation.
Here, the cost is measured by the number of $\order{1}$-local quantum gates.
We do not assume the geometrical locality, i.e., $\order{1}$-local gates between distant qubits are available with the same cost as geometrically-local gates.
In the time-dependent PFs, each local time evolution under $h_X^\gamma (t)$ can be implemented by $\order{1}$ $k$-qubit quantum gates.
We note that, even though we use the time-ordered product for the generalized PF, the local time evolution given by
\begin{equation}
    \mcl{T} \exp \left( -i \int_0^t \dd \tau h_X^\gamma (\tau) \right)
\end{equation}
is classically computable, and hence it can be decomposed into local gates.
The gate count per query to the time-dependent PFs is proportional to the number of local terms in the Hamiltonian $H(t)$.
As a result, the total gate count amounts to $\order{Nr}$ for finite- and short-ranged interacting systems and $\order{N^kr}$ for $k$-local long-ranged interacting systems.
We summarize these results in Table \ref{Table:comparison_algorithms}.

We remark the $\Gamma$-dependency of the cost, which is present for the standard and the generalized PFs in general but absent for the standard PF with uniform coefficients.
The number of terms in PFs, $\Gamma$, in Eq. (\ref{Eq_Pre:H_t_decomp}) depends on the range of interactions.
The number $\Gamma$ can be $\order{1}$ for generic Hamiltonians with finite-range interactions.
For instance, we can decompose a 1d Hamiltonian with nearest-neighbor interactions $H(t)=\sum_{i=1}^{N-1} h_{i,i+1}(t)$ into $H(t)=H_1(t)+H_2(t)$ with $H_1(t)=\sum_{i: \text{odd}} h_{i,i+1}(t)$ and $H_2(t)=\sum_{i: \text{even}} h_{i,i+1}(t)$.
The time-evolution operators under $H_1(t)$ or $H_2(t)$ can be implemented by $\order{N}$ local quantum gates.
In general, we can decompose the set of finite-range interactions into $\Gamma \in \order{1}$ groups so that each of them is composed of mutually-commuting terms, and we can perform such an implementation.
For short-ranged interacting systems, $\Gamma$ can be regarded as $\order{1}$ since the exponentially small interactions can be truncated by Lieb-Robinson bound \cite{Lieb1972-uo}.
Therefore, the number $\Gamma$ is irrelevant to the cost for Hamiltonians with finite- and short-range interactions.

\begin{figure}
    \includegraphics[height=4cm, width=8.5cm]{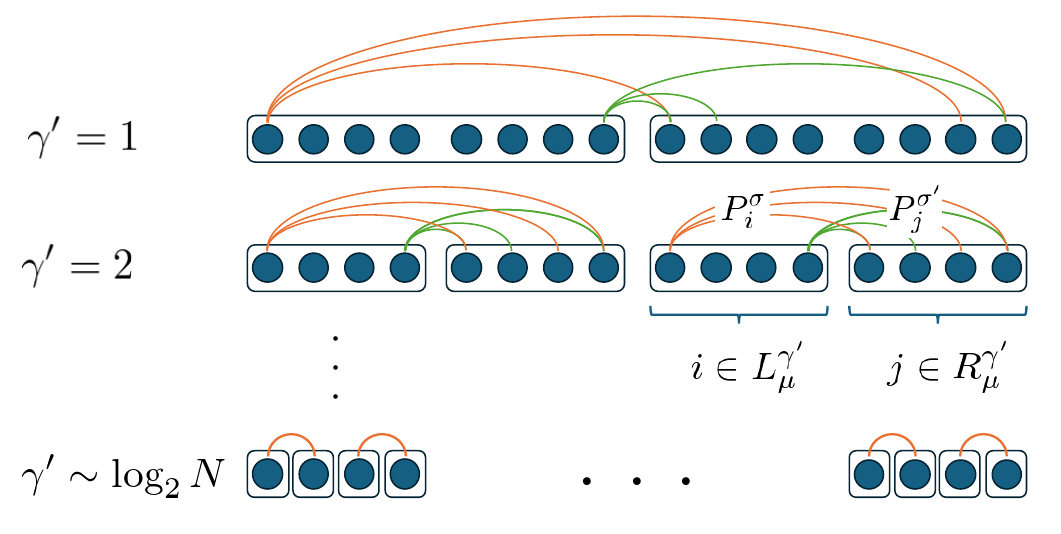}
    \caption{Decomposition of a Hamiltonian $H(t)=\sum_\gamma H_\gamma (t)$ when the interactions are long-ranged. This decomposition is based on Ref.~\cite{Low-prxq2023-long_range} and can be generalized to generic dimension $d$.} 
    \label{Fig_Long_Range}
\end{figure}

On the other hand, the quantity $\Gamma$ becomes significant for long-ranged interacting systems.
Let us consider the $2$-local case, whose Hamiltonian can be generally written by
\begin{equation}
    H(t) = \sum_{\substack{i,j \in \Lambda \\ i \neq j}} \sum_{\sigma,\sigma^\prime=1}^3 h_{ij}^{\sigma \sigma^\prime}(t) P_i^{\sigma} P_j^{\sigma^\prime} + \sum_{i \in \Lambda} \sum_{\sigma=1}^3 h_i^\sigma(t)P_i^\sigma
\end{equation}
with some coefficients $h_{ij}^{\sigma \sigma^\prime}(t), h_i^\sigma(t) \in \bbR$.
The symbol $P_i^{\sigma}$ denotes the Pauli operators $X,Y,Z$ at the site $i$ respectively for $\sigma=1,2,3$.
The decomposition $H(t)= \sum_{\gamma=1}^\Gamma H_\gamma (t)$ such that each local term in $H_\gamma (t)$ commutes with one another can be organized by a divide-and-conquer method \cite{Low-prxq2023-long_range}.
Figure \ref{Fig_Long_Range} shows the decomposition of the two-body interactions when the spatial dimension $d$ is $1$.
At the $\gamma^\prime$-th stage for $\gamma^\prime=1,\cdots,\lceil \log_2 N \rceil$, we split the lattice $\Lambda$ into $2^{\gamma^\prime}$ blocks with the size $N/2^{\gamma^\prime}$.
We consider pairwise interactions between adjacent blocks $L_\mu^{\gamma^\prime}$ and $R_\mu^{\gamma^\prime}$ separately for each $\sigma,\sigma^\prime=1,2,3$.
Namely, we choose
\begin{equation}\label{Eq_Exam:decomp_long_range_2body}
    H_{\gamma^\prime}^{\sigma\sigma^\prime}(t) = \sum_{\mu=1}^{2^{\gamma^\prime-1}} \sum_{i \in L_\mu^{\gamma^\prime}} \sum_{j \in R_\mu^{\gamma^\prime}} h_{ij}^{\sigma\sigma^\prime} (t) P_i^\sigma P_j^{\sigma^\prime}
\end{equation}
as each term $H_\gamma(t)$ with the index $\gamma=(\gamma^\prime,\sigma,\sigma^\prime)$.
We set the last term by $H_\Gamma (t) = \sum_i \sum_\sigma h_i^\sigma(t) P_i^\sigma$.
The number of terms amounts to $\Gamma = 9 \lceil \log_2 N \rceil + 1 \in \order{\log N}$.
Based on the same strategy, the partition number $\Gamma$ can be $\order{N^{k-1} \log N}$ for Hamiltonians with $k$-local long-range interactions on a $d$-dimensional lattice.
Therefore, the standard and generalized PFs are generally efficient for Hamiltonians with $2$-local long-range interactions.
On the other hand, the standard PF with uniform coefficients, which is common with the Lie-Suzuki-Trotter or Forest-Ruth-Suzuki formulas, is applicable to those with generic $k$-local long-range interactions owing to the $\Gamma$-independence of the cost.

\subsection{Comparison with the previous results}

We first compare our gate count estimates with the previously obtained ones for the time-dependent PFs.
The cost of the standard PF based on the $1$-norm scaling error, Eq. (\ref{Eq_Pre:1_norm_error_t_dep}), is shown in the first row of Table \ref{Table:comparison_algorithms} \cite{Wiebe2010-mu}, which mainly focused on the Lie-Suzuki-Trotter formula.
Our result for the standard PF with uniform coefficients completely overwhelms it with respect to the system size $N$, giving the $N$ times smaller cost.
This improvement comes from the substantial error reduction from Eq. (\ref{Eq_Cost:error_scaling_1_norm}) to Eq. (\ref{Eq_Cost:error_scaling_com_uni}) based on the error bound with the commutator scaling, in which the locality of the Hamiltonian plays a central role.
Our results for the standard and generalized PFs in generic forms are also superior to the one by the $1$-norm scaling error for Hamiltonians with finite-, short-, and $2$-local long-range interactions.
Ref. \cite{childs-prl2019-pf} has recently derived the cost of the standard PF for a generic order $p \in \bbN$ based on the commutator-scaling error.
Our result reproduces its result for finite- and short-ranged interactions.
The advantage of our formalism lies in the applicability to generic local Hamiltonians. 
Our error bound is expressed directly by nested commutators among $\{H_\gamma(t)\}_\gamma$ and their time derivative, thereby allowing us to obtain its scaling as Eqs. (\ref{Eq_Cost:error_scaling_com_Std})-(\ref{Eq_Cost:error_scaling_com_uni}).
We arrive at the size-efficient cost brought by the commutator scaling also for Hamiltonians with long-range interactions, while it is not achieved in Ref. \cite{childs-prl2019-pf}.

We also compare the results with linear combination of unitaries (LCU) \cite{Berry-prl2015-LCU} and quantum singular-value transformation (QSVT) \cite{Gilyen2019-qsvt}, known as promising post-Trotter algorithms.
These algorithms have exponentially better dependence in the inverse error $1/\varepsilon$ than PFs.
While they were originally invented for time-independent systems, they have been partially extended for time-dependent cases, such as LCU for generic time-dependent Hamiltonians \cite{Low2018-dyson,Kieferova2019-dyson} and QSVT for time-periodic and -quasiperiodic Hamiltonians \cite{Mizuta_Quantum_2023,Mizuta_2023_multi}.
Their costs are shown in Table \ref{Table:comparison_algorithms}
While the post-Trotter algorithms keep the exponential advantage in the inverse error $1/\varepsilon$, our results show that the time-dependent PFs have better scaling in the system size $N$ for generic local Hamiltonians with finite-, short-, and long-range interactions.
Such an advantage in the system size has already been understood for time-independent PFs due to their commutator-scaling errors \cite{childs2021-trotter}.
By contrast, it has been unclear for time-dependent cases since the conventional analysis based on the $1$-norm scaling error cannot predict it.
Our analysis has shown that the time-dependent PFs are superior to the post-Trotter algorithms, LCU and QSVT, with respect to the system size for a generic order $p$ and Hamiltonians with generic local interactions.

\section{Time-dependent multi-product formula}\label{Sec:MPF}

A multi-product formula (MPF) is a linear combination of PFs \cite{Chin2010-mpf}, and a promising algorithm that can have both the efficiency in the system size like PF and the one in the accuracy like LCU.
For time-independent cases, a $p$th-based order-$m$ MPF $M(t)$ is defined by a linear combination of the $p$th-order PF $T(t)=e^{-iHt}+\order{t^{p+1}}$,
\begin{equation}\label{Eq_MPF:MPF_t_indep}
    M(t) = \sum_{j=1}^J c_j [ T(t/k_j) ]^{k_j},
\end{equation}
satisfying the order condition $M(t)=e^{-iHt} + \order{t^{m+1}}$.
The set of coefficients $\{ c_j \in \bbR \}_{j=1}^J$ and the set of natural numbers $\{ k_j \in \bbN \}_{j=1}^J$ with $J \in \bbN$ are determined by the order condition.
With a particular choice of $\{ c_j \}_j$ and $\{ k_j \}_j$, called the well-conditioned solution, MPF can exponentially improve the gate complexity in the allowable error compared to the PF for time-independent Hamiltonians \cite{low2019-mpf,aftab2024-mpf,mizuta2025-mpf}.

In contrast, MPFs for time-dependent Hamiltonians are poorly understood.
Ref.~\cite{watkins-2024-clock} has provided the construction of time-dependent MPFs, and obtained their error bound and cost.
Although they imply the existence of the commutator scaling for time-dependent cases, any formula explicitly expressed by commutators has not been derived yet.
Thus, it still remains an open problem whether the time-dependent MPFs can simulate time-dependent systems exponentially faster than the time-dependent PFs with respect to the inverse error $1/\varepsilon$ while keeping the system-size efficiency brought by the commutator scaling.

In this section, we give an error bound having the commutator scaling for the time-dependent MPF in a similar strategy for the time-dependent PFs.
We clarify the relationship between the time-independent and -dependent MPFs in Section~\ref{Subsec:MPF_relation}.
We give the error bound by evaluating the one for the time-independent case with infinite-dimension in Section~\ref{Subsec:MPF_error}, while its detailed derivation is provided in Appendix~\ref{A_Sec:MPF}.
Section~\ref{Subsec:MPF_Examples} discusses the computational cost of the time-dependent MPF.

\subsection{Relation between time-independent and time-dependent MPFs}\label{Subsec:MPF_relation}

We begin by introducing time-dependent MPFs and show that they have time-independent counterparts on the infinite-dimensional Floquet-Hilbert space as well as the PFs.
As a natural extension of Eq.~(\ref{Eq_MPF:MPF_t_indep}), we define the standard and generalized MPFs for time-dependent Hamiltonians. 

\begin{definition}\label{Def_MPF:MPF_t_dep}
\textbf{(Time-dependent MPFs)}

The $p$th-based order-$m$ standard and generalized MPFs $\bar{N}(t,0)$ and $N(t,0)$ for time-dependent Hamiltonians are defined respectively by
\begin{eqnarray}
    \bar{N}(t,0) &=& \sum_{j=1}^J c_j \prod_{l=1}^{k_j} \bar{S}\left(\frac{lt}{k_j},\frac{(l-1)t}{k_j} \right), \label{Eq_MPF:Standard_MPF}\\
    N(t,0) &=& \sum_{j=1}^J c_j \prod_{l=1}^{k_j} S \left(\frac{lt}{k_j},\frac{(l-1)t}{k_j} \right), \label{Eq_MPF:Generalized_MPF}
\end{eqnarray}
satisfying the order condition,
\begin{eqnarray}
    \bar{N}(t,0) &=& U(t,0) + \order{t^{m+1}}, \\
    N(t,0) &=& U(t,0) + \order{t^{m+1}}.
\end{eqnarray}
The operators $\bar{S}(\cdot,\cdot)$ and $S(\cdot,\cdot)$ are the $p$th-order standard and generalized PFs by Eqs. (\ref{Eq_Pre:Standard_PF}) and (\ref{Eq_Pre:Generalized_PF}).
We denote the standard MPF constructed by the standard PFs in the uniform case $\bar{S}_p(\cdot,\cdot)$ by $\bar{N}_p(\cdot,\cdot)$.

\end{definition}

As we will discuss later, the set of the coefficients $\{c_j\}$ and the integers $\{ k_j \}$ is common with the time-independent cases in Eq.~(\ref{Eq_MPF:MPF_t_indep}).
The time-dependent MPFs can be implemented on quantum circuits by the LCU method \cite{Berry-prl2015-LCU}.
We evaluate their error bounds for proving their efficiency.
The strategy is the same as the one for the time-dependent PFs in Section~\ref{Sec:relation}.
First, we show the correspondence between the time-dependent and -independent MPFs as follows.

\begin{theorem}\label{Thm_MPF:MPF_relation}
\textbf{(Relation of MPFs)}

Let $\bar{M}^F(t)$ and $M^F(t)$ be time-independent MPFs on the Floquet-Hilbert space, defined by
\begin{eqnarray}
    \bar{M}^F(t) &=& \sum_{j=1}^J c_j [ \bar{T}^F(t/k_j) ]^{k_j}, \\
    M^F(t) &=& \sum_{j=1}^J c_j [ T^F(t/k_j) ]^{k_j}.
\end{eqnarray}
The operators $\bar{T}^F(t)$ and $T^F(t)$ are the $p$th-order time-independent PFs for the standard and generalized PFs by Eqs. (\ref{Eq_Rel:T_F_Std}) and (\ref{Eq_Rel:T_F_Gen}).
The time-dependent MPFs are expressed by
\begin{eqnarray}
    \bar{N}(t,0) &=& \sum_{l \in \bbZ} e^{-il\omega t} \braket{l|\bar{M}^F(t)|0}, \label{Eq_MPF:Rel_MPF_Std}\\
    N(t,0) &=& \sum_{l \in \bbZ} e^{-il\omega t} \braket{l|M^F(t)|0}. \label{Eq_MPF:Rel_MPF_Gen}
\end{eqnarray}
\end{theorem}

\textbf{Proof.---}
The proof is similar to the one for Theorem~\ref{Thm_Rel:Relation}.
We focus on the standard MPF $\bar{M}(t,0)$.
Using the translation symmetry of the Floquet Hamiltonian $H^F$ [See Eq. (\ref{Eq_Pre:tr_sym_H_F})] and the absolute convergence, we obtain
\begin{eqnarray}
    && \sum_{l \in \bbZ} e^{-il \omega t} \bra{l}[ \bar{T}^F(t/k_j) ]^{k_j} \ket{0} \nonumber \\
    && \quad = \sum_{\substack{l_1,\cdots,l_{k_j} \in \bbZ \\ l_0=0}} e^{-il_{k_j} \omega t} \prod_{k=1}^{k_j} \braket{l_k|\bar{T}^F(t/k_j)|l_{k-1}} \nonumber \\
    && \quad = \prod_{k=1}^{k_j} \sum_{l_1,\cdots,l_{k_j} \in \bbZ}  e^{-il_k \omega (k/k_j)t} \braket{l_k|\bar{T}^F(t/k_j)|0} \nonumber \\
    && \quad = \prod_{k=1}^{k_j} \bar{S}(kt/k_j, (k-1)t/k_j).
\end{eqnarray}
We use Theorem~\ref{Thm_Rel:Relation} in the last equality.
We obtain Eq. (\ref{Eq_MPF:Rel_MPF_Std}) by taking the summation with the weight $c_j$.
The proof for the generalized MPF is parallel to this discussion. $\quad \square$

This theorem states that the errors of the time-dependent MPFs can be evaluated via those for time-independent MPFs like 
\begin{equation}\label{Eq_MPF:Rel_MPF_error}
    U(t,0)-\bar{N}(t,0) = \sum_{l \in \bbZ} e^{-il\omega t} \braket{l|\left[e^{-iH^Ft}-\bar{M}^F(t)\right]|0}.
\end{equation}
Not only it allows the error analysis of time-independent MPF for time-dependent one as well as PF, but also it says that the coefficients $\{c_j\}$ and $\{k_j\}$ constructed for time-independent cases qualify as those for time-dependent cases.
Namely, using the coefficients for the well-conditioned MPF \cite{low2019-mpf}, the time-dependent MPFs can be efficiently implemented by LCU with a small overhead.

\subsection{Error bounds of time-dependent MPF}\label{Subsec:MPF_error}

We discuss the error bounds of the time-dependent MPFs while the detailed derivation is given in Appendix~\ref{A_Sec:MPF}.
We suppose that $H(t)=\sum_\gamma H_\gamma(t)$ is a smooth Hamiltonian such that each $H_\gamma(t)$ belongs to class $C^\infty$ here. 
This is in contrast to the time-dependent PFs, where class $C^{p+2}$ is sufficient for the order $p$ as discussed in Section \ref{Subsec:Setup}.
The Hamiltonian $H(t)$ is supposed to be $k$-local and $g$-extensive [See Eq. (\ref{Eq_Cost:extensiveness})].
The degree of the local time-dependency $f$ is defined by a number satisfying Eq. (\ref{Eq_Cost:time_dependency}) for any $q \in \bbN$, reflecting the change in differentiability.
We prove the following bound on the standard MPF error.

\begin{theorem}\label{Thm_MPF:error_MPF}
\textbf{(Error of the standard MPF)}

Let $\epsilon$ be an arbitrary fixed value in $(0,1)$.
We define a quantity expressed by nested commutators,
\begin{equation}\label{Eq_MPF:alpha_MPF_Std}
    \bar{\alpha}_{\mr{com},p}^\mr{MPF}(\tau) = \max_{\substack{q \in \bbN: \\ p \leq q \leq p_0(N,\epsilon)}} \left( \left[ \bar{\alpha}_{\mr{com},q} (\tau)\right]^{\frac1{q+1}}\right), 
\end{equation}
where $p_0(N,\epsilon)$ and $\bar{\alpha}_{\mr{com},q} (\tau)$ are respectively defined by $p_0(N,\epsilon)=\lceil \log (2N/\epsilon) \rceil$ and Eq. (\ref{Eq_Sum:alpha_com_Std}).
When the time $t$ is small enough to satisfy
\begin{equation}\label{Eq_MPF:requirement}
    t \in \order{\min \left( \frac{1}{p_0(N,\epsilon)(g+\Gamma f)}, \frac{1}{\max_{\tau \in [0,t]}\bar{\alpha}_{\mr{com},p}^\mr{MPF} (\tau)}\right)},
\end{equation}
the error of the standard MPF is bounded by
\begin{eqnarray}
    && \norm{U(t,0)-\bar{N}(t,0)} \nonumber \\
    && \quad \leq \norm{\vec{c}}_1 \left( 2V_p \max_{\tau \in [0,t]}\bar{\alpha}_{\mr{com},p}^\mr{MPF} (\tau) t\right)^{m+1} + \norm{\vec{c}}_1 \|\vec{k}\|_1 \epsilon. \nonumber \\
    && \label{Eq_MPF:MPF_error_bound_Std}
\end{eqnarray}
\end{theorem}

When simulating large evolution time $t$, the requirement Eq. (\ref{Eq_MPF:requirement}) should be satisfied for each Trotter step $t/r$.
Since the Trotter number $r$ for the MPF will be large enough for it, this requirement does not matter in practice.
We also note that there is an option of $\epsilon \in (0,1)$.
For achieving the desirable error $\varepsilon$, it will be chosen as $\epsilon \in o(\varepsilon)$.
While we provide the proof of this theorem in Appendix \ref{A_Sec:MPF}, it is essentially parallel to the time-dependent PFs in Section \ref{Sec:Error}.
The error of the time-dependent MPFs can be expressed by a time-independent counterpart by Eq. (\ref{Eq_MPF:Rel_MPF_error}) owing to Theorem \ref{Thm_MPF:MPF_relation}.
Then, we apply the analysis for the time-independent MPF $\bar{M}^F(t)$ \cite{aftab2024-mpf,mizuta2025-mpf}.

We next identify the scaling of $\bar{\alpha}_{\mr{com},p}^\mr{MPF}(\tau)$, which plays a central role in characterizing the commutator scaling of the standard MPF.
For generic local Hamiltonians, we obtain the following upper bound.

\begin{lemma}\label{Lem_MPF:alpha_MPF_bound}
\textbf{}

The quantity $\bar{\alpha}_{\mr{com},p}^\mr{MPF} (\tau)$ defined by Eq. (\ref{Eq_MPF:alpha_MPF_Std}) is bounded by
\begin{eqnarray}
    && \bar{\alpha}_{\mr{com},p}^\mr{MPF} (\tau) \nonumber \\
    && \quad \leq 2p[(2kg+\Gamma f)^p g N]^{\frac1{p+1}}+ 2e (2kg+\Gamma f)p_0(N,\epsilon) \nonumber \\
    && \quad \in \order{ [(g+\Gamma f)^p g N]^{\frac1{p+1}} + (g+\Gamma f)\log (N/\epsilon) }. \nonumber \\
    && \label{Eq_MPF:alpha_MPF_bound}
\end{eqnarray}
\end{lemma}

\textbf{Proof.---}
Using the upper bound of $\bar{\alpha}_{\mr{com},q} (\tau)$ by Theorem \ref{Thm_Cost:bound_commutators}, we obtain
\begin{eqnarray}
    \left[ \bar{\alpha}_{\mr{com},q} (\tau)\right]^{\frac1{q+1}} &\leq& [ q! (2kg+\Gamma f)^q Ng ]^{\frac1{q+1}} \nonumber \\
    &\leq& (q+1) \left(\frac{Ng}{2kg+\Gamma f}\right)^{\frac1{q+1}} (2kg+ \Gamma f). \nonumber \\
    &&
\end{eqnarray}
Since the function $x a^{\frac1x}$ ($a>0$ : a constant) monotonically decreases in $ x < \log a$ and increases in $\log a < x$, the right-hand side becomes the largest at either $q=p$ or $q=p_0(N,\epsilon)$.
Thus, we have
\begin{eqnarray}
    \frac{\bar{\alpha}_{\mr{com},p}^\mr{MPF}(\tau)}{2kg+\Gamma f} &\leq& (p+1) \left(\frac{Ng}{2kg+\Gamma f}\right)^{\frac1{p+1}} \nonumber \\
    && \quad + [p_0(N,\epsilon)+1] \left(\frac{Ng}{2kg+\Gamma f}\right)^{\frac{1}{p_0(N,\epsilon)+1}}. \nonumber \\
    &&
\end{eqnarray}
By using the relation, $N^{1/p_0(N,\epsilon)} \leq e^{\log N/\log (3N/\epsilon)} < e$, we obtain Eq. (\ref{Eq_MPF:alpha_MPF_bound}). $\quad \square$

The error bounds for the generalized MPF $N(t,0)$ and the standard one in the uniform case $\bar{N}_p(t,0)$ are obtained in the same way.
The quantity $\bar{\alpha}_{\mr{com},p}^\mr{MPF}(\tau)$ in the error bound Eq. (\ref{Eq_MPF:MPF_error_bound_Std}) is replaced by
\begin{eqnarray}
    \alpha_{\mr{com},p}^\mr{MPF}(\tau) &=& \max_{\substack{q \in \bbN: \\ p \leq q \leq p_0(N,\epsilon)}} \left( \left[ \alpha_{\mr{com},q} (\tau)\right]^{\frac1{q+1}}\right), \\
    \bar{\alpha}_{\mr{com},p}^\mr{uni,MPF}(\tau) &=& \max_{\substack{q \in \bbN: \\ p \leq q \leq p_0(N,\epsilon)}} \left( \left[ \bar{\alpha}_{\mr{com},q}^\mr{uni} (\tau)\right]^{\frac1{q+1}}\right),
\end{eqnarray}
respectively for $N(t,0)$ and $\bar{N}_p(t,0)$, in which $\alpha_{\mr{com},q} (\tau)$ and $\bar{\alpha}_{\mr{com},q}^\mr{uni} (\tau)$ are defined by Eqs. (\ref{Eq_Sum:alpha_com_Gen}) and (\ref{Eq_Err:alpha_com_uni}).
As well as the discussion for PFs in Section \ref{Sec:Examples}, their scalings can be evaluated by substituting $2\Gamma f$ [for $N(t,0)$] or $f$ [for $\bar{N}_p(t,0)$] into $\Gamma f$.
We note that this replacement also applies to the requirement on the time, Eq. (\ref{Eq_MPF:requirement}).

Ref. \cite{watkins-2024-clock} has recently shown the error bound of the standard MPF in the uniform case $\bar{N}_p(t,0)$, especially composed of the second order Lie-Suzuki-Trotter formula.
It obeys the $1$-norm scaling, which amounts to
\begin{equation}\label{Eq_MPF:MPF_1_norm}
 \norm{U(t,0)-\bar{N}_p(t,0)} \in \order{\left[ (Ngt+Nft)^p Ngt\right]^{\frac{m+1}{p+1}}}.
\end{equation}
By contrast, combining Theorem \ref{Thm_MPF:error_MPF} and Lemma \ref{Lem_MPF:alpha_MPF_bound} for the uniform case results in the following system-size dependence,
\begin{equation}\label{Eq_MPF:MPF_com_scaling}
     \norm{U(t,0)-\bar{N}_p(t,0)} \in \order{\left[ (gt+ft)^p Ngt\right]^{\frac{m+1}{p+1}}},
\end{equation}
where we neglect logarithmic terms.
It is polynomially better than the $1$-norm scaling.
In particular, while increasing the order $p$ does not improve the system-size dependence of the $1$-norm scaling, Eq. (\ref{Eq_MPF:MPF_1_norm}), the commutator-scaling error by Eq. (\ref{Eq_MPF:MPF_com_scaling}) decreases.

\subsection{Cost of time-dependent MPFs}\label{Subsec:MPF_Examples}

We evaluate the cost of Hamiltonian simulation by using the time-dependent MPFs for local Hamiltonians.
The algorithm runs in a similar manner to the one with the time-independent MPFs as follows \cite{low2019-mpf}.
We focus on the standard MPF.
First, we split the large evolution time $t$ into $r$ parts like PF.
The Trotter number $r$ is determined so that the $r$-times repetition of the MPF for the small time $t/r$ approximates the time evolution within a desirable error as
\begin{equation}
    \norm{U(t,0)- \prod_{r'=0}^{r-1} \bar{N}\left(\frac{(r^\prime+1)t}{r},\frac{r't}r\right)} \in \order{\varepsilon}.
\end{equation}
It is sufficient to make the error at each step bounded by
\begin{equation}\label{Eq_MPF:r_determine}
    \norm{U\left(\frac{(r'+1)t}{r},\frac{r't}{r}\right)-\bar{N}\left(\frac{(r'+1)t}{r},\frac{r't}{r}\right)} \leq \frac{\varepsilon}{2r},
\end{equation}
for every $r'=1,\cdots,r$.
Each MPF for the time $t/r$ is implemented by LCU \cite{Berry-prl2015-LCU}, where we use the controlled operation of the time-dependent PF,
\begin{equation}
    C[\bar{S}(t_2,t_1)] = \ket{0}\bra{0} \otimes I + \ket{1}\bra{1} \otimes \bar{S}(t_2,t_1),
\end{equation}
with certain
$t_1,t_2 \in \bbR$.
The number of queries to $C[\bar{S}(t_2,t_1)]$ amounts to
\begin{equation}
    \order{r  \norm{\vec{c}}_1  \| \vec{k} \|_1}.
\end{equation}
The factor $\norm{\vec{c}}_1$ represents the cost of quantum amplitude amplification \cite{Brassard_2002_qaa}, which is required to implemente the LCU deterministically.
The last factor $\| \vec{k} \|_1$ is the number of queries for each time evolution in the linear combination.
The number of ancilla qubits for implementing the LCU is $\lceil \log_2 J \rceil$.

We evaluate the computational cost with determining the Trotter number $r$, the order $m$, and the number of terms $J$.
We set the value $\epsilon \in (0,1)$ by $\epsilon = \varepsilon / 4 \norm{\vec{c}}_1 \| \vec{k}\|_1 r$ in Theorem \ref{Thm_MPF:error_MPF}.
Equation (\ref{Eq_MPF:r_determine}) suggests that the scaling of the proper Trotter number $r$ can be obtained by solving
\begin{equation}
    \begin{cases}
        \displaystyle
        \norm{\vec{c}}_1 \frac{[(gt+\Gamma ft)^p Ngt]^{\frac{m+1}{p+1}}}{r^m} \in \order{\varepsilon}, \\
        \displaystyle
        \norm{\vec{c}}_1 \frac{[(gt+\Gamma ft) \log \, (N \norm{\vec{c}}_1 \| \vec{k} \|_1 r/\varepsilon)]^{m+1}}{r^m} \in \order{\varepsilon}.
    \end{cases}
\end{equation}
These inequalities are the essentially the same as those required in the time-independent MPF (See Theorem 6 in Ref. \cite{mizuta2025-mpf}), where $N^{\frac1{p+1}}$ and $g$ are replaced respectively by $\{ Ng (g+\Gamma f)^{-1} \}^{\frac1{p+1}}$ and $g+\Gamma f$, reflecting the time-dependency.
We set the order $m$ by
\begin{equation}\label{Eq_MPF:m}
    m = \left\lceil \log \left( \frac{N(g+\Gamma f)t}{\varepsilon}\right) \right\rceil.
\end{equation}
When the coefficients $\{ c_j \}$ and the integers $\{ k_j \}$ are organized as the well-conditioned MPF \cite{low2019-mpf}, the number $J$ is proportional to $m$ and the coefficients scale as
\begin{equation}
    \norm{\vec{c}}_1, \| \vec{k} \|_1 \in \poly{J} \subset \mr{polylog} (N,gt,\Gamma ft,1/\varepsilon).
\end{equation}
Then, there exists a Trotter number such that
\begin{eqnarray}
    r &\in& \Theta \biggl( [(g+\Gamma f)^p g N ]^{\frac1{p+1}} t \biggr. \nonumber \\
    && \qquad \biggl. + (g+\Gamma f)t \log^2\left( \frac{N(g+ \Gamma f)t}{\varepsilon}\right)\biggr). \label{Eq_MPF:Trotter_number_MPF}
\end{eqnarray}
The second term except for $\Gamma f t$ in the above equation can be absorbed into the first term or the poly-logarithmic factors in the symbol $\Otilde{\cdot}$.
Consequently, the query complexity in the controlled standard PF $C[\bar{S}(t_2,t_1)]$ amounts to
\begin{equation}\label{Eq_MPF:Cost_Std}
    \Otilde{ [(g+\Gamma f)^p g N ]^{\frac1{p+1}}t + \Gamma f t}.
\end{equation}

The number of $\order{1}$-qubit gates is obtained by multiplying the query complexity with the number of gates for the controlled PF, which is proportional to the number of local interactions in the Hamiltonian.
The number of ancilla qubits is determined by $\order{\log J}$ with the scaling of $J \propto m$ by Eq. (\ref{Eq_MPF:m}).
The computational resources for the generalized MPF $S(t,0)$ or the standard MPF in the uniform case $\bar{S}_p(t,0)$ are obtained similarly, in which we replace $\Gamma f$ in Eq. (\ref{Eq_MPF:Cost_Std}) by $2\Gamma f$ or $f$.
We summarize these results in Table \ref{Table:comparison_algorithms}.
As well as PFs, the difference among the time-dependent MPFs appear in the applicability to Hamiltonians with long-range interactions due to the number $\Gamma$ in their cost.
The standard MPF in the uniform case $\bar{N}_p(t,0)$, which is the most commonly used, can be employed for generic local Hamiltonians.
On the other hand, the standard MPF $\bar{N}(t,0)$ and the generalized MPF $N(t,0)$ are efficient for Hamiltonians with finite- or short-range interactions and some classes of long-range interactions such that the number of partitions $\Gamma$ can be small (See Section \ref{Sec:Examples}).

The time-dependent MPFs achieve the polylogarithmic dependency in $1/\varepsilon$, overcoming the shortcoming of the PFs.
While such efficiency in the allowable error has already been predicted by the $1$-norm scaling \cite{watkins-2024-clock}, we prove that the time-dependent MPFs are efficient also in the system size.
To be precise, when we focus on the system-size dependence, their gate counts achieve the scaling $\Otilde{N^{1+1/(p+1)}}$ (for finite-range interactions) or $\Otilde{N^{k+1/(p+1)}g}$ (for $k$-local long-range interactions).
These are as large as those for the $p$th-order time-dependent PFs obtained based on the commutator-scaling errors in Section \ref{Sec:Examples}, and smaller than those for the post-Trotter algorithms like LCU and QSVT.
This fact gives an evidence of the commutator scaling in the time-dependent MPFs.
To conclude, the time-dependent MPFs are simultaneously efficient in the system size, the evolution time, and the inverse error like the time-independent MPFs \cite{mizuta2025-mpf}.
We note that MPFs require distant local gates and some ancilla qubits in contrast to PFs. 

We make comparison also with the Haah-Hastings-Kothari-Low (HHKL) algorithm \cite{Haah2021-time-dep}, which can be efficient in $N,t,1/\varepsilon$ like MPF.
This algorithm achieves the near-optimal gate count $\Otilde{Nt}$ for Hamiltonians with finite- or short-range interactions.
The time-dependent MPFs have a little worse scaling in the system size $N$.
However, the HHKL algorithm strongly relies on the Lieb-Robinson bound \cite{Lieb1972-uo} and cannot keep its efficiency in the inverse error $1/\varepsilon$ for long-range interactions \cite{Tran-PRX2019-hhkl}.
Since the time-dependent MPFs only exploit the locality via the commutator scaling, it serves as a versatile method for generic local Hamiltonians as well as time-independent cases.

\section{Conclusion and Discussion}\label{Sec:Discussion}

In this paper, we derive the explicit error bounds for the product and multi-product formulas (PFs and MPFs) for generic smooth time-dependent Hamiltonians $H(t)=\sum_{\gamma=1}^\Gamma H_\gamma(t)$.
The error bounds are expressed by the commutators among $\{ H_\gamma (t) \}$ and their time derivatives, which decreases the cost of time-dependent Hamiltonian simulation in the system size.
While the additional error term arising from the time derivatives are inherent in time-dependent cases, we show that various time-dependent local Hamiltonians can be simulated as efficiently as time-independent cases with exploiting this commutator scaling.
The derivation of the error bound itself is insightful.
We use Floquet theory and discover that generic time-dependent PFs are mapped to time-independent one defined on an infinite-dimensional space.
This enables us to evaluate the error bound by the technique for the time-independent PFs, although we have to confirm the convergence of the summation over the infinite-dimensional space for getting a meaningful error bound.
By the careful calculation based on the translation symmetry appearing in Floquet theory, we obtain the error bound explicitly expressed by the commutators of $\{H_\gamma(t)\}$ and their derivatives.

We leave some future directions.
Our results will lead to the acceleration of various quantum algorithms relying on time-dependent PFs, such as time-independent Hamiltonian simulation exploiting an interaction picture \cite{Low2018-dyson,bosse-2024_int_pic,Sharma-2024-int_pic}.
Since the interaction picture under integrable Hamiltonians maintains the locality and the range of interactions, the commutator scaling will improve the PF error and thereby the cost of Hamiltonian simulation.
Recently, the error analysis on PFs has diverged to various directions, such as the low-energy subspace simulation~\cite{Sahinoglu2021-ng-low-energy,Gong2024-wr-low-energy,Hejazi2024-xs-low-energy,mizuta-2025-low-energy} and the connections to thermalization~\cite{Heyl2019-wf-chaos,Sieberer2019-zy-chaos}, and their extensions to time-dependent Hamiltonians will be interesting.
In particular, concerning to the former example, the time-independent PFs have been proven to host substantial decrease in their errors when the initial states belong to low-energy subspaces \cite{Sahinoglu2021-ng-low-energy}, and the commutator scaling plays a significant role there \cite{Hejazi2024-xs-low-energy,mizuta-2025-low-energy}.
If one can find the time-dependent analogue based on our formalism, it will accelerate quantum algorithms for adiabatic state preparation~\cite{Wu-PRL2002-adiabatic,Aspuru-Guzik2005-adiabatic,Albash-RMP2018-adiabatic}, with which we can prepare various desirable energy eigenstates.

\begin{center}
    \textit{Note added}
\end{center}
After the appearance of this manuscript, a preprint whose author is one of us \cite{mizuta2025-mpf} pointed out that the analysis of the time-independent MPF relying on the convergence of BCH formula \cite{aftab2024-mpf} is insufficient for the size-efficient cost reflecting the commutator scaling.
The analysis of the time-independent MPFs in the first version of our preprint had the same problem.
We modify the proofs in Appendix \ref{A_Sec:MPF} while the resulting cost does not essentially change.

\section*{Acknowledgment}

We thank Y. Ito for giving us some information about smooth Hamiltonians in Appendix \ref{A_Sec:Smooth}.
K. M. is supported by JST PRESTO Grant No. JPMJPR235A and JSPS KAKENHI Grant No. JP24K16974.
T. N. I. is supported by JST PRESTO Grant No. JPMJPR2112 and by JSPS KAKENHI Grant No. JP21K13852.
This work is supported by MEXT
Quantum Leap Flagship Program (MEXTQLEAP)
Grant No. JPMXS0118067394, JPMXS0120319794, and
JST COI-NEXT program Grant No. JPMJPF2014.

\appendix
\onecolumngrid
\begin{center}
\bf{\large Appendix}
\end{center}

\section{Properties of smooth Hamiltonians}\label{A_Sec:Smooth}

\renewcommand{\thetheorem}{\thesection\arabic{theorem}}
\setcounter{theorem}{0}

\subsection{Extrapolation of smooth Hamiltonians}

As discussed in Section \ref{Subsec:Setup}, our results apply to generic smooth Hamiltonians.
This is based on the fact that there exists a time-periodic extrapolation for such a smooth Hamiltonian, and here we explicitly construct it.
We first begin with defining the smoothness of time-dependency.

\begin{definition}
\textbf{(Smoothness)}

A time-dependent Hamiltonian $H(t)$ is said to be of class $C^p$ in the closed interval $[0,t]$ if it has a $q$-th derivative $H^{(q)}(t)$, which is bounded and continuous in $[0,t]$, for every $q = 0,1,\cdots,p$.
Note that the derivatives at the edges of $[0,t]$ are defined by the left and the right derivatives respectively.

\end{definition}

We consider the $p$-th order PF for a smooth Hamiltonian $H(t)=\sum_\gamma H_\gamma(t)$.
While $H_\gamma (t)$ is not necessarily time-periodic, it is known that its time-periodic extrapolation $H^\mr{ex}(t)$ can be organized by using so-called a bump function.

\begin{theorem}
\textbf{(Extrapolation)}

When $H_\gamma (t)$ is a Hamiltonian of class $C^{p+2}$ in the interval $[t_0,t+t_0]$, there exists a Hamiltonian $H^\mr{ex}_\gamma (t)$ such that the following conditions,
\begin{enumerate}[(i)]
    \item $H^\mr{ex}(\tau)$ is a class-$C^{p+2}$ Hamiltonian in $\tau \in \bbR$, satisfying the periodicity $H^\mr{ex}_\gamma (\tau+T) = H^\mr{ex}_\gamma (\tau)$ for any $\tau \in \bbR$ with the $\gamma$-independent period $T=2t$,
    \item $H^\mr{ex}_\gamma(\tau)=H_\gamma(\tau)$ for any $\tau \in [t_0,t+t_0]$,
    \item The Fourier series for $H^\mr{ex}_\gamma(t)$ and its derivatives shows the uniform and absolute convergence in $[t_0,t+t_0]$ with
    \begin{equation}\label{EqA_Smo:Converge_derivative}
        \qquad \dv[p^\prime]{t} H^\mr{ex}_\gamma (t) = \lim_{M \to \infty} \sum_{|m| \leq M} (-im\omega)^{p^\prime} H^\mr{ex}_{\gamma m} e^{-im\omega t},
    \end{equation}
\end{enumerate}
are satisfied.
\end{theorem}

\textbf{Proof.---} We set $t_0=0$ without loss of generality.
Using a bump function given by
\begin{equation}
    b (t) = \begin{cases}
        e^{-1/t} & (t>0) \\
        0 & (t \leq 0),
    \end{cases}
\end{equation}
we define a scalar function $c(t)$ by
\begin{equation}
    c(t) = \frac{\int_0^t \dd \tau b(\tau)b(1-\tau)}{\int_0^1 \dd \tau b(\tau)b(1-\tau)}.
\end{equation}
The function $c(t)$ is of class $C^\infty$ in $t \in \bbR$ and satisfies $c(t) = 0$ for any $t \leq 0$, $c(t) =1$ for any $t \geq 1$, and $c^{(p^\prime)}(0)=c^{(p^\prime)}(1)=0$ for any $p^\prime \in \bbN$.
The extrapolation $H^\mr{ex}_\gamma(t)$ is explicitly organized by
\begin{equation}\label{EqA_Smo:extrapolation}
    H^\mr{ex}_\gamma(\tau) = \begin{cases}
        H_\gamma (\tau) & (\tau \in [0,t]) \\
        \sum_{k=0}^{p+2} \frac{H^{(k)}_\gamma(t) \tau^k}{k!}c\left(7-\frac{6\tau}t \right) & (\tau \in (t,\frac43 t)) \\
        0 & (\tau \in [\frac43t, \frac53 t]) \\
        \sum_{k=0}^{p+2} \frac{H^{(k)}_\gamma(0) \tau^k}{k!}c\left(\frac{6\tau}t - 11\right) & (\tau \in (\frac53 t, 2t)).
    \end{cases}
\end{equation}
The period $T=2t$ is independent of $\gamma$, and $H^\mr{ex}_\gamma(\tau)$ for $\tau \notin [0,2t]$ is by the periodic extension $H^\mr{ex}_\gamma(\tau+2t)=H^\mr{ex}_\gamma(\tau)$.
Confirming the continuity at $\tau=0,t,\frac43 t, \frac53 t, 2t$, it is easy to check the satisfaction of (i) and (ii).

About the condition (iii), the Fourier coefficient of $H^\mr{ex}_\gamma (t)$ is
\begin{equation}
    H^\mr{ex}_{\gamma m} = \frac1T \int_0^T \dd \tau H_\gamma^\mr{ex}(\tau) e^{im\omega \tau}. 
\end{equation}
Repeating the integration by part, we obtain
\begin{eqnarray}
    \norm{H^\mr{ex}_{\gamma m}} &=& \norm{\frac{i^{p+2}}{(m\omega)^{p+2}T} \int_0^T \dd \tau  H^{\mr{ex} \, (p+2)}_\gamma (\tau) e^{im\omega \tau}}\nonumber \\
    &\leq& \frac{1}{(|m|\omega)^{p+2}} \max_{\tau \in [0,2t]} \left( \norm{H^\mr{ex}_\gamma (\tau)} \right)  \label{EqA_Smo:Fourier_decay}
\end{eqnarray}
for each $m \neq 0$, and it decays as $\norm{H^\mr{ex}_{\gamma m}} \in o(|m|^{-p-1})$.
This implies the uniform and absolute convergence of Eq. (\ref{EqA_Smo:Converge_derivative}), which validates the condition (iii). $\quad \square$

When $H(t)$ is a smooth Hamiltonian such that each $H_\gamma(t)$ is of class $C^\infty$, the Fourier coefficient $H_{\gamma m}^\mr{ex}$ decays faster than every polynomial of $m$.
\subsection{Decay in the Floquet-Hilbert space}

We here discuss the decay of operators on the Floquet-Hilbert space, which will be used for proving the error bounds.
It is characterized by the smoothness, inheriting the decaying property of the Fourier components.
First, we show the following lemma. which allows us to easily evaluate the decay in the product of operators.

\begin{lemma}\label{LemA_smooth:product_decay}
\textbf{}

When two bounded operators $A^F,B^F$ acting on the Floquet-Hilbert space satisfy
\begin{equation}
    \norm{\braket{l|A^F|l'}} \in o \left( |l-l'|^{-p-1}\right), \quad  \norm{\braket{l|B^F|l'}} \in o \left( |l-l'|^{-p-1}\right), 
\end{equation}
the following inequality is satisfied,
\begin{equation}\label{EqA_smooth:AB_decay}
    \norm{\braket{l|A^FB^F|l'}} \in o\left(|l-l'|^{-p-1}\right).
\end{equation}
\end{lemma}
\textbf{Proof.---}
Without loss of generality, we suppose
\begin{equation}
    \norm{\braket{l|A^F|l'}} \leq \frac{a}{(1+|l-l'|)^{p+1}}, \quad  \norm{\braket{l|B^F|l'}} \leq \frac{b}{(1+|l-l'|)^{p+1}},
\end{equation}
with some constants $a,b \geq 0$.
Inserting the completeness $\sum_{l''} \ket{l''}\bra{l''}=1$, we obtain
\begin{eqnarray}
    \norm{\braket{l|A^FB^F|l'}} &\leq& \sum_{l'' \in \bbZ} \norm{\braket{l|A^F|l''}} \times \norm{\braket{l''|B^F|l'}} \nonumber \\
    &\leq&\sum_{l'' \in \bbZ} \frac{a}{(1+|l-l''|)^{p+1}} \times \frac{b}{(1+|l''-l'|)^{p+1}} \nonumber \\
    &\leq& \frac{C_p ab}{(1+|l-l'|)^{p+1}},
\end{eqnarray}
where $C_p$ denotes a positive constant dependent on $p$.
The last inequality can be confirmed by an elementary calculation \cite{Gong2022-bound}.
This completes the proof of Eq. (\ref{EqA_smooth:AB_decay}). $\quad \square$

Combining the decay of the Fourier components by the smoothness and this lemma, we obtain the decay of the time-evolution operator in the Floquet-Hilbert space as follows.

\begin{lemma}\label{LemA_smooth:evolution_decay}
\textbf{}

When the Hamiltonian $H(t)=\sum_m H_m e^{-im\omega t}$ has polynomially-decaying Fourier coefficients as $\norm{H_m} \in o(|m|^{-p-1})$, we have
\begin{equation}\label{EqA_smooth:decay_evol}
    \norm{ \braket{l|e^{-iH^\mr{Add}t}|l'}} \in o(|l-l'|^{-p-1}), \quad \norm{ \braket{l|e^{-iH^Ft}|l'}} \in o(|l-l'|^{-p-1}),
\end{equation}
under $|l-l'| \to \infty$, where $H^\mr{Add}$ and $H^F$ are respectively defined by Eqs. (\ref{Eq_Pre:H_add}) and (\ref{Eq_Pre:H_F}).
\end{lemma}

\textbf{Proof.---}
Since $H^\mr{Add}=\sum_m \mr{Add}_m \otimes H_m$ is a Toeplitz matrix \cite{Gray2006-da-toeplitz}, its norm is bounded by
\begin{equation}
    \norm{H^\mr{Add}} \leq \max_{\tau \in [0,T]} \left( \norm{H(\tau)} \right).
\end{equation}
This allows us to employ the Taylor series expansion for $e^{-iH^{\mr{Add}}t}$.
The matrix element $\braket{l|e^{-iH^\mr{Add}t}|l'}$ is bounded by
\begin{eqnarray}
    \norm{ \braket{l|e^{-iH^\mr{Add}t}|l'} } &\leq& \sum_{n=0}^\infty \frac{|t|^n}{n!} \norm{\braket{l|(H^\mr{Add})^n|l'}}.
\end{eqnarray}
We have $\norm{\braket{l|H^\mr{Add}|l'}}=\norm{H_{l-l'}} \in o(|l-l'|^{-p-1})$.
By repeating Lemma \ref{LemA_smooth:product_decay}, there exists a constant $C_p$ such that 
\begin{equation}
    \norm{\braket{l|(H^\mr{Add})^n|l'}} \leq \frac{[C_p  \max_{\tau \in [0,T]} \left( \norm{H(\tau)} \right)]^n}{(1+|l-l'|)^{p+1}}.
\end{equation}
As a result, we obtain
\begin{equation}
    \norm{ \braket{l|e^{-iH^\mr{Add}t}|l'} } \leq \frac{e^{C_pt  \max_{\tau \in [0,T]} \left( \norm{H(\tau)} \right)}}{(1+|l-l'|)^{p+1}} \in o\left( |l-l'|^{-p-1}\right)
\end{equation}
for any finite time $t$, which proves the first half of Eq. (\ref{EqA_smooth:decay_evol}).

For the matrix element $\braket{l|e^{-iH^Ft}|l'}$, we should be careful of the unboundedness of $H^F$.
Using the interaction picture in $H^\mr{LP}$, it is expressed by
\begin{eqnarray}
    \braket{l|e^{-iH^Ft}|l'} &=& \bra{l} e^{iH^\mr{LP}t} \mcl{T} \exp \left( -i \int_0^t \dd \tau e^{-iH^\mr{LP}\tau} H^\mr{Add} e^{iH^\mr{LP}\tau}\right) \ket{l'} \nonumber \\
    &=& e^{il\omega t} \sum_{n=0}^\infty \int_0^t\dd \tau_n \cdots \int_0^{\tau_2} \dd \tau_1 \bra{l}\prod_{n'=1}^n e^{-iH^\mr{LP}\tau_{n'}} H^\mr{Add} e^{iH^\mr{LP}\tau_{n'}} \ket{l'}.
\end{eqnarray}
The second line comes from the boundedness of $e^{-iH^\mr{LP}\tau} H^\mr{Add} e^{iH^\mr{LP}\tau}$.
We can apply Lemma \ref{LemA_smooth:product_decay} owing to the relation $\| \braket{l|e^{-iH^\mr{LP}\tau} H^\mr{Add} e^{iH^\mr{LP}\tau}|l'}\|=\| H_{l-l'}e^{-i(l-l')\omega \tau}\| = \|H_{l-l'}\|$, and this immediately leads to the latter part of Eq. (\ref{EqA_smooth:decay_evol}). $\quad \square$

\section{Proof of the error bounds not mentioned in the main text}\label{A_Sec:Generalized_PF}

\renewcommand{\thetheorem}{\thesection\arabic{theorem}}
\setcounter{theorem}{0}

\subsection{Proof of the generalized PF error bound}
In this section, we describe the proof of Theorem \ref{Thm_Err:main_thm_Gen}, which states the upper bound on the generalized PF error, $\norm{U(t,0)-S(t,0)}$.

\textbf{Proof of Theorem \ref{Thm_Err:main_thm_Gen}.---}
The proof is completely parallel to that for the standard PF in Section \ref{Sec:Error}.
The time-independent PF in the Floquet-Hilbert space, $T^F(t)$, by Eq. (\ref{Eq_Rel:T_F_Gen}) is expressed by
\begin{equation}
    T^F(t) \equiv \prod_{v=1}^{V_p} \prod_{\tilde{\gamma}=1}^{2\Gamma} e^{-i \tilde{H}_{v \tilde{\gamma}} \tilde{\alpha}_{v \tilde{\gamma}} t},
\end{equation}
where $\tilde{H}_{v \tilde{\gamma}}$ and $\tilde{\alpha}_{v \tilde{\gamma}}$ are respectively given by
\begin{equation}\label{EqA_Gen:H_tilde_Gen}
    (\tilde{H}_{v\tilde{\gamma}}, \tilde{\alpha}_{v \tilde{\gamma}}) = \begin{cases}
        (H_{\pi_v(\gamma)}^F,\alpha_{v \gamma}) & (\tilde{\gamma}=2\gamma-1) \\
        (-H^\mr{LP},\beta_{v(\gamma+1)}-\beta_{v\gamma}-\alpha_{v\gamma}) & (\tilde{\gamma}=2\gamma)
    \end{cases}
\end{equation}
instead of Eq. (\ref{Eq_Err:H_tilde_def}).
The coefficient $\tilde{\alpha}_{v \gamma}$ has an absolute value smaller than $1$ owing to Eqs. (\ref{Eq_Pre:coef_1})-(\ref{Eq_Pre:coef_3}).
In a similar manner to Eq. (\ref{Eq_Err:t_dep_error_norm_Delta}), we obtain
\begin{equation}
    \norm{U(t,0)-S(t,0)} \leq \int_0^t \dd \tau \norm{\sum_{l \in \bbZ} \braket{l|\Delta^F(\tau)|0}},
\end{equation}
where the operator $\Delta^F(\tau)$ is equal to Eq. (\ref{Eq_Err:Delta_F_expansion}) with substituting Eq. (\ref{EqA_Gen:H_tilde_Gen}) as $(\tilde{H}_{v\tilde{\gamma}},\tilde{\alpha}_{v\tilde{\gamma}})$.

We next examine how the operators $C_1(H^F)$ and $C_2(\tilde{H}_{v'\tilde{\gamma}'})$ in $\Delta^F(\tau)$ form the nested commutators and the time derivative.
As a counterpart of Lemma \ref{Lem_Err:Form_commutators}, we prove the following lemma.
\begin{lemma}
\textbf{}

Let $\tilde{H}_{v\tilde{\gamma}}$ be an operator defined by Eq. (\ref{EqA_Gen:H_tilde_Gen}).
The commutators with $\tilde{H}_{v\tilde{\gamma}}$ are given by
\begin{eqnarray}
    \ad_{\tilde{H}_{v\tilde{\gamma}}} H_{\gamma'}^\mr{Add} &=& \sum_{m \in \bbZ} \mr{Add}_m \otimes (D_{v\tilde{\gamma}}(t)H_{\gamma'}(t))_m, \quad D_{v\tilde{\gamma}}(t) = \begin{cases}
    \displaystyle  \ad_{H_{\pi_{v}(\gamma)}(t)} + i \dv{t}  & (\tilde{\gamma}=2\gamma-1) \\
    \displaystyle i \dv{t} & (\tilde{\gamma}=2\gamma) 
    \end{cases},\\
    \ad_{\tilde{H}_{v\tilde{\gamma}}}
    H^\mr{LP} &=&
    \begin{cases}
        \displaystyle \sum_{m \in \bbZ} \mr{Add}_m \otimes \left( i \dv{t} H_{\pi_{v}(\gamma)}(t) \right)_m & (\tilde{\gamma}=2\gamma-1)\\
        \displaystyle 0 & (\tilde{\gamma}=2\gamma)
    \end{cases},
\end{eqnarray}
for $\gamma=1,\cdots,\Gamma$.
\end{lemma}

\textbf{Proof.---}
It is easy to check this lemma.
For an odd integer $\tilde{\gamma}=2\gamma-1$, we have 
\begin{eqnarray}
    \ad_{\tilde{H}_{v\tilde{\gamma}}} H_{\gamma'}^\mr{Add} &=& [H_{\pi_{v}(\gamma)}^{\mr{Add}},H_{\gamma'}^{\mr{Add}}]-[H^\mr{LP},H_{\gamma'}^{\mr{Add}}] \nonumber \\
    &=& \sum_{m \in \bbZ} \mr{Add}_m \otimes \left( \ad_{H_{\pi_v(\gamma)}} H_{\gamma'}(t) + i \dv{t} H_{\pi_{v}(\gamma)}(t) H_{\gamma'}(t)\right)_m, 
\end{eqnarray}
in which the second line comes from Lemma \ref{Lem_Err:Form_commutators}.
The same goes also for the other cases. $\quad \square$

We also need to show the counterpart of Lemma \ref{Lem_Err:truncated}.
With using the truncated generalized PF $S_{\prec v \gamma}(t,0) = \prod_{v'\gamma' \prec v\gamma} U_{\pi_{v'}(\gamma')}(\beta_{v'\gamma'}t+\alpha_{v'\gamma'}t,\beta_{v'\gamma'}t)$ like Eq. (\ref{Eq_Err:truncated_Standard_PF}), we have
\begin{equation}\label{EqA_Gen:truncated_Gen}
    \sum_{l \in \bbZ} e^{-il \omega \tau_{v\tilde{\gamma}}(\tau,\tau_1)} \braket{l|e^{-i \tilde{H}_{v\tilde{\gamma}} \tilde{\alpha}_{v\tilde{\gamma}} \tau_1} T^F_{\prec v \tilde{\gamma}}(\tau)|0} = \begin{cases}
        U_{\pi_v(\gamma)}(\beta_{v\gamma}\tau+\alpha_{v\gamma} \tau_1,\beta_{v\gamma}\tau) S_{\prec v\gamma}(\tau,0) & (\tilde{\gamma}=2\gamma-1) \\
        S_{\prec v \gamma} (\tau,0) & (\tilde{\gamma}=2\gamma)
    \end{cases},
\end{equation}
where the time $\tau_{v\tilde{\gamma}}(\tau,\tau_1)$ is equal to $\beta_{v\gamma}\tau+\alpha_{v\gamma} \tau_1$ for $\tilde{\gamma}=2\gamma-1$ and $\beta_{v(\gamma+1)}\tau_1+(\beta_{v\gamma}+\alpha_{v\gamma})(\tau-\tau_1)$ for $\tilde{\gamma}=2\gamma$.
The time $\tau_{v\tilde{\gamma}}(\tau,\tau_1)$ belongs to $[0,\tau]$ for any $\tau_1 \in [0,\tau]$.
This relation can be confirmed easily like Lemma \ref{Lem_Err:truncated} with using Eq. (\ref{Eq_Rel:Time_Evol_Floquet_gamma}).
Since Eq. (\ref{EqA_Gen:truncated_Gen}) gives a unitary operator, it is irrelevant to the error bound.
As a result, the quantity characterizing the generalized PF error is bounded by 
\begin{eqnarray}
\norm{\sum_{l \in \bbZ} \braket{l|\Delta^F(\tau)|0}} &\leq&  \tau^{p-1} \int_0^\tau \dd \tau_1 \sum_{v=1}^{V_p} \sum_{\tilde{\gamma}=1}^{2\Gamma}
    \sum_{\substack{q_{v\tilde{\gamma}}, \cdots,q_{(V_p,2\Gamma)} \geq 0: \\ q_{v\tilde{\gamma}} + \cdots + q_{(V_p,2\Gamma)} = p,
    \\ q_{v\tilde{\gamma}} \neq 0}} \left( \sum_{\gamma'=1}^\Gamma \norm{C_{v\tilde{\gamma},V_p 2\Gamma}^{\{q\},\gamma'}(\tau_{v\tilde{\gamma}} (\tau,\tau_1))} + \norm{C_{v\tilde{\gamma},V_p 2\Gamma}^{\{q\},\mr{LP}}(\tau_{v\tilde{\gamma}} (\tau,\tau_1))}\right) \nonumber \\
&& \qquad + \tau^{p-1} \int_0^\tau \dd \tau_1 \sum_{v'=1}^{V_p} \sum_{\tilde{\gamma}':\text{odd}} \sum_{v\tilde{\gamma} \preceq v'\tilde{\gamma}'} 
    \sum_{\substack{q_{v\tilde{\gamma}}, \cdots,q_{v'\tilde{\gamma}'} \geq 0: \\ q_{v\tilde{\gamma}} + \cdots + q_{v'\tilde{\gamma}'} = p,
    \\ q_{v\tilde{\gamma}} \neq 0}}   \norm{C_{v\tilde{\gamma},v'\tilde{\gamma}'}^{\{q\},\frac{\tilde{\gamma}'+1}2}(\tau_{v\tilde{\gamma}} (\tau,\tau_1))} \nonumber \\
&& \qquad \qquad + \tau^{p-1}\int_0^\tau \dd \tau_1 \sum_{v'=1}^{V_p} \sum_{\tilde{\gamma}'=1}^{2\Gamma} \sum_{v\tilde{\gamma} \preceq v'\tilde{\gamma}'} 
    \sum_{\substack{q_{v\tilde{\gamma}}, \cdots,q_{v'\tilde{\gamma}'} \geq 0: \\ q_{v\tilde{\gamma}} + \cdots + q_{v'\tilde{\gamma}'} = p,
    \\ q_{v\tilde{\gamma}} \neq 0}}  \norm{C_{v\tilde{\gamma},v'\tilde{\gamma}'}^{\{q\},\mr{LP}}(\tau_{v\tilde{\gamma}} (\tau,\tau_1))},
\end{eqnarray}
in which we properly replace the operator $\bar{D}_{v\tilde{\gamma}}(t)$, the coefficient $\tilde{\alpha}_{v\tilde{\gamma}}$, and the time $\tau_{v\tilde{\gamma}}(\tau,\tau_1)$ by those for the generalized PF.
We note that the range of the index $\tilde{\gamma}'$ in the third line is different from the original one, Eq. (\ref{Eq_Err:Delta_F_bound_start}).
This is because $\tilde{H}_{v' \tilde{\gamma
}'}$ appearing in the second term of Eq. (\ref{Eq_Err:Delta_F_expansion}) contains $H^\mr{LP}$ also for odd $\tilde{\gamma}'$ according to Eq. (\ref{EqA_Gen:H_tilde_Gen}).
Finally, repeating the evaluation from Eq. (\ref{Eq_Err:Delta_F_bound_start}) to Eq. (\ref{Eq_Err:sum_LP_b}), we obtain the error bound of the generalized PF,
\begin{equation}
    \norm{U(t,0)-S(t,0)} \leq 5 (V_p t)^{p+1} \max_{\tau \in [0,t]} \alpha_{\mr{com},p}(\tau),
\end{equation}
where the quantity $\alpha_{\mr{com},p}(\tau)$ is defined by Eq. (\ref{Eq_Sum:alpha_com_Gen}).
The change in the coefficient from Eq. (\ref{Eq_Err:PF_bound_theorem}) reflects that in the range of $\tilde{\gamma}'$ discussed above. $\quad \square$

\subsection{Error bounds of the time-dependent PFs with $\beta_{11} \neq 0$}\label{SubsecA:Error_beta_nonzero}

In the main text, we focus on the standard or generalized PFs with setting $\beta_{11}=0$ in Eqs. (\ref{Eq_Pre:Standard_PF}) or (\ref{Eq_Pre:Generalized_PF}).
This is just for simplifying the calculation in Sections \ref{Sec:relation} and \ref{Sec:Error}, and hence we briefly discuss the case with $\beta_{11} \neq 0$ here.

We focus on the standard PF $\bar{S}(t,0)$ with $\beta_{11} \neq 0$.
The problem of the calculation in Section \ref{Sec:Error} is that the time-independent PF $\bar{T}^F(t)$ defined by Eq. (\ref{Eq_Rel:T_F_Std}) cannot approximate the time evolution $e^{-iH^Ft}$ in the case of $\beta_{11} \neq 0$.
Namely, the corresponding PF $\bar{T}(t)$ by Eq. (\ref{Eq_Err:PF_equiv}) cannot satisfy the order condition $\bar{T}(t)=e^{-iHt}+\order{t^{p+1}}$ for a Hamiltonian $H=\sum_{\gamma=0}^\Gamma H_\gamma$, which is required in Theorem \ref{Thm_Err:Main_theorem}.
To see this, let us consider the total evolution time under $H_\gamma$ included in the PF $\bar{T}(t)$.
The evolution time for each $\gamma \in \{1,2,\cdots, \Gamma\}$ amounts to
\begin{equation}
    \sum_{v=1}^{V_p} \sum_{\gamma': \pi_v(\gamma')=\gamma} \alpha_{v\gamma'} t= t,
\end{equation}
which directly follows from the assumption, Eq. (\ref{Eq_Pre:coef_1}).
On the other hand, the total evolution time for $\gamma=0$ is
\begin{equation}
    \sum_{v=1}^{V_p} \sum_{\gamma=1}^\Gamma (\beta_{v(\gamma+1)}-\beta_{v\gamma})t =  \beta_{V_p(\Gamma+1)} t - \beta_{11}t = t-\beta_{11}t, 
\end{equation}
which cannot be equal to $t$ except for $\beta_{11}=0$.
Thus, the PF $\bar{T}(t)$ cannot approximate $e^{-iHt}=e^{-i\sum_{\gamma=0}^\Gamma H_\gamma t}$ even within an error $\order{t}$ under $\beta_{11} \neq 0$, implying the impossibility of satisfying the order condition $\bar{T}(t)=e^{-iHt}+\order{t^{p+1}}$.
This problem is easily resolved by modifying the time-independent PF $\bar{T}^F(t)$ on the Floquet-Hilbert space $\mcl{H}_\mr{FT}$.
Instead of Eq. (\ref{Eq_Rel:T_F_Std}), we set it by
\begin{equation}\label{EqA_beta:modified_T_F_Std}
    \bar{T}^F(t) = \left[ \prod_{v=1}^{V_p}  \prod_{\gamma=1}^\Gamma \left( e^{iH^\mr{LP} (\beta_{v(\gamma+1)}-\beta_{v\gamma})t} e^{-iH_{\pi_v(\gamma)}^\mr{Add} \alpha_{v\gamma} t}\right) \right] e^{iH^\mr{LP} \beta_{11}t}.
\end{equation}
The correspondence between the time-independent and -dependent PFs, i.e., the relation $\bar{S}(t,0)=\sum_l e^{-il\omega t}\braket{l|\bar{T}(t)|0}$, still holds under this modification due to $e^{iH^\mr{LP}\beta_{11}t}\ket{0}=\ket{0}$.
Then, the corresponding PF $\bar{T}(t)$ is Eq. (\ref{Eq_Err:PF_equiv}) multiplied by $e^{iH_0 \beta_{11}t}$ on the right.
The evolution time under $H_\gamma$ in $\bar{T}(t)$ is $t$ for every $\gamma=0,1,\cdots,\Gamma$, and hence the satisfaction of the order condition $\bar{T}(t)=e^{-iHt}+\order{t^{p+1}}$ is allowed.

The error bound of the time-dependent PFs should be modified from Theorem \ref{Thm_Err:Main_theorem} in the case of $\beta_{11} \neq 0$ as well.
Section \ref{Sec:Error} employs the error analysis of the time-dependent PF \cite{childs2021-trotter} for $\bar{T}^F(t)$.
It is sufficient to replace the expression Eqs. (\ref{Eq_Err:T_F_rewrite}) and (\ref{Eq_Err:H_tilde_def}) by
\begin{equation}\label{EqA_beta:T_F_rewrite_modified}
    \bar{T}^F(t) \equiv \prod_{v=0}^{V_p} \prod_{\tilde{\gamma}=1}^{2\Gamma} e^{-i \tilde{H}_{v \tilde{\gamma}} \tilde{\alpha}_{v \tilde{\gamma}} t}, \quad (\tilde{H}_{v\tilde{\gamma}}, \tilde{\alpha}_{v \tilde{\gamma}}) = \begin{cases}
        (H_{\pi_v(\gamma)}^\mr{Add},0) & \text{if $v=0$, $\tilde{\gamma}=2\gamma-1$} \\
        (-H^\mr{LP}, \beta_{11} \delta_{\gamma,1}) & \text{if $v=0$, $\tilde{\gamma}=2\gamma$} \\
        (H_{\pi_v(\gamma)}^\mr{Add},\alpha_{v \gamma}) & \text{if $v \neq 0$, $\tilde{\gamma}=2\gamma-1$} \\
        (-H^\mr{LP},\beta_{v(\gamma+1)}-\beta_{v\gamma}) & \text{if $v \neq 0$, $\tilde{\gamma}=2\gamma$}
    \end{cases},
\end{equation}
where $\gamma$ is an integer in $\{1,2,\cdots,\Gamma\}$ for reproducing the modified PF by Eq. (\ref{EqA_beta:modified_T_F_Std}).
It shares the structure with the original expression by Eqs. (\ref{Eq_Err:T_F_rewrite}) and (\ref{Eq_Err:H_tilde_def}) in that they are organized by the time-evolution operators under $H_\gamma^\mr{Add}$ or $H^\mr{LP}$ and that their coefficients satisfy $|\tilde{\alpha}_{v\tilde{\gamma}}| \leq 1$.
On the other hand, the number of the layers $V_p$ is replaced by $V_p+1$.
As a result, we obtain the following error bound valid for the case with $\beta_{11} \neq 0$.

\begin{corollary}\label{CorA_beta:Std_error_beta_nonzero}
\textbf{(Standard PF error in the case of $\beta_{11} \neq 0$)}

Suppose that the coefficients $\{\alpha_{v\gamma}, \beta_{v\gamma}\}$ are chosen so that a time-independent PF,
\begin{equation}
    \bar{T}(t) = \left[ \prod_{v=1}^{V_p} \prod_{\gamma=1}^\Gamma \left( e^{-iH_0 (\beta_{v(\gamma+1)}-\beta_{v\gamma})t} e^{-i H_{\pi_v(\gamma)} \alpha_{v\gamma}t} \right)\right] e^{-iH_0 \beta_{11}t}
\end{equation}
can satisfy the order condition $\bar{T}(t)=e^{-iHt}+\order{t^{p+1}}$ for the Hamiltonian $H=\sum_{\gamma=0}^\Gamma H_\gamma$.
The standard PF $\bar{S}(t,0)$ by Eq. (\ref{Eq_Pre:Standard_PF}) has an error bounded by
\begin{equation}\label{EqA_beta:Error_Std_modified}
    \norm{U(t,0)-\bar{S}(t,0)} \leq 4 [ (V_p+1) t ]^{p+1} \max_{\tau \in [0,t]} \bar{\alpha}_{\mr{com},p}(\tau).
\end{equation}
\end{corollary}

\textbf{Proof.---}
We follow the calculation in the proof of Theorem \ref{Thm_Err:Main_theorem}.
The explicit error bound is immediately obtained by replacing $V_p$ in Eq. (\ref{Eq_Err:PF_bound_theorem}) with $V_p+1$. $\quad \square$

We realize that the upper bound Eq. (\ref{EqA_beta:Error_Std_modified}) excessively overestimates the PF error since Eq. (\ref{EqA_beta:T_F_rewrite_modified}) contains redundant copies of $H_{\pi_v(\gamma)}^\mr{Add}$ for $v=0$.
However, the above error bound is sufficient for ensuring that the scalings of the error bound or the computational cost in the case of $\beta_{11} \neq 0$ are exactly the same as those for $\beta_{11}=0$, discussed in the main text.
When the standard PF has uniform coefficients as Definition \ref{Def_Pre:uniform}, the quantity $\bar{\alpha}_{\mr{com},p}(\tau)$ in Eq. (\ref{EqA_beta:Error_Std_modified}) is replaced by $\bar{\alpha}_{\mr{com},p}^{\mr{uni}}(\tau)$ like Corollary \ref{Cor_Err:error_uniform}.

The same discussion goes also for the generalized PF $S(t,0)$.
In the corresponding time-independent PF for a Hamiltonian $H=\sum_{\gamma=1}^\Gamma H_\gamma + (\Gamma-1) H_0$ by Eq. (\ref{Eq_Err:T_F_ctrpart_Gen}), the total evolution time under $H_0$ amounts to
\begin{equation}
    -\sum_{v=1}^{V_p} \sum_{\gamma=1} (\beta_{v(\gamma+1)}-\beta_{v\gamma} -\alpha_{v\gamma}) t = (\Gamma-1+\beta_{11})t.
\end{equation}
We use the assumption Eq. (\ref{Eq_Pre:coef_1}) for the above summation of the coefficients $\{\alpha_{v\gamma}\}$.
The evolution time does not coincides with $t$ unless $\beta_{11} \neq 0$, implying the impossibility of the order condition. 
In a similar manner to the standard PF, it is sufficient to employ the modified time-independent PF,
\begin{equation}
    T^F(t) = \left[ \prod_{v=1}^{V_p} \prod_{\gamma=1}^\Gamma \left( e^{iH^\mr{LP} (\beta_{v(\gamma+1)}-\beta_{v\gamma}-\alpha_{v\gamma})t} e^{-i H_{\pi_v(\gamma)}^F \alpha_{v\gamma}t} \right)\right] e^{iH^\mr{LP} \beta_{11}t}
\end{equation}
instead of Eq. (\ref{Eq_Rel:T_F_Gen}), which satisfies the relation, Eq. (\ref{Eq_Rel:Rel_PF_Gen}).
As a result, we obtain the error bound of the generalized PF $S(t,0)$ in the case of $\beta_{11} \neq 0$ as follows.

\begin{corollary}
\textbf{(Generalized PF error in the case of $\beta_{11} \neq 0$)}

Suppose that the coefficients $\{\alpha_{v\gamma}, \beta_{v\gamma}\}$ are chosen so that a time-independent PF,
\begin{equation}
    \bar{T}(t) = \left[ \prod_{v=1}^{V_p} \prod_{\gamma=1}^\Gamma \left( e^{iH_0 (\beta_{v(\gamma+1)}-\beta_{v\gamma}t-\alpha_{v\gamma})} e^{-i H_{\pi_v(\gamma)} \alpha_{v\gamma}t} \right)\right] e^{iH_0 \beta_{11}t}
\end{equation}
can satisfy the order condition $T(t)=e^{-iHt}+\order{t^{p+1}}$ for the Hamiltonian $H=\sum_{\gamma=1}^\Gamma H_\gamma + (\Gamma-1) H_0$.
The generalized PF $S(t,0)$ by Eq. (\ref{Eq_Pre:Generalized_PF}) has an error bounded by
\begin{equation}
    \norm{U(t,0)-\bar{S}(t,0)} \leq 5 [ (V_p+1) t ]^{p+1} \max_{\tau \in [0,t]} \alpha_{\mr{com},p}(\tau).
\end{equation}
\end{corollary}

\textbf{Proof.---} It is sufficient to replace $V_p$ by $V_p+1$ in Theorem \ref{Thm_Err:main_thm_Gen} like the proof of Corollary \ref{CorA_beta:Std_error_beta_nonzero}. $\quad \square$

\section{Proof of time-dependent MPF error}\label{A_Sec:MPF}

\renewcommand{\thetheorem}{\thesection\arabic{theorem}}
\setcounter{theorem}{0}

\renewcommand{\thefigure}{\thesection\arabic{figure}}
\setcounter{figure}{0}

\begin{figure*}
    \centering
    \includegraphics[width=\linewidth]{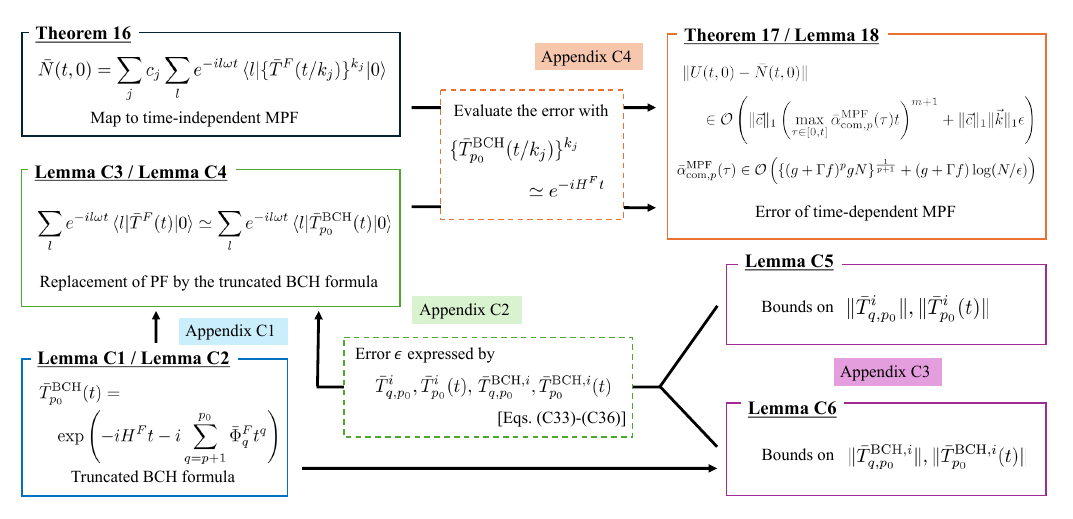}
    \caption{Proof sketch for an upper bound on the time-dependent MPF error.}
    \label{Fig:Proof_MPF}
\end{figure*}

In this section, we provide the detailed derivation of the error for the time-dependent multi-product formula (MPF), discussed in Section \ref{Sec:MPF}.
The time-dependent MPFs are related to the time-independent ones by Theorem \ref{Thm_MPF:MPF_relation}.
As a result, their errors are also related to time-independent ones like Eq. (\ref{Eq_MPF:Rel_MPF_error}).
Throughout this appendix, we mainly focus on the standard MPF $\bar{N}(t)$ by Eq. (\ref{Eq_MPF:Standard_MPF}).
The derivation of the time-dependent MPF error is sketched in Fig.~\ref{Fig:Proof_MPF}, which is parallel to the time-independent case \cite{mizuta2025-mpf}.
First, we introduce the truncated Baker-Campbell-Hausdorff (BCH) formula for approximating the time-independent PF and discuss its properties in Appendix \ref{SubsecA:truncated_BCH}.
We next show that it can approximate the time-dependent MPF with an arbitrarily small error in Appendix \ref{SubsecA:err_truncated_BCH}, while some inequalities used in this part are proven in Appendix \ref{SubsecA:T_q_expansion}.
Finally, by evaluating the error between the exact time-evolution operator and the approximate one composed of the truncated BCH formulas in Appendix \ref{SubsecA:err_proof_MPF}, we arrive at the upper bound of the time-dependent MPF error shown as Theorem \ref{Thm_MPF:error_MPF}.

As discussed in the beginning of Section \ref{Sec:MPF}, the Hamiltonian $H(t)$ is composed of the terms $H_\gamma(t)$ in class $C^\infty$.
It is supposed to be $k$-local and have the extensiveness $g$ defined by Eq. (\ref{Eq_Cost:extensiveness}).
We also assume the existence of the degree of time-dependency $f$ satisfying Eq. (\ref{Eq_Cost:time_dependency}) for any $q \in \bbN$.

\subsection{Baker-Campbell-Hausdorff (BCH) formula and its properties}\label{SubsecA:truncated_BCH}

In order to evaluate the time-independent MPF error $\bar{M}^F(t)-e^{-iH^Ft}$, we first consider the error $\{ T (t/k_j) \}^{k_j} - e^{-iH^Ft}$.
It can be calculated based on the Baker-Campbell-Hausdorff (BCH) formula like $e^{At}e^{Bt}=e^{(A+B)t+[B,A]t^2 + \order{t^3}}$.
The order-$p_0$ ($p_0 \in \bbN$) BCH formula for the time-independent PF $\bar{T}^F(t)$ is defined by
\begin{equation}\label{EqA_MPF:BCH_p0}
    \bar{T}_{p_0}^{\text{BCH}}(t) = \exp \left( -iH^Ft -i \sum_{q=p+1}^{p_0} \bar{\Phi}_q^F t^q \right) = \exp \left( -i \sum_{q=1}^{p_0} \bar{\Phi}_q^F t^q \right),
\end{equation}
where the order-$q$ operator $\bar{\Phi}_q^F$ is defined by $\bar{\Phi}_1^F=H^F$ and 
\begin{equation}\label{EqA_MPF:Phi_q_F}
    \bar{\Phi}_q^F = (-i)^{q-1} \sum_{\substack{q_{11},\cdots,q_{V_p2\Gamma} \geq 0; \\ q_{11}+\cdots+q_{V_p2\Gamma} = q}} \left( \prod_{v=1}^{V_p} \prod_{\tilde{\gamma}=1}^{2\Gamma} \frac{(\tilde{\alpha}_{v\tilde{\gamma}})^{q_{v\tilde{\gamma}}}}{q_{v\tilde{\gamma}}!}\right) \phi_q (\underbrace{\tilde{H}_{V_p2\Gamma},\cdots,\tilde{H}_{V_p2\Gamma}}_{q_{V_p2\Gamma}}, \cdots,\underbrace{\tilde{H}_{v\tilde{\gamma}},\cdots,\tilde{H}_{v\tilde{\gamma}}}_{q_{v\tilde{\gamma}}},\cdots,\underbrace{\tilde{H}_{11},\cdots,\tilde{H}_{11}}_{q_{11}}),
\end{equation}
\begin{equation}\label{EqA_MPF:phi_q_com}
    \phi_q (H_1,\cdots,H_q) = \sum_{\sigma \in \mcl{S}_q} \frac{(-1)^{d_\sigma} d_\sigma! (q-1-d_\sigma)!}{q^2 (q-1)!} [H_{\sigma(1)},\cdots,[H_{\sigma(q-1)},H_{\sigma(q)}]]],
\end{equation}
for $q \geq 2$.
The symbol $\mcl{S}_q$ is the symmetric group of $q$ elements and $d_\sigma$ is the number of adjacent pairs such that $\sigma(i+1)<\sigma(i)$, i.e., $d_\sigma = |\{ i \, | \, \sigma(i+1)<\sigma(i) \}|$.
We have $\bar{\Phi}_q^F=0$ for $q=2,3,\cdots,p$ due to the order condition of the PF, $\bar{T}^F(t)=e^{-iH^Ft}+\order{t^{p+1}}$.
At the same time, the order condition for the order-$p_0$ BCH formula implies $\bar{T}_{p_0}^{\text{BCH}}(t)=\bar{T}^F(t)+\order{t^{p_0+1}}$.
Each order-$q$ term $\bar{\Phi}_q^F$ in the BCH formula has the following property.

\begin{lemma}\label{LemA_MPF:Phi_q_bound}
\textbf{}

Each term in the BCH formula, $\bar{\Phi}_q^F$ by Eq. (\ref{EqA_MPF:Phi_q_F}) is written in the form of
\begin{equation}\label{EqA_MPF:Phi_q_F_form}
    \bar{\Phi}_q^F = \sum_{m \in \bbZ} \mr{Add}_m \otimes \bar{\Phi}_{q,m}, \quad \bar{\Phi}_{q,m} = \frac1T \int_0^T \dd t \bar{\Phi}_q(t) e^{im\omega t}.
\end{equation}
The operator $\bar{\Phi}_q(t)$ is bounded by
\begin{equation}\label{EqA_MPF:Phi_q_t_bound}
    \norm{\bar{\Phi}_q(t)} \leq (V_p)^q \bar{\alpha}_{\mr{com},q-1}(t) \leq (q-1)! (V_p)^q (2kg+\Gamma f)^{q-1} Ng,
\end{equation}
where $\bar{\alpha}_{\mr{com},q}(t)$ is defined by Eq. (\ref{Eq_Sum:alpha_com_Std}).
\end{lemma}

\textbf{Proof.---}
By the definition $\bar{\Phi}_1^F=H^F$, we have $\bar{\Phi}_1(t)=H(t)$ and $\norm{\bar{\Phi}_1(t)} \leq Ng$, which completes the proof for $q=1$.
We focus on the case with $q \geq 2$.
Equations (\ref{EqA_MPF:Phi_q_F}) and (\ref{EqA_MPF:phi_q_com}) say that $\bar{\Phi}_q^F$ is composed of nested commutators among $\{ H_\gamma^\mr{Add} \}$ and $H^\mr{LP}$.
Since they have the form like Eq. (\ref{Eq_Err:Multi_Com_repr}) as discussed in Section \ref{Subsec:form_commutators}, the operator $\bar{\Phi}_q^F$ is also written in the same form.
We next evaluate the upper bound on $\norm{\bar{\Phi}_q(t)}$.
Each $q$-fold nested commutator $[\tilde{H}_{v_q\tilde{\gamma}_q}, \cdots,[\tilde{H}_{v_2\tilde{\gamma}_2},\tilde{H}_{v_1\tilde{\gamma}_1}]]$ in the operator $\bar{\Phi}_q^F$ comes from the permutation $\sigma \in S_q$ such that $(\sigma(1),\cdots,\sigma(q))=(v_q\tilde{\gamma}_q,\cdots,v_1\tilde{\gamma}_1)$ under fixed indices $q_{11},\cdots,q_{V_p2\Gamma}$ satisfying $q_{11}+\cdots+q_{V_p2\Gamma}=q$.
Let us denote the set of such permutations by $S_q'$, and then we have $|S_q'|=q_{11}!\cdots q_{V_p2\Gamma}!$.
The absolute value of the nested commutator $[\tilde{H}_{v_q\tilde{\gamma}_q}, \cdots,[\tilde{H}_{v_2\tilde{\gamma}_2},\tilde{H}_{v_1\tilde{\gamma}_1}]]$ is bounded by
\begin{equation}
    \left| \left( \prod_{v=1}^{V_p} \prod_{\tilde{\gamma}=1}^{2\Gamma} \frac{(\tilde{\alpha}_{v\tilde{\gamma}})^{q_{v\tilde{\gamma}}}}{q_{v\tilde{\gamma}}!}\right) \sum_{\sigma \in S_q'} \frac{(-1)^{d_\sigma} d_\sigma !(q-1-d_\sigma)!}{q^2 (q-1)!}\right| \leq \frac{1}{q_{11}!\cdots q_{V_p2\Gamma}!} \frac{|S_q'|}{q^2} = \frac1{q^2}.
\end{equation}
When we move to the representation in the time domain, the commutator $\ad_{\tilde{H}_{v\tilde{\gamma}}}$ is replaced by $\bar{D}_{v\tilde{\gamma}}(t)$ as shown in Lemma \ref{Lem_Err:Form_commutators}.
We should be careful of the parity of the index $\tilde{\gamma}_1$.
For an odd integer $\tilde{\gamma}_1=2\gamma_1-1$, the nested commutator $[\tilde{H}_{v_q\tilde{\gamma}_q}, \cdots,[\tilde{H}_{v_2\tilde{\gamma}_2},\tilde{H}_{v_1\tilde{\gamma}_1}]]$ gives rise to the term $\prod_{q'=2}^p \bar{D}_{v_{q'}\tilde{\gamma}_{q'}}(t) H_{\gamma_1}(t)$ in the operator $\bar{\Phi}_q(t)$ following Eq. (\ref{Eq_Err:Com_gamma_Add}).
On the other hand, when $\tilde{\gamma}_1$ is even, the commutator $[\tilde{H}_{v_2\tilde{\gamma}_2},\tilde{H}_{v_1\tilde{\gamma}_1}]$ survives only for an odd integer $\tilde{\gamma}_2=2\gamma_2-1$, and then it gives $i \dv{t} H_{\gamma_2}(t)$ according to Eq. (\ref{Eq_Err:Com_gamma_LP}).
Therefore, the operator $\bar{\Phi}_q^F(t)$ is bounded by
\begin{eqnarray}
    \norm{\bar{\Phi}_q(t)} &\leq& \frac1{q^2} \sum_{v_1=1}^{V_p} \sum_{\gamma_1=1}^\Gamma \sum_{v_2\tilde{\gamma}_2,\cdots,v_q\tilde{\gamma}_q=11}^{V_p 2\Gamma} \norm{\left(\prod_{q'=2}^p \bar{D}_{v_{q'}\tilde{\gamma}_{q'}}(t)\right) H_{\gamma_1}(t)} \nonumber \\
    && \qquad  \qquad  \qquad 
    + \frac1{q^2} \sum_{v_1=1}^{V_p} \sum_{\gamma_1=1}^\Gamma \sum_{v_2=1}^{V_p} \sum_{\gamma_2=1}^\Gamma \sum_{v_3\tilde{\gamma}_3,\cdots,v_q\tilde{\gamma}_q=11}^{V_p 2\Gamma} \norm{\left(\prod_{q'=3}^p \bar{D}_{v_{q'}\tilde{\gamma}_{q'}}(t)\right) \dv{t} H_{\gamma_2}(t)} \nonumber \\
    &\leq& \frac2{q^2} (V_p)^q \sum_{\gamma_1=1}^\Gamma \sum_{\gamma_2,\cdots,\gamma_p=1}^{\Gamma+1} \norm{\left( \prod_{q=2}^p \bar{\mcl{D}}_{\gamma_q}(t)\right) H_{\gamma_1}(t)} \leq (V_p)^q \bar{\alpha}_{\mr{com},q-1}(t),
\end{eqnarray}
which completes the proof of the first inequality in Eq. (\ref{EqA_MPF:Phi_q_t_bound}).
The second inequality in Eq. (\ref{EqA_MPF:Phi_q_t_bound}) immediately comes from Theorem \ref{Thm_Cost:bound_commutators}. $\quad \square$

We next consider the locality, extensiveness, and the strength of time-dependency of the operator $\bar{\Phi}_q(t)$ as a Hamiltonian.
The operator $\bar{\Phi}_q(t)$ is written in the form of
\begin{equation}
    \bar{\Phi}_q(t) = \sum_{X \subset \Lambda} h_X^{\bar{\Phi}_q}(t),
\end{equation}
where each $h_X^{\bar{\Phi}_q}(t)$ denotes a local operator nontrivially acting on the domain $X$.
In a similar manner to Eq. (\ref{Eq_Cost:extensiveness}), we define the extensiveness $g(\bar{\Phi}_q)$ by a number such that the relation,
\begin{equation}\label{EqA_MPF:g_Phi_q}
    \max_{i \in \Lambda} \sup_{\tau \in [0,t]} \sum_{X \ni i} \norm{h_X^{\bar{\Phi}_q}(\tau)} \leq g(\bar{\Phi}_q),
\end{equation}
holds.
We also define the strength of time-dependency $f(\bar{\Phi}_q)$ by a number such that the relation,
\begin{equation}\label{EqA_MPF:f_Phi_q}
    \max_{i \in \Lambda} \sup_{\tau \in [0,t]}\sum_{X \ni i} \norm{\dv[q']{\tau} h_X^{\bar{\Phi}_q}(\tau)} \leq [ f(\bar{\Phi}_q) ]^{q'} g(\bar{\Phi}_q),
\end{equation}
holds for every $q' \in \bbN$.
We characterize these quantities based on the counterparts of the Hamiltonian $H(t)$ as follows.

\begin{lemma}\label{LemA_MPF:Phi_q_extensive}
\textbf{}

The operator $\bar{\Phi}_q(t)$ defined by Eq. (\ref{EqA_MPF:Phi_q_F_form}) is at most $qk$-local.
In addition, its extensiveness and strength of time-dependency can be chosen as
\begin{equation}\label{EqA_MPF:gf_Phi_values}
    g(\bar{\Phi}_q) = q! (V_p)^q (2kg+\Gamma f)^{q-1} g, \quad f(\bar{\Phi}_q) = qf,
\end{equation}
which satisfy the relations Eqs. (\ref{EqA_MPF:g_Phi_q}) and (\ref{EqA_MPF:f_Phi_q}).
\end{lemma}

\textbf{Proof.---}
The case with $q = 1$ is trivial and we consider $q \geq 2$.
The operator $\bar{\Phi}_q(t)$ is composed of $q$-fold nested commutators among $H_\gamma (t)$ and their time derivative.
Since each $H_\gamma (t)$ is $k$-local and the time derivative preserves the locality, the operator $\bar{\Phi}_q(t)$ is at most $qk$-local.

We evaluate the quantities $g(\bar{\Phi}_q)$ and $f(\bar{\Phi}_q)$.
Let us expand the operator $\prod_{q'=2}^q \bar{\mcl{D}}_{\gamma_{q'}}(\tau) H_{\gamma_1}(\tau)$ by
\begin{equation}\label{EqA_MPF:prod_D_H_local}
    \left[\prod_{q'=2}^q \bar{\mcl{D}}_{\gamma_{q'}}(\tau) \right] H_{\gamma_1}(\tau) = \sum_{X \subset \Lambda} \bar{h}_X^{\gamma_1,\cdots,\gamma_q}(\tau),
\end{equation}
with a local term $\bar{h}_X^{\gamma_0,\cdots,\gamma_q}(\tau)$ nontrivially acting on a domain $X$.
The discussion on the locality says that $\bar{h}_X^{\gamma_0,\cdots,\gamma_q}(\tau)$ can be nonzero only for a domain $X$ such that $|X| \leq qk$.
As well as the discussion in the proof of Lemma \ref{LemA_MPF:Phi_q_bound}, the local terms in $\bar{\Phi}_q(t)$ is covered by $(V_p)^q$ copies of the nested commutators among $\{ H_\gamma(t)\}$ and their time derivative with a coefficient smaller than $2/q^2 \leq 1$, which leads to the relations,
\begin{eqnarray}
    \sum_{X \ni i} \norm{h_X^{\bar{\Phi}_q}(\tau)} &\leq& (V_p)^q \sum_{X \ni i}\sum_{\gamma_1=1}^\Gamma \sum_{\gamma_2,\cdots,\gamma_q=1}^{\Gamma+1} \norm{\bar{h}_X^{\gamma_1,\cdots,\gamma_q}(\tau)}, \\
    \sum_{X \ni i} \norm{\dv[q']{\tau} h_X^{\bar{\Phi}_q}(\tau)} &\leq& (V_p)^q \sum_{X \ni i}\sum_{\gamma_1=1}^\Gamma \sum_{\gamma_2,\cdots,\gamma_q=1}^{\Gamma+1} \norm{\dv[q']{\tau} \bar{h}_X^{\gamma_1,\cdots,\gamma_q}(\tau)}.
\end{eqnarray}
We aim to find the quantities $g(\bar{\Phi}_q)$ and $f(\bar{\Phi}_q)$ such that the inequalities 
\begin{equation}\label{EqA_MPF:g_f_Phi_q_local}
    (V_p)^q \sum_{X \ni i}\sum_{\gamma_1=1}^\Gamma \sum_{\gamma_2,\cdots,\gamma_q=1}^{\Gamma+1} \norm{\bar{h}_X^{\gamma_1,\cdots,\gamma_q}(\tau)} \leq g(\bar{\Phi}_q), \quad  (V_p)^q \sum_{X \ni i}\sum_{\gamma_1=1}^\Gamma \sum_{\gamma_2,\cdots,\gamma_q=1}^{\Gamma+1} \norm{\dv[q']{\tau} \bar{h}_X^{\gamma_1,\cdots,\gamma_q}(\tau)} \leq [ f(\bar{\Phi}_q) ]^{q'} g(\bar{\Phi}_q)
\end{equation}
are satisfied.

We determine the quantities $g(\bar{\Phi}_q)$ and $f(\bar{\Phi}_q)$ recursively.
Let us consider the extensiveness for $q+1$.
The local terms of Eq. (\ref{EqA_MPF:prod_D_H_local}) for $q+1$ are composed of the commutators $[h_{X'}^{\gamma_{q+1}}(\tau),\bar{h}_X^{\gamma_1,\cdots,\gamma_q}(\tau)]$, which come from the application of $\bar{D}_{\gamma_{q+1}}(\tau)$ for $\gamma_{q+1}=1,\cdots,\Gamma$, and the time derivative $\Gamma \dv{\tau} \bar{h}_X^{\gamma_1,\cdots,\gamma_q}(\tau)$, which comes from that for $\gamma_{q+1}=\Gamma+1$.
As a result, the extensiveness is evaluated by
\begin{eqnarray}
    && (V_p)^{q+1} \sum_{X \ni i}\sum_{\gamma_1=1}^\Gamma \sum_{\gamma_2,\cdots,\gamma_{q+1}=1}^{\Gamma+1} \norm{\bar{h}_X^{\gamma_1,\cdots,\gamma_{q+1}}(\tau)} \nonumber \\
    && \qquad \leq (V_p)^{q+1} \sum_{\gamma_1=1}^\Gamma \sum_{\gamma_2,\cdots,\gamma_q=1}^{\Gamma+1} \left( \sum_{\substack{X,X' \subset \Lambda; \\ X \cup X' \ni i}}  \sum_{\gamma_{q+1}=1}^\Gamma  \norm{[h_{X'}^{\gamma_{q+1}}(\tau),\bar{h}_X^{\gamma_1,\cdots,\gamma_q}(\tau)]} + \sum_{X \ni i} \norm{\Gamma\dv{\tau} \bar{h}_X^{\gamma_1,\cdots,\gamma_q}(\tau)} \right) \nonumber \\
    && \qquad \leq (V_p)^{q+1} \sum_{\gamma_1,\gamma_{q+1}=1}^\Gamma \sum_{\gamma_2,\cdots,\gamma_q=1}^{\Gamma+1} \sum_{\substack{X,X' \subset \Lambda; \\ X \cup X' \ni i}} \norm{[h_{X'}^{\gamma_{q+1}}(\tau),\bar{h}_X^{\gamma_1,\cdots,\gamma_q}(\tau)]} + V_p\Gamma f(\bar{\Phi}_q) g(\bar{\Phi}_q). \label{EqA_MPF:sum_Phi_q_local}
\end{eqnarray}
The second term comes from the existence of $f(\bar{\Phi}_q)$ saturating Eq. (\ref{EqA_MPF:g_f_Phi_q_local}).
In the first term, only the contributions from $X,X' \in \Lambda$ such that $X \cup X' \ni i$ and $X \cap X' \neq \phi$ survive.
Since $X$ or $X'$ includes the site $i$, Eq. (\ref{EqA_MPF:sum_Phi_q_local}) is further bounded by    
\begin{eqnarray}
    [\text{Eq. (\ref{EqA_MPF:sum_Phi_q_local})}] &\leq& (V_p)^{q+1} \sum_{\gamma_1,\gamma_{q+1}=1}^\Gamma \sum_{\gamma_2,\cdots,\gamma_q=1}^{\Gamma+1} \left( \sum_{X \ni i} \sum_{X'; X \cap X' \neq \phi} + \sum_{X' \ni i} \sum_{X; X \cap X' \neq \phi}\right) \norm{[h_{X'}^{\gamma_{q+1}}(\tau),\bar{h}_X^{\gamma_1,\cdots,\gamma_q}(\tau)]} + V_p\Gamma f(\bar{\Phi}_q) g(\bar{\Phi}_q) \nonumber \\
    &\leq& (V_p)^{q+1} \sum_{\gamma_1,\gamma_{q+1}=1}^\Gamma \sum_{\gamma_2,\cdots,\gamma_q=1}^{\Gamma+1} \left( \sum_{X \ni i} \sum_{j \in X} \sum_{X' \ni j} + \sum_{X' \ni i} \sum_{j \in X'}\sum_{X \ni j}\right) 2 \norm{h_{X'}^{\gamma_{q+1}}(\tau)} \norm{\bar{h}_X^{\gamma_1,\cdots,\gamma_q}(\tau)} + V_p\Gamma f(\bar{\Phi}_q) g(\bar{\Phi}_q) \nonumber \\
    &\leq& V_p (qk+k) 2g \cdot g(\bar{\Phi}_q) + V_p\Gamma f(\bar{\Phi}_q) g(\bar{\Phi}_q). \label{EqA_MPF:sum_Phi_q_local_2}
\end{eqnarray}
In the last inequality, we use the fact that the domains of the local terms, $X$ and $X'$, respectively contain at most $qk$ and $k$ sites due to the locality.
We also use $g(\bar{\Phi}_q)$ that saturates Eq. (\ref{EqA_MPF:g_f_Phi_q_local}).
Thus, one can choose the extensiveness $g(\bar{\Phi}_{q+1})$ by
\begin{equation}\label{EqA_MPF:g_Phi_q_recursive}
    g(\bar{\Phi}_{q+1}) = V_p \{ 2(q+1)kg+\Gamma f(\bar{\Phi}_q) + \Gamma f\} g(\bar{\Phi}_q),
\end{equation}
which is strictly larger than Eq. (\ref{EqA_MPF:sum_Phi_q_local_2}).
The quantity $\Gamma f$ is added just for solving this recursive relation.

The strength of time-dependency is evaluated in a similar way.
Considering the Leibnitz rule, the time derivative of the local terms is bounded by
\begin{eqnarray}
    && (V_p)^{q+1} \sum_{X \ni i}\sum_{\gamma_1=1}^\Gamma \sum_{\gamma_2,\cdots,\gamma_{q+1}=1}^{\Gamma+1} \norm{\dv[q']{\tau}\bar{h}_X^{\gamma_1,\cdots,\gamma_{q+1}}(\tau)} \nonumber \\
    && \qquad \leq (V_p)^{q+1} \sum_{\gamma_1=1}^\Gamma \sum_{\gamma_2,\cdots,\gamma_q=1}^{\Gamma+1} \left( \sum_{\substack{X,X' \subset \Lambda; \\ X \cup X' \ni i}}  \sum_{\gamma_{q+1}=1}^\Gamma  \norm{\dv[q']{\tau}[h_{X'}^{\gamma_{q+1}}(\tau),\bar{h}_X^{\gamma_1,\cdots,\gamma_q}(\tau)]} + \sum_{X \ni i} \norm{\Gamma\dv[q'+1]{\tau} \bar{h}_X^{\gamma_1,\cdots,\gamma_q}(\tau)} \right) \nonumber \\
    && \qquad \leq (V_p)^{q+1} \sum_{\gamma_1,\gamma_{q+1}=1}^\Gamma \sum_{\gamma_2,\cdots,\gamma_q=1}^{\Gamma+1} \sum_{\substack{X,X' \subset \Lambda; \\ X \cup X' \ni i}} \sum_{q''=0}^{q'} \frac{q'!}{q''!(q'-q'')!} \norm{\left[  h_{X'}^{\gamma_{q+1} (q'')}(\tau), \bar{h}_X^{\gamma_1,\cdots,\gamma_q (q'-q'')}(\tau)\right]}+ V_p \Gamma [ f(\bar{\Phi}_q) ]^{q'+1} g(\bar{\Phi}_q) \nonumber \\
    && \qquad \leq \sum_{q''=0}^{q'} \frac{q'!}{q''!(q'-q'')!} V_p (qk+k) 2 f^{q''} g \cdot [ f(\bar{\Phi}_q) ]^{q'-q''} g(\bar{\Phi}_q) +  V_p \Gamma [ f(\bar{\Phi}_q) ]^{q'+1} g(\bar{\Phi}_q) \nonumber \\
    && \qquad \leq [ f(\bar{\Phi}_q) + f]^{q'} V_p [ 2(q+1)kg+\Gamma f(\bar{\Phi}_q)] g(\bar{\Phi}_q) \leq [ f(\bar{\Phi}_q) + f]^{q'} g(\bar{\Phi}_{q+1}).
\end{eqnarray}
Thus, one can choose the strength of time-dependency by
\begin{equation}
    f(\bar{\Phi}_{q+1}) = f(\bar{\Phi}_q) + f.
\end{equation}
We immediately obtain $f(\bar{\Phi}_q)$ with the initial condition $f(\bar{\Phi}_1)=f$.
We substitute this into Eq. (\ref{EqA_MPF:g_Phi_q_recursive}) and solve the resulting recursive relation $g(\bar{\Phi}_{q+1})=(q+1) V_p (2kg+\Gamma f) g(\bar{\Phi}_q)$ with $g(\bar{\Phi}_1)=V_p g$.
This results in $g(\bar{\Phi}_q)=q!(V_p)^q (2kg+\Gamma f)^{q-1} g$. $\quad \square$

\subsection{Error by the truncated Baker-Campbell-Hausdorff formula}\label{SubsecA:err_truncated_BCH}

The next step for evaluating the error bound is to interleave the truncated BCH formula between the exact time-evolution operator and the time-dependent MPF.
Although the power of the time-independent PF $[ \bar{T}^F(t/k_j) ]^{k_j}$ in the MPF is not expressed in a simple way, that for the truncated BCH formula can be easily calculated as
\begin{equation}\label{EqA_MPF:power_BCH}
    [ \bar{T}_{p_0}^\mr{BCH}(t/k_j) ]^{k_j} = \exp \left( -iH^Ft - i \sum_{q=p+1}^{p_0} \bar{\Phi}_q^F \frac{t^q}{(k_j)^{q-1}}\right).
\end{equation}
Thus, the evaluation of the error $[ \bar{T}_{p_0}^\mr{BCH}(t/k_j) ]^{k_j}-e^{-iH^Ft}$ is easier than that of $[ \bar{T}_F(t/k_j) ]^{k_j} - e^{-iH^Ft}$ included in Eq. (\ref{Eq_MPF:Rel_MPF_error}).
We leave the calculation of the former one in Appendix \ref{SubsecA:err_proof_MPF}, and we begin with considering the error caused by adopting $\bar{T}_{p_0}^\mr{BCH}(t/k_j)$ instead of $\bar{T}_F(t/k_j)$ in this section.

First, we focus on the error by replacing the PF for a single step, i.e.,
\begin{equation}
    \sum_{l \in \bbZ} e^{-il\omega t}\braket{l|\bar{T}^F(t)|0} - \sum_{l \in \bbZ} e^{-il\omega t} \braket{l|\bar{T}_{p_0(N,\epsilon)}^\mr{BCH}(t)|0}.
\end{equation}
We prove the following upper bound by using the locality of Hamiltonians.

\begin{lemma}\label{LemA_MPF:truncated_BCH_error}
\textbf{(Truncated BCH)}

Let $\epsilon$ be an arbitrary fixed value in $(0,1)$.
Suppose that the time $t$ is small enough to satisfy
\begin{equation}\label{EqA_MPF:BCH_t_condition}
    |t| \leq \frac{1}{4e^3 V_p p_0(N,\epsilon)(2kg+\Gamma f)} \in \order{\frac{1}{(g+\Gamma f)\log (N/\epsilon)}},
\end{equation}
where $p_0(N,\epsilon)= \lceil \log (2N/\epsilon) \rceil$ denotes the truncation order.
The error of replacing the time-independent PF $\bar{T}^F(t)$ by the truncated BCH formula $\bar{T}_{p_0}^\mr{BCH}(t)$ is bounded by
\begin{equation}\label{EqA_MPF:error_BCH}
    \norm{\sum_{l \in \bbZ} e^{-il\omega t}\braket{l|\left[ \bar{T}^F(t)- \bar{T}_{p_0(N,\epsilon)}^\mr{BCH}(t)\right]|0}} \leq \epsilon.
\end{equation}
\end{lemma}

\textbf{Remark.---}
We note that the truncation at the order $p_0(N,\epsilon)$ plays an essential role in the MPF.
If we sets $p_0(N,\epsilon) \to \infty$, the error Eq. (\ref{EqA_MPF:error_BCH}) becomes exactly zero, which seems to be preferable compared to the truncated case.
However, such a situation demands the convergence of the BCH formula Eq. (\ref{EqA_MPF:BCH_p0}).
According to its convergence radius, no bound tighter than $|t| \in \order{[N(g+\Gamma f)]^{-1}}$ is known in general, like the time-independent case \cite{mizuta2025-mpf}.
It is incompatible with the complexity respecting the commutator scaling since we use the time-dependent MPF for the small time $t/r \in \tilde{\Omega}(N^{-1/(p+1)})$ characterized by the Trotter number $r$ in Eq. (\ref{Eq_MPF:Trotter_number_MPF}).
On the other hand, this lemma exploits the asymptotic convergence up to the order $p_0(N,\epsilon)$ under the locality of Hamiltonians, allowing us to use the truncated BCH formula under the condition looser than the convergence radius, i.e., Eq. (\ref{EqA_MPF:BCH_t_condition}).
The time under Eq. (\ref{EqA_MPF:BCH_t_condition}) is compatible with the Trotter number by Eq. (\ref{Eq_MPF:Trotter_number_MPF}), whose system-size dependence reflects the locality via the commutator scaling.
The necessity of this truncation is present also in the time-independent MPF, and see Refs. \cite{aftab2024-mpf,mizuta2025-mpf} for its detailed discussion.

\textbf{Proof.---}
Let $p_0$ denote $p_0(N,\epsilon)$ for simplicity.
For the $N$-site lattice $\Lambda = \{1,2,\cdots,N\}$, we consider its subsystem,
\begin{equation}\label{EqA_MPF:subsystem}
    \Lambda_i = \begin{cases}
        \Lambda & (i=0) \\
        \{ i+1, \cdots, N \} & (i=1,\cdots,N-1) \\
        \phi & (i=N)
    \end{cases}.
\end{equation}
Let us define the subsystem Hamiltonians,
\begin{equation}\label{EqA_MPF:subsys_H}
    H^{\Lambda_i}(t) = \sum_{X \subset \Lambda_i} h_X(t), \quad H^{\Lambda_i}_\gamma (t) = \sum_{X \subset \Lambda_i} h_X^\gamma(t).
\end{equation}
The extensiveness and the strength of $H^{\Lambda_i}(t)$, which are counterparts of Eqs. (\ref{Eq_Cost:extensiveness}) and (\ref{Eq_Cost:time_dependency}), are respectively smaller than those of $H(t)$ by definition.
We consider the PFs and the BCH formula for the subsystem Hamiltonian $H^{\Lambda_i}(t)$, for which we attach the superscript $\Lambda_i$.
For instance, the standard PF is defined by
\begin{equation}\label{EqA_MPF:subsys_standard_PF}
    \bar{S}^{\Lambda_i}(t,0) = \prod_{v=1}^{V_p} \sum_{\gamma=1}^{\Gamma} e^{-i H_{\pi_v(\gamma)}^{\Lambda_i} (\beta_{v\gamma} t) \alpha_{v\gamma}t},
\end{equation}
and each term in the BCH formula $\bar{\Phi}_q^{F,\Lambda_i}$ is given in a similar manner to Eq. (\ref{EqA_MPF:Phi_q_F}).
We also define the subsystem PF $\bar{T}^{F,\Lambda_i}(t)$ and the subsystem BCH formula  and the subsystem BCH formula $\bar{T}^{\mr{BCH},\Lambda_i}_{p_0}(t)$ as well.
The time-independent PF $\bar{T}^F(t)$ and the BCH formula $\bar{T}^{\mr{BCH}}_{p_0}(t)$ are respectively decomposed into
\begin{eqnarray}
    \bar{T}^F(t) &=& e^{i H^\mr{LP} t} \prod_{i=1}^N \left[ \bar{T}^{F,\Lambda_i}(t)^\dagger \bar{T}^{F,\Lambda_{i-1}}(t) \right], \\
    \bar{T}^{\mr{BCH}}_{p_0}(t) &=& e^{i H^\mr{LP}t} \prod_{i=1}^N \left[ \bar{T}^{\mr{BCH},\Lambda_i}_{p_0}(t)^\dagger \bar{T}^{\mr{BCH},\Lambda_{i-1}}_{p_0}(t) \right]
\end{eqnarray}
since we have $\bar{T}^{F,\Lambda_0}(t)=\bar{T}^F(t)$, $\bar{T}^{\mr{BCH},\Lambda_i}_{p_0}(t)=\bar{T}^{\mr{BCH}}_{p_0}(t)$, and $\bar{T}^{F,\Lambda_N}(t)=\bar{T}^{\mr{BCH},\Lambda_N}_{p_0}(t)=e^{iH^\mr{LP}t}$.

We express the error in Eq. (\ref{EqA_MPF:error_BCH}) by the above decompositions.
First, it is easy to check that the translation symmetry like Theorem \ref{Lem_Rel:transl_sym_PF} holds also for the operators $\bar{T}^{F,\Lambda_i}(t)$.
By calculation like Eq. (\ref{Eq_Err:t_dep_error_Delta_1}), we obtain
\begin{equation}\label{EqA_MPF:prod_T_dag_T}
    \sum_{l \in \bbZ} e^{-il\omega t} \braket{l|\bar{T}^F(t)|0} = \prod_{i=1}^N \left( \sum_{l \in \bbZ} \braket{l|\bar{T}^{F,\Lambda_i}(t)^\dagger \bar{T}^{F,\Lambda_{i-1}}(t)|0}\right).
\end{equation}
We note that the above relation is ensured by the absolute convergence of $\sum_{l \in \bbZ} \braket{l|\bar{T}^{F,\Lambda_i}(t)^\dagger \bar{T}^{F,\Lambda_{i-1}}(t)|0}$, which is easily confirmed by Lemma \ref{LemA_smooth:product_decay} and Lemma \ref{LemA_smooth:evolution_decay}.
The translation symmetry of $\bar{T}^{F,\Lambda_i}(t)$ also implies the unitarity of the operator in the above formula by
\begin{eqnarray}
    && \left( \sum_{l \in \bbZ} \braket{l|\bar{T}^{F,\Lambda_i}(t)^\dagger \bar{T}^{F,\Lambda_{i-1}}(t)|0}\right)^\dagger \left( \sum_{l \in \bbZ} \braket{l|\bar{T}^{F,\Lambda_i}(t)^\dagger \bar{T}^{F,\Lambda_{i-1}}(t)|0}\right) \nonumber \\
    && \qquad \qquad  = \sum_{l,l' \in \bbZ} \bra{l-l'} \bar{T}^{F,\Lambda_{i-1}}(t)^\dagger \bar{T}^{F,\Lambda_i}(t) \ket{l}  \bra{l} \bar{T}^{F,\Lambda_i}(t)^\dagger \bar{T}^{F,\Lambda_{i-1}}(t) \ket{0} = I. \label{EqA_MPF:unitarity_T_dag_T}
\end{eqnarray}
Similar calculation goes also for the subsystem BCH formula, $\bar{T}_{p_0}^\mr{BCH}(t)$, leading to the decomposition like Eq. (\ref{EqA_MPF:prod_T_dag_T}) and the unitarity like Eq. (\ref{EqA_MPF:unitarity_T_dag_T}).
As a result, the left-hand side of the error, Eq. (\ref{EqA_MPF:error_BCH}), is bounded by
\begin{equation}\label{EqA_MPF:sum_T_F_BCH}
    \norm{\sum_{l \in \bbZ} e^{-il\omega t}\braket{l|\left[ \bar{T}^F(t)- \bar{T}_{p_0}^\mr{BCH}(t)\right]|0}} \leq \sum_{i=1}^N \norm{\sum_{l \in \bbZ} \braket{l|\left[ \bar{T}^{F,\Lambda_i}(t)^\dagger \bar{T}^{F,\Lambda_{i-1}}(t) - \bar{T}^{\mr{BCH},\Lambda_i}_{p_0}(t)^\dagger \bar{T}^{\mr{BCH},\Lambda_{i-1}}_{p_0}(t) \right]|0}}.
\end{equation}

With a technique similar to the time-independent case \cite{mizuta2025-mpf}, we can expand the right-hand side with respect to $t$ under the condition Eq. (\ref{EqA_MPF:BCH_t_condition}) as we will prove as Lemma \ref{LemA_MPF:T_q_F_expansion} and Lemma \ref{LemA_MPF:T_q_BCH_expansion} in Appendix \ref{SubsecA:T_q_expansion}.
First, about the operator $\bar{T}^{F,\Lambda_i}(t)$, we can show the relation,
\begin{equation}\label{EqA_MPF:TT_F_expansion}
    \sum_{l \in \bbZ} \bra{l}\bar{T}^{F,\Lambda_i}(t)^\dagger \bar{T}^{F,\Lambda_{i-1}}(t) \ket{0} = \sum_{q=0}^\infty \bar{T}_{q,p_0}^i t^q + \bar{T}^{i}_{p_0}(t), \quad \bar{T}^{i}_{p_0}(t) \in \order{t^{p_0+1}}.
\end{equation}
Each of the operators $\bar{T}_{q,p_0}^i$ and $\bar{T}^{i}_{p_0}(t)$ is bounded by
\begin{equation}\label{EqA_MPF:TT_F_expansion_bound}
    \norm{\bar{T}_{q,p_0}^i}t^q \leq \frac{e}2 [2p_0 V_p(2kg+\Gamma f)t]^q, \quad
    \norm{\bar{T}^{i}_{p_0}(t)} \leq e^2[2p_0V_p (2kg+\Gamma f)t]^{p_0+1}.
\end{equation}
As well, we have a similar expansion for the operator $\bar{T}^{\mr{BCH},\Lambda_i}_{p_0}(t)$, given by
\begin{equation}\label{EqA_MPF:TT_BCH_expansion}
    \sum_{l \in \bbZ} \bra{l} \bar{T}^{\mr{BCH},\Lambda_i}_{p_0}(t)^\dagger \bar{T}^{\mr{BCH},\Lambda_{i-1}}_{p_0}(t) \ket{0} = \sum_{q=0}^\infty \bar{T}_{q,p_0}^{\mr{BCH},i} t^q + \bar{T}_{p_0}^{\mr{BCH},i}(t),
\end{equation}
where the operators $\bar{T}_{q,p_0}^{\mr{BCH},i}$ and $\bar{T}_{p_0}^{\mr{BCH},i}(t)$ are respectively bounded by
\begin{equation}\label{EqA_MPF:TT_BCH_expansion_bound}
    \norm{\bar{T}_{q,p_0}^{\mr{BCH},i}} t^q \leq \frac{1}{2e^2} \{4e^2 p_0 V_p (2kg+\Gamma f)t\}^q, \quad 
    \norm{\bar{T}_{p_0}^{\mr{BCH},i}(t)} \leq e^{3/2} [2e^2 p_0 V_p (2kg+\Gamma f)t]^{p_0+1}.
\end{equation}
We attach their detailed derivation in Appendix \ref{SubsecA:T_q_expansion} and continue the proof of Lemma \ref{LemA_MPF:truncated_BCH_error}.

Owing to the order condition $\bar{T}^{\mr{BCH},\Lambda_i}_{p_0}(t) = \bar{T}^{F,\Lambda_i}(t) + \order{t^{p_0+1}}$, we have $\bar{T}_{q,p_0}^i=\bar{T}_{q,p_0}^{\mr{BCH},i}$ for every $q=0,1,\cdots,p_0$.
Equation (\ref{EqA_MPF:sum_T_F_BCH}) can be bounded by
\begin{eqnarray}
    \norm{\sum_{l \in \bbZ} e^{-il\omega t}\braket{l|\left[ \bar{T}^F(t)- \bar{T}_{p_0}^\mr{BCH}(t)\right]|0}} &\leq& \sum_{i=1}^N  \left[ \sum_{q=p_0+1}^\infty \left( \norm{\bar{T}_{q,p_0}^i}t^q +\norm{\bar{T}_{q,p_0}^{\mr{BCH},i}} t^q \right) + \norm{\bar{T}^i_{p_0}(t)} + \norm{\bar{T}_{p_0}^{\mr{BCH},i}(t)}\right] \nonumber \\
    &\leq& N\sum_{n=p_0+1}^\infty \left[ \frac{e}2 (2e^3)^{-q} + \frac{1}{2e^2} e^{-q}\right] + e^2N (2e^3)^{-p_0-1} + e^{3/2} N (2e)^{-p_0-1} \nonumber \\
    &\leq& N \left( \frac{e+e^{-2}}2 \sum_{q=p_0+1}^\infty e^{-q} + \frac{e^{-2}+e^{1/2}}{2} e^{-p_0}\right) \leq 2N e^{-p_0}.
\end{eqnarray}
Since the integer $p_0=p_0(N,\epsilon)$ is set to $\lceil \log (2N/\epsilon) \rceil$, this becomes smaller than $\epsilon$. $\quad \square$

The aim of this section is to evaluate
\begin{equation}
    \bar{N}(t,0) -\sum_{j=1}^J c_j \sum_{l \in \bbZ} e^{-il\omega t} \braket{l|[T_{p_0(N,\epsilon)}^{\mr{BCH}} (t/k_j) ]^{k_j}|0} = \sum_{j=1}^J c_j \sum_{l \in \bbZ} e^{-il\omega t} \braket{l|\left\{ [\bar{T}^F (t/k_j) ]^{k_j} - [\bar{T}_{p_0(N,\epsilon)}^{\mr{BCH}} (t/k_j) ]^{k_j} \right\}|0},
\end{equation}
which represents the error by replacing the repeated PF $[\bar{T}^F(t/k_j)]^{k_j}$ with the corresponding truncated BCH formula $[\bar{T}_{p_0(N,\epsilon)}^{\mr{BCH}} (t/k_j) ]^{k_j}$.
It can be expressed by the single-step error discussed in Lemma \ref{LemA_MPF:truncated_BCH_error}, which results in the following statement.

\begin{lemma}\label{LemA_MPF:MPF_BCH_error}
\textbf{}

The standard MPF $\bar{N}(t,0)$ is approximated as
\begin{equation}
    \norm{\bar{N}(t,0)-\sum_{j=1}^J c_j \sum_{l \in \bbZ} e^{-il\omega t} \braket{l|[\bar{T}_{p_0(N,\epsilon)}^{\mr{BCH}} (t/k_j) ]^{k_j}|0}} \leq \norm{\vec{c}}_1 \| \vec{k} \|_1 \epsilon,
\end{equation}
under the same condition as Lemma \ref{LemA_MPF:truncated_BCH_error}.
\end{lemma}

\textbf{Proof.---}
The error of replacing $[\bar{T}^F(t/k_j)]^{k_j}$ in the time-dependent MPF $\bar{N}(t,0)$ by the truncated BCH formula $[ \bar{T}_{p_0}^\mr{BCH}(t/k_j) ]^{k_j}$ is equal to 
\begin{eqnarray}
    && \bar{N}(t,0) - \sum_j c_j \sum_{l \in \bbZ}e^{-il\omega t}\braket{l|\{ \bar{T}_{p_0}^\mr{BCH}(t/k_j) \}^{k_j}|0} \nonumber \\
    && \quad = \sum_j c_j \sum_{k_j'=1}^{k_j} \sum_{l \in \bbZ} e^{-il\omega t} \bra{l} [ \bar{T}^F(t/k_j) ]^{k_j - k_j'} \left(  \bar{T}^F(t/k_j) - \bar{T}_{p_0}^\mr{BCH}(t/k_j)\right) [  \bar{T}_{p_0}^\mr{BCH}(t/k_j) ]^{k_j'-1}\ket{0} \nonumber \\
    && \quad = \sum_j c_j \sum_{k_j'=1}^{k_j}  \sum_{l \in \bbZ} e^{-il\omega t} \bra{l} [ \bar{T}^F(t/k_j) ]^{k_j - k_j'} \ket{0} \nonumber \\
    && \qquad \quad \times \sum_{l' \in \bbZ} e^{-il'\omega tk_j'/k_j}\bra{l'}\left[\bar{T}^F(t/k_j) - \bar{T}_{p_0}^\mr{BCH}(t/k_j)  \right] \ket{0} \sum_{l'' \in \bbZ} e^{-il''\omega t (k_j'-1)/k_j} \bra{l''}[ \bar{T}_{p_0}^\mr{BCH}(t/k_j) ]^{k_j'-1}\ket{0},
\end{eqnarray}
where we use the completeness, the translation symmetry, and the absolute convergence like Eq. (\ref{Eq_Err:t_dep_error_Delta_1}).
With calculation similar to Eq. (\ref{EqA_MPF:unitarity_T_dag_T}), we can check that the two operators appearing in the above formula,
\begin{equation}
    \sum_{l \in \bbZ} e^{-il\omega t} \bra{l} [ \bar{T}^F(t/k_j) ]^{k_j - k_j'} \ket{0}, \qquad \sum_{l'' \in \bbZ} e^{-il''\omega t (k_j'-1)/k_j} \bra{l''}[ \bar{T}_{p_0}^\mr{BCH}(t/k_j) ]^{k_j'-1}\ket{0},
\end{equation}
are both unitary.
As a result, we obtain
\begin{eqnarray}
    \norm{\bar{N}(t,0)-\sum_j c_j \sum_{l \in \bbZ}e^{-il\omega t}\braket{l|[ \bar{T}_{p_0}^\mr{BCH}(t/k_j) ]^{k_j}|0}} &\leq& \sum_j |c_j| \sum_{k_j'=1}^{k_j} \norm{\sum_{l' \in \bbZ} e^{-il'\omega tk_j'/k_j}\bra{l'}\left[\bar{T}^F(t/k_j)-\bar{T}_{p_0}^\mr{BCH}(t/k_j) \right] \ket{0}} \nonumber \\
    &\leq& \norm{\vec{c}}_1 \|\vec{k}\|_1 \epsilon,
\end{eqnarray}
where we use Lemma \ref{LemA_MPF:truncated_BCH_error}. $\quad \square$

Considering the triangle inequality, the error of the time-dependent MPF is bounded by
\begin{equation}\label{EqA_MPF:Evol_MPF_error_triangle}
    \norm{U(t,0)-\bar{N}(t,0)} \leq \norm{U(t,0)-\sum_{j=1}^J c_j \sum_{l \in \bbZ}e^{-il\omega t}\braket{l|[ \bar{T}_{p_0}^\mr{BCH}(t/k_j) ]^{k_j}|0}} + \norm{\vec{c}}_1 \|\vec{k}\|_1 \epsilon.
\end{equation}
We will evaluate the first term and complete the error bound in Appendix \ref{SubsecA:err_proof_MPF}.

\subsection{Proof of the relations for the subsystem operators}\label{SubsecA:T_q_expansion}

In this section, we prove the relations Eqs. (\ref{EqA_MPF:TT_F_expansion})-(\ref{EqA_MPF:TT_BCH_expansion_bound}), provided by Lemma \ref{LemA_MPF:T_q_F_expansion} and Lemma \ref{LemA_MPF:T_q_BCH_expansion}.
First of all, we clarify the notations of some operators nontrivially acting on the 
$\Lambda_i$.
Using the subsystem Hamiltonian defined in Eq. (\ref{EqA_MPF:subsys_H}), the time-independent subsystem PF on the Floquet-Hilbert space like Eq. (\ref{Eq_Rel:T_F_Std}) is precisely given by 
\begin{equation}\label{EqA_MPF:subsys_T_F}
    \bar{T}^{F,\Lambda_i}(t) = \prod_{v=1}^{V_p} \prod_{\tilde{\gamma}=1}^{2\Gamma} e^{-i\tilde{H}_{v\tilde{\gamma}}^{\Lambda_i}\tilde{\alpha}_{v\tilde{\gamma}}t}, \quad (\tilde{H}_{v\tilde{\gamma}}^{\Lambda_i}, \tilde{\alpha}_{v \tilde{\gamma}}) = \begin{cases}
        (H_{\pi_v(\gamma)}^{\mr{Add},\Lambda_i}\alpha_{v \gamma}) & (\tilde{\gamma}=2\gamma-1) \\
        (-H^\mr{LP},\beta_{v(\gamma+1)}-\beta_{v\gamma}) & (\tilde{\gamma}=2\gamma),
    \end{cases}
\end{equation}
in which the operator $H_\gamma^{\mr{Add},\Lambda_i}$ is defined by
\begin{equation}
    H_\gamma^{\mr{Add},\Lambda_i} = \sum_{m \in \bbZ} \mr{Add}_m \otimes H_{\gamma m}^{\Lambda_i}, \quad H_{\gamma m}^{\Lambda_i} = \frac1T \int_0^T H_\gamma^{\Lambda_i}(t) e^{im\omega t} \dd t.
\end{equation}
Similarly, the subsystem BCH formula $\bar{T}_{p_0}^{\text{BCH},\Lambda_i}(t)$ is precisely written by
\begin{equation}\label{EqA_MPF:subsys_BCH_def}
    \bar{T}_{p_0}^{\text{BCH},\Lambda_i}(t) = \exp \left( -iH^{F,\Lambda_i}t -i \sum_{q=p+1}^{p_0} \bar{\Phi}_q^{F,\Lambda_i} t^q \right).
\end{equation}
The Floquet Hamiltonian $H^{F,\Lambda_i}$ is defined for the subsystem Hamiltonian $H^{\Lambda_i}(t)$ like Eq. (\ref{Eq_Pre:H_F}) and the operator $\bar{\Phi}_q^{F,\Lambda_i}$ is obtained by replacing all of $\{ \tilde{H}_{v\tilde{\gamma}} \}$ in Eq. (\ref{EqA_MPF:Phi_q_F}) with $\{ \tilde{H}_{v\tilde{\gamma}}^{\Lambda_i}\}$.
We first show the relations about the subsystem PF $\bar{T}^{F,\Lambda_i}(t)$, Eqs. (\ref{EqA_MPF:TT_F_expansion}) and (\ref{EqA_MPF:TT_F_expansion_bound}), as follows.

\begin{lemma}\label{LemA_MPF:T_q_F_expansion}
\textbf{}

Let $\bar{T}^{F,\Lambda_i}(t)$ denotes the subsystem PF on the Floquet-Hilbert space defined by Eq. (\ref{EqA_MPF:subsys_T_F}).
When the time $t$ is small enough to satisfy
\begin{equation}\label{EqA_MPF:t_condition_TT_F}
    t < \frac1{2p_0V_p(2kg+\Gamma f)} \in \order{[p_0(g+\Gamma f)]^{-1}}
\end{equation}
for a certain natural number $p_0 \in \bbN$, the operator $\sum_{l \in \bbZ} \bra{l}\bar{T}^{F,\Lambda_i}(t)^\dagger \bar{T}^{F,\Lambda_{i-1}}(t) \ket{0}$ is expanded by an absolutely convergent series,
\begin{equation}\label{EqA_MPF:TT_F_series}
    \sum_{l \in \bbZ} \bra{l}\bar{T}^{F,\Lambda_i}(t)^\dagger \bar{T}^{F,\Lambda_{i-1}}(t) \ket{0} = \sum_{q=0}^\infty \bar{T}_{q,p_0}^i t^q + \bar{T}^{i}_{p_0}(t).
\end{equation}
The operators $\bar{T}_{q,p_0}^i$ and $\bar{T}^i_{p_0}(t)$ on the Hilbert space $\mcl{H}$ satisfy the following relations,
\begin{eqnarray}
    \norm{\bar{T}_{q,p_0}^i}t^q &\leq& \frac{e}2 (4p_0V_pkgt+2p_0V_p\Gamma ft)^q \in \order{[p_0(g+\Gamma f)t]^p}, \label{EqA_MPF:T_q_p0_bound} \\
    \norm{\bar{T}^{i}_{p_0}(t)} &\leq& e^2(4p_0V_pkgt+2p_0V_p\Gamma ft)^{p_0+1} \in \order{[p_0(g+\Gamma f)t]^{p_0+1}}. \label{EqA_MPF:T_p0_t_bound}
\end{eqnarray}
\end{lemma}

\textbf{Proof.---}
Since $\bar{T}^F(t)$ is a time-dependent PF defined by Eq. (\ref{Eq_Rel:T_F_Std}), we can apply the calculation for the time-independent case to the coefficient $\bar{T}^{F,i}_q$ (See Lemma 9 in Ref. \cite{mizuta2025-mpf}).
We define two kinds of time-dependent Hamiltonians $h_i^F(\tau)$ and $\tilde{H}^{F,\Lambda_i}(\tau)$ for the time $\tau \in [0,2V_p\Gamma t)$, which act on the Floquet-Hilbert space $\mcl{H}_\mr{FH}$.
They are piecewise constant and defined by
\begin{equation}\label{EqA_MPF:h_F_tau_setting}
    h_i^F(\tau) = \tilde{\alpha}_{v\tilde{\gamma}}( \tilde{H}_{v\tilde{\gamma}}^{\Lambda_{i-1}} -\tilde{H}_{v\tilde{\gamma}}^{\Lambda_i}), \quad \tilde{H}^{F,\Lambda_i}(\tau) = \tilde{\alpha}_{v\tilde{\gamma}} \tilde{H}_{v\tilde{\gamma}}^{\Lambda_i}, \quad  \frac{\tau}t \in [(v-1)2\Gamma+\tilde{\gamma}-1,(v-1)2\Gamma+\tilde{\gamma}),
\end{equation}
for each $v=1,\cdots,V_p$ and each $\tilde{\gamma}=1,\cdots,2\Gamma$.
The operator $\bar{T}^{F,\Lambda_{i-1}}(t)$ defined by Eq. (\ref{EqA_MPF:subsys_T_F}) is equal to the time-evolution operator under the Hamiltonian $\tilde{H}^{F,\Lambda_{i-1}}(\tau)$ over the time $\tau \in [0,2V_p\Gamma t]$.
Considering the interaction picture in the Hamiltonian $\tilde{H}^{F,\Lambda_{i}}(\tau)$, it can be also represented by
\begin{equation}\label{EqA_MPF:TT_F_int_pic}
    \bar{T}^{F,\Lambda_{i}}(t)^\dagger \bar{T}^{F,\Lambda_{i-1}}(t) = \mcl{T} \exp \left( -i \int_0^{2V_p\Gamma t} \dd \tau U^{F,\Lambda_i}(\tau)^\dagger h_i^F(\tau) U^{F,\Lambda_i}(\tau)\right), \quad
    U^{F,\Lambda_i}(\tau) = \mcl{T} \exp \left( - i \int_0^\tau \dd \tau' \tilde{H}^{F,\Lambda_{i}}(\tau')\right).
\end{equation}

\textit{Deriving the series expansion, Eq. (\ref{EqA_MPF:TT_F_series}).---}
We first derive the coefficient of the series expansion $\bar{T}_{q,p_0}^i$.
Applying the Dyson series expansion to the operator $\bar{T}^{F,\Lambda_i}(t)^\dagger \bar{T}^{F,\Lambda_{i-1}}(t)$, we obtain
\begin{eqnarray}
    && \sum_{l \in \bbZ} \bra{l}\bar{T}^{F,\Lambda_i}(t)^\dagger \bar{T}^{F,\Lambda_{i-1}}(t)\ket{0} \nonumber \\
    && \qquad = \sum_{l \in \bbZ} \sum_{q'=0}^\infty \frac{(-i)^{q'}}{q'!} \int_0^{2V_p\Gamma t} \dd \tau_1 \cdots \int_0^{2V_p\Gamma t} \dd \tau_{q'} \bra{l} \mcl{T}_\rightarrow \left[ \prod_{q''=1}^{q'} U^{F,\Lambda_i}(\tau_{q''})^\dagger h_i^F(\tau_{q''}) U^{F,\Lambda_i}(\tau_{q''})\right] \ket{0}, \label{EqA_MPF:TT_sum_Dyson}
\end{eqnarray}
where the symbol $\mcl{T}_\rightarrow$ denotes the time-ordered product, in which $\tau_1,\cdots,\tau_{q'}$ are aligned in the descending order.
Owing to the smoothness of the Hamiltonian discussed in Appendix \ref{A_Sec:Smooth}, we have at least $\norm{\braket{l|U^{F,\Lambda_i}(\tau)|0}} \in \order{|l|^{-2}}$ and $\norm{\braket{l|h_i^F(\tau)|0}} \in \order{|l|^{-2}}$ for $|\tau| < \infty$.
Using Lemma \ref{LemA_smooth:product_decay}, there exists a certain positive constant $C_2$ such that
\begin{equation}
    \norm{\frac{(-i)^{q'}}{q'!} \int_0^{2V_p\Gamma t} \dd \tau_1 \cdots \int_0^{2V_p\Gamma t} \dd \tau_{q'} \bra{l} \mcl{T}_\rightarrow \left[ \prod_{q''=1}^{q'} U^{F,\Lambda_i}(\tau_{q''})^\dagger h_i^F(\tau_{q''}) U^{F,\Lambda_i}(\tau_{q''})\right] \ket{0}} \leq \frac{(2V_p \Gamma t)^{q'} (C_2)^{q'}}{q'!(1+|l|)^2}.
\end{equation}
Thus, the infinite series by Eq. (\ref{EqA_MPF:TT_sum_Dyson}) is absolutely convergent with respect to $l$ and $q'$, and we can change the order of taking the summation freely.
As a result, in a similar manner to the calculation for Eq. (\ref{Eq_Err:t_dep_error_Delta_1}), we obtain
\begin{equation}
    [\text{Eq. (\ref{EqA_MPF:TT_sum_Dyson})}] = \sum_{q'=0}^\infty \frac{(-it)^{q'}}{q'!} \int_0^{2V_p\Gamma} \dd s_1 \cdots \int_0^{2V_p\Gamma} \dd s_{q'}\,   \mcl{T}_\rightarrow \left[ \prod_{q''=1}^{q'} \left( \sum_{l \in \bbZ} \bra{l} U^{F,\Lambda_i}(s_{q''}t)^\dagger h_i^F(s_{q''}t) U^{F,\Lambda_i}(s_{q''}t) \ket{0}\right)\right].
\end{equation}
We use the translation symmetry $\braket{l|U^{F,\Lambda_i}(\tau)^\dagger h_i^F(\tau) U^{F,\Lambda_i}(\tau)|l'} = \braket{l-l''|U^{F,\Lambda_i}(\tau)^\dagger h_i^F(\tau) U^{F,\Lambda_i}(\tau)|l'-l''}$, which is easily obtained by Eqs. (\ref{Eq_Pre:tr_sym_H_Add})-(\ref{Eq_Pre:tr_sym_H_F}), and change the variables in the integral by $\tau_{q''}=s_{q''}t$ in the above.

We next consider the expansion of the operator $U^{F,\Lambda_i}(st)^\dagger h_i^F(st) U^{F,\Lambda_i}(st)$ in the integral.
Let $v(s)$ and $\tilde{\gamma}(s)$ denote the indices such that $s \in [(v(s)-1)2\Gamma + \tilde{\gamma}(s)-1,(v(s)-1)2\Gamma + \tilde{\gamma}(s))$.
It can be expressed by
\begin{equation}\label{EqA_MPF:U_h_U_setting}
    U^{F,\Lambda_i}(st)^\dagger h_i^F(st) U^{F,\Lambda_i}(st) = \left(\prod_{v'\tilde{\gamma}'= v(s)\tilde{\gamma}(s)}^{11} e^{i\tilde{\alpha}_{v'\tilde{\gamma}'}^s t \ad_{\tilde{H}_{v'\tilde{\gamma}'}^{\Lambda_i}}} \right) h_i^F(st), \quad \tilde{\alpha}_{v\tilde{\gamma}}^s = \begin{cases}
        \tilde{\alpha}_{v(s)\tilde{\gamma}(s)} (s-\lfloor s \rfloor) & \text{if $v\tilde{\gamma}=v(s)\tilde{\gamma}(s)$} \\
        \tilde{\alpha}_{v\tilde{\gamma}} & \text{otherwise} 
    \end{cases},
\end{equation}
and the coefficient $\tilde{\alpha}_{v\tilde{\gamma}}^s$ is bounded by $|\tilde{\alpha}_{v\tilde{\gamma}}^s| \leq |\tilde{\alpha}_{v\tilde{\gamma}}| \leq 1$.
We can employ Eq. (\ref{Eq_Err:adjoint_product_exp}) with the replacement of the order $p$ by $p_0$, which is the result of the Taylor's theorem.
This leads to the series expansion,
\begin{equation}\label{EqA_MPF:U_h_U_exp_Xi}
    \sum_{l \in \bbZ} \bra{l} U^{F,\Lambda_i}(st)^\dagger h_i^F(st) U^{F,\Lambda_i}(st) \ket{0} = \sum_{q=0}^{p_0} \Xi_q^i(s) t^q + \Xi_{p_0}^i(t;s),
\end{equation}
where the operators $\Xi_q^i(s)$ and $\Xi_{p_0}^i(t;s)$ are respectively defined by
\begin{eqnarray}
    \Xi_q^i (s) &=& i^q \sum_{\substack{q_{11},\cdots,q_{v(s)\tilde{\gamma}(s)} \geq 0, \\ q_{11}+\cdots+q_{v(s)\tilde{\gamma}(s)} = q}} \sum_{l \in \bbZ} \bra{l} \left(\prod_{v'\tilde{\gamma}'=v(s)\tilde{\gamma}(s)}^{11} \frac{(\tilde{\alpha}_{v'\tilde{\gamma}'}^s \ad_{\tilde{H}_{v'\tilde{\gamma}'}^{\Lambda_i}})^{q_{v'\tilde{\gamma}'}}}{q_{v'\tilde{\gamma}'}!} \right) h_i^F(st)\ket{0}, \label{EqA_MPF:Xi_q_i}\\
    \Xi_{p_0}^i(t;s) &=& (it)^{p_0+1} \sum_{v\tilde{\gamma} \preceq v(s)\tilde{\gamma}(s)}\sum_{\substack{q_{v\tilde{\gamma}},\cdots,q_{v(s)\tilde{\gamma}(s)} \geq 0, \\
    q_{v\tilde{\gamma}} \neq 0, \\ q_{v\tilde{\gamma}}+\cdots+q_{v(s)\tilde{\gamma}(s)} = p_0+1}} \int_0^1 \dd u \frac{u^{q_{v\tilde{\gamma}}-1}}{(q_{v\tilde{\gamma}}-1)!q_{v(\tilde{\gamma}+1)}!\cdots q_{v(s)\tilde{\gamma}(s)}!} \nonumber \\
    && \quad \sum_{l \in \bbZ}  \bra{l} \left( \prod_{v'\tilde{\gamma}'=v(\tilde{\gamma}+1)}^{11} e^{i \tilde{\alpha}_{v'\tilde{\gamma}'}^s t \ad_{\tilde{H}_{v'\tilde{\gamma}'}^{\Lambda_i}}} \right) e^{i \tilde{\alpha}_{v\tilde{\gamma}}^s ut \ad_{\tilde{H}_{v\tilde{\gamma}}^{\Lambda_i}}}\left(\prod_{v'\tilde{\gamma}'=v(s)\tilde{\gamma}(s)}^{v\tilde{\gamma}} ( \tilde{\alpha}_{v'\tilde{\gamma}'}^s\ad_{\tilde{H}_{v'\tilde{\gamma}'}^{\Lambda_i}})^{q_{v'\tilde{\gamma}'}}\right) h_i^F(st) \ket{0}. \label{EqA_MPF:Xi_p0_i}
\end{eqnarray}
The operator $\Xi_q^i (s)$ is independent of the time $t$ since $h_i^F(\tau)$ is determined by the time $\tau$ modulo $t$ according to Eq. (\ref{EqA_MPF:h_F_tau_setting}).
We note that the above calculation executes the reordering of the infinite summation with respect to $l \in \bbZ$.
It is justified by the absolute convergence of the operators $\Xi_q^i (s)$ and $\Xi_{p_0}^i(t;s)$ with respect to $l \in \bbZ$.
We use the assumption $h_X^\gamma (t) \in C^\infty$ for this purpose.
Since the Fourier coefficients like $\braket{l|H_\gamma^\mr{Add}|l'}$ and $\braket{l|h_i^F(st)|l'}$ decay faster than every polynomial of $l$ as discussed at the end of Appendix \ref{A_Sec:Smooth}, each term in the summation appearing in Eq. (\ref{EqA_MPF:Xi_q_i}) rapidly decay at least as $o(|l|^{-2})$.
This implies the absolute convergence of $\Xi_q^i (s)$ and so does also for $\Xi_{p_0}^i (t;s)$.
If we only assume $h_X^\gamma (t) \in C^{p+2}$ like the time-dependent PF, the above series may not be absolutely convergent for $q=p+1,\cdots,p_0$, which involve differentiation more than $p$ times. 
The reason why we employ the Taylor's theorem in Eq. (\ref{EqA_MPF:U_h_U_exp_Xi}) rather than the infinite series expansion in contrast to Eq. (\ref{EqA_MPF:TT_sum_Dyson}) is the absolute convergence, too.
Since $\tilde{H}_{v\tilde{\gamma}}^{\Lambda_i}$ contains $H^\mr{LP}$, which is an unbounded operator, each term in Eq. (\ref{EqA_MPF:Xi_q_i}) can have a norm whose scaling is $\order{|l|^q}$.
We cannot ensure the absolute convergence of Eq. (\ref{EqA_MPF:U_h_U_exp_Xi}) if $q$ can be infinite, which leads to the truncation at the order $p_0$.

We substitute the expansion Eq. (\ref{EqA_MPF:U_h_U_exp_Xi}) into Eq. (\ref{EqA_MPF:TT_sum_Dyson}) and rearrange the terms by the order in time $t$.
The coefficient $\bar{T}_{q,p_0}^i$ comes from the collection of the products of $\{ \Xi_q^i(s)\}$ and it is given by
\begin{equation}\label{EqA_MPF:T_q_p0_def}
    \bar{T}_{q,p_0}^i = \sum_{n=0}^q \sum_{\substack{0 \leq q_1,\cdots,q_n \leq p_0, \\ q_1+\cdots+q_n=q-n}}\frac{(-i)^n}{n!} \int_0^{2V_p\Gamma} \dd s_1 \cdots \int_0^{2V_p\Gamma} \dd s_n\,   \mcl{T}_\rightarrow \left[ \prod_{n'=1}^n \Xi_{q_{n'}}^i(s_{n'}) \right].
\end{equation}
The symbol $\mcl{T}_\rightarrow$ means the time-ordered product that aligns $s_1,\cdots,s_n$ in the descending order.
The operator $\bar{T}_{p_0}^i(t)$ comes from the remaining terms involving the Taylor's reminder $\Xi_{p_0}^i(t;s)$, and it is given by
\begin{equation}\label{EqA_MPF:T_p0_t_def}
    \bar{T}_{p_0}^i(t) = \sum_{n=0}^\infty \frac{(-it)^n}{n!} \int_0^{2V_p\Gamma} \dd s_1 \cdots \int_0^{2V_p\Gamma} \dd s_n\,   \mcl{T}_\rightarrow \left[ \prod_{n'=1}^n \left(\sum_{q=0}^{p_0} \Xi_q^i(s_{n'}) t^q + \Xi_{p_0}^i(t;s_{n'})\right) - \prod_{n'=1}^n \left(\sum_{q=0}^{p_0} \Xi_q^i(s_{n'}) t^q \right)\right].
\end{equation}

\textit{Upper-bounding the coefficients.---}
We next derive the upper bounds on $\bar{T}_{q,p_0}^i$ and $\bar{T}_{p_0}^i(t)$ respectively defined by Eqs. (\ref{EqA_MPF:T_q_p0_def}) and (\ref{EqA_MPF:T_p0_t_def}).
We begin with evaluating $\Xi_q^i(s)$ and $\Xi_{p_0}^i(t;s)$ in them, which are respectively defined by Eqs. (\ref{EqA_MPF:Xi_q_i}) and (\ref{EqA_MPF:Xi_p0_i}).
When the integer $\tilde{\gamma}(s)$ is even, we have $\norm{\Xi_q^i(s)}=0$ due to $h_i^F(st)=0$ by Eqs. (\ref{Eq_Err:H_tilde_def}) and (\ref{EqA_MPF:h_F_tau_setting}).
When the integer $\tilde{\gamma}(s)$ is odd with $\tilde{\gamma}(s)=2\gamma(s)-1$, it is bounded by
\begin{eqnarray}
    \norm{\Xi_q^i(s)} &\leq& \sum_{\substack{q_{11},\cdots,q_{v(s)\tilde{\gamma}(s)} \geq 0, \\ q_{11}+\cdots+q_{v(s)\tilde{\gamma}(s)} = q}} \norm{\sum_{l \in \bbZ} \bra{l} \left(\prod_{v'\tilde{\gamma}'=v(s)\tilde{\gamma}(s)}^{11} \frac{\left(  \ad_{\tilde{H}_{v'\tilde{\gamma}'}^{\Lambda_i}}\right)^{q_{v'\tilde{\gamma}'}}}{q_{v'\tilde{\gamma}'}!} \right) h_i^F(st)\ket{0}} \nonumber \\
    &=& \sum_{\substack{q_{11},\cdots,q_{v(s)\tilde{\gamma}(s)} \geq 0, \\ q_{11}+\cdots+q_{v(s)\tilde{\gamma}(s)} = q}} \norm{ \left(\prod_{v'\tilde{\gamma}'=v(s)\tilde{\gamma}(s)}^{11} \frac{[ \bar{D}_{v'\tilde{\gamma}'}^{\Lambda_i}(0)]^{q_{v'\tilde{\gamma}'}}}{q_{v'\tilde{\gamma}'}!} \right) (H_{\pi_{v(s)}(\gamma(s))}^{\Lambda_{i-1}}(0)-H_{\pi_{v(s)}(\gamma(s))}^{\Lambda_{i}}(0))} \nonumber \\
    &\leq& (V_p)^q \sum_{\gamma_1,\cdots,\gamma_q=1}^{\Gamma+1} \sum_{X \subset \Lambda; X \ni i} \norm{\left( \prod_{q'=1}^q \bar{\mcl{D}}_{\gamma_{q'}}^{\Lambda_i} (0) \right) h_X^{\pi_{v(s)}(\gamma(s))}(0)}, \label{EqA_MPF:Xi_q_nested_com}
\end{eqnarray}
where the operators $\bar{D}_{v'\tilde{\gamma}'}^{\Lambda_i}(t)$ and $\bar{\mcl{D}}_{\gamma_{q'}}^{\Lambda_i}(t)$ are defined by replacing $H_\gamma(t)$ with $H_\gamma^{\Lambda_i}(t)$ respectively in Eq. (\ref{Eq_Err:Dbar_v_gamma_def}) and (\ref{Eq_Sum:alpha_com_Std}).
In the above calculation, the second line comes from the fact that the nested commutators among $\{ H_\gamma^\mr{Add} \}$ and $H^\mr{LP}$ are converted to those among $\{H_\gamma (t)\}$ and their time derivatives, as discussed in Lemma \ref{Lem_Err:Form_commutators}.
In the last inequality, we use the fact that the summation over $q_{11},\cdots,q_{v(s)\tilde{\gamma}(s)}$ such that $q_{11}+\cdots+q_{v(s)\tilde{\gamma}(s)}=q$ is covered by $(V_p)^q$ copies of the summation over $\gamma_1,\cdots,\gamma_q$, as discussed in Eqs. (\ref{Eq_Err:Delta_F_bound_start})-(\ref{Eq_Err:sum_LP_b}).
We also take into account that
\begin{equation}\label{EqA_MPF:H_Lambda_i_diff}
    H_\gamma^{\Lambda_{i-1}}(t)-H_{\gamma}^{\Lambda_i}(t) = \sum_{X \subset \Lambda_i; X \ni i} h_X^\gamma
\end{equation}
is covered by all the terms $\{h_X^\gamma (t)\}$ nontrivially acting on the site $i$.
When we integrate $\norm{\Xi_q(s)}$ over $s \in (0,2V_p\Gamma)$, the index $\pi_{v(s)}(\gamma (s))$ reproduces every index $\gamma$ for $V_p$ times, and hence we have 
\begin{eqnarray}
    \int_0^{2V_p\Gamma} \dd s \norm{\Xi_q(s)} &\leq& (V_p)^{q+1} \sum_{\gamma=1}^\Gamma \sum_{\gamma_1,\cdots,\gamma_q=1}^{\Gamma+1} \sum_{X \subset \Lambda; X \ni i} \norm{\left( \prod_{q'=1}^q \bar{\mcl{D}}_{\gamma_{q'}}^{\Lambda_i} (0) \right) h_X^\gamma (0)} \nonumber \\
    &\leq& (V_p)^{q+1} q! (2kg+\Gamma f)^q g \nonumber \\
    &\leq& [ q V_p (2kg+\Gamma f) ]^{q+1}. \label{EqA_MPF:Xi_q_integral}
\end{eqnarray}
The second inequality comes from Theorem \ref{Thm_Cost:bound_commutators}, which is valid also for the subsystem Hamiltonian $H^\mr{\Lambda_i}(t)$ since it shares the locality, the extensiveness by Eq. (\ref{Eq_Cost:extensiveness}), and the scale of time-dependency by Eq. (\ref{Eq_Cost:time_dependency}) with $H(t)$.

We next evaluate the upper bound of the operator $\Xi_{p_0}^i(t;s)$ given by Eq. (\ref{EqA_MPF:Xi_p0_i}).
We trivially have $\norm{\Xi_{p_0}^i(t;s)}=0$ when $\tilde{\gamma}(s)$ is even.
When $\tilde{\gamma}(s)$ is odd by $\tilde{\gamma}(s)=2\gamma(s)-1$, the calculation similar to Lemma \ref{Lem_Err:truncated} ensures the existence of a certain time $\tau_{s,u} \in [0,t]$ and a unitary operator $\bar{U}_{v\tilde{\gamma}'}(\tau_{s,u})$ such that
\begin{eqnarray}
    && \sum_{l \in \bbZ}  \bra{l} \left( \prod_{v'\tilde{\gamma}'=v(\tilde{\gamma}+1)}^{11} e^{i \tilde{\alpha}_{v'\tilde{\gamma}'}^s t \ad_{\tilde{H}_{v'\tilde{\gamma}'}^{\Lambda_i}}} \right) e^{i \tilde{\alpha}_{v\tilde{\gamma}}^s ut \ad_{\tilde{H}_{v\tilde{\gamma}}^{\Lambda_i}}}\left(\prod_{v'\tilde{\gamma}'=v(s)\tilde{\gamma}(s)}^{v\tilde{\gamma}} ( \tilde{\alpha}_{v'\tilde{\gamma}'}^s \ad_{\tilde{H}_{v'\tilde{\gamma}'}^{\Lambda_i}})^{q_{v'\tilde{\gamma}'}}\right) h_i^F(st)\ket{0} \nonumber \\
    && \quad = \bar{U}_{v\tilde{\gamma}'}(\tau_{s,u})^\dagger \left[ \left(\prod_{v'\tilde{\gamma}'=v(s)\tilde{\gamma}(s)}^{v\tilde{\gamma}} \bar{D}_{v'\tilde{\gamma}'}(\tau_{s,u}) \right) \alpha_{v(s)\gamma(s)}[H_{\pi_{v(s)}(\gamma(s))}^{\Lambda_{i-1}}(\tau_{s,u})-H_{\pi_{v(s)}(\gamma(s))}^{\Lambda_{i}}(\tau_{s,u})] \right] \bar{U}_{v\tilde{\gamma}'}(\tau_{s,u}).
\end{eqnarray}
The norm of the operator $\Xi_{p_0}^i(t;s)$ is bounded by
\begin{eqnarray}
    \norm{\Xi_{p_0}^i(t;s)} 
    &\leq& t^{p_0+1} \max_{\tau \in [0,t]} \left( \sum_{v\tilde{\gamma} \preceq v(s)\tilde{\gamma}(s)}\sum_{\substack{q_{v\tilde{\gamma}},\cdots,q_{v(s)\tilde{\gamma}(s)} \geq 0, \\ q_{v\tilde{\gamma}}+\cdots+q_{v(s)\tilde{\gamma}(s)} = p_0+1, \\ q_{v\tilde{\gamma}}\neq 0}} \sum_{X \subset \Lambda; X \ni i} \norm{\left(\prod_{v'\tilde{\gamma}'=v(s)\tilde{\gamma}(s)}^{v\tilde{\gamma}} \bar{D}_{v'\tilde{\gamma}'}(\tau) \right)h_X^{\pi_{v(s)}(\gamma(s))}(\tau)}\right) \nonumber \\
    &\leq& (V_p t)^{p_0+1} \max_{\tau \in [0,t]} \left( \sum_{\gamma_1,\cdots,\gamma_{p_0+1}=1}^{\Gamma+1} \sum_{X \subset \Lambda; X \ni i} \norm{\left( \prod_{q'=1}^{p_0+1} \bar{\mcl{D}}_{\gamma_{q'}} (\tau) \right) h_X^{\pi_{v(s)}(\gamma(s))}(\tau)}\right),
\end{eqnarray}
where the second inequality is obtained in parallel to Eq. (\ref{EqA_MPF:Xi_q_nested_com}).
Its integral is bounded by
\begin{equation}
    \int_0^{2V_p \Gamma} \dd s \norm{\Xi_{p_0}^i(t;s)} \leq t^{p_0+1} [ (p_0+1)V_p(2kg+\Gamma f) ]^{p_0+2}.
\end{equation}
as well as Eq. (\ref{EqA_MPF:Xi_q_integral}).

We are ready to prove the upper bounds on the operators $\bar{T}_{q,p_0}^i$ and $\bar{T}^i_{p_0}(t)$ .
Using the expression of $\bar{T}_{q,p_0}^i$ by Eq. (\ref{EqA_MPF:T_q_p0_def}), its norm is bounded by
\begin{eqnarray}
    \norm{\bar{T}_{q,p_0}^i} &\leq& \sum_{n=0}^q \sum_{\substack{0 \leq q_1,\cdots,q_n \leq p_0, \\ q_1+\cdots+q_n=q-n}}\frac{1}{n!}  \prod_{n'=1}^n \left( \int_0^{2V_p\Gamma} \dd s \norm{\Xi_{q_{n'}}^i(s)} \right) \nonumber \\
    &\leq& \sum_{n=0}^q \sum_{\substack{0 \leq q_1,\cdots,q_n \leq p_0, \\ q_1+\cdots+q_n=q-n}}\frac{1}{n!}  \prod_{n'=1}^n \left[ q_{n'}V_p(2kg+\Gamma f) \right]^{q_{n'}+1} \nonumber \\
    &\leq& 2^{q-1}[p_0 V_p (2kg +\Gamma f) ]^q \sum_{n=0}^q \frac{1}{n!} \leq \frac{e}2 [2p_0 V_p (2kg +\Gamma f) ]^q.
\end{eqnarray}
We use Eq. (\ref{EqA_MPF:Xi_q_integral}) in the second inequality and use $q_{n'} \leq p_0$ in the third inequality.
This completes the proof of Eq. (\ref{EqA_MPF:T_q_p0_bound}).
We note that this bound also implies the absolute convergence of the series Eq. (\ref{EqA_MPF:TT_F_series}) with respect to $q$ under the condition Eq. (\ref{EqA_MPF:t_condition_TT_F}).
On the other hand, the norm of the operator $\bar{T}^i_{p_0}(t)$ expressed by Eq. (\ref{EqA_MPF:T_p0_t_def}) is bounded by
\begin{eqnarray}
    \norm{\bar{T}^i_{p_0}(t)} &\leq& \sum_{n=0}^\infty \frac{t^n}{n!}  \sum_{n'=1}^n\frac{n!}{n'!(n-n')!} \left(\int_0^{2V_p\Gamma} \dd s\norm{\Xi_{p_0}^i(t;s)}\right)^{n'} \left(\sum_{q=0}^{p_0} \int_0^{2V_p\Gamma} \dd s \norm{\Xi_q^i(s)} t^q\right)^{n-n'} \nonumber \\
    &\leq& \sum_{n=0}^\infty \frac{t^n}{n!}  \sum_{n'=1}^n \frac{n!}{n'!(n-n')!} \left\{ t^{p_0+1} [(p_0+1)V_p (2kg+\Gamma f)]^{p_0+2} \right\}^{n'} \left( \sum_{q=0}^{p_0} t^q [qV_p(2kg+\Gamma f)]^{q+1} \right)^{n-n'}. \nonumber \\
    &&
\end{eqnarray}
The condition on the time $t$ by Eq. (\ref{EqA_MPF:t_condition_TT_F}) implies the relation,
\begin{equation}
    \sum_{q=0}^{p_0}   [qV_p(2kg+\Gamma f)t]^q \leq \sum_{q=0}^{p_0} [p_0 V_p (2kg+\Gamma f)t]^q \nonumber \\
    \leq \sum_{q=0}^\infty 2^{-q} = 2.
\end{equation}
As a result, we obtain
\begin{eqnarray}
    \norm{\bar{T}^i_{p_0}(t)} &\leq& \sum_{n=0}^\infty \frac{[2p_0 V_p (2kg +\Gamma f) t]^n}{n!}  \sum_{n'=1}^n \frac{n!}{n'!(n-n')!} \left\{ [2p_0 V_p (2kg +\Gamma f) t]^{p_0+1}\right\}^{n'} \nonumber \\
    &\leq& \sum_{n=0}^\infty \frac{2^n}{n!} [2p_0 V_p (2kg +\Gamma f) t]^{p_0+1}  = e^2 [2p_0 V_p (2kg +\Gamma f) t]^{p_0+1},
\end{eqnarray}
which completes the proof of Eq. (\ref{EqA_MPF:T_p0_t_bound}). $\quad \square$

We briefly explain the meaning of Lemma \ref{LemA_MPF:T_q_F_expansion}.
The subsystem PFs $\bar{T}^{F,\Lambda_i}(t)$ and $\bar{T}^{F,\Lambda_{i-1}}(t)$ are global operators on $\Lambda_i$ and $\Lambda_{i-1}$, whose sizes can be as large as the system size $N$.
On the other hand, the operators $\bar{T}^{F,\Lambda_i}(t)$ and $\bar{T}^{F,\Lambda_{i-1}}(t)$ have different actions only around the site $i$ by construction, and hence $\bar{T}^{F,\Lambda_i}(t)^\dagger\bar{T}^{F,\Lambda_{i-1}}(t)$ becomes quasi-local around the site $i$.
This is why the upper bounds of $\norm{\bar{T}_{q,p_0}^i}$ and $\norm{\bar{T}^i_{p_0}(t)}$, appearing in its series expansion, are determined by the local scale of the energy $g$ and that of time-dependency $f$ without explicitly depending on the size $N$ except for logarithmic corrections by $p_0(N,\epsilon)$.
This property originates from the locality of Hamiltonians, and hence it lies also in the subsystem BCH formula $\bar{T}^{\mr{BCH},\Lambda_i}_{p_0}(t)$.
As a counterpart of Lemma \ref{LemA_MPF:T_q_F_expansion}, we prove the following lemma corresponding to Eq. (\ref{EqA_MPF:TT_BCH_expansion_bound}) in Appendix \ref{SubsecA:err_truncated_BCH}.

\begin{lemma}\label{LemA_MPF:T_q_BCH_expansion}
\textbf{}

Let $\bar{T}^{\mr{BCH},\Lambda_i}_{p_0}(t)$ denote the subsystem BCH formula on the Floquet-Hilbert space defined by Eq. (\ref{EqA_MPF:subsys_BCH_def}) and suppose that the time $t$ is small enough to satisfy
\begin{equation}\label{EqA_MPF:t_condition_T_BCH}
    |t| < \frac{1}{4e^2 p_0 V_p (2kg+\Gamma f)} \in \order{[p_0(g+\Gamma f)]^{-1}},
\end{equation}
for a certain natural number $p_0 \in \bbN$. The operator $\sum_{l \in \bbZ} \braket{l|\bar{T}^{\mr{BCH},\Lambda_i}_{p_0}(t)^\dagger \bar{T}^{\mr{BCH},\Lambda_{i-1}}_{p_0}(t)|0}$ is expanded by an absolutely convergent series,
\begin{equation}\label{EqA_MPF:TT_sum_BCH_expansion}
    \sum_{l \in \bbZ} \bra{l} \bar{T}^{\mr{BCH},\Lambda_i}_{p_0}(t)^\dagger \bar{T}^{\mr{BCH},\Lambda_{i-1}}_{p_0}(t) \ket{0} = \sum_{q=0}^\infty \bar{T}_{q,p_0}^{\mr{BCH},i} t^q + \bar{T}_{p_0}^{\mr{BCH},i}(t).
\end{equation}
where the operators $\bar{T}_{q,p_0}^{\mr{BCH},i}$ and $\bar{T}_{p_0}^{\mr{BCH},i}(t)$ are bounded by
\begin{eqnarray}
    \norm{\bar{T}_{q,p_0}^{\mr{BCH},i}} t^q &\leq& \frac{1}{2e^2} [4e^2 p_0 V_p (2kg+\Gamma f)t]^q \in \order{[p_0(g+\Gamma f)t]^q}, \label{EqA_MPF:T_q_p0_BCH_bound}\\
    \norm{\bar{T}_{p_0}^{\mr{BCH},i}(t)} &\leq& e^{3/2} [2e^2 p_0 V_p (2kg+\Gamma f)t]^{p_0+1}\in \order{[p_0(g+\Gamma f)t]^{p_0+1}}. \label{EqA_MPF:T_p0_t_BCH_bound}
\end{eqnarray}
\end{lemma}

\textbf{Proof.---}
The proof is parallel to the one for Lemma \ref{LemA_MPF:T_q_F_expansion}.
Namely, we first execute the series expansion in the time $t$ and obtain the expressions of the operators $\bar{T}_{q,p_0}^{\mr{BCH},i}$ and $\bar{T}_{p_0}^{\mr{BCH},i}(t)$.
Then, we evaluate the upper bounds of them using the locality, the extensiveness, and the strength of time-dependency discussed in Appendix \ref{SubsecA:truncated_BCH}.
In the calculation below, we need to ensure the absolute convergence of the infinite series.
However, it is essentially the same as the discussion in Lemma \ref{LemA_MPF:T_q_F_expansion} in that the assumption on the smoothness is sufficient to ensure it, and hence we omit the discussion for the absolute convergence.

\textit{Deriving the series expansion, Eq. (\ref{EqA_MPF:TT_sum_BCH_expansion}).---}
The operator $\bar{T}^{\mr{BCH},\Lambda_i}_{p_0}(t)^\dagger \bar{T}^{\mr{BCH},\Lambda_{i-1}}_{p_0}(t)$ can be recast by a time-evolution operator under the interaction picture like Eq. (\ref{EqA_MPF:TT_F_int_pic}).
Regarding the operators $\sum_{q=1}^{p_0} \bar{\Phi}_q^{F,\Lambda_i} t^{q-1}$ and $\sum_{q=1}^{p_0} (\bar{\Phi}_q^{F,\Lambda_{i-1}}-\bar{\Phi}_q^{F,\Lambda_i}) t^{q-1}$ respectively as an unperturbed Hamiltonian and an interaction Hamiltonian, we obtain the relation,
\begin{eqnarray}
    && \sum_{l \in \bbZ} \bra{l} \bar{T}^{\mr{BCH},\Lambda_i}_{p_0}(t)^\dagger \bar{T}^{\mr{BCH},\Lambda_{i-1}}_{p_0}(t) \ket{0} \nonumber \\
    && \qquad = \sum_{l \in \bbZ} \sum_{n=0}^\infty \frac{(-i)^n}{n!}\int_0^t \dd \tau_1 \cdots \int_0^t \dd \tau_n \bra{l} \mcl{T}_\rightarrow \left[ \prod_{n'=1}^n \left( e^{i\tau_{n'} \sum_{q=1}^{p_0} t^{q-1} \ad_{\bar{\Phi}_q^{F,\Lambda_i}}} \sum_{q=1}^{p_0} (\bar{\Phi}_q^{F,\Lambda_{i-1}}-\bar{\Phi}_q^{F,\Lambda_i})t^{q-1} \right) \right]\ket{0} \nonumber \\
    && \qquad = \sum_{n=0}^\infty \frac{(-i)^n}{n!}\int_0^1 \dd s_1 \cdots \int_0^1 \dd s_n\mcl{T}_\rightarrow \left[ \prod_{n'=1}^n \left( \sum_{l \in \bbZ}\bra{l} e^{is_{n'} \sum_{q=1}^{p_0} t^q \ad_{\bar{\Phi}_q^{F,\Lambda_i}}} \sum_{q=1}^{p_0} (\bar{\Phi}_q^{F,\Lambda_{i-1}}-\bar{\Phi}_q^{F,\Lambda_i})t^q \ket{0}\right) \right]. \label{EqA_MPF:TT_BCH_Dyson}
\end{eqnarray}
The second equality comes from the symmetry of the operator $\bar{\Phi}_q^{F,\Lambda_i}$, which is equal to $H^{F,\Lambda_i}$ for $q=1$ and is zero or composed of nested commutators on the Floquet-Hilbert space for $q \geq 2$.
We have the translation symmetry,
\begin{equation}\label{EqA_MPF:Phi_q_symmetry}
    \braket{l-l''|\bar{\Phi}_q^{F,\Lambda_i}|l'-l''} = \braket{l|\bar{\Phi}_q^{F,\Lambda_i}|l'}+l''\omega \delta_{q1},
\end{equation}
by Lemma \ref{Lem_Err:Form_commutators}, and the second equality follows from the calculation similar to Eq. (\ref{Eq_Err:t_dep_error_Delta_1}).

We next expand the operators appearing in Eq. (\ref{EqA_MPF:TT_BCH_Dyson}).
Using the Taylor's theorem on an operator exponential by Eq. (\ref{Eq_Err:Taylor_theorem}) up to the order $p_0$, we obtain
\begin{eqnarray}
    && \sum_{l \in \bbZ}\bra{l} e^{is \sum_{q=1}^{p_0} t^q \ad_{\bar{\Phi}_q^{F,\Lambda_i}}} \sum_{q=1}^{p_0} (\bar{\Phi}_q^{F,\Lambda_{i-1}}-\bar{\Phi}_q^{F,\Lambda_i})t^q \ket{0} \nonumber \\
    &=& \sum_{l \in \bbZ}  \bra{l}\left[ \sum_{n=0}^{p_0} \frac{\left( is\sum_{q=1}^{p_0} t^q \ad_{\bar{\Phi}_q^{F,\Lambda_i}}\right)^n}{n!} + \int_0^s \dd s' e^{is' \sum_{q=1}^{p_0} t^q \ad_{\bar{\Phi}_q^{F,\Lambda_i}}} \frac{\left( is'\sum_{q=1}^{p_0} t^q \ad_{\bar{\Phi}_q^{F,\Lambda_i}}\right)^{p_0+1}}{(p_0+1)!}\right] \sum_{q=1}^{p_0}(\bar{\Phi}_q^{F,\Lambda_{i-1}}-\bar{\Phi}_q^{F,\Lambda_i})t^q \ket{0}. \nonumber \\
    &&
\end{eqnarray}
The above series can be arranged in terms of the time $t$ and expressed by
\begin{equation}\label{EqA_MPF:Psi_q_def_expansion}
    \sum_{l \in \bbZ}\bra{l} e^{is \sum_{q=1}^{p_0} t^q \ad_{\bar{\Phi}_q^{F,\Lambda_i}}} \sum_{q=1}^{p_0} (\bar{\Phi}_q^{F,\Lambda_{i-1}}-\bar{\Phi}_q^{F,\Lambda_i})t^q \ket{0} = \sum_{q=1}^\infty \bar{\Psi}_q^i (s) t^q + \bar{\Psi}_{p_0}^i (t;s), 
\end{equation}
where we define the operators $\bar{\Psi}_q^i (s)$ and $\bar{\Psi}_{p_0}^i (t;s)$ respectively by
\begin{eqnarray}
    \bar{\Psi}_q^i(s) &=& \sum_{n=0}^{p_0} \frac{(is)^n}{n!} \sum_{\substack{1 \leq q_0,\cdots,q_n \leq p_0, \\ q_0+\cdots+q_n=q}} \sum_{l \in \bbZ}\bra{l} \left(\prod_{n'=1}^n \ad_{\bar{\Phi}_{q_{n'}}^{F,\Lambda_i}}\right) \left( \bar{\Phi}_{q_0}^{F,\Lambda_i}- \bar{\Phi}_{q_0}^{F,\Lambda_{i-1}} \right) \ket{0}, \label{EqA_MPF:Psi_q_def}\\
    \bar{\Psi}_{p_0}^i (t;s) &=& \int_0^s \dd s' \sum_{l \in \bbZ} \bra{l}e^{is' \sum_{q=1}^{p_0} t^q \ad_{\bar{\Phi}_q^{F,\Lambda_i}}} \frac{\left( is'\sum_{q=1}^{p_0} t^q \ad_{\bar{\Phi}_q^{F,\Lambda_i}}\right)^{p_0+1}}{(p_0+1)!} \sum_{q=1}^{p_0}(\bar{\Phi}_q^{F,\Lambda_{i-1}}-\bar{\Phi}_q^{F,\Lambda_i})t^q \ket{0}. \label{EqA_MPF:Psi_p0_t_def}
\end{eqnarray}
We note that $\bar{\Psi}_q^i(s)$ is equal to zero for $q > (p_0)^2 + p_0$.
We substitute the series expansion Eq. (\ref{EqA_MPF:Psi_q_def_expansion}) into Eq. (\ref{EqA_MPF:TT_BCH_Dyson}) and again arrange it in terms of the time $t$ like Eqs. (\ref{EqA_MPF:T_q_p0_def}) and (\ref{EqA_MPF:T_p0_t_def}).
Consequently, we obtain the series expansion in the form of Eq. (\ref{EqA_MPF:TT_sum_BCH_expansion}), in which the operators $\bar{T}_{q,p_0}^{\mr{BCH},i}$ and $\bar{T}_{p_0}^{\mr{BCH},i}(t)$ are respectively given by 
\begin{eqnarray}
    \bar{T}_{q,p_0}^{\mr{BCH},i} &=& \sum_{n=0}^\infty \frac{(-i)^n}{n!} \sum_{\substack{ q_1,\cdots,q_n \geq 1, \\ q_1+\cdots+q_n=q}}\int_0^1 \dd s_1 \cdots \int_0^1 \dd s_n \mcl{T}_\rightarrow \left[ \prod_{n'=1}^n \bar{\Psi}_{q_{n'}}^i (s_{n'}) \right], \label{EqA_MPF:T_q_p0_BCH_expression}\\
    \bar{T}_{p_0}^{\mr{BCH},i}(t) &=& \sum_{n=0}^\infty \frac{(-i)^n}{n!}\int_0^1 \dd s_1 \cdots \int_0^1 \dd s_n\mcl{T}_\rightarrow \left[ \prod_{n'=1}^n \left(\sum_{q=1}^\infty \bar{\Psi}_q^i(s_{n'}) t^q + \bar{\Psi}_{p_0}^i(t;s_{n'})\right) - \prod_{n'=1}^n \left(\sum_{q=1}^\infty \bar{\Psi}_q^i(s_{n'}) t^q \right) \right]. \nonumber \\
    && \label{EqA_MPF:T_p0_t_BCH_expression}
\end{eqnarray}

\textit{Upper-bounding the norm of $\bar{T}_{q,p_0}^{\mr{BCH},i}$}.---
We next consider the upper bound on the norm of the operator $\bar{T}_{q,p_0}^{\mr{BCH},i}$.
Focusing on the operator $\bar{\Psi}_q^i(s)$ given by Eq. (\ref{EqA_MPF:Psi_q_def}), the commutator appearing in it is expressed by
\begin{eqnarray}
    \sum_{l \in \bbZ} \braket{l|\ad_{\bar{\Phi}_{q'}^{F,\Lambda_i}} (\bar{\Phi}_q^{F,\Lambda_i}-\bar{\Phi}_q^{F,\Lambda_{i-1}})|0} &=& \sum_{l,l' \in \bbZ} \left( \braket{l|\bar{\Phi}_{q'}^{F,\Lambda_i}|l'}\braket{l'|(\bar{\Phi}_q^{F,\Lambda_i}-\bar{\Phi}_q^{F,\Lambda_{i-1}})|0}-\braket{l|(\bar{\Phi}_q^{F,\Lambda_i}-\bar{\Phi}_q^{F,\Lambda_{i-1}})|l'}\braket{l'|\bar{\Phi}_{q'}^{F,\Lambda_i}|0}\right) \nonumber \\
    &=& \sum_{l,l' \in \bbZ} \left[ \braket{l|\bar{\Phi}_{q'}^{F,\Lambda_i}|0}, \braket{l'|(\bar{\Phi}_q^{F,\Lambda_i}-\bar{\Phi}_q^{F,\Lambda_{i-1}})|0} \right] - \delta_{q1}\sum_{l' \in \bbZ} l' \omega \braket{l'|(\bar{\Phi}_q^{F,\Lambda_i}-\bar{\Phi}_q^{F,\Lambda_{i-1}})|0} \nonumber \\
    &=& \left[ \bar{\Phi}_{q'}^{\Lambda_i}(0), \bar{\Phi}_{q}^{\Lambda_i}(0) - \bar{\Phi}_{q}^{\Lambda_{i-1}}(0) \right] + \delta_{q1}\left( i\dv{t} \bar{\Phi}_{q}^{\Lambda_i}(0) - i\dv{t} \bar{\Phi}_{q}^{\Lambda_{i-1}}(0) \right). \label{EqA_MPF:com_Phi_Phi}
\end{eqnarray}
We use the translation symmetry of $\bar{\Phi}_{q}^{F,\Lambda_i}$ by Eq. (\ref{EqA_MPF:Phi_q_symmetry}) for the second equality, and use the operator $\bar{\Phi}_{q}^{\Lambda_i}(t)$ defined by Eq. (\ref{EqA_MPF:Phi_q_F_form}) for the last equality.
Since the commutator $[\bar{\Phi}_{q'}^{F,\Lambda_i},\bar{\Phi}_q^{F,\Lambda_i}-\bar{\Phi}_q^{F,\Lambda_{i-1}}]$ has the translation invariance like $\bar{\Phi}_q^{F,\Lambda_i}-\bar{\Phi}_q^{F,\Lambda_{i-1}}$, the above calculation can be repeated for the nested commutators in Eq. (\ref{EqA_MPF:Psi_q_def}).
We obtain the relation,
\begin{equation}\label{EqA_MPF:An_tau_def}
    \bar{\Psi}_q^i(s) = \sum_{n=0}^{p_0} \frac{(is)^n}{n!} \sum_{\substack{1 \leq q_0,\cdots,q_n \leq p_0, \\ q_0+\cdots+q_n=q}} \left. \mcl{A}_n(\tau) \left[ \bar{\Phi}_{q_0}^{\Lambda_i}(\tau)- \bar{\Phi}_{q_0}^{\Lambda_{i-1}}(\tau) \right] \right|_{\tau=0}, \quad 
    \mcl{A}_n(\tau) = \prod_{n'=1}^n \left( \ad_{\bar{\Phi}_{q_{n'}}^{\Lambda_i}(\tau)}+\delta_{q_{n'}1} i \dv{\tau} \right),
\end{equation}
for the operator $\bar{\Psi}_q^i(s)$.

We consider the norm of each operator appearing in the summation, Eq. (\ref{EqA_MPF:An_tau_def}).
As discussed in Appendix \ref{SubsecA:truncated_BCH}, every term of the BCH expansion $\bar{\Phi}_q^{\Lambda_i}(\tau)$ is reproduced by $(V_p)^q$ copies of $\prod_{q'=2}^q \bar{D}_{\gamma_{q'}}^{\Lambda_i}(\tau) H_{\gamma_1}^{\Lambda_i}(\tau)$.
Each term with certain indices $\gamma_1,\cdots,\gamma_q$ for the subsystem $\Lambda_i$ shares the same coefficient with the one with the same indices for the subsystem $\Lambda_{i-1}$ by definition.
As a result, we obtain
\begin{eqnarray}
    && \norm{\mcl{A}_n(\tau) \left[ \bar{\Phi}_{q_0}^{\Lambda_i}(\tau)- \bar{\Phi}_{q_0}^{\Lambda_{i-1}}(\tau) \right]} \nonumber \\
    && \qquad \leq (V_p)^{q_0}\sum_{\gamma_1=1}^\Gamma \sum_{\gamma_2,\cdots,\gamma_{q_0}=1}^{\Gamma + 1} \norm{\mcl{A}_n(\tau)  \left[ \left(\prod_{q=2}^{q_0} \bar{\mcl{D}}_{\gamma_q}^{\Lambda_i}(\tau)\right) H_{\gamma_1}^{\Lambda_i}(\tau) - \left(\prod_{q=2}^{q_0} \bar{\mcl{D}}_{\gamma_q}^{\Lambda_{i-1}}(\tau)\right) H_{\gamma_1}^{\Lambda_{i-1}}(\tau) \right]} \nonumber \\
    && \qquad \leq (V_p)^{q_0}\sum_{\gamma_1=1}^\Gamma \sum_{\gamma_2,\cdots,\gamma_{q_0}=1}^{\Gamma + 1} \norm{\mcl{A}_n(\tau) \left(\prod_{q=2}^{q_0} \bar{\mcl{D}}_{\gamma_q}^{\Lambda_i}(\tau)\right) \left[ H_{\gamma_1}^{\Lambda_i}(\tau) - H_{\gamma_1}^{\Lambda_{i-1}}(\tau)\right] } \nonumber \\
    && \qquad \qquad + \sum_{q'=2}^{q_0} (V_p)^{q_0}\sum_{\gamma_1=1}^\Gamma \sum_{\gamma_2,\cdots,\gamma_{q_0}=1}^{\Gamma + 1} \norm{\mcl{A}_n(\tau) \left(\prod_{q=q'+1}^{q_0} \bar{\mcl{D}}_{\gamma_q}^{\Lambda_i}(\tau)\right) \left[ \bar{\mcl{D}}_{\gamma_{q'}}^{\Lambda_i}(\tau) - \bar{\mcl{D}}_{\gamma_{q'}}^{\Lambda_{i-1}}(\tau)\right] \left(\prod_{q=2}^{q'-1} \bar{\mcl{D}}_{\gamma_q}^{\Lambda_{i-1}}(\tau)\right)H_{\gamma_1}^{\Lambda_{i-1}}(\tau) }. \nonumber \\
    && \label{EqA_MPF:An_Phi_q_diff}
\end{eqnarray}
The last inequality comes from the triangular inequality.
As discussed in Eq. (\ref{EqA_MPF:H_Lambda_i_diff}), all the local terms of $H_{\gamma_1}^{\Lambda_i}(\tau) - H_{\gamma_1}^{\Lambda_{i-1}}(\tau)$ are covered by the set of $\{ h_X^{\gamma_1}(\tau)\}$ such that the domain $X$ includes the site $i$.
Similarly, the operator $\bar{\mcl{D}}_{\gamma_{q'}}^{\Lambda_i}(\tau) - \bar{\mcl{D}}_{\gamma_{q'}}^{\Lambda_{i-1}}(\tau)$ is composed of $\ad_{h_X^{\gamma_{q'}}(\tau)}$ with $X \ni i$ for $\gamma_{q'}=1,\cdots,\Gamma$, while it vanishes for $\gamma_{q'}=\Gamma +1$.
Thus, Eq. (\ref{EqA_MPF:An_Phi_q_diff}) is further bounded by
\begin{eqnarray}
    [\text{Eq. (\ref{EqA_MPF:An_Phi_q_diff})}] &\leq& (V_p)^{q_0}\sum_{X \ni i}\sum_{\gamma_1=1}^\Gamma \sum_{\gamma_2,\cdots,\gamma_{q_0}=1}^{\Gamma + 1} \norm{\mcl{A}_n(\tau) \left(\prod_{q=2}^{q_0} \bar{\mcl{D}}_{\gamma_q}^{\Lambda_i}(\tau)\right) h_X^{\gamma_1}(\tau)} \nonumber \\
    && \quad + \sum_{q'=2}^{q_0} (V_p)^{q_0} \sum_{X \ni i}\sum_{\gamma_1,\gamma_{q'}=1}^\Gamma \sum_{\gamma_2,\cdots,\gamma_{q_0}=1}^{\Gamma + 1} \norm{\mcl{A}_n(\tau) \left(\prod_{q=q'+1}^{q_0} \bar{\mcl{D}}_{\gamma_q}^{\Lambda_i}(\tau)\right) \left[ h_X^{\gamma_{q'}}(\tau) , \left(\prod_{q=2}^{q'-1} \bar{\mcl{D}}_{\gamma_q}^{\Lambda_i}(\tau)\right)H_{\gamma_1}^{\Lambda_{i-1}}(\tau)\right]}. \nonumber \\
    && \label{EqA_MPF:An_Phi_q_diff_2}
\end{eqnarray}
Using the locality, the extensiveness, and the strength of time-dependency of the operator $\bar{\Phi}_q(\tau)$ by Lemma \ref{LemA_MPF:Phi_q_extensive}, we can obtain the following relations,
\begin{equation}\label{EqA_MPF:An_D_h_bound}
    \sum_{X \ni i}\sum_{\gamma_1=1}^\Gamma \sum_{\gamma_2,\cdots,\gamma_{q_0}=1}^{\Gamma + 1} \norm{\mcl{A}_n(\tau) \left(\prod_{q=2}^{q_0} \bar{\mcl{D}}_{\gamma_q}^{\Lambda_i}(\tau)\right) h_X^{\gamma_1}(\tau)} \leq \prod_{n'=1}^n \left[ (2k g(\bar{\Phi}_{q_{n'}})+f) \sum_{n''=0}^{n'-1} q_{n''}\right] (q_0-1)! (2kg+\Gamma f)^{q_0-1} g, 
\end{equation}
and
\begin{eqnarray}
    && \sum_{X \ni i}\sum_{\gamma_1,\gamma_{q'}=1}^\Gamma \sum_{\gamma_2,\cdots,\gamma_{q_0}=1}^{\Gamma + 1} \norm{\mcl{A}_n(\tau) \left(\prod_{q=q'+1}^{q_0} \bar{\mcl{D}}_{\gamma_q}^{\Lambda_i}(\tau)\right) \left[ h_X^{\gamma_{q'}}(\tau) , \left(\prod_{q=2}^{q'-1} \bar{\mcl{D}}_{\gamma_q}^{\Lambda_i}(\tau)\right)H_{\gamma_1}^{\Lambda_{i-1}}(\tau)\right]} \nonumber \\
    && \qquad \qquad \qquad \qquad\qquad \qquad\qquad \qquad \qquad \leq \prod_{n'=1}^n \left[ (2k g(\bar{\Phi}_{q_{n'}})+f) \sum_{n''=0}^{n'-1} q_{n''}\right] (q_0-1)! (2kg+\Gamma f)^{q_0-1} g. \label{EqA_MPF:An_D_h_D_H_bound}
\end{eqnarray}
The quantity $g(\bar{\Phi}_q)$ denotes the extensiveness of $\bar{\Phi}_q(\tau)$ given by Eq. (\ref{EqA_MPF:gf_Phi_values}).
Since their derivation is too long and essentially parallel to the calculation in Section \ref{Subsec:bound_commutator}, we attach it at the end of this section and continue the proof of Lemma \ref{LemA_MPF:T_q_BCH_expansion} here.

Using the inequalities Eqs. (\ref{EqA_MPF:An_D_h_bound}) and (\ref{EqA_MPF:An_D_h_D_H_bound}), Eq. (\ref{EqA_MPF:An_Phi_q_diff_2}) is bounded by
\begin{eqnarray}
    \norm{\mcl{A}_n(\tau) \left[ \bar{\Phi}_{q_0}^{\Lambda_i}(\tau)- \bar{\Phi}_{q_0}^{\Lambda_{i-1}}(\tau) \right]} &\leq& (V_p)^{q_0}[1+(q_0-1)] \prod_{n'=1}^n \left[ (2k g(\bar{\Phi}_{q_{n'}})+f) \sum_{n''=0}^{n'-1} q_{n''}\right] (q_0-1)! (2kg+\Gamma f)^{q_0-1} g \nonumber \\
    &\leq& (V_p)^{q_0} q_0! (2kg+\Gamma f)^{q_0-1} g \prod_{n'=1}^n \left\{ \left[ q_{n'}! (V_p)^{q_{n'}} (2kg+\Gamma f)^{q_{n'}-1} 2kg +f \right]q\right\} \nonumber \\
    &\leq& (p_0V_p)^q (2kg+\Gamma f)^{q-1} g (2q)^n. \label{EqA_MPF:An_Phi_diff_bound}
\end{eqnarray}
In the last inequality, we consider the indices $q_0,q_1,\cdots,q_n$ such that $q_0,\cdots,q_n \leq p_0$ and $q_0+\cdots+q_n = q$, which are valid relations in the summation Eq. (\ref{EqA_MPF:An_tau_def}), and employ the resulting inequality $q_0! \cdots q_n ! \leq (p_0)^{q_0+\cdots+q_n} = (p_0)^q$.
The operator $\bar{\Psi}_q^i (s)$ expressed by Eq. (\ref{EqA_MPF:Psi_q_def}) is bounded by
\begin{eqnarray}
    \norm{\bar{\Psi}_q^i(s)} &\leq& \sum_{n=0}^{p_0} \frac{s^n}{n!} \sum_{\substack{1 \leq q_0,\cdots,q_n \leq p_0, \\ q_0+\cdots+q_n=q}}  (p_0 V_p)^q (2kg+\Gamma f)^{q-1} g (2q)^n \nonumber \\
    &\leq& \frac12 \{ 2p_0V_p(2kg+\Gamma f)\}^q \sum_{n=0}^\infty \frac{(2sq)^n}{n!} = \frac12 \{ 2e^{2s} p_0 V_p(2kg+\Gamma f)\}^q. \label{EqA_MPF:Psi_q_bound}
\end{eqnarray}
Finally, substituting this relation into Eq. (\ref{EqA_MPF:T_q_p0_BCH_expression}), we arrive at the upper bound of the operator $\bar{T}_{q,p_0}^{\mr{BCH},i}$ give by
\begin{eqnarray}
    \norm{\bar{T}_{q,p_0}^{\mr{BCH},i}}t^q &\leq& \sum_{n=0}^\infty \frac{1}{n!} \sum_{\substack{1 \leq q_1,\cdots,q_n \leq p_0, \\ q_1+\cdots+q_n=q}}\int_0^1 \dd s_1 \cdots \int_0^1 \dd s_n \prod_{n'=1}^n \left\{ \frac12[ 2e^{2s_{n'}} p_0V_p (2kg+\Gamma f)]^{q_{n'}} \right\} t^q \nonumber \\
    &\leq& [ 2e^2 p_0 V_p (2kg+\Gamma f)t]^q \sum_{n=0}^\infty \frac{2^{q-n-1}}{n!} = \frac1{2e^2} [ 4e^2 p_0 V_p (2kg+\Gamma f)t]^q.
\end{eqnarray}
This completes the proof of Eq. (\ref{EqA_MPF:T_q_p0_BCH_bound}).

\textit{Upper-bounding the norm of $\bar{T}_{p_0}^{\mr{BCH},i}(t)$.---}
We consider the norm of the higher order term $\bar{T}_{p_0}^{\mr{BCH},i}(t)$. 
Using Eq. (\ref{EqA_MPF:T_p0_t_BCH_expression}), it is bounded by
\begin{equation}\label{EqA_MPF:T_p0_BCH_t_bound_0}
    \norm{\bar{T}_{p_0}^{\mr{BCH},i}(t)} \leq \sum_{n=0}^\infty \frac1{n!} \sum_{n'=1}^n \frac{n!}{n'!(n-n')!} \left( \int_0^1 \dd s \sum_{q=1}^\infty \norm{\bar{\Psi}_q^i(s)}t^q \right)^{n-n'} \left( \int_0^1 \dd s \norm{\bar{\Psi}_{p_0}^i (t;s)}\right)^{n'}.
\end{equation}
The inequality Eq. (\ref{EqA_MPF:Psi_q_bound}) and the assumption Eq. (\ref{EqA_MPF:t_condition_T_BCH}) imply an upper bound,
\begin{equation}\label{EqA_MPF:Psi_q_t_integral_bound}
    \int_0^1 \dd s \sum_{q=1}^\infty \norm{\bar{\Psi}_q^i(s)}t^q \leq \sum_{q=1}^\infty \frac12 [ 2e^2 p_0 V_p(2kg+\Gamma f) t]^q \leq \sum_{q=1}^\infty 2^{-n-1} = \frac12,
\end{equation}
and hence it is sufficient to evaluate the upper bound on $\norm{\bar{\Psi}_{p_0}^i (t;s)}$.

Let us define an operator $\bar{U}_{p_0}^{\bar{\Phi}}(st)$ by
\begin{equation}
    \bar{U}_{p_0}^{\bar{\Phi}}(st) = \sum_{l \in \bbZ} e^{-il\omega st} \braket{l|e^{-is \sum_{q=1}^{p_0} \bar{\Phi}_q^{F,\Lambda_i} t^q}|0}.
\end{equation}
Then, using the symmetry of $\bar{\Phi}_q^{F,\Lambda_i}$ by Eq. (\ref{EqA_MPF:Phi_q_symmetry}), the operator $\bar{U}_{p_0}^{\bar{\Phi}}(st)$ is a unitary operator like Eq. (\ref{EqA_MPF:unitarity_T_dag_T}) and gives an expression of $\bar{\Psi}_{p_0}^i (t;s)$ by
\begin{eqnarray}
    \bar{\Psi}_{p_0}^i (t;s) &=& \int_0^s \dd s' \bar{U}_{p_0}^{\bar{\Phi}}(s't)^\dagger \left[ \sum_{l \in \bbZ} e^{-il\omega s' t} \bra{l}\frac{\left( is'\sum_{q=1}^{p_0} t^q \ad_{\bar{\Phi}_q^{F,\Lambda_i}}\right)^{p_0+1}}{(p_0+1)!} \sum_{q=1}^{p_0}(\bar{\Phi}_q^{F,\Lambda_{i-1}}-\bar{\Phi}_q^{F,\Lambda_i})t^q \ket{0} \right] \bar{U}_{p_0}^{\bar{\Phi}}(s't) \nonumber \\
    &=& \int_0^s \dd s' \bar{U}_{p_0}^{\bar{\Phi}}(s't)^\dagger \left[ \frac{(is')^{p_0+1}}{(p_0+1)!} \sum_{q=p_0+1}^{p_0(p_0+2)} t^q \sum_{\substack{1 \leq q_0,\cdots,q_{p_0+1} \leq p_0; \\ q_0+\cdots+q_{p_0+1}=q}}  \left. \mcl{A}_{p_0+1}(\tau) \left( \bar{\Phi}_{q_0}^{\Lambda_i}(\tau)- \bar{\Phi}_{q_0}^{\Lambda_{i-1}}(\tau) \right) \right|_{\tau=s't} \right]\bar{U}_{p_0}^{\bar{\Phi}}(s't). \nonumber \\
    &&
\end{eqnarray}
The first equality comes from the calculation similar to Eq. (\ref{Eq_Err:t_dep_error_Delta_1}), in which we use the translation symmetry Eq. (\ref{EqA_MPF:Phi_q_symmetry}).
The second equality is obtained in a similar manner to Eqs. (\ref{EqA_MPF:com_Phi_Phi}) and (\ref{EqA_MPF:An_tau_def}) while the time $\tau=0$ is replaced by $\tau=s't \in [0,t]$ due to the phase factor $e^{-il\omega s't}$ in the first line.
Employing the upper bound Eq. (\ref{EqA_MPF:An_Phi_diff_bound}) under $n=p_0+1$, the operator $\bar{\Psi}_{p_0}^i (t;s)$ has a norm bounded by
\begin{eqnarray}
    \norm{\bar{\Psi}_{p_0}^i (t;s)} &\leq& \frac{s^{p_0+1}}{(p_0+1)!} \sum_{q=p_0+1}^\infty t^q \sum_{\substack{1 \leq q_0,\cdots,q_{p_0+1} \leq p_0; \\ q_0+\cdots+q_{p_0+1}=q}} (p_0V_p)^q (2kg+\Gamma f)^{q-1} g (2q)^{p_0+1} \nonumber \\
    &\leq& \frac12 s^{p_0+1} \sum_{q=p_0+1}^\infty [ 2e^2 p_0 V_p (2kg + \Gamma f)t]^q \leq [ 2e^2 s p_0 V_p(2kg+\Gamma f)t ]^{p_0+1}. \label{EqA_MPF:Psi_p0_t_bound}
\end{eqnarray}
We use the relation $(2q)^{p_0+1}/(p_0+1)! \leq e^{2q}$ and the assumption Eq. (\ref{EqA_MPF:t_condition_T_BCH}) respectively for the second and last inequalities.
Summarizing the relations, Eqs. (\ref{EqA_MPF:T_p0_BCH_t_bound_0}), (\ref{EqA_MPF:Psi_q_t_integral_bound}), and (\ref{EqA_MPF:Psi_p0_t_bound}), we finally obtain
\begin{eqnarray}
    \norm{\bar{T}_{p_0}^{\mr{BCH},i}(t)} &\leq& \sum_{n=0}^\infty \frac1{n!} \sum_{n'=1}^n \frac{n!}{n'!(n-n')!} \left( \frac12 \right)^{n-n'} [ 2e^2 p_0 V_p(2kg+\Gamma f)t ]^{p_0+1} \nonumber \\
    &\leq& e^{3/2} [ 2e^2 p_0 V_p(2kg+\Gamma f)t ]^{p_0+1},
\end{eqnarray}
which completes the proof of Eq. (\ref{EqA_MPF:T_p0_t_BCH_bound}). $\quad \square$

\textbf{Proof of Eq. (\ref{EqA_MPF:An_D_h_bound}).---} 
We first consider an upper bound of
\begin{equation}\label{EqA_MPF:D_h_sum}
    \sum_{X \ni i}\sum_{\gamma_1=1}^\Gamma \sum_{\gamma_2,\cdots,\gamma_{q_0}=1}^{\Gamma + 1} \norm{ \left(\prod_{q=2}^{q_0} \bar{\mcl{D}}_{\gamma_q}^{\Lambda_i}(\tau)\right) h_X^{\gamma_1}(\tau)},
\end{equation}
which corresponds to the case, $n=0$.
In a similar manner to the discussion in Section \ref{Subsec:bound_commutator}, the operator $\prod_{q=2}^{q_0} \{ \bar{\mcl{D}}_{\gamma_q}^{\Lambda_i}(\tau) \} h_X^{\gamma_1}(\tau)$ is expanded by
\begin{equation}\label{EqA_MPF:nested_com_h_h_h}
    \Gamma^{q_0-1-r}\sum_{X_2,\cdots,X_r \subset \Lambda_i} [h_{X_r}^{\gamma_r (n_r)}(\tau), \cdots, [h_{X_2}^{\gamma_2 (n_2)}(\tau),h_X^{\gamma_1 (n_1)}(\tau)]],
\end{equation}
such that $r+n_1+\cdots+n_r=q_0-1$.
Its summation has a norm bounded by
\begin{eqnarray}
    && \Gamma^{q_0-1-r} \sum_{\gamma_1}^\Gamma \sum_{X \ni i}\sum_{X_2,\cdots,X_r \subset \Lambda_i} \norm{[h_{X_r}^{\gamma_r (n_r)}(\tau), \cdots, [h_{X_2}^{\gamma_2 (n_2)}(\tau),h_X^{\gamma_1 (n_1)}(\tau)]]} \nonumber \\
    && \qquad \leq 2^{r-1} \Gamma^{q_0-1-r} \sum_{\gamma_1,\cdots,\gamma_r=1}^\Gamma \sum_{X \ni i}\norm{h_X^{\gamma_1(n_1)}(\tau)} \prod_{r'=r}^2 \left( \sum_{\gamma_{r'}=1}^\Gamma \sum_{\substack{X_{r'} \subset \Lambda_i; \\ X_{r'}\cap Y_{r'} \neq \phi}}\norm{h_{X_{r'}}^{\gamma_{r'}(n_{r'})}(\tau)} \right) \nonumber \\
    && \qquad \leq 2^{r-1} \Gamma^{q_0-1-r} f^{n_1} g \prod_{r'=2}^r \left[ (r'-1)k f^{n_{r'}} g \right] = (r-1)! (2kg)^{r-1} (\Gamma f)^{q_0-1-r} g.
\end{eqnarray}
As discussed in Section \ref{Subsec:bound_commutator}, Eq. (\ref{EqA_MPF:D_h_sum}) is bounded by at most $(q_0-1)!/\{r! (q_0-r-1)!\} \times r^{q_0-1-r}$ copies of Eq. (\ref{EqA_MPF:nested_com_h_h_h}).
The former factor comes from the number of ways to pick up $q_0-1-r$ indices giving the time derivative $\mcl{D}_{\Gamma+1}(\tau)=\Gamma \dv{\tau}$.
The latter one means the maximal number of copies when the Leibnitz rule is taken into account.
As a result, Eq. (\ref{EqA_MPF:D_h_sum}) is bounded by
\begin{eqnarray}
    \sum_{X \ni i}\sum_{\gamma_1=1}^\Gamma \sum_{\gamma_2,\cdots,\gamma_{q_0}=1}^{\Gamma + 1} \norm{\left(\prod_{q=2}^{q_0} \bar{\mcl{D}}_{\gamma_q}^{\Lambda_i}(\tau)\right) h_X^{\gamma_1}(\tau)} 
    &\leq& \sum_{r=0}^{q_0-1} \frac{(q_0-1)!}{r! (q_0-1-r)!} r^{q_0-1-r}  (r-1)! (2kg)^{r-1} (\Gamma f)^{q_0-r-1} g \nonumber \\
    &\leq& (q_0-1)! (2kg+\Gamma f)^{q_0-1} g. \label{EqA_MPF:D_h_i_bound}
\end{eqnarray}

We next consider the case where the operator $\mcl{A}_n(\tau)$ is applied.
The operator $\mcl{A}_n(\tau) \prod_{q=2}^{q_0} [ \bar{\mcl{D}}_{\gamma_q}^{\Lambda_i}(\tau) ] h_X^{\gamma_1}(\tau)$ is composed of nested commutators in the form of
\begin{equation}\label{EqA_MPF:An_nested_com_h_h}
    \sum_{X_2,\cdots,X_{r+r'} \in \Lambda_i}[h_{X_{r+r'}}^{\bar{\Phi}_{q_{r'}}^{\Lambda_i}(n_{r+r'})}(\tau),[\cdots,[h_{X_{r+1}}^{\bar{\Phi}_{q_1}^{\Lambda_i}(n_{r+1})}(\tau),[h_{X_r}^{\gamma_r (n_r)}(\tau), \cdots, [h_{X_2}^{\gamma_2 (n_2)}(\tau),h_X^{\gamma_1 (n_1)}(\tau)]]]]]
\end{equation}
with some set of the integers $n_1,\cdots,n_{r+r'}$.
The operator $h_X^{\bar{\Phi}_q (n')}(\tau)$ denotes the $n'$-th derivative of $h_X^{\bar{\Phi}_q}(\tau)$, which arises from the time derivative in $\mcl{A}_n(\tau)$.
The number $r'$ satisfies $r' \leq n$ since the operator $\mcl{A}_n(\tau)$ applies at most $n$-fold commutators to Eq. (\ref{EqA_MPF:nested_com_h_h_h}).
The summation of Eq. (\ref{EqA_MPF:An_nested_com_h_h}) over $X \ni i$ and $\gamma_1,\cdots,\gamma_r$ has a norm bounded by
\begin{eqnarray}
    && \sum_{\gamma_1,\cdots,\gamma_r=1}^\Gamma \sum_{X \ni i} \sum_{X_2,\cdots,X_{r+r'} \in \Lambda_i} \norm{[h_{X_{r+r'}}^{\bar{\Phi}_{q_{r'}}^{\Lambda_i}(n_{r+r'})}(\tau),[\cdots,[h_{X_{r+1}}^{\bar{\Phi}_{q_1}^{\Lambda_i}(n_{r+1})}(\tau),[h_{X_r}^{\gamma_r (n_r)}(\tau), \cdots, [h_{X_2}^{\gamma_2 (n_2)}(\tau),h_X^{\gamma_1 (n_1)}(\tau)]]]]]} \nonumber \\
    && \qquad \leq 2^{r+r'-1} \sum_{\gamma_1=1}^\Gamma \sum_{X \ni i} \norm{h_X^{\gamma_1 (n_1)}} \prod_{r''=r}^2 \left( \sum_{\gamma_{r''}=1}^\Gamma \sum_{\substack{X_{r''} \subset \Lambda_i; \\ X_{r''} \cap Y_{r''} \neq \phi}} \norm{h_{X_{r''}}^{\gamma_{r''}(n_{r''})}(\tau)}\right)  \prod_{r''=r'}^1 \left(  \sum_{\substack{X_{r+r''} \subset \Lambda_i; \\ X_{r+r''} \cap Y_{r+r''} \neq \phi}} \norm{h_{X_{r+r''}}^{\bar{\Phi}_{q_{r''}}(n_{r+r''})}(\tau)}\right) \nonumber \\
    && \qquad \leq 2^{r+r'-1} f^{n_1} g \prod_{r''=r}^2 \left[  (r''-1)kf^{n_{r''}} g\right] \prod_{r''=r'}^1 \left[ \left( r+ \sum_{r'''=1}^{r''-1} q_{r'''} \right)k [f(\bar{\Phi}_{q_{r''}})]^{n_{r+r''}} g(\bar{\Phi}_{q_{r''}}) \right]. \label{EqA_MPF:com_h_h_bound}
\end{eqnarray}
In the last inequality, we use the fact that $Y_{r+r''}$ contains at most $rk+\sum_{r'''<r'} q_{r'''}k$ sites since each $\bar{\Phi}_q(\tau)$ is $qk$-local.
We also use the extensiveness and the strength of time-dependency for the operator $\bar{\Phi}_q(\tau)$ given by Lemma \ref{LemA_MPF:Phi_q_extensive}.
Let $C_{r,r',n_1,\cdots,n_{r+r'}}^n$ denote the number of copies of Eq. (\ref{EqA_MPF:An_nested_com_h_h}) in the left-hand side of Eq. (\ref{EqA_MPF:An_D_h_bound}).
For instance, we have $C_{r,r',n_1,\cdots,n_{r+r'}}^0 \leq (q_0-1)!/[r! (q_0-r-1)!] \times r^{q_0-1-r} \times \Gamma^{q_0-1-r}$ for all the possible numbers $r,r',n_1,\cdots,n_r$, as discussed in Eq. (\ref{EqA_MPF:D_h_i_bound}).
Then, using the relation Eq. (\ref{EqA_MPF:com_h_h_bound}), we obtain
\begin{eqnarray}
    && \sum_{X \ni i}\sum_{\gamma_1=1}^\Gamma \sum_{\gamma_2,\cdots,\gamma_{q_0}=1}^{\Gamma + 1} \norm{\mcl{A}_n(\tau) \left(\prod_{q=2}^{q_0} \bar{\mcl{D}}_{\gamma_q}^{\Lambda_i}(\tau)\right) h_X^{\gamma_1}(\tau)} \nonumber \\
    && \qquad \leq \sum_{r,r',n_1,\cdots,n_{r+r'}} C_{r,r',n_1,\cdots,n_{r+r'}}^n   2^{r+r'-1} g \prod_{r''=r}^2 \left[  (r''-1)kf^{n_{r''}} g\right] \prod_{r''=r'}^1 \left[  \left(\sum_{r'''=0}^{r''-1} q_{r'''} \right) k [f(\bar{\Phi}_{q_{r''}})]^{n_{r+r''}} g(\bar{\Phi}_{q_{r''}}) \right]. \nonumber \\
    && \label{EqA_MPF:An_D_h_bound_relation}
\end{eqnarray}

We evaluate the above upper bound in a recursive way.
Supposing that the operator $\mcl{A}_n(\tau) \prod_{q=2}^{q_0} [ \bar{\mcl{D}}_{\gamma_q}^{\Lambda_i}(\tau) ] h_X^{\gamma_1}(\tau)$ is expanded by $C_{r,r',n_1,\cdots,n_{r+r'}}^n$ copies of Eq. (\ref{EqA_MPF:An_nested_com_h_h}), we focus on the operator $\mcl{A}_{n+1}(\tau) \prod_{q=2}^{q_0} [ \bar{\mcl{D}}_{\gamma_q}^{\Lambda_i}(\tau) ] h_X^{\gamma_1}(\tau)$.
When the index $q_{n+1}$ is equal to $1$, it is expanded by $C_{r,r',n_1,\cdots,n_{r+r'}}^n$ copies of copies of
\begin{eqnarray}
    && \left( \ad_{\bar{\Phi}_1^{\Lambda_i}(\tau)} + i \dv{\tau}\right) \sum_{X_2,\cdots,X_{r+r'} \in \Lambda_i}[h_{X_{r+r'}}^{\bar{\Phi}_{q_{r'}}^{\Lambda_i}(n_{r+r'})}(\tau),[\cdots,[h_{X_{r+1}}^{\bar{\Phi}_{q_1}^{\Lambda_i}(n_{r+1})}(\tau),[h_{X_r}^{\gamma_r (n_r)}(\tau), \cdots, [h_{X_2}^{\gamma_2 (n_2)}(\tau),h_X^{\gamma_1 (n_1)}(\tau)]]]]] \nonumber \\
    && \qquad = \sum_{X_2,\cdots,X_{r+r'+1} \in \Lambda_i}[h_{X_{r+r'+1}}^{\bar{\Phi}_1^{\Lambda_i}}(\tau),[h_{X_{r+r'}}^{\bar{\Phi}_{q_{r'}}^{\Lambda_i}(n_{r+r'})}(\tau),[\cdots,[h_{X_{r+1}}^{\bar{\Phi}_{q_1}^{\Lambda_i}(n_{r+1})}(\tau),[h_{X_r}^{\gamma_r (n_r)}(\tau), \cdots, [h_{X_2}^{\gamma_2 (n_2)}(\tau),h_X^{\gamma_1 (n_1)}(\tau)]]]]]] \nonumber \\
    && \qquad \qquad + i\sum_{l=0}^{r+r'} \sum_{X_2,\cdots,X_{r+r'} \in \Lambda_i}[h_{X_{r+r'}}^{\bar{\Phi}_{q_{r'}}^{\Lambda_i}(n_{r+r'}+\delta_{l,r+r'})}(\tau),[\cdots,[h_{X_{r+1}}^{\bar{\Phi}_{q_1}^{\Lambda_i}(n_{r+1}+\delta_{l,r+1})}(\tau),[ \cdots, [h_{X_2}^{\gamma_2 (n_2+\delta_{l,2})}(\tau),h_X^{\gamma_1 (n_1+\delta_{l,1})}(\tau)]]]]]. \nonumber \\
    && \label{EqA_MPF:A_n_plus_1_com}
\end{eqnarray}
The second term comes from the Liebnitz rule.
We consider the summation of the norm over $X \ni i$ and $\gamma_1,\cdots,\gamma_r$ in a similar manner to Eq. (\ref{EqA_MPF:com_h_h_bound}), and then it is bounded by
\begin{eqnarray}
    && \sum_{\gamma_1,\cdots,\gamma_r=1}^\Gamma [\text{norm of Eq. (\ref{EqA_MPF:A_n_plus_1_com})}] \nonumber \\
    && \qquad \leq 2^{r+r'} f^{n_1}g \prod_{r''=r}^2 \left[  (r''-1)kf^{n_{r''}} g\right] \prod_{r''=r'}^1 \left[ \left( \sum_{r'''=0}^{r''-1} q_{r'''} \right)k [f(\bar{\Phi}_{q_{r''}})]^{n_{r+r''}} g(\bar{\Phi}_{q_{r''}}) \right] \times |Y_{r+r'}|g(\bar{\Phi}_1^{\Lambda_i}) \nonumber \\
    && \qquad \qquad + \sum_{l=0}^{r+r'} 2^{r+r'-1} f^{n_1+\delta_{l,1}} g \prod_{r''=r}^2 \left[  (r''-1)kf^{n_{r''}+\delta_{l,r''}} g\right] \prod_{r''=r'}^1 \left[ \left(  \sum_{r'''=0}^{r''-1} q_{r'''} \right)k [f(\bar{\Phi}_{q_{r''}})]^{n_{r+r''}+\delta_{l,r+r''}} g(\bar{\Phi}_{q_{r''}}) \right] \nonumber \\
    && \qquad \leq \left[ 2 \left( r+ \sum_{r'''=1}^{r''} q_{r'''}\right) kg(\bar{\Phi}_1^{\Lambda_i}) + rf + \sum_{r''=1}^r f(\bar{\Phi}_{q_{r''}})\right] \nonumber \\
    && \qquad \qquad \times 2^{r+r'-1} f^{n_1}g \prod_{r''=r}^2 \left[  (r''-1)kf^{n_{r''}} g\right] \prod_{r''=r'}^1 \left[ \left( \sum_{r'''=0}^{r''-1} q_{r'''} \right)k [f(\bar{\Phi}_{q_{r''}})]^{n_{r+r''}} g(\bar{\Phi}_{q_{r''}}) \right] \nonumber \\
    && \qquad \leq \left[ (2kg(\bar{\Phi}_1^{\Lambda_i})+f) \sum_{n'=1}^n q_{n'}\right] 2^{r+r'-1} f^{n_1}g \prod_{r''=r}^2 \left[  (r''-1)kf^{n_{r''}} g\right] \prod_{r''=r'}^1 \left[ \left( \sum_{r'''=0}^{r''-1} q_{r'''} \right)k [f(\bar{\Phi}_{q_{r''}})]^{n_{r+r''}} g(\bar{\Phi}_{q_{r''}}) \right]. \nonumber \\
    && \label{EqA_MPF:An_plus_1_bound}
\end{eqnarray}
In the last inequality, we use the relation $f(\bar{\Phi}_{q}) = qf$ by Lemma \ref{LemA_MPF:Phi_q_extensive}.
We also use $r+\sum_{r'''=1}^{r''} q_{r''} \leq \sum_{n'=0}^n q_{n'}$.
This comes from the fact that $r$ and $r'$ are at most $q_0$ and $n$ respectively so that the corresponding term Eq. (\ref{EqA_MPF:An_nested_com_h_h}) can appear in $\mcl{A}_n(\tau)\prod_{q=2}^{q_0} \{ \bar{\mcl{D}}_{\gamma_q}^{\Lambda_i}(\tau) \} h_X^{\gamma_1}(\tau)$.

The inequality Eq. (\ref{EqA_MPF:An_plus_1_bound}) holds also for $q_{n+1} \neq 1$ when we replace $g(\bar{\Phi}_1^{\Lambda_i})$ with $g(\bar{\Phi}_{q_{n+1}}^{\Lambda_i})$ with similar calculation, while it overestimates the upper bound due to the absence of the time derivative in Eq. (\ref{EqA_MPF:A_n_plus_1_com}).
In any case, this relation with the common number of the copies $C_{r,r',n_1,\cdots,n_{r+r'}}^n$ implies the recursive relation, 
\begin{equation}\label{EqA_MPF:D_h_bound_recursive}
    [\text{r.h.s. of Eq. (\ref{EqA_MPF:An_D_h_bound_relation}) for $n+1$}] \leq \left( [2kg(\bar{\Phi}_{n+1}^{\Lambda_i})+f] \sum_{n'=0}^n q_{n'}\right) \times [\text{r.h.s. of Eq. (\ref{EqA_MPF:An_D_h_bound_relation}) for $n$}].  
\end{equation}
By repeating this relation until the bound for $n=0$ appears, we obtain
\begin{equation}
    [\text{r.h.s. of Eq. (\ref{EqA_MPF:An_D_h_bound_relation}) for $n$}] \leq \prod_{n'=0}^{n-1} \left( [2kg(\bar{\Phi}_{n'}^{\Lambda_i})+f] \sum_{n''=0}^{n'-1} q_{n''}\right) (q_0-1)! (2kg+\Gamma f)^{q_0-1} g,
\end{equation}
where we use Eq. (\ref{EqA_MPF:D_h_i_bound}).
This completes the proof of Eq. (\ref{EqA_MPF:An_D_h_bound}). $\quad \square$

\textbf{Proof of Eq. (\ref{EqA_MPF:An_D_h_D_H_bound}).---}
The proof is parallel to the one for Eq. (\ref{EqA_MPF:An_D_h_bound}).
We first consider the case with $n=0$, i.e., the upper bound on
\begin{equation}\label{EqA_MPF:D_h_D_h}
    \sum_{X \ni i}\sum_{\gamma_1,\gamma_{q'}=1}^\Gamma \sum_{\gamma_2,\cdots,\gamma_{q_0}=1}^{\Gamma + 1} \norm{ \left(\prod_{q=q'+1}^{q_0} \bar{\mcl{D}}_{\gamma_q}^{\Lambda_i}(\tau)\right) \left[ h_X^{\gamma_{q'}}(\tau) , \left(\prod_{q=2}^{q'-1} \bar{\mcl{D}}_{\gamma_q}^{\Lambda_{i-1}}(\tau)\right)H_{\gamma_1}^{\Lambda_{i-1}}(\tau)\right]}
\end{equation}
In a similar manner to Eq. (\ref{EqA_MPF:D_h_sum}), the above equation is bounded by at most $(q_0-1)!/[r!(q_0-1-r)!] \times r^{q_0-1-r} \times \Gamma^{q_0-1-r}$ copies of
\begin{equation}\label{EqA_MPF:com_h_h_ver2}
    \sum_{\gamma_1,\cdots,\gamma_r=1}^\Gamma \sum_{X \ni i} \sum_{\substack{X_2,\cdots,X_{r'} \subset \Lambda_{i-1}, \\ X_{r'+1},\cdots,X_r \subset \Lambda_i}} \norm{[h_{X_{\gamma_r}}^{\gamma_{q_r}(n_r)}(\tau),\cdots,[h_X^{\gamma_1 (n_1)}(\tau),[h_{X_{r'}}^{\gamma_{r'} (n_{r'})}(\tau),[\cdots,[h_{X_3}^{\gamma_{3} (n_{3})}(\tau),h_{X_{2}}^{\gamma_{2} (n_{2})}(\tau)]]]]]}
\end{equation}
with some integers $r,r',n_1,\cdots,n_r$ satisfying $r' < r \leq q_0$ and $r+n_1+\cdots+n_r=q_0-1$.
This is different from Eq. (\ref{EqA_MPF:nested_com_h_h_h}) in that the domain $X$ containing the site $i$ appears in the middle of the nested commutator.
Let $Z_{r''}$ denote the union $Z_{r''} = X_2 \cup X_3 \cup \cdots \cup X_{r''-1}$ for $r''=2,3,\cdots,r'$, which contains at most $(r''-2)k$ sites.
Note that it is different from the union $Y_{r''} = X \cup X_2 \cup X_3 \cup \cdots \cup X_{r''-1}$.
Considering the set of the domains $X,X_2,\cdots,X_r$ such that the nested commutators among the operators defined on them can survive, Eq. (\ref{EqA_MPF:com_h_h_ver2}) is bounded by
\begin{eqnarray}
    [\text{Eq. (\ref{EqA_MPF:com_h_h_ver2})}] &\leq& 2^{r-1} \sum_{\gamma_1,\cdots,\gamma_{r'}=1}^\Gamma \sum_{X \ni i} \norm{h_X^{\gamma_1 (n_1)}(\tau)} \times \sum_{\substack{X_2,\cdots,X_{r'} \subset \Lambda_{i-1}; \\ X_{r''} \cap Z_{r''} \neq \phi, (r''=2,\cdots,r') ,\\ X \cap (X_{r'}\cup Z_{r'}) \neq \phi}} \prod_{r''=2}^{r'} \norm{h_{X_{r''}}^{\gamma_{r''}(n_{r''})}(\tau)} \nonumber \\
    && \qquad \times \prod_{r''=r}^{r'+1} \left\{ \sum_{\gamma_{r''}=1}^\Gamma \sum_{\substack{X_{r''} \subset \Lambda_i; \\ X_{r''} \cap Y_{r''} \neq \phi}} \norm{h_{X_{r''}}^{\gamma_{r''}(n_{r''})}(\tau)}\right\} \nonumber \\
    &\leq& 2^{r-1} \frac{(r-1)!}{r'!} (kg)^{r-r'} f^{n_{r'+1}+\cdots+n_r} \sum_{\gamma_1,\cdots,\gamma_{r'}=1}^\Gamma \sum_{X \ni i} \norm{h_X^{\gamma_1 (n_1)}(\tau)} \sum_{\substack{X_2,\cdots,X_{r'} \subset \Lambda_{i-1}; \\ X_{r''} \cap Z_{r''} \neq \phi, (r''=2,\cdots,r') ,\\ X \cap (X_{r'}\cup Z_{r'}) \neq \phi}} \prod_{r''=2}^{r'} \norm{h_{X_{r''}}^{\gamma_{r''}(n_{r''})}(\tau)}. \nonumber \\
    && \label{EqA_MPF:S_r_intro}
\end{eqnarray}
The second inequality is obtained by the same calculation as Eq. (\ref{EqA_MPF:com_h_h_bound}), in which we use the extensiveness and the strength of the time-dependency for $r''=r'+1,\cdots,r$.

In order to evaluate the upper bound of Eq. (\ref{EqA_MPF:S_r_intro}), we define the quantity $S_{r'}$ by
\begin{equation}
    S_{r'} = \max_{X \subset \Lambda; |X| \leq k} \left( \sum_{\gamma_2,\cdots,\gamma_{r'}=1}^\Gamma \sum_{\substack{X_2,\cdots,X_{r'} \subset \Lambda_{i-1}; \\ X_{r''} \cap Z_{r''} \neq \phi, (r''=2,\cdots,r') ,\\ X \cap (X_{r'}\cup Z_{r'}) \neq \phi}} \prod_{r''=2}^{r'} \norm{h_{X_{r''}}^{\gamma_{r''}(n_{r''})}(\tau)} \right),
\end{equation}
for $r'\geq 2$ and compute its upper bound.
The relation $X \cap (X_{r'}\cup Z_{r'})$ implies that there exists a certain domain $X_{r'''} \in \{X_2,\cdots,X_{r'}\}$ such that $X \cap X_{r'''} \neq \phi$, and hence $S_{r'}$ is bounded by
\begin{eqnarray}
    S_{r'} &\leq& \sum_{r'''=2}^{r'} \sum_{\gamma_2,\cdots,\gamma_{r'}=1}^\Gamma \sum_{\substack{X_2,\cdots,X_{r'} \subset \Lambda_{i-1}; \\ X_{r''} \cap Z_{r''} \neq \phi, (r''=2,\cdots,r') ,\\ X \cap X_{r'''} \neq \phi}} \prod_{r''=2}^{r'} \norm{h_{X_{r''}}^{\gamma_{r''}(n_{r''})}(\tau)} \nonumber \\
    &\leq& \sum_{r'''=2}^{r'} \sum_{\gamma_2,\cdots,\gamma_{r'}=1}^\Gamma \sum_{\substack{X_{r'''} \subset\Lambda_{i-1}; \\ X_{r'''} \cap X \neq \phi}} \norm{h_{X_{r'''}}^{\gamma_{r'''}(n_{r'''})}(\tau)} \sum_{\substack{X_2,\cdots,X_{r'''-1} \subset \Lambda_{i-1}; \\ X_{r''} \cap Z_{r''} \neq \phi, (r''=2,\cdots,r'''-1) ,\\ X_{r'''}\cap (X_{r'''-1} \cup Z_{r'''-1}) \neq \phi}} \prod_{r''=2}^{r'''-1} \norm{h_{X_{r''}}^{\gamma_{r''}(n_{r''})}(\tau)} \nonumber \\
    && \qquad \qquad\qquad\qquad\qquad\qquad\qquad\qquad\qquad\qquad \times \prod_{r''=r'''+1}^{r'} \left( \sum_{\substack{X_{r''} \subset \Lambda_{i-1}; \\X_{r''}\cap Z_{r''}\neq \phi}} \norm{h_{X_{r''}}^{\gamma_{r''}(n_{r''})}(\tau)}\right).
\end{eqnarray}
For $r''' \geq 3$, the summation over $X_2,\cdots,X_{r'''-1}$ can be bounded by $S_{r'''-1}$ since the domain $X_{r'''}$ contains at most $k$ sites due to the locality.
The corresponding part for $r'''=2$ is equal to $1$, and hence we set $S_1=1$ so that the bound by $S_{r'''-1}$ can work for any $r'''$.
The summation over $X_{r'''+1},\cdots,X_{r'}$ in Eq. (\ref{EqA_MPF:S_r_intro}) is evaluated in a similar manner to Eq. (\ref{EqA_MPF:com_h_h_bound}).
As a result, the quantity $S_{r'}$ is bounded by
\begin{eqnarray}
    S_{r'} &\leq& \sum_{r'''=2}^{r'} kf^{n_{r'''}} g S_{r'''-1} \times \prod_{r''=r'''+1}^{r'} \left[ (r''-1)k f^{n_{r''}} g\right] = \sum_{r'''=2}^{r'} \frac{(r'-1)!}{(r'''-1)!}(kg)^{r'-r''}f^{n_{r'''}+\cdots+n_{r'}}S_{r'''-1}.
\end{eqnarray}
This results in the recursive relation,
\begin{equation}
    \frac{S_{r'}}{r'! f^{n_2+\cdots+n_{r'}} (kg)^{r'}} \leq \frac1{r'} \sum_{r'''=2}^{r'} \frac{S_{r'''-1}}{(r'''-1)!f^{n_2+\cdots+n_{r'''-1}} (kg)^{r'''-1}}
\end{equation}
for $r' \geq 2$.
With the initial condition $S_1=1$, we can easily show that this implies the satisfaction of $S_{r'} \leq r'! f^{n_2+\cdots+n_{r'}} (kg)^{r'-1}$ by induction.

We go back to the evaluation of Eq. (\ref{EqA_MPF:S_r_intro}).
Substituting $S_{r'} \leq r'! f^{n_2+\cdots+n_{r'}} (kg)^{r'-1}$ into Eq. (\ref{EqA_MPF:S_r_intro}), we arrive at
\begin{eqnarray}
    [\text{Eq. (\ref{EqA_MPF:S_r_intro})}] &\leq& 2^{r-1} \frac{(r-1)!}{r'!} (kg)^{r-r'} f^{n_{r'+1}+\cdots+n_r} \sum_{\gamma_1,\cdots,\gamma_{r'}=1}^\Gamma \sum_{X \ni i} \norm{h_X^{\gamma_1 (n_1)}(\tau)} S_{r'} \nonumber \\
    &\leq& (r-1)! (2kg)^{r-1} f^{q_0-1-r}g.
\end{eqnarray}
Finally, we obtain
\begin{eqnarray}
    [\text{Eq. (\ref{EqA_MPF:D_h_D_h})}] &\leq& \sum_{r=0}^{q_0-1} \frac{(q_0-1)!}{r!(q_0-1-r)!} r^{q_0-1-r} \Gamma^{q_0-1-r} (r-1)! (2kg)^{r-1} f^{q_0-1-r}g \nonumber \\
    &\leq& (q_0-1)! (2kg+\Gamma f)^{q_0-1} g,
\end{eqnarray}
which completes the proof of Eq. (\ref{EqA_MPF:An_D_h_D_H_bound}) for $n=0$.
The upper bound when the operator $\mcl{A}_n(\tau)$ is applied is essentially the same as Eq. (\ref{EqA_MPF:An_D_h_bound}).
The locality, the extensiveness, and the strength of time-dependency appearing in Eq. (\ref{EqA_MPF:com_h_h_ver2}) are common with Eq. (\ref{EqA_MPF:nested_com_h_h_h}).
We can construct the recursive relation for the upper bound like Eq. (\ref{EqA_MPF:D_h_bound_recursive}).
This results in the inequality, Eq. (\ref{EqA_MPF:An_D_h_D_H_bound}). $\quad \square$

\subsection{Error bounds of time-dependent MPF}\label{SubsecA:err_proof_MPF}

Here, we complete the proof of Theorem \ref{Thm_MPF:error_MPF}, which gives the bound on the time-dependent MPF errors.
We prove the following theorem using the results of Lemma \ref{LemA_MPF:truncated_BCH_error} and Lemma \ref{LemA_MPF:MPF_BCH_error}.

\begin{theorem}
\textbf{(Formal version of Theorem \ref{Thm_MPF:error_MPF})}

Let $\bar{T}(t)=e^{-iHt}+\order{t^{p+1}}$ be a $p$th-order time-independent PF for a Hamiltonian $H=\sum_{\gamma=0}^\Gamma H_\gamma$, defined by Eq. (\ref{Eq_Err:PF_equiv}).
We set the quantities $\{c_j,k_j\}_{j=1}^J$ so that the time-independent MPF $\bar{M}(t)=\sum_{j=1}^J c_j \{\bar{T}(t/k_j)\}^{k_j}$ can satisfy the order condition, $\bar{M}(t)=e^{-iHt}+\order{t^{m+1}}$.
For the time $t$ small enough to satisfy
\begin{equation}\label{EqA_MPF:t_condition_MPF}
    t \leq \min \left( \frac{1}{4e^3 V_p p_0(N,\epsilon)(2kg+\Gamma f)},  \frac1{4V_p \max_{\tau \in [0,\tau]}\bar{\alpha}_{\mr{com},p}^\mr{MPF} (\tau)} \right),
\end{equation}
the standard MPF $\bar{N}(t,0)$ has an error bounded by
\begin{equation}\label{EqA_MPF:MPF_error}
   \norm{U(t,0)-\bar{N}(t,0)} \leq \norm{\vec{c}}_1 \left( 2V_p \max_{\tau \in [0,t]}\alpha_{\mr{com},p}^\mr{MPF} (\tau) t\right)^{m+1} + \norm{\vec{c}}_1 \|\vec{k}\|_1 \epsilon.
\end{equation}
\end{theorem}

\textbf{Proof.---}
We begin the proof with Eq. (\ref{EqA_MPF:Evol_MPF_error_triangle}), which tells us the error bound by the triangle inequality via the truncated BCH formula.
We have already computed the difference between the standard MPF and its approximation by the truncated BCH formula in Lemma \ref{LemA_MPF:truncated_BCH_error} or Lemma \ref{LemA_MPF:MPF_BCH_error}, and the assumption $t \leq \{ 4e^3 V_p p_0(N,\epsilon)(2kg+\Gamma f) \}^{-1}$ in Eq. (\ref{EqA_MPF:t_condition_MPF}) is made for applying them.
The remaining task is to evaluate the norm of
\begin{equation}\label{EqA_MPF:Evol_BCH_diff}
    U(t,0)-\sum_j \sum_{l \in \bbZ} c_j e^{-il\omega t} \braket{l|[ \bar{T}_{p_0}^\mr{BCH}(t/k_j) ]^{k_j} |0} =  \sum_j \sum_{l \in \bbZ} c_j e^{-il\omega t} \braket{l|\left\{ e^{-iH^Ft}-[ \bar{T}_{p_0}^\mr{BCH}(t/k_j) ]^{k_j} \right\} |0}.
\end{equation}
The right-hand side comes from the expression of the time evolution by Eq. (\ref{Eq_Pre:Time_Evol_Floquet}) and the relation $\sum_{j=1}^J c_j=1$, which is always demanded for the order condition $\bar{M}(t)=e^{-iHt}+\order{t}$.

The power of the truncated BCH formula $[\bar{T}_{p_0}^\mr{BCH}(t/k_j) ]^{k_j}$ expressed by Eq. (\ref{EqA_MPF:power_BCH}) can be seen as a time-evolution operator under the Hamiltonian $H^F+\sum_{q=p+1}^{p_0} \bar{\Phi}_q^F t^{q-1}/(k_j)^{q-1}$.
Regarding $H^F$ and $\sum_{q=p+1}^{p_0} \bar{\Phi}_q^F t^{q-1}/(k_j)^{q-1}$ respectively as unperturbed and perturbed Hamiltonians in the interaction picture, it can be represented by the Dyson series,
\begin{equation}
    [\bar{T}_{p_0}^\mr{BCH}(t/k_j) ]^{k_j} = e^{-iH^Ft} + e^{-iH^Ft} \sum_{n=1}^\infty (-i)^n \int_0^t \dd \tau_n \cdots \int_0^{\tau_2} \dd \tau_1 \prod_{n'=1}^n \left( e^{i \tau_{n'} \ad_{H^F} } \sum_{q=p+1}^{p_0} \frac{\bar{\Phi}_q^F t^{q-1}}{(k_j)^{q-1}} \right).
\end{equation}
This series is absolutely convergent with respect to $n$ since each $\bar{\Phi}_q^F$ is a Toeplitz matrix given by Eq. (\ref{EqA_MPF:Phi_q_F_form}) and its norm is bounded by $\norm{\bar{\Phi}_q^F} \leq \max_{\tau \in [0,T]} (\norm{\bar{\Phi}_q(\tau)}) < \infty$.
Thus, the difference between $e^{-iH^Ft}$ and $[\bar{T}_{p_0}^\mr{BCH}(t/k_j) ]^{k_j}$ in Eq. (\ref{EqA_MPF:Evol_BCH_diff}) is calculated as follows,
\begin{eqnarray}
    && \sum_j c_j \left\{ e^{-iH^Ft}-[ \bar{T}_{p_0}^\mr{BCH}(t/k_j) ]^{k_j} \right\} \nonumber \\
    && \qquad = - \sum_j c_j \sum_{n=1}^\infty (-i)^n \int_0^1 \dd s_n \cdots \int_0^{s_2} \dd s_1 \prod_{n'=1}^n \left(e^{-iH^Ft(s_{n'+1}-s_{n'})}  \sum_{q=p+1}^{p_0} \frac{\bar{\Phi}_{q}^F t^{q}}{(k_j)^{q-1}} \right) e^{-iH^Fts_1}, \label{EqA_MPF:Evol_BCH_diff_1}
\end{eqnarray}
where we set $s_{n+1}=1$ and use $\sum_j c_j=1$.
We arrange the terms with respect to the time $t$, which results in
\begin{eqnarray}
    && \sum_j c_j \left\{ e^{-iH^Ft}-[ \bar{T}_{p_0}^\mr{BCH}(t/k_j) ]^{k_j} \right\} \nonumber \\
    && \quad = - \sum_j c_j \sum_{n=1}^\infty (-i)^n \sum_{q=pn+1}^{(p_0-1)n+1}\sum_{\substack{p+1 \leq q_1,\cdots,q_n \leq p_0; \\ q_1+\cdots+q_n=q+n-1}} \frac{t^{q+n-1}}{(k_j)^{q-1}} \int_0^1 \dd s_n \cdots \int_0^{s_2} \dd s_1 \prod_{n'=1}^n \left( e^{-iH^Ft(s_{n'+1}-s_{n'})} \bar{\Phi}_{q_{n'}}^F \right) e^{-iH^Fts_1} \nonumber \\
    && \quad =  \sum_{q=p+1}^\infty t^q \sum_j \frac{c_j}{(k_j)^{q-1}}\sum_{n=\left\lceil \frac{q-1}{p_0} \right\rceil}^{\left\lfloor \frac{q-1}{p} \right\rfloor} (-it)^{n-1} \sum_{\substack{p+1 \leq q_1,\cdots,q_n \leq p_0; \\ q_1+\cdots+q_n=q+n-1}} \int_0^1 \dd s_n \cdots \int_0^{s_2} \dd s_1 \prod_{n'=1}^n \left( e^{-iH^Ft(s_{n'+1}-s_{n'})} \bar{\Phi}_{q_{n'}}^F \right) e^{-iH^Fts_1}. \nonumber \\
    && \label{EqA_MPF:Evol_BCH_diff_2}
\end{eqnarray}
The calculation in Eqs. (\ref{EqA_MPF:Evol_BCH_diff_1}) and (\ref{EqA_MPF:Evol_BCH_diff_2}) is exactly the same as that for the time-independent MPF \cite{mizuta2025-mpf}, in which $H^\mr{LP}, H_1^\mr{Add},\cdots,H_\Gamma^\mr{Add}$ are associated with $H_0,H_1,\cdots,H_\Gamma$ in the time-independent case via the correspondence of $\bar{T}^F(t)$ [Eq. (\ref{Eq_Rel:T_F_Std})] and $\bar{T}(t)$ [Eq. (\ref{Eq_Err:PF_equiv})].
The quantities $\{ c_j , k_j \}$ established for the time-independent MPF $\bar{M}(t)$ to satisfy the order condition $\bar{M}(t)=e^{-iHt}+\order{t^{m+1}}$ are chosen so that the terms with $q=p+1,p+2,\cdots,m$ in Eq. (\ref{EqA_MPF:Evol_BCH_diff_2}) can vanish (See the end of Appendix \ref{SubsecA:err_proof_MPF} for the detailed choice).
Therefore, with the same choice of $\{c_j,k_j\}$, the error for the time-dependent case is expressed by
\begin{eqnarray}
    [\text{Eq. (\ref{EqA_MPF:Evol_BCH_diff})}] &=& \sum_{l \in \bbZ} \sum_{q=m+1}^\infty t^q \sum_j \frac{c_j}{(k_j)^{q-1}}\sum_{n=\left\lceil \frac{q-1}{p_0} \right\rceil}^{\left\lfloor \frac{q-1}{p} \right\rfloor} (-it)^{n-1} \nonumber \\
    && \, \times \sum_{\substack{p+1 \leq q_1,\cdots,q_n \leq p_0; \\ q_1+\cdots+q_n=q+n-1}} \int_0^1 \dd s_n \cdots \int_0^{s_2} \dd s_1 e^{-il\omega t} \bra{l} \left[  \prod_{n'=1}^n \left( e^{-iH^Ft(s_{n'+1}-s_{n'})} \bar{\Phi}_{q_{n'}}^F \right) e^{-iH^Fts_1} \right] \ket{0}. \nonumber \\
    && \label{EqA_MPF:Evol_BCH_diff_3}
\end{eqnarray}

We note that the above series is absolutely convergent with respect to $l$ and $q$, which can be easily ensured by the smoothness of the Hamiltonian as we do for Eq. (\ref{EqA_MPF:TT_sum_Dyson}).
Changing the order of the summation with respect to $l$ and $q$, we focus on the operator,
\begin{equation}
    \sum_{l \in \bbZ} e^{-il\omega t} \bra{l} \left[  \prod_{n'=1}^n \left( e^{-iH^Ft(s_{n'+1}-s_{n'})} \bar{\Phi}_{q_{n'}}^F \right) e^{-iH^Fts_1} \right] \ket{0}.
\end{equation}
We insert the completeness $\sum_{l \in \bbZ} \ket{l}\bra{l}=1$ and use the translation symmetry of $H^F$ by Eq. (\ref{Eq_Pre:tr_sym_H_F}) and $\bar{\Phi}_q^F$ by Eq. (\ref{EqA_MPF:Phi_q_symmetry}).
This results in the relation,
\begin{eqnarray}
    && \sum_{l \in \bbZ} e^{-il\omega t} \braket{l|\prod_{n'=1}^n \left(e^{-iH^Ft(s_{n'+1}-s_{n'})}   \bar{\Phi}_{q_{n'}}^F \right) e^{-iH^Fts_1} |0} \nonumber \\
    && \quad = \sum_{\substack{l_1,\cdots,l_{n+1} \in \bbZ, \\ m_1,\cdots,m_n \in \bbZ}} e^{-il_{n+1}\omega t} \prod_{n'=1}^n \left( \braket{l_{n'+1}|e^{-iH^Ft(s_{n'+1}-s_{n'})}|l_{n'}+m_{n'}} \braket{l_{n'}+m_{n'}|\bar{\Phi}_{q_{n'}}^F|l_{n'}}\right) \braket{l_1|e^{-iH^Fts_1}|0} \nonumber \\
    && \quad =  \sum_{\substack{l_1,\cdots,l_{n+1} \in \bbZ, \\ m_1,\cdots,m_n \in \bbZ}} \prod_{n'=1}^n \left( e^{-il_{n'+1} \omega t s_{n'+1}}\braket{l_{n'+1}|e^{-iH^Ft(s_{n'+1}-s_{n'})}|0} \{\bar{\Phi}_{q_{n'}}(t)\}_{m_{n'}}e^{-im_{n'}\omega t s_{n'}} \right) e^{-il_1 \omega t s_1}\braket{l_1|e^{-iH^Fts_1}|0} \nonumber \\
    && \quad = \prod_{n'=1}^n \left[ U(s_{n'+1}t,s_{n'}t) \bar{\Phi}_{q_{n'}}(s_{n'}t) \right] U(s_1 t,0). 
\end{eqnarray}
We use the expression of $\bar{\Phi}_q^F$ by Eq. (\ref{EqA_MPF:Phi_q_F_form}) in the second and the last equalities, and use that of the time evolution $U(t,0)$ by Eq. (\ref{Eq_Pre:Time_Evol_Floquet}) in the last equality.
The unitary operators $U(s_{n'+1}t,s_{n'}t)$ and $U(s_1t,0)$ do not affect the norm.
We can use the relation $\norm{\bar{\Phi}_q(s_{n'}t)} \leq \max_{\tau \in [0,t]} (\norm{\bar{\Phi}_q(\tau)})$ since we have $0 \leq s_{n'} \leq 1$ in Eq. (\ref{EqA_MPF:Evol_BCH_diff_3}).
As a result, the norm of Eq. (\ref{EqA_MPF:Evol_BCH_diff}) is bounded by
\begin{eqnarray}
    \norm{U(t,0)-\sum_j c_j \sum_{l \in \bbZ}\braket{l|[ \bar{T}_{p_0}^\mr{BCH}(t/k_j) ]^{k_j}|0}} &\leq& \norm{\vec{c}}_1 \sum_{q=m+1}^\infty  \sum_{n=\left\lceil \frac{q-1}{p_0} \right\rceil}^{\left\lfloor \frac{q-1}{p} \right\rfloor} t^{q+n-1} \sum_{\substack{p+1 \leq q_1,\cdots,q_n \leq p_0; \\ q_1+\cdots+q_n=q+n-1}} \frac1{n!} \prod_{n'=1}^n \max_{\tau \in [0,t]} \left( \norm{\bar{\Phi}_{q_{n'}} (\tau)}\right) \nonumber \\
    &\leq& \norm{\vec{c}}_1 \sum_{q=m+1}^\infty \sum_{n=1}^\infty \frac{t^{q+n-1}}{n!} \left( V_p \max_{\tau \in [0,t]}\bar{\alpha}_{\mr{com},p}^\mr{MPF} (\tau)\right)^{q+n-1} \sum_{\substack{ 1 \leq q_1,\cdots,q_n \leq p_0; \\ q_1+\cdots+q_n=q+n-1}} 1 \nonumber \\
    &\leq& \frac{\norm{\vec{c}}_1}{4} \sum_{q=m+1}^\infty \left( 2V_p \max_{\tau \in [0,t]}\alpha_{\mr{com},p}^\mr{MPF} (\tau) t\right)^{q} \sum_{n=1}^\infty \frac{1}{n!} \left( 2V_p \max_{\tau \in [0,t]}\alpha_{\mr{com},p}^\mr{MPF} (\tau) t\right)^{n-1} \nonumber \\
    &\leq& \norm{\vec{c}}_1 \left( 2V_p \max_{\tau \in [0,t]}\alpha_{\mr{com},p}^\mr{MPF} (\tau) t\right)^{m+1}.
\end{eqnarray}
In the second inequality, we use Lemma \ref{LemA_MPF:Phi_q_bound} for upper-bounding $\norm{\bar{\Phi}_q(\tau)}$ by the quantity  $\bar{\alpha}_{\mr{com},p}(\tau)$ and use the definition of the quantity $\bar{\alpha}_{\mr{com},p}^{\mr{MPF}}(\tau)$ by Eq. (\ref{Eq_MPF:alpha_MPF_Std}).
The last inequality comes from the assumption Eq. (\ref{EqA_MPF:t_condition_MPF}). 
The error bound Eq. (\ref{EqA_MPF:MPF_error}) is immediately obtained by Eq. (\ref{EqA_MPF:Evol_MPF_error_triangle}). $\quad \square$

\textbf{Choice of $\{c_j,k_j\}$.---}
We discuss the concrete choice of the coefficients $\{ c_j \}$ and the integers $\{ k_j \}$.
We always impose $\sum_{j=1}^J c_j = 1$ so that at least the order condition $\bar{N}(t,0)=e^{-iHt}+\order{t}$ can be satisfied.
On the other hand, we demand that the term with $q \leq m$ in Eq. (\ref{EqA_MPF:Evol_BCH_diff_2}) should vanish for $\bar{N}(t,0)=e^{-iHt}+\order{t^{m+1}}$.
Thus, it is sufficient to find $\{c_j,k_j\}$ such that
\begin{equation}\label{EqA_MPF:cj_kj_eq}
    \sum_{j=1}^J c_j = 1, \quad \sum_{j=1}^J \frac{c_j}{(k_j)^{q-1}} = 0, \quad \text{for $q=p+1,p+2,\cdots,m$}  
\end{equation}
is satisfied.
Or if the time-dependent PF $\bar{S}(t,0)$ has an even order $p$ and the corresponding time-dependent PF $\bar{T}^F(t)$ satisfies the symmetric condition $\bar{T}^F(-t)^\dagger = \bar{T}^F(t)$, the terms with even $q$ in Eq. (\ref{EqA_MPF:Evol_BCH_diff_2}) are all zero due to $\bar{\Phi}_{q_{n'}}^F = 0$ for even $q_{n'}$.
In that case, it is sufficient to impose
\begin{equation}\label{EqA_MPF:cj_kj_eq_symmetric}
    \sum_{j=1}^J c_j = 1, \quad \sum_{j=1}^J \frac{c_j}{(k_j)^{q-1}} = 0, \quad \text{for $q=p+1,p+3,\cdots,2 \lfloor m/2 \rfloor -1$}. 
\end{equation}
For example, this is the case for the even-order Lie-Suzuki-Trotter formula $\bar{S}_p(t,0)$ since the time-dependent Lie-Suzuki-Trotter formula $\bar{T}^F(t)$, given by Theorem \ref{Thm_Rel:Coincidence_LST}, satisfies the symmetric condition.
The conditions Eqs. (\ref{EqA_MPF:cj_kj_eq}) or (\ref{EqA_MPF:cj_kj_eq_symmetric}) are exactly the same as those for the time-independent MPF \cite{low2019-mpf}.
While we focus on the standard MPF $\bar{N}(t,0)$ here, they are valid also for the generalized MPF $N(t,0)$.

The well-conditioned MPF is a MPF satisfying $\norm{\vec{c}}_1, \| \vec{k} \|_1 \in \poly{m}$, which can be efficiently implemented on quantum circuits by LCU.
Such a solution is obtained when we consider PFs satisfying the symmetric condition \cite{low2019-mpf}.
We set $J=m/2$ for even $m$ and demand
\begin{equation}\label{EqA_MPF:cj_kj_eq_well_symmetric}
    \sum_{j=1}^J c_j = 1, \quad \sum_{j=1}^J \frac{c_j}{(k_j)^{q-1}} = 0, \quad \text{for $q=3,5,\cdots, 2J+1$}. 
\end{equation}
Although it is redundant by the conditions for $q=3,5,\cdots,p-1$ compared to Eq. (\ref{EqA_MPF:cj_kj_eq_symmetric}), it allows a good choice of $\{k_j\}$ given by
\begin{equation}\label{EqA_MPF:well_conditioned_kj}
    k_j = \left\lceil K \left| \sin \left( \frac{\pi (2j-1)}{8m}\right)\right|^{-1} \right\rceil, \quad \| \vec{k} \|_1 \in \order{m^2 \log m}. \\
\end{equation}
The scale factor $K < 2\sqrt{2}m/\pi$ is a number such that $\{k_j\}$ becomes mutually different integers.
The set of the coefficients $\{c_j\}_j$ is given by
\begin{equation}\label{EqA_MPF:well_conditioned_cj}
    c_j = \prod_{i \neq j} \frac{1}{1-(k_i/k_j)^2}, \quad \norm{\vec{c}}_1 \in \order{\log m}.
\end{equation}
When we compose of the MPF with PFs which do not necessarily satisfy the symmetric condition, we set $J=m$ and demand
\begin{equation}
    \sum_{j=1}^J c_j = 1, \quad \sum_{j=1}^J \frac{c_j}{(k_j)^{q-1}} = 0, \quad \text{for $q = 2,3,\cdots,m$}.
\end{equation}
This reduces to Eq. (\ref{EqA_MPF:cj_kj_eq_well_symmetric}) when we replace $k_j$ with $(k_j)^2$, and hence we can use a squared value of Eq. (\ref{EqA_MPF:well_conditioned_kj}) as $k_j$.
The choice of $\{c_j\}$ is just the same as the symmetric case.

\section{PF for non-unitary dynamics}\label{A_Sec:non_unitary}

\renewcommand{\thetheorem}{\thesection\arabic{theorem}}
\setcounter{theorem}{0}

While we concentrate on evaluating the explicit error for time-dependent hermitian Hamiltonians $H(t)$, our results can be easily extended to the case when they are non-hermitian as well as time-independent cases \cite{childs2021-trotter}.
Such cases are useful for solving general linear differential equations \cite{Berry2014-linear_diff}, simulating open quantum systems \cite{Kliesch-prl2011-open,Kamakari-prxq2022-open}, or preparing thermal states via imaginary time evolution \cite{Motta2019-po-thermal}.

Let us consider a time-dependent Hamiltonian $H(t) = \sum_{\gamma=1}^\Gamma H_\gamma (t)$, where each term $H_\gamma (t)$ is of class $C^{p+2}$ but not necessarily hermitian.
We note that, regardless of hermiticity of the Hamiltonian, the decay of the transition amplitude $\braket{l|e^{-iH^Ft}|0}$ and the resulting absolute convergence of Eq. (\ref{Eq_Pre:Time_Evol_Floquet}) are maintained \cite{Mizuta_Quantum_2023}.
This means that Floquet analysis in the main text holds also for non-hermitian cases.
What differs from hermitian cases is the absence of unitary operators in the error representation.
Reviewing the discussion in Sections \ref{Sec:relation}, Eq. (\ref{Eq_Err:t_dep_error_Delta}) implies that the error of the standard PF is bounded by
\begin{equation}\label{EqA_NonU:error_bound_int}
    \norm{U(t,0)-\bar{S}(t,0)} \leq \int_0^t \dd \tau \norm{U(t,\tau)} \cdot \norm{\bar{S}(\tau,0) \sum_{l \in \bbZ} \braket{l|\bar{\Delta}^F(\tau)|0}}.
\end{equation}
The operator $\Delta^F(\tau)$ is defined by Eq. (\ref{Eq_Err:Delta_F_expansion}) with replacing the conjugate of matrices by their inverse.
The norm of the time-evolution operator $U(t,\tau)$ is bounded by
\begin{equation}
    \norm{U(t,\tau)} \leq \exp \left( \int_\tau^t \dd \tau_1 \norm{\mr{Im} H(\tau_1)}\right),
\end{equation}
where $\mr{Im} H$ denotes $(H-H^\dagger)/2i$.
On the other hand, as shown in Eq. (\ref{Eq_Err:Proof_Thm_1v}), the operator $\bar{S}(\tau,0) \sum_{l \in \bbZ} \braket{l|\bar{\Delta}^F(\tau)|0}$ includes non-unitary operators, each of which has the following norm,
\begin{eqnarray}
    \norm{\bar{S}(\tau,0) \bar{S}_{\prec v\tilde{\gamma}}(\tau,0)^\dagger} \cdot \norm{\bar{S}_{\prec v\tilde{\gamma}}(\tau,0)} &\leq& \norm{\prod_{(v',\gamma')=(v,\gamma)}^{(V_p,\Gamma)} e^{-i H_{\pi_{v'}(\gamma')}(\beta_{v'\gamma'}\tau)\alpha_{v'\gamma'}\tau}} \cdot \norm{\prod_{(v',\gamma')=(1,1)}^{(v,\gamma-1)} e^{-i H_{\pi_{v'}(\gamma')}(\beta_{v'\gamma'}\tau)\alpha_{v'\gamma'}\tau}} \nonumber \\
    &\leq& \exp \left( V_p \tau \sum_{\gamma=1}^\Gamma \max_{\tau \in [0,t]}\norm{\mr{Im}H_\gamma(\tau)} \right), 
\end{eqnarray}
or 
\begin{equation}
    \norm{\bar{S}(\tau,0) \bar{S}_{\prec v\tilde{\gamma}}(\tau,0)^\dagger e^{iH_{\pi_v(\gamma)}(\beta_{v\gamma}\tau)\alpha_{v\gamma}\tau_1}} \cdot \norm{e^{-iH_{\pi_v(\gamma)}(\beta_{v\gamma}\tau)\alpha_{v\gamma}\tau_1} \bar{S}_{\prec v\tilde{\gamma}}(\tau,0)} \leq \exp \left( V_p \tau \sum_{\gamma=1}^\Gamma \max_{\tau \in [0,t]}\norm{\mr{Im}H_\gamma(\tau)} \right).
\end{equation}
Following the discussion from Eq. (\ref{Eq_Err:Delta_F_bound_start}) to Eq. (\ref{Eq_Err:sum_LP_b}), we obtain an upper bound,
\begin{eqnarray}
    \norm{\bar{S}(\tau,0) \sum_{l \in \bbZ} \braket{l|\bar{\Delta}^F (\tau)|0}} &\leq& 4 (V_p)^{p+1} \tau^p  \exp \left( V_p \tau \sum_{\gamma=1}^\Gamma \max_{\tau \in [0,t]}\norm{\mr{Im}H_\gamma(\tau)} \right) \max_{\tau_1 \in [0,\tau]} \bar{\alpha}_{\mr{com},p}(\tau_1).
\end{eqnarray}
Finally, using the relation Eq. (\ref{EqA_NonU:error_bound_int}), we arrive at the following error bound for non-unitary dynamics.

\begin{lemma}\label{LemA_NonU:error_NonU}
\textbf{(Standard PF error for non-unitary dynamics)}

When the Hamiltonian $H(t)$ is not necessarily hermitian, the $p$th-order standard PF $\bar{S}(t,0)$ has an error bounded by
\begin{equation}
\norm{U(t,0)-\bar{S}(t,0)} \leq 4 (V_pt)^{p+1}  \exp \left( V_pt \sum_{\gamma=1}^\Gamma \max_{\tau \in [0,\tau]}\norm{\Im H_{\gamma}(\tau)} + \int_0^t \dd \tau \norm{\mr{Im}H(\tau)}\right) \max_{\tau \in [0,\tau]} \bar{\alpha}_{\mr{com},p}(\tau).
\end{equation}
\end{lemma}

This bound is parallel to the one for time-independent cases \cite{childs2021-trotter} in that they have additional exponential factors in the time $t$.
While we hereby concentrate on the standard PF for non-unitary dynamics, such extensions of other variants of time-dependent PFs are straightforward.
We note that this extension also applies to the time-dependent MPF.
Namely, the error bounds of the generalized PF or the time-dependent MPFs for non-unitary dynamics can be obtained by upper-bounding the non-unitary time-evolution operators appearing in their error representations.
Such error bounds will have some additional exponential factors like Lemma \ref{LemA_NonU:error_NonU}.

\bibliography{bibliography.bib}

\end{document}